\documentclass[useAMS,usenatbib]{mn2e}
\usepackage{times}
\usepackage{url}
\usepackage{graphicx}
\usepackage[usenames]{color}
\usepackage{graphics,graphicx}
\usepackage{deluxetable}
\usepackage[usenames]{color}
\DeclareGraphicsExtensions{.pdf,.png,.jpg,.mps,.eps,.ps}
\usepackage{amsmath}
\usepackage{natbib}

\def\unit #1{\,{\rm #1}}

\newcommand\kms{\rm \,\unit{km\,s^{-1}}}

\newcommand\cmsqi{\rm \,\unit{cm^{-2}}}

\newcommand\kev{\rm \,\unit{keV}}

\newcommand\funit{\rm \,erg\,cm^{-2}\,s^{-1}}

\newcommand\lunit{\rm \,erg \,s^{-1}}

\newcommand\xiunit{\rm \,erg\,cm\,s^{-1}}

\newcommand\nh{\rm N_{H}}

\newcommand\nhwa{N^{\rm WA}_{\rm H}}

\newcommand\ks{\, \rm ks}

\newcommand\cd{\,\rm \chi^2/dof}

\newcommand\ev{\unit{\, eV}}

\newcommand\lgxi{\rm \, log\xi}

\newcommand\asca{{\it ASCA}}
\newcommand\suzaku{{\it Suzaku}}

\newcommand\chandra{{\it Chandra}}

\newcommand\xmm{{\it XMM-Newton}}

\newcommand\ion[2]{#1$\;${\scshape{#2}}}%

\usepackage{ulem}
\usepackage{graphicx}

\title[Warm aborbers in X-rays]{Warm absorbers in X-rays (WAX), a comprehensive high resolution grating spectral study of a sample of Seyfert galaxies: \\ I. A global view and frequency of occurrence of warm absorbers.}
\author[Laha et al.]{Sibasish Laha$^{1}$\thanks{laha@iucaa.ernet.in}, Matteo Guainazzi$^{2}$\thanks{matteo.guainazzi@sciops.esa.int }, Gulab \ C.\ Dewangan$^{1}$, Susmita Chakravorty$^{3}$, \newauthor and Ajit \ K.\ Kembhavi$^{1}$\\
$^{1}$Inter University Centre for Astronomy and Astrophysics, Post bag 4, Ganeshkhind, Pune, India\\
$^{2}$European Space Astronomy Centre of ESA, PO Box 78, Villanueva de la Cañada, 28691 Madrid\\
$^3$Laboratoire d’Astrophysique, Universit´e Joseph Fourier, CNRS UMR 5571, Grenoble, France}

\date{\today}
\begin{document}




\maketitle
\label{firstpage}

\begin{abstract}

	We present results from a homogeneous analysis of the broadband $0.3-10\kev$ CCD resolution as well as of soft X-ray high-resolution grating spectra of a hard X-ray flux-limited sample of 26 Seyfert galaxies observed with {\it XMM-Newton}. Our goal is to characterise warm absorbers (WAs) along the line-of-sight to the active nucleus. We significantly detect WAs in $65\%$ of the sample sources. {\color{black} Our results are consistent with WAs being present in at least half of the Seyfert galaxies in the nearby Universe, in agreement with previous estimates .} {\color{black} We find a gap in the distribution of the ionisation parameter in the range $0.5<\log\xi<1.5$ which we interpret as a thermally unstable region for WA clouds. This may indicate that the warm absorber flow is probably constituted by a clumpy distribution of discrete clouds rather than a continuous medium. The distribution of the WA column densities for the sources with broad Fe K$\alpha$ lines are similar to those sources which do not have broadened emission lines. Therefore the detected broad Fe K$\alpha$ emission lines are bonafide and not artifacts of ionised absorption in the soft X-rays.} The WA parameters show no correlation among themselves, with the exception of the ionisation parameter versus column density. The shallow slope of the $\log\xi$ versus $\log v_{\rm out}$ linear regression ($0.12\pm 0.03$) is inconsistent with the scaling laws predicted by radiation or magneto-hydrodynamic-driven winds. Our results suggest also that WA and Ultra Fast Outflows (UFOs) do not represent extreme manifestation of the same astrophysical system.

\end{abstract}


\section{Introduction}

The X-ray and $\gamma$-ray emission of an active galaxy mostly originate from the innermost regions of the central engine, known as the Active Galactic Nuclei (AGN). It is commonly believed that the AGN is powered by accretion of plasma onto a super-massive black hole. The $0.1-10 \kev$ X-ray emission spectrum of radio quiet (RQ) AGN can be well approximated by a power-law with a photon index in the range of $1.6\le \Gamma\le2.4$,  suggesting Comptonisation as the main underlying physical process \citep{1991ApJ...380L..51H,1994ApJ...434..570T}. Superposed on this continuum, several spectral components have been discovered
    along the history of X-ray astronomy. Most of these are believed
    to be due to reprocessing by an optically thick and geometrically
    thin accretion disk \citep{george-fabian}. Spectral distortions,
    most notably observed in the profile of emission lines such as the
    iron K${\alpha}$, suggest that the bulk of the X-ray emission comes
    from a region just a few gravitational radii from the black hole's
    event horizon \citep[for e.g., see the review by][for a different interpretative scenario]{2009A&ARv..17...47T}.
Extrapolating this non-thermal component to energies lower than
$\sim$1~keV unveils a ``soft X-ray excess".
The physical origin of this component has so far proven elusive \citep[see e.g.,][]{2004MNRAS.349L...7G,2004A&A...422...85P,2006MNRAS.365.1067C, 2005A&A...432...15P,2006MNRAS.371...81S,2007ApJ...671.1284D}. { Redwards of the soft X-rays lie the unobservable region of the AGN spectra. The energy range of $13.6-100\ev$ is obscured from us due to the Galactic extinction. Below $13.6\ev$ the big blue bump (BBB) is the most important and dominant spectral feature. It typically spans the energy range from $\rm1\mu m$-$\rm 3nm$ or $\rm \log(\nu/Hz)\sim 14.5-17$ and is a major contributor to the AGN bolometric luminosity \citep{2013MNRAS.431..210C,2007MNRAS.381.1235V,1994ApJS...95....1E,1989ApJ...347...29S}. The most commonly accepted origin of the BBB is the multi-colored blackbody emission from a geometrically thin and optically thick accretion disk \citep{1973A&A....24..337S}.} 

The $0.1-10\kev$ spectra of many AGN are affected by obscuration by partially ionised material along our line
of sight and intrinsic to the source, first detected by \cite{1984ApJ...281...90H} using {\it Einstein} data. Such X-ray absorbing clouds
have been named warm absorbers (WA).
These warm absorbers imprint their signatures on the X-ray spectrum as
narrow absorption lines and edges, from elements at a wide range of
ionisation stages \citep[see e.g.,][]{2000A&A...354L..83K,
2000ApJ...535L..17K,2005A&A...431..111B}. 
The high resolution
grating X-ray spectra from the \xmm{} and \chandra{} observatories have profoundly improved
our understanding of these discrete absorption and emission features.

Absorption lines and photo-absorption edges are
sensitive diagnostics of the ionisation structure and kinematics of
the gas. The measured blue-shift of the absorption lines with respect to the
systemic velocity implies that these absorbers are outflowing with
moderate velocities in the range of $\sim 100-1000 \kms$. 
In some AGNs, high velocity ($\sim \rm 0.1c$) outflows have also been detected \citep[e.g.,][]{2003ApJ...593L..65R,2004A&A...413..921D,2005ApJ...618L..87D,2007ApJ...670..978B}. 
The resulting mass outflow
rate can be a substantial fraction of the accretion rate required to
power the AGN. Thus, warm absorbers can be dynamically important and
the knowledge of their state, location, geometry and dynamics would
help in understanding the central engines of AGN
\citep{1995AAS...186.4501M}.

\subsection{Previous studies of warm-absorbed AGN samples}
\label{sect_previousstudies}

Spectroscopic measurements of AGN with the X-ray CCD detectors onboard ASCA allowed the first systematic studies of sizable samples of warm absorbed AGN.
\cite{1997MNRAS.286..513R} presented a study of X-ray spectral properties along with warm absorber properties for a sample of 24 type 1 AGN, using the medium resolution \asca{} data. Almost half of the objects in this sample show K-shell absorption edges of \ion{O}{VII} and \ion{O}{VIII}, and were interpreted as signatures of warm absorbers.
Reynolds found a trend for less ionised absorption to be present in more luminous objects. 
A similar fraction of warm absorbed sources was found in a sample of 18 Seyfert~1--1.5 galaxies observed by ASCA \citep{1998ApJS..114...73G}, as well as in a sample of $\simeq$40 Palomar-Green Quasi Stellar objects 
(QSOs) observed by the XMM-Newton EPIC CCD cameras \citep{2005A&A...432...15P}.
   
\cite{2005A&A...431..111B} compiled the results of previous studies of a number of Seyfert galaxies observed with the RGS (Reflection Grating Spectrometer) onboard {\it XMM-Newton}. They suggested that warm absorbers in nearby Seyfert galaxies and QSOs originate in outflows from the dusty torus. The kinetic luminosity of these outflows accounts for well under $1\%$ of the bolometric luminosities of AGN. The authors showed that the amount of matter processed through an AGN outflow system, averaged over the lifetime of the AGN, is probably large enough to have a significant influence on the evolution of the host galaxy. \cite{2007MNRAS.379.1359M} analysed a sample of 15 type 1 sources using the high resolution Chandra grating data. Ten out of the 15 sources show signatures due to intrinsic absorption from H- or He-like Oxygen, with nine out of the ten sources requiring multiple ionisation components. The ionisation parameter of the warm absorbing gas spanned almost four orders of magnitude ($\xi\sim 10^{0-4}\xiunit$), and the column density 3 orders of magnitude ($\nh\sim 10^{20-23}\cmsqi$). These authors found an apparent gap in the distribution of outflow velocities between $300-500\kms$, whose origin was unclear. \cite{2012ApJ...745..107W} studied a sample of 48 Seyfert 1-1.5 galaxies in the X-rays and hard X-rays using the {\it Suzaku} and {\it XMM-Newton} telescopes. The sources were detected in the hard X-rays using the {\it Swift} Burst alert telescope (BAT) in the $14-195\kev$ band. They detected significantly \ion{O}{VII} and \ion{O}{VIII} edges in $52\%$ of their sample. The ionised column densities of sources with \ion{O}{VII} and \ion{O}{VIII} detections cluster in a narrow range around $\rm \nh \sim 10^{21} \cmsqi$, while sources without strong detection have column densities an order of magnitude lower. Taking into account the inhomogeneous coverage of their sample in the soft X-ray band, they conclude that up to $80\%$ of nearby AGN could host warm absorbers.

\cite{2010A&A...521A..57T} (hereafter T10) have carried out an extensive sample study of ultra fast outflows (UFOs), which are highly ionised outflowing clouds of gas with velocities a fraction ($\sim 0.1-0.3$) of the speed of light. Their signatures in the spectra are in the form of narrow absorption lines in the energy range $\sim 6-9 \kev$. The authors conclude that these outflowing clouds are present in about $\sim 35\%$ of Seyfert galaxies in the nearby universe. \cite{2013MNRAS.430.1102T} (hereafter T13) carried out a detailed comparison between the UFOs and the WAs and conclude that possibly both the types of ionised outflows belong to a single extended stratified ouflow. They also suggest that the UFOs may be launched from the innermost parts of the accretion disk. 


\subsection{Motivation of our study}

Our study aims at characterising the properties of warm absorbing gas in a well defined, flux-limited sample of nearby AGN. At variance with most of the studies discussed in
Sect.~\ref{sect_previousstudies}, we apply a homogeneous data reduction and analysis procedure on all the sources in our sample,
use simultaneous UV and X-ray measurements to obtain the Spectral Energy Distribution (SED) of each object, employ state-of-the-art photoionisation codes
to model the spectra of the warm absorbing gas, and fit simultaneously CCD spectra in the broad 0.3--10~keV energy band and high-resolution grating spectra in the 0.4--2~keV
energy band to achieve a self-consistent astrophysical description of the spectra. We made use of data obtained with the XMM-Newton satellite \citep{2001A&A...365L...1J}. For each source we could combine the large collecting area of the XMM-Newton optics, the high quantum efficiency of the EPIC-pn \citep{2001A&A...365L..18S}, the high resolution RGS \citep{2001A&A...365L...7D} detectors, and the simultaneous optical photometry obtained with the Optical Monitor \citep{2001A&A...365L..36M}. This broadband coverage allowed us 
to self-consistently calculate the SED of each object at the same time as the warm absorber was studied. 
The WAX-AGN ({\it warm absorbers in X-ray AGN}, WAX hereafter) project aims at addressing the following important questions:

\begin{enumerate}

\item How fundamental are ionised outflows in AGN?

\item What are the distribution of warm absorber properties (column density, ionisation parameter, outflow velocity, launching radius) in the parent population of nearby Seyfert galaxies ?

\item What is the outflow acceleration mechanism?

\item Do ionised outflows play an important role on the host galaxy chemical history and evolution?

\end{enumerate}

\noindent The first two questions are specifically addressed in this paper.
 
The paper is organised as follows: Section~\ref{sect_sample}
deals with the sample selection. Section~\ref{sect_data} describes the data reprocessing and spectral extraction which involves the data from all the three types of instruments
aboard {\it XMM-Newton}. Sections~\ref{sect_epic}, \ref{subsec-CLOUDY} and ~\ref{section:RGS-analysis} describe the spectral analysis.
Sections.~\ref{sect_results} and~\ref{sect_discussion} discusses the results, which is followed by summary and conclusion.

\section{The Sample selection}
\label{sect_sample}

The WAX sample was selected as a sub-sample of the {\it CAIXA} (Catalogue of AGN In XMM Archive) sample \citep{2009A&A...495..421B}. {\it CAIXA} consists of all radio quiet X-ray unobscured ($\rm N_H<2\times 10^{22} \cmsqi$) AGN observed by \xmm{}  at the boresight, whose data were public as of March 2007, and have at least 200 background-subtracted counts in the 0.3--2~keV, as well as in the 2--10~keV energy band. Sources, whose EPIC-pn spectra were affected by pile-up larger than $1\%$ were not included in CAIXA. To ensure that only radio quiet objects were included in the {\it CAIXA} catalogue, \cite{2009A&A...495..421B} collected radio data at $6$ cm ($4.85$ GHz) and 20 cm (1.4 GHz) from the literature and calculated the K-corrected radio-loudness parameter $R$ \citep{2007A&A...467..519P}. All quasars with log$R>1$ were excluded. For the Seyferts, however, known to be on an average ``radio louder'', log$R>2.4$ was used as the boundary. At the end of the selection procedure CAIXA consists of 156 radio-quiet AGN.  

We cross-correlated the CAIXA sources with the RXTE X-ray Sky Survey \citep[XSS:][]{2004A&A...418..927R}, and retained only those sources with an XSS $3-8\kev$ count rate $\ge$ 1~s$^{-1}$. The hard X-ray selection avoids biases against sources whose spectrum is affected by strong opacity in the soft X-ray band (which is the most interesting band for the purposes of any warm absorber studies). This cross-correlation yielded a hard X-ray flux-limited sample of 34 sources. From this parent flux-limited sample we additionally removed three sources (NGC~7314, MCG-5-23-16, and NGC~526A) because their spectra exhibit column density of cold (neutral) photoabsorbing gas $>$5$\times 10^{21}$~cm$^{-2}$ . Such a column density substantially affects the signal-to-noise ratio in the soft X-ray energy band, where features due to warm absorbers are expected to be the most prominent. The XMM-Newton observations of four other sources in the {\color{black}remaining sample of 31 sources} (5C3.100, H1846-786, RXJ1424.9+4214, E1346+266) were affected by high particle background, making the signal-to-noise ratio in the RGS spectra too poor for the sake of our study. Furthermore, H0557-385 was in a very low flux state during most of the XMM-Newton observations. Only one of the {\it XMM-Newton} observations caught this target at a flux level high enough to potentially warrant a good SNR in the soft X-ray band \citep{2009MNRAS.394L...1L}. However, its exposure time is too short and hence the SNR of its RGS spectrum is again too poor. The final sample of 26 sources with their XMM-Newton EPIC-pn and RGS spectra will be referred to as the ``WAX sample'' throughout this work. {\color{black}The WAX sample is therefore $\sim 76\%$ complete with respect to RXTE/XSS sources with a count rate $>1$ count $\rm s^{-1}$, and at least $84\%$ complete with respect to the sub-sample of X-ray unobscured AGN}. When multiple observations of the same source are available, we selected only the longest exposure data for our analysis to avoid any bias related to the flux and/or spectral state of the source. 

The WAX sample consists of a total of 26 sources of which nine Seyfert 1 or Broad Line Seyferts, five are Seyfert 1.2, 11 are Seyfert 1.5, and one is LINER (the classification is according to the NASA Extragalactic Database, NED). All the sources are in the local universe with a maximum redshift of $0.063$ and a median redshift of $z=0.02$ (see Fig.~\ref{sample_2}). The blackhole masses range from $\rm log(M_{BH}/M_{\odot})=6.5-9.0$  (see Fig.~\ref{sample_2}).
The average RGS exposure time is $\sim 60$~ks. Readers are referred to Tables \ref{basic-info} and \ref{obs-info} for a list of the main properties of the sources in the WAX sample.

\section{Data reprocessing, filtering and spectral extraction}
\label{sect_data}

All sources in the WAX sample were simultaneously observed by all the instruments on-board XMM-Newton. For broad band spectral analysis we used only the EPIC-pn data as its signal to noise ratio (SNR) is higher compared to the EPIC-MOS cameras \citep{2001A&A...365L..27T}. The details of the observation done using the {\it XMM-Newton} satellite are given in Table~\ref{obs-info}. We used SASv11 for data reprocessing and spectral extraction, using the most update calibration files at the time the data were reduced (2011).

Calibrated and concatenated event lists for the EPIC-pn camera were generated using the SAS task {\tt epchain}.
We extracted {full} field-of-view, single event light curves at energies $E \ge 10$~keV to estimate the level of particle background during each observation. Good Time Intervals (GTI) for the
accumulation of scientific products were defined as those corresponding to a count rate $\le$1~counts s$^{-1}$ above $10\kev$.
Spectra were accumulated using single events only, because they have the best spectral resolution. Single events
constitute the bulk of the events at energies lower than $2 \kev$. The source spectra were extracted from circular regions of 45 arc-seconds radius centred on the source X-ray centroid, while the background spectra were extracted 
from appropriate nearby circular regions on the same CCD free of serendipitous contaminating sources. We
created the ancillary response file (ARF) and the redistribution matrix file (RMF) using the SAS tasks {\tt arfgen} and {\tt rmfgen}.
The EPIC-pn spectral data were grouped with a minimum of 20 counts per
energy bin and at most 5 energy bins per resolution element, using the
{\tt specgroup} command in the SAS. We used ISIS version 1.6.2-12 \citep{2000ASPC..216..591H} for
our spectral fitting.  We performed spectral fits on the EPIC-pn spectra using the $0.3-10 \kev$ energy band, where the instrument is well calibrated.

The RGS data were reprocessed using the SAS task {\tt rgsproc} assuming the source optical coordinates as reference for the calculation of the position-dependent wavelength scale. We filtered particle background using background light curves extracted from CCD9, which is most sensitive to proton events, and where the effective area to X-ray photons is the lowest. All intervals with a background count rate $<0.8 \rm \, cts/s$ were selected, to maximise the SNR. 

We reprocessed the OM data using the SAS task {\tt omichain}. The OM
camera simultaneously observed the sources along with EPIC
and RGS cameras. We considered fluxes measured by any one of the following filters:
UVW1 ($\rm 2120 \AA$), UVM2 ($\rm 2300 \AA$), or UVW2 ($\rm 2900 \AA$). These UV energy bands ($4-7\ev$) lie in the region of our interest where the AGN ``Big-Blue-Bump'' (BBB) peaks.
Observed UV fluxes were corrected for the Galactic reddening assuming \cite{1999PASP..111...63F} redenning law with $\rm R_v= 3.1$ (see Table~\ref{OM}). 

We checked for the spectral variability in the WAX sources, and found that the 2--10~keV versus 0.3--2~keV hardness ratio of 10 sources varied during the observation. In a future paper, we will carry out time resolved spectroscopy of those sources. In this paper we instead focus on the properties of their time-averaged spectra.
All errors quoted on the fitted parameters
reflect the $90\%$ confidence interval for one interesting parameter corresponding to $\Delta \chi^2=2.7$ \citep{1976ApJ...208..177L}. Throughout this paper, we consider the `best-fit' as that corresponding to the global minimum of the $\chi^2$ surface in the given energy band with the given set of models.

\section{X-ray spectral analysis: Medium resolution}
\label{sect_epic}

The X-ray emission from AGN is very complex with a number of physical phenomena contributing to the emitted flux and in most cases overlapping with each other in the energy space \citep[see ][for a review]{2004ASSL..308..187R}. A typical Seyfert spectrum in the $0.3-10 \kev$ band can be fit with a combination of the following components: a powerlaw, most likely due to unsaturated thermal Comptonization of accretion disk photons in a hot corona \citep{1979rpa..book.....R, 1991ApJ...380L..51H}, a soft-excess (SE), neutral Compton-reflection continuum, fluorescent and recombination iron emission lines, and warm absorption and emission in the soft X-rays, along with Galactic and intrinsic neutral absorption. Emission lines can be ``narrow'' (unresolved at the resolution of the detectors) or ``broad'', in the latter case due to either Keplerian motions in the torus or broad line regions clouds \citep{2004ApJ...604...63Y, 2006AN....327.1039N} or by a combination of special and general relativistic effects if the lines are emitted in the innermost regions of a X-ray illuminated accretion disk \citep{1989MNRAS.238..729F,1991A&A...247...25M}.
We fit the EPIC-pn spectra with an astrophysically-motivated combination of spectral components.
For all the sources we follow a common procedure for modeling the broad band spectra as enumerated below. 

We start with the $2-10 \kev$ EPIC-pn data as this part of the continuum is not affected by the SE. We fit the data with a powerlaw absorbed by Galactic and intrinsic neutral columns and narrow/broad Gaussian lines to describe the Fe K$\alpha$ and Fe K$\beta$ emission. In many cases we find that the rest frame Fe K$\alpha$ profiles are resolved. Also the Fe K$\alpha$ line energy in some cases is $>6.4 \kev$ indicating that the lines are emitted from an ionised disk. In most cases the fit is satisfactory (See Table \ref{2-10}). 

Extrapolating the best fit $2-10 \kev$ model down to $0.3 \kev$ uncovered soft excess emission in all cases except for MR~2251-178. We used blackbody models to describe this excess. The blackbody temperatures are in the range $0.01-0.3 \kev$ as typically measured in Seyfert~1s \citep[0.1-0.25 keV, ][]{2004MNRAS.349L...7G}. We interpret residual excess emission above $\sim$7~keV as a signature of Compton-reflection from optically thick gas in the nuclear environment \citep{1994MNRAS.268..405N}. The Xspec model {\tt pexmon} \citep{2007MNRAS.382..194N} was used to describe this excess emission. {\tt pexmon} treats self-consistently the fluorescent emission lines from iron and nickel associated with Compton-reflection. When using {\tt pexmon}, we have therefore removed the phenomenological Gaussian profiles originally included in our model. In principle, {\tt pexmon} was designed to describe reflection from a plane-parallel infinite slab, such as an accretion disk as seen from the primary X-ray source. If reflection occurs in an azimuthally symmetric distribution of clouds (the most likely geometry of the reflecting AGN ``torus") the parameters obtained from {\tt pexmon} may not correctly describe the system \citep[see an extensive discussion of this point in][]{2009MNRAS.397.1549M}. However, we still used {\tt pexmon} due to its simplicity and flexibility. We tied the normalization and the spectral index in the {\tt pexmon} model to the normalization and the spectral index of the primary power-law continuum (assuming therefore a plane-parallel infinite geometry and an isotropic primary source).
When the energy at the centroid of the iron line indicates ionisation (E$>6.4 \kev$), we needed a separate Gaussian because {\tt pexmon} describes only reflection off cold matter. We have not attempted fitting these ionized emission lines with physical models, due to their weakness and the lack of spectral ranges within the EPIC-pn sensitive bandpass where continuum components associated with these lines are expected to contribute. For 11 sources in our sample \cite{2010A&A...524A..50D} detected broad Fe K$\alpha$ emission lines. For those sources we removed the broad Gaussian lines and used more physical {\tt diskline} \citep{1989MNRAS.238..729F} or {\tt laor} \citep{1991ApJ...376...90L} models which describe the broadened and skewed profile of an emission line emitted by an X-ray illuminated relativistic accretion disk. The models {\tt diskline} and {\tt Laor} are characterised by four important parameters: the central line energy of emission, the inner radius, the emissivity profile, and the inclination of the disk. The inner radius of the disk decides the effect of the strong gravity regime on the emission line. The emissivity profile of the disk on the other hand determines the amount of flux emitted as a function of the radius of the disk. The outer radius of these model is kept fixed to $400 \rm R_G$. In all cases {\tt diskline} profiles improve the fit with respect to Gaussian profiles, except for Mrk766 and NGC~3516 where we have used a broad Gaussian and a {\tt Laor} profile, respectively.

Once all these components are included in the best-fit model, narrow features are still visible in the residuals, primarily below 2~keV. We interpreted these features as due to ionized gas along the line of sight (warm absorber), or scattering of the primary continuum by ionized gas into the line of sight (warm emitter).
Warm absorbers were modeled using multiplicative table models generated using a photo-ionisation code. We followed an iterative process to finally generate a WA model.
In the following section, we describe briefly the generation of the warm absorber model grids.

\section{CLOUDY grids}
\label{subsec-CLOUDY}

We used the photoionisation simulation code CLOUDY \citep[version 08.00, ][]{1998PASP..110..761F} to calculate the
absorption of continuum photons by a partially ionized medium.
CLOUDY uses an extensive atomic database to predict the absorption
and emission spectrum through and from a cloud. The clouds are
assumed to have a uniform spherical distribution around the central
source and are photoionised by the source. For a cloud we approximate a plane parallel geometry by making
the distance of the cloud very large compared to the thickness of
the cloud.  CLOUDY performs the simulations by dividing a cloud into
thin concentric shells referred to as zones. The thickness of the
zones are chosen small enough for the physical conditions across each zone to be nearly constant. For each zone the simulations are
carried out by simultaneously solving the equations that account for
ionisation and thermal balance. The model predicts the absorption
and emission from such clouds in thermal and ionisation
equilibrium. {\color{black} We have assumed a Solar abundance for the chemical composition of the gas \citep{1998SSRv...85..161G}. The turbulent velocity was frozen to a value of $100\kms$ while generating the model. We discuss our choice of turbulent velocity in appendix \ref{subsec:turbulent-velocity}}.

The physical condition of the photoionised gas are dependant on the SED in the energy band $13.6\ev-10 \kev$ \citep{1999ApJ...517..108N,2013ApJ...777....2L}. Hence the input continuum required by CLOUDY should be as close to the observationally obtained SED as possible for each source (in the said band). A brief description of the procedure for generating the SED is given in section \ref{real-cont}.

\subsection{Generating the SED }
\label{real-cont}

In order to characterize the UV-to-X-ray SED, we need to specify three components: the ``UV bump'', the X-ray ``soft-excess'' and the hard X-ray power-law component.

We assumed that the UV bump in the SED of Seyferts is characterised by \cite{1973A&A....24..337S} thin disk emission which has a multi-temperature disk black body spectrum. The inner disk temperature is determined by the black hole mass obtained from previous studies assuming a $10\%$ accretion rate with respect to the Eddington rate, and an accretion disk extending to 6 R$\rm _g$ (gravitational radius).
The typical values of the temperature $\rm kT_{in}$ of the inner radius of the disk-blackbody ranges from $5-30\ev$ (see Table~\ref{OM}).
The integrated model flux was then obtained from the OM observational data points, once corrected for Galactic reddening as described in Sect.~\ref{sect_data}.

The $0.3-10\kev$ X-ray band spectral shape is determined from the model which best fits the EPIC-pn data. This model is then extrapolated to lower energies down to the energy where it meets the UV bump. We use the {\tt nthcomp} model in XSPEC \citep{1995MNRAS.273..837M} to characterize the X-ray powerlaw and {\tt bbody} model for the SE.
We did not include other components in the X-ray SED such as the Compton reflection hump or the Fe K$\alpha$ line, because they have a negligible impact on the warm absorber's properties due to their comparative weakness. However to obtain the correct SED in the X-ray we followed an iterative procedure. In the first iteration we fitted the dataset with a generic CLOUDY WA model and obtained the best fit $0.3-10\kev$ continuum. This continuum is then used along with the UV bump as mentioned above and we generated the 'realistic' WA model. This model was then used to fit the EPIC-pn X-ray data. See table \ref{EPIC-pn} for the best fit broad band continuum parameters.

{Figure \ref{sed} shows the broad band SEDs for individual sources as mentioned above}. This $ 1\ev- 10\kev$ continuum shape for each source was fed to CLOUDY as the ionizing continuum. For each source, we produced a grid of CLOUDY table models with a range $\xi$ ($10^{0-4}$~erg~s$^{-1}$) and $\nhwa$ ($10^{20-24}$~cm$^{-2}$) following \cite{2006PASP..118..920P}. 

\section{X-ray spectral analysis: high resolution}
\label{section:RGS-analysis}

After the determination of the best-fit X-ray broadband model (Sect.~\ref{sect_epic}) and the UV-to-X-ray SED (Sect.~\ref{real-cont}), a simultaneous fit was made of the EPIC-pn, RGS1 and RGS2 spectra. The RGS spectra were fitted in the energy range of $0.4-2 \kev$ where they are well calibrated. We allowed the relative normalisation to vary among the datasets of the three instruments. We used the Cash statistics \citep{1979ApJ...228..939C} as a goodness-of-fit test, because it properly treats spectra whose count distribution is not (or cannot be) well approximated by a Gaussian distribution.
We added warm absorber components and narrow Gaussian emission lines to the baseline broadband model calculated with the EPIC-pn spectra alone, whenever their inclusion improved the value of the Cash statistic by at least $11$, corresponding to the $99\%$ confidence level for 3 interesting parameters \citep{1976ApJ...208..177L}. Further, we eye-inspected the residuals for any unaccounted features. The errors on the warm absorber parameters were calculated using the Levenberg-Marquardt grid search method (in {\tt ISIS}) with all the discrete as well as continuum parameters free to vary in the simultaneous fit. The best fit WA parameters from the simultaneous fits of the EPIC-pn and the RGS are listed in Table \ref{RGS-table}. For the sources where we did not detect statistically significant WA components we calculated the corresponding upper/lower limits on their ionisation and column density parameters following a procedure similar to those sources showing WA detections. We added a CLOUDY WA component to the continuum best-fit model anf fit simultaneously the EPIC and RGS spectra. Using the grid search method used in ISIS (as mentioned above) we calculated the $90\%$ confidence ranges on the WA parameters. This yielded strict upper/lower limits on the WA parameters for five sources and unconstrained limits for four sources. The results are reported in Table \ref{RGS-table}.  

{\color{black} We used CLOUDY WA table models to fit narrow and weak FeK absorption features at energies $>6.4\kev$. However, in no source we could find any evidence for a high $\xi$ and high velocity absorption phase, which would be the ultra fast outflows (UFOs) found by {\it e.g.,} \cite{2010A&A...521A..57T} in some of the WAX sources. This is evident from the residuals of the best fit plot using EPIC-pn data (Fig.~\ref{epicpn_zoomed}). There are no visible absorption features in the energy range $\sim6-8 \kev$ after we obtain the best fit. We discuss further this issue in Section \ref{subsec:UFO-detection}. We confirmed the lack of UFO detections by fitting the data with XSTAR table models \citep{2004ApJS..155..675K} encompassing the ionisation parameter range of $1<\lgxi<5$ and column density of $\rm 21 <log\nh<24$.}

\section{Results}
\label{sect_results}

\subsection{The broadband $0.3-10 \kev$ spectral shape}

{ The broad band continuum as well as discrete spectral properties of the sources obtained in the WAX sample are typical of Seyfert 1 galaxies. The best-fit parameters resulting from the spectral modeling of the $0.3-10\kev$ EPIC-pn data are shown in Table~\ref{EPIC-pn}. In Fig.~\ref{sample_1} we show the distribution of $\alpha_{\rm OX}$, photon index, and absorption-corrected $0.3-10\kev$ luminosity, where $\alpha_{OX}$ is given by the expression $0.385 \log\left[{\frac{f_{\nu}(2500\AA)}{f_{\nu}(2 keV)}}\right]$ \citep{1979ApJ...234L...9T}.
The best fit spectral slope of the $0.3-10 \kev$ continuum for the WAX sample ranged from $\Gamma=1-2.5$, with a mean value of 1.8 and a standard deviation of $0.2$, consistent with previous studies (Bianchi et al. 2009). Slopes as flat as $\Gamma=1$ found in our sample are, however, unusual in Seyfert galaxies. There are two such cases, NGC~3516 and UGC~3973, where the sources exhibited a very low flux state during their XMM-Newton observations. This low flux state is often associated with (disk) reflection-dominated phases in the light bending scenario proposed by Miniutti \& Fabian (2004).

All the sources in the sample exhibit narrow Fe K$\alpha$ emission lines. These lines are mostly neutral, but, in some cases we additionally detect ionised Fe K$\alpha$ emission lines with rest frame central energies lying in the range $\rm6.4 <E<6.96 \kev$. Such ionized lines were found in 16 sources with equivalent widths ranging from $15-185\ev$ (see Table \ref{2-10}). The narrow neutral Fe K$\alpha$ lines were modeled simultaneously with the distant neutral reflection component. We find that values of the reflection coefficient R (ratio of the reflected and the primary flux) obtained in our study are consistent with those obtained by \cite{2012ApJ...745..107W} in the broad band study with \xmm{} and \suzaku{} observatories. This reflection component is statistically required for 17 sources in the sample. For the sources Fairall~9, NGC~4051, IRAS~050278+1626, Mrk~279, and MCG--6--30--15 the reflection coefficient is large ($\rm 3<R<5$) indicating a more complex geometry than a simple plane-parallel infinite slab subtending an isotropic source. 

For 11 sources in our sample we found broad Fe K$\alpha$ emission line profiles which were also detected by \cite{2010A&A...524A..50D} in their sample. Table \ref{diskline} enumerates all the best fit parameters obtained for the broad line components. They are emitted mostly by ionised disks as the best fit rest frame line energies are $>6.4\kev$. The high emissivity profiles in all cases indicate that the bulk of the emission comes from a few gravitational radii. The inclination angles of $<50$ degrees are consistent with those expected for Seyfert 1 galaxies. As a special case where \xmm{} captured the source NGC~3516 in a low flux state, we detected a large equivalent width of the broad Fe K$\alpha$ emission line ($\sim 1560 \ev$), in agreement with the light bending scenario. In the case of Mrk~766 the broad emission profile could not be modeled either with {\tt diskline} or {\tt laor} models, whereupon we used a broad Gaussian to describe it. In our sample not all sources exhibiting broad Fe K$\alpha$ emission lines have statistical detection of warm absorbers, for e.g., ESO~511-G030 and ARK~120. Also, the neutral reflection coefficient is not correlated with the presence of broad emission lines, indicating that the two emission features arise from distinctly different regions.

The soft-excess is nearly ubiquitous (25/26) in our sample, confirming the finding of \cite{2012ApJ...745..107W} who detected soft-excess in 45 out of 48 sources in their sample. The blackbody temperatures required to describe the soft-excess ranges from $0.04-0.32\kev$, which is similar to the range found in previous studies \citep[e.g., 0.1-0.25 keV, ][]{2004MNRAS.349L...7G}. {The $0.3-10\kev$ unabsorbed flux for each source are listed in Table \ref{Luminosity}.}

\subsection{The EPIC-pn and RGS joint fit}
 \label{subsec:joint-fit}
 
 The best fit parameters obtained from the simultaneous fit of high and medium resolution spectra are reported in Table \ref{RGS-table}. The WAX sample is constituted by 26 X-ray type 1 AGN out of which 17 ($65\%$) sources have at least one warm absorber component. In 11 sources we detect upto three components. The WAX sample probes the column density of $\rm N_H=10^{20-23.5}\cmsqi$, ionisation parameter of $\rm log\xi=-1 \,to\,3.2$, and outflow velocity of $0-10^4 \kms$. While the sample is by construction flux-limited (with a completeness fraction of $84\%$), the observations discussed in this paper were taken in the framework of observational programs with largely diverse scientific goals. Consequently, the statistical quality of the spectra presented in this paper is very different for different sources. It is in principle possible that the detection fraction reflects more the statistical quality of the data rather than the intrinsic properties of the sample. In Fig.~\ref{fig:SNR-RGS} right panel, we plot the number of detected warm absorbed components as a function of the net count rates in the RGS spectrum. We use the latter quantity as a proxy for the statistical quality of the XMM-Newton observations. {\color{black}We find that even for high SNR objects a WA is not present. This may indicate that the absence of WA is not solely driven by the statistical quality of the data.}

 In Figure \ref{xi-NH-distribution} we compare the distribution of column density and ionization parameter for all the sources in the WAX sample. The green histograms correspond to all the warm absorber components detected in WAX; the blue histograms indicate ionization parameter (column density) lower (upper) limit for the sources where no warm absorber component was detected. Upper/lower limits were calculated following the procedure described in Sec.~\ref{section:RGS-analysis} and are reported in Table \ref{RGS-table}. The distribution of the WA measurements and that of the censored data are different. However, a KS test could not be performed to confirm the result as the sample size of the no-WA sources is too small to derive any meaningful statistical conclusion. {\color{black}From Fig. \ref{xi-NH-distribution} we find that there is gap in the $\log\xi$ space in the range $\sim 0.5-1.5$, which we will discuss in Section \ref{subsec:xi-gap}.}\\

\subsection{Correlation analysis}
\label{subsec:corr-analysis} 
We tested for correlations between the three WA parameters: $\log\xi$, $\log\nh$ and $ v_{\rm out}$ and also between these and the continuum parameters of the sources. The results of the non-parametric Spearman rank correlation test are reported in Table \ref{corr}.  We find that the WA column density is correlated with the ionisation parameter with a correlation coefficient of 0.64 with $>99.99\%$ confidence. We obtain correlations at a confidence level $\sim 95\%$ between the outflow velocity of the absorbers and the column density or the ionisation parameter (see Table \ref{corr}). To test the linear dependence of the parameters in the log-log space we carried out linear regression analysis between the WA parameters following the method presented by \cite{1996ApJ...470..706A} and using their public code to carry out the analysis. In this method the errors in both variables defining a data point are taken into account, as is any intrinsic scatter that may be present in the data, in addition to the scatter produced by the random variables. We quote the results of the analysis in Table \ref{Table:lin-reg} and Figure \ref{fig:linear-reg-xi-V}. A detailed comparison of these results with those obtained by T13 is reported in Section \ref{subsec:UFO-WA}. To deal with censored data (lower limits)on the outflow velocity we used a method adapted from \cite{1985ApJ...293..178S}. We generated 1000 Monte-Carlo simulations, where each of the censored data points $v^{*}_{out}$ was replaced by a randomly extracted number from a flat distribution in the range $v^{*}_{\rm out}$ to $30,000\kms$. For each of the simulated datasets we calculated the best-fit parameters, and the correlation coefficient. The results in Table \ref{Table:lin-reg} represents the average of the values obtained in the simulations. In addition we note that there are three data points where the outflow velocity is consistent with $\rm \sim zero \kms$. In comparing the WAX sample results with the prediction of outflow models, we have ignored those data points from the analysis.

The correlation between the WA parameters and the continuum parameters is not straight forward, as we need to consider the inherent degeneracy in the WA parameters for a particular source. For e.g., if we consider a source which has three WA components, then for the given value of its blackhole mass there will be three values of $\xi$ in the $\rm M_{BH}$-$\log\xi$ space. A possible solution to the problem would be to take average of the WA parameters for a given source, as in \cite{2005A&A...431..111B}. However given the unknown structure of the WA clouds, different components of the WA for the same source may actually reflect different properties of the clouds which are stratified in nature. Therefore the average values may not be an unbiased estimate of the outflow properties. We have instead considered correlation between warm absorbers and source and continuum parameters after selecting one component for each source (e.g., that with the highest/lowest column density, ionization parameter, outflow velocity). In all the cases we found that there are no strong correlations between WA parameters and source and continuum parameters (See Table \ref{Tab:corr-high-and-low}). {We have also tested for correlations in the $\log\xi$ - $\log v_{\rm out}$ separately for each source having multiple WA components. This is motivated from the fact that different sources may have different driving mechanisms for the WA clouds. If for a particular source a single driving mechanism is in operation, we should expect a correlation. We still find no correlation between these two parameters, indicating that even for a single source the outflow physics is not determined by a single process.} {Interestingly we do not find any correlation between the black hole mass and the corresponding WA outflow velocity. A correlation would vindicate the direct role of the black hole as the primary driver of the WA outflows. The lack of a correlation indicates that the simple virial arguments may not be used for estimating the distances of the WA from the central source, as the black hole may not have a direct influence on the outflows}.

\subsection{UFOs in the WAX sample}
\label{subsec:UFO-detection}
T10 in their analysis found the presence of ultra-fast-outflows (UFOs) in 15 sources out of which six sources are also in the WAX list. These are MRK~509, ARK~120, UGC~3973, NGC~4051, MRK~766, and IC~4329A. We do not detect UFOs in any of the sources of our sample. The lack of detection of UFOs in the six sources could be due to the different statistical thresolds and methods used in the spectral analysis. We investigate this discrepancy in this section. 

 T10 fitted the EPIC-pn spectra from $4-10\kev$ only. However, the soft X-ray WA which they ignored in their analysis may also show some opacity in the hard X-rays especially near the Fe K band $\sim6-9 \kev$ if their ionisation parameter as well as the column density are high enough. Limiting the fit to the hard X-ray band may lead to an erroneous estimate of the standard WA contribution to the opacity above $6 \kev$, where UFO observational features are detected.

 {\color{black}Secondly, limiting the spectral fitting to the hard X-ray band yields larger sytematic uncertainties in the determination of the underlying continuum. They might affect the detection of weak absorption features at the CCD resolution in a band where additional spectral complexity (Compton reflection, partial covering obscuration, recombination lines from highly ionised plasma) may take place at least in some sources. Figure \ref{epicpn_zoomed} shows that there are no statistically significant negative residuals after a full band modeling has been done as enumerated in Sections \ref{sect_epic} and \ref{section:RGS-analysis}.}


 Lastly, we also note that none of the previous individual studies done on the six common sources between T10 and our work found any UFO. See for e.g., \cite{2007A&A...461..135S,2004MNRAS.351..193V,2011MNRAS.411..607G,2008RMxAC..32..123K,2003ApJ...582...95M,2005A&A...432..453S} for MRK~509, ARK~120, UGC~3973, NGC~4051, MRK~766, and IC~4329A, respectively. \cite{2013A&A...549A..72P} in a long 600 ks \xmm{} campaign of the source MRK~509 found no statistically significant highly ionised, high velocity outflows.

\subsection{WA parameter ranges probed by WAX}
\label{subsec:WA-simulations}

As mentioned in Section \ref{subsec:joint-fit}, the WAX sample probes the column density of $\rm N_H=10^{20-23.5}\cmsqi$, ionisation parameter of $\rm log\xi=-1 \,to\,3.2$, and outflow velocity of $0-0.03c$. {\color{black} In this Section, we discuss the possibility that the lack of the shallow WA components with low ionisation parameters and low column density in the ranges $\log\xi=0.5-1.5$ and $\log\nh\sim 20$ is driven by the statistical quality of our data. The lack of components with low ionisation parameter and high column density is a consequence of the WAX sample definition. We carried out simulations of the bright WAX source Ark~120 as well as a less luminous source ESO~511-G030, showing no WA features. We added a CLOUDY WA model component with $\rm log\xi=0.5$ with $\rm N_H=10^{20}\cmsqi$ to the best fit continuum of AKN~120 and ESO~511-G030 and generated fake EPIC-pn as well as RGS datasets using the {\tt fakeit} command in ISIS. We fitted these fake datasets with a model consisting of continuum parameters only to statistically detect the presence of the WA in the data. We found that the fit worsened by $\Delta C\ge 50$ for both the cases. This indicates that the non-detection of such WA components in the WAX sample is not instrument limited.

In the right panel of Figure \ref{fig:linear-reg-xi-V} we compare the column density versus the ionisation parameter relation for the warm absorber components for WAX data points with the extrapolation of the correlation of the same parameters for the UFOs in T13 (dashed line). We observe a relative dearth of measurements in the central part of this plot. To investigate possible observational biases, we have performed simulations similar to those described above  assuming pairs of values $\rm log\xi=2.5$ with $\rm N_H=10^{21.5}\cmsqi$ and $\rm log\xi=3.5$ with $\rm N_H=10^{23}\cmsqi$ for both the sources. In all cases we found that the quality of the fit of the simulated datasets worsened with respect to the real data by $\rm\Delta C\ge 1000$ indicating that the observed gap in the column density versus ionisation parameter plane is not an instrumental bias.}

\section{Discussion}
\label{sect_discussion}
In this paper, we have performed a systematic high resolution spectral study of a sample of Seyferts and studied the WA properties in detail. The sample consists of bright nearby Seyfert 1 galaxies which have \xmm{} data. We discuss below the findings of our study.

\subsection{Detection fraction of warm absorbers}
The best fit parameters for the WA components are shown in Table \ref{RGS-table}. We detect warm absorber components in 17 sources out of 26 WAX sources. In four sources the values of the WA column density and the ionisation parameters cannot be constrained. The quality of these data is therefore insufficient to address whether these sources host a WA. For the remaining five sources the lower limit on the ionisation parameter and the upper limit on the column density are inconsistent with the distribution of these quantities in detected WAX components (see Figure \ref{xi-NH-distribution}). We interpret these objects as {\it bona fide} WA-bare AGN.

{\color{black} The WA detection fraction in the WAX sample is $77\%$ (17 out of 22 sources with definite detections) with a statistical error of $\pm 9\%$, which is calculated using a binomial distribution for the occurrence of WA in the parent sample. Taking into account further uncertainties yielded by the 4 objects whose spectra do not have sufficient statistical quality, a further systematic error of [$+3\%, -{14}\%$] has to be added to the estimate of the detection fraction. Our sample is only complete at $76\%$, so an error of $1\%$ is added futher. This gives an upper limit of detection of $\sim 90\%$. On the other hand the most conservative lower limit on the occurence of WA is $\sim 50\%$ (17 out of 34 sources). This indicates that at least $50\%$ of the Seyferts in nearby universe should have WA.}

Our study yields a detection fraction which is broadly consistent with the previous sample studies on AGN WA with {\it ASCA}, {\it Chandra}, and {\it XMM-Newton}. \cite{2005A&A...431..111B} obtained a detection fraction of $\sim 50\%$. The study by \cite{2012ApJ...745..107W} on the other hand has put the upper limit to $\sim 52\%$ although they claim that WA are ubiquitous in AGN. The comparative advantage of our study is that we have investigated a flux limited sample of AGN with a known (and comparably small) incompleteness, as opposed to the sparse collections or largely incomplete samples used in previous studies. Lacking any direct estimate of the X-ray WA covering fraction, $C_f$, we assume hereafter a value comparable to the detection fraction in our sample, $C_f\simeq 0.75$. However we need to keep in mind a couple of caveats in the context of calculating a conservative upper limit on the detection fraction. Firstly, the source Mrk 279 does not show any WA in WAX analysis. We have used a single exposure ($\sim 60$ ks) dataset. \cite{2010A&A...512A..25C} studied the same source and have combined 3 other datasets to effectively have $\sim 160 \ks$ exposure. They could detect two weak WA componets. However for no other WAX sources this is the case. Secondly, WA features could be transient, as recently discovered in Mkn~335 \citep{2013ApJ...766..104L} 

\subsection{The driving mechanisms of the WA}

There are primarily three existing models for driving mechanisms of the WA outflows. The first scenario is the radiatively driven wind where the radiation pressure is mainly accounted for by the UV absorption lines \citep{2004ApJ...616..688P}. However, if the cloud is highly ionised then the UV and X-ray line opacities may not be sufficient to drive the clouds. In such a case Compton scattering can provide the necessary momentum to the outflow \citep{2003MNRAS.345..657K,2010MNRAS.402.1516K}. In the radiatively driven wind scenario a simple scaling relation holds: $v_{out} \propto \xi^{1/2}$. 

The second scenario is that the WA are accelerated by magneto hydrodynamically driven (MHD) processes \citep{1982MNRAS.199..883B,2004ApJ...615L..13E, 2010ApJ...715..636F}. This possibility stems from the fact that some of the most important AGN nuclear phenomenon like the accretion disk viscosity and the existence of a corona can be explained by MHD mechanisms. \cite{2010ApJ...715..636F} suggested that MHD winds could be fundamental to the AGN outflows and suggested few scaling relations between obervable WA parameters ($\nh$, $\xi$ and $v$). These relations can be reduced roughly to $v_{\rm out} \propto \xi$ \citep{2012ASPC..460..181K}. However, in this model there are certain limitations, for e.g., the authors consider the disk as a boundary only, while in a realistic situation the accretion physics should also be taken in to account.

The third scenario is that of the thermally driven wind \citep{1983ApJ...271...70B,2001ApJ...561..684K, 2006ApJ...652L.117N}. The flux of the central source irradiates the accretion disk and heats it up to some larger radius where the heating produced locally is negligible compared to the radiative heating. The rate of heating is proportional to the flux of the incident radiation but the cooling rate is dependent on collisional cooling, whereby it declines with density as we move away from the equatorial plane of the accretion disk. The AGN central radiation is sufficiently hard to heat the gas to $10^7$ K by the Compton process. This temperature is very high compared to the internal temperatures of the outer part of the accretion disk. Hence, if one focusses on a thin layer above the outer parts of the accretion disk, the gas forms a tenuous plasma with the thickness exceeding that of the disk. If the sound speed in this heated gas is higher than the escape speed from the system at a given radius, the gas will escape resulting in a wind which is thermally driven. As enumerated by \cite{1983ApJ...271...89B}, for a Compton heated wind we would observe: (1) an extended source of scattered X-rays, (2) X-ray polarisation, (3) X-ray line emission due to reprocessing of incident X-rays by the disk, and (4) in the UV the spectrum will differ from the simple blackbody scaling. (1), (2), and (4) are beyond the observational capability of our sample. We discuss here briefly the possibility that the emission lines ubiquitously associated to WA features in the WAX sample could be the signature of a hot disk atmosphere in this scenario. {The He like triplet lines of \ion{O}{VII} are the most commonly detected emission lines in the energy band of RGS and they serve as the most important tool in plasma diagnostics of the emitting medium. The three emission lines: {\it resonant} (r), {\it intercombination} (i), and {\it forbidden} (f) for the \ion{O}{VII} He like ion have rest frame energies of $0.574,\, 0.568,\, 0.561\kev$ respectively. The {\it resonant} line is mostly emitted by a collisionally ionised plasma while the forbidden line is emitted by colder low density  photo-ionised plasma. Figure \ref{OVII} shows the distribution of the detected line centroid energies of the \ion{O}{VII} triplet. We find that the distribution is mostly clustered around the rest frame energy of the {\it forbidden} line indicating that the emission lines are possibly emitted from a photo-ionised plasma instead of a collisionally ionised, high-temperature thermal plasma. If these emission lines are emitted by the same clouds which show warm absorption, we can rule out the origin of WA from a thermally driven wind. However a detailed study of the warm emission lines will be carried out in a future study.}

From our study we find the linear regression slope between the two parameters is $v_{\rm out} \propto \xi^{0.12\pm 0.03}$. Our measurements on the whole sample do not follow the simple scaling relations predicted by either radiation-driven or MHD-accelerated winds. The WA winds expand as it moves out and hence its volume density decreases which has to be taken into account in a geometrical framework to consider any driving mechanism. We cannot rule out the presence of other mechanisms or the combination of the above mentioned ones. Even for individual sources with multiple components of WA we find that the linear regression slope between the $\log\xi$ and $\log v_{\rm out}$ do not conform with the predictions of the above driving mechanisms.


\subsection{The gap in the $\xi$ distribution}
\label{subsec:xi-gap}

{\color{black} From Fig. \ref{xi-NH-distribution} we find that there is a gap in the distribution of the ionisation parameter in the range $\log\xi=0.5-1.5$. This gap has been previously observed in many of the studies on individual sources \citep{2001A&A...365L.168S,2003ApJ...597..832K,2005ApJ...620..165K}. For the first time we have confirmed the absence of particular ionic states in high resolution data, for a uniformly analysed sample of 26 Seyfert galaxies. The gap could therefore be a universal feature which disfavours a continuous distribution of ionisation parameter \citep[see for e.g.,][]{2005ApJ...620..165K}. However, as a caveat we should remember that the exact values of the fitted ionisation parameter depend on the assumed values of the cloud metallicity and the source SED. Nevertheless, such a gap in the distribution of $\log\xi$ is intriguing. There are two possible reasons for it. Firstly, it may be due to the absence of strong ionic transitions in that energy range with weak lines being undetected in our data. Secondly, it could be that clouds inherently fail to have those states, for e.g, if the clouds are thermally unstable in that range of ionisation parameter. 

The first reason can be ruled out from the simulations carried out in Section \ref{subsec:WA-simulations} which showed that a WA component of $\log\xi=0.5$ and $\log\nh=20$ is detectable by our present instrument and data quality. Therefore, the only reason for this gap could be thermal instability of the gas in that regime of ionisation parameter.

\citet{2007ApJ...663..799H} studied the absorption measure distribution (AMD), defined as d$\nh$/d$\log\xi$, for two bright sources (IRAS~13349+2438 and NGC~3783) showing warm absorption. Both the sources showed a double peak distribution of the AMD with a deep minimum in the range $0.75<\log\xi<1.75$, which is nearly same as the range in which we find a gap in the distribution of the ionisation parameter. \citet{2007ApJ...663..799H} believe that this minimum is due to thermal instability. A systematic stability curve analysis will be carried out in a future paper.

 One other interesting feature is the large distribution range of the ionisation parameters for individual sources. For sources like MRK~704, NGC~4051, NGC~3227, MCG-6-30-15, IC~4329A, we have detected ionisation parameters ranging over almost three decades. \citet{2009ApJ...703.1346B} pointed out that the AMD slopes for such sources could be a tool for density diagnostic of the clouds. A systematic study of AMD for WAX sources will be taken up by us in the future.}

\subsection{Relation between outflows and relativistic spectral features}
 It has been suggested that the broad profiles of the Fe K${\alpha}$ fluorescent line detected in the X-ray spectra of nearby Seyfert galaxies \citep{2007MNRAS.382..194N,2010A&A...524A..50D} could be an artifact of fitting multiple layers of partial covering, ionised absorption with a relativistically smeared disk reflection continuum. To test this hypothesis, we plot the WA column density distributions for sources where our analysis suggests the presence/absence of a relativistically broadened K$\alpha$ profile (Fig. \ref{Fe-line-NH-distrib}). For sources with multiple WA components, we plot the highest column density, as this is likely to yield the largest spectral impact on iron line modeling. The distributions are indistinguishable. This rules out that an adequate modelling of the WA opacity is the main driver underlying the detection of relativistically broadened spectral components, in particular of the Fe K$\alpha$ red wing \citep{1989MNRAS.238..729F}.

\subsection{Are WA and UFOs different manifestations of the same physical system?}
\label{subsec:UFO-WA}
 It is still a matter of debate whether the UFOs and the WA are different manifestations of the same physical system or not. T13 have concluded that in the parameter space spanned by $\log\xi$, $\log\nh$ and $\log v_{\rm out}$, WA and the UFOs lie at the extreme ends of the same correlation, pointing to a common nature of the two types of outflows. To test this hypothesis, we compare the locus occupied by the WAX (UFO) points in the parameter spaces defined by the warm absorber observables with the extrapolation of the regression fit on the UFO (WAX) data points.

Figure \ref{fig:linear-reg-xi-V} left panel shows the linear regression fit in the $\log\xi$-$\log v_{\rm out}$ space for both the WAX sample and the UFOs (T13). We find that the UFOs do not lie on the extrapolated linear regression fit of the WAX sample. On the other hand we find that the regression analysis of the UFOs predict a large span of $\log\xi$-$\log v_{\rm out}$ space, which includes the WAX data points. The right panel of figure \ref{fig:linear-reg-xi-V} shows the linear regression fit in the $\log\xi$-$\log \nh$ space where we find that the UFOs do not lie on the extrapolated linear regression fit of the WAX sample. We therefore conclude that our results are inconsistent with the warm absorbers and UFOs occupying extreme loci of the same correlation between the wind observables. They do not directly indicate a common origin between UFOs and the WAs.

\section{Summary and conclusions}
  In this paper we analyse for the first time a hard X-ray flux-limited sample of nearby Seyfert galaxies, using broad-band EPIC-pn and high resolution RGS X-ray data. The primary goals of our study are: a) to estimate the incident rate of ionised outflows in nearby, X-ray bright Seyfert galaxies; b) To constrain their acceleration mechanism; and c) their relation with high-velocity UFOs.

The main conclusions of the paper are:

\begin{itemize}

	\item  WA do not appear ubiquitously in our sample. {\color{black} The detection fraction of WA in the WAX parent population is constrained between $65-90\%$. However we could put a strict lower limit on the detection fraction $\sim 50\%$ which indicates that at least half of all the Seyfert galaxies in the nearby Universe exhibits warm ionised outflows.}

		\item {\color{black}We find a gap in the distribution of the ionisation parameter in the range $0.5<\log\xi<1.5$ which we interpret as a thermally unstable region for WA clouds. This may point towards discrete and clumpy configuration of WA clouds as against a continuous medium.}

\item   WA parameters ($\xi$, $\nh$ and $v_{\rm out}$) do not show any significant correlation between themselves except for $\log\xi$, $\log\nh$. They also do not correlate with any of the X-ray broad band continuum parameters.

\item  Our results are inconsistent with the correlation between the outflow velocity and the ionisation parameter predicted either by Compton-scattering driven (King 2003, $\xi \propto v_{\rm out}^{0.5}$) or MHD driven (Fuzukawa et al. 2010, $\xi \propto v_{\rm out}$) winds. The linear regression between $\log\xi$ and $\log v_{\rm out}$ on the whole sample is $0.12\pm 0.03$. A similar discrepancy occurs if the independent components of each individual source are separately considered.

\item  The presence of WA are not related to the presence of relativistic Fe K$\alpha$ features as we find sources without WA but with a broad Fe K$\alpha$ line. The detection of the broad line features in the X-rays are not influenced by the WA modeling.

\item The UFOs and the WA may not be the same physical system. The correlations among gas observables obtained in our paper do not extend to the UFO regime. However, given the poor statistical quality of UFO data points, the linear regression line in the UFO parameter space encompasses the WA data points as well.

\end{itemize}





\begin{table*}

{\footnotesize
\centering
  \caption{WAX source basic properties.} \label{basic-info}
  \begin{tabular}{llllllllllll} \hline\hline 

 No. & Object&  Seyfert & V   & Redshift & Lum dist &  $\rm M_{BH}$ & $\rm N_H^{Gal}$ & \\ 
     &   name         &  type& mag &  &  (Mpc)     &    $\rm log(M/ M_{\odot})$      &    $\times 10^{20} \cmsqi$ & \\ \hline \\ 

 1.& NGC4593 &  1  & 11.67 &   0.009 &   42.0   &   6.77  &  1.9     \\ \\

 2.& MRK 704 & 1.2  & 15.38 &   0.029 &   127.2  &   7.62 &  2.9    \\ \\

 3.& ESO511-G030 & 1  & 13.30 & 0.022&   97.0  &  8.4  &  4.69    \\  \\

 4.& NGC 7213 &  1.5 & 11.01 &    0.0058 &  21.2   &   8.3  &  1.06    \\ \\

 5.& AKN 564 &  LINER  & 14.55 &   0.024   &  98.6   &   6.7  &   5.34    \\  \\

 6.& MRK 110  & 1 & 15.6  &   0.035  & 151     &  7.4  &   1.30 \\ \\

 7.& ESO198-G024 &  1 & 15.36 &  0.045  & 192     &  8.1  &  2.93  \\  \\

 8.& Fairall-9 & 1  &  13.5  &  0.047  & 199     &  8.6 &  3.16   \\ \\

19.& UGC 3973 & 1.5 & 13.9  &  0.022   & 94.4   &  8.1 & 5.27 \\   \\

10.& NGC 4051 & 1.5 &  10.83 & 0.002   & 12.7    &  6.3   &  1.32 &  \\   \\

11.& MCG-2-58-22 & 1.5  & 14.6 &  0.046 & 194     & 7.36   &  2.91 \\   \\

12.& NGC 7469 & 1.2 & 13.0    &  0.016 & 62.7    & 7.1   &  4.45\\   \\

13.& MRK 766 & 1.5 & 13.7    &  0.0129& 57.7    & 7.5 & 1.71  \\   \\

14.& MRK 590 & 1.2 & 13.85   &  0.026 & 107    &  8.6 &  2.65  \\   \\

15.& IRAS05078+1626 & 1.5 & 15.6&0.017& 74.3   & 7.86   &  2.20 \\   \\

16.& NGC3227 & 1.5 & 11.1    &  0.003  & 20.4   & 7.4  &  2.15  \\   \\

17.& MR2251-178 & 1 & 14.36&  0.063& 271    &  8.5   &   2.42  \\   \\

18.& MRK 279 & 1.5   & 14.57& 0.030   & 129    &  7.5  & 1.52  \\   \\

19.& ARK 120 &  1  & 15.3& 0.032   & 138    &  8.4 & 9.78  \\   \\

20.& MCG+8-11-11& BLSY & 15   & 0.02    &85.8    &  8.48   & 1.84  \\   \\

21.& MCG-6-30-15 & 1.5 & 13.7& 0.007   &35.8    & 7.0  & 3.92  \\   \\

22.& MRK 509 &  1.2 & 3.0  & 0.034   & 141    & 8.3  & 4.25   \\   \\

23.& NGC 3516 & 1.5 & 12.5 & 0.008   & 37.5   & 7.7  & 3.45    \\   \\

24.& NGC 5548 & 1.5  & 13.3 & 0.017   &74.5    & 8.1  & 1.55   \\   \\

25.& NGC 3783 & 1  & 12.64 &0.009   & 44.7   &  7.8  & 9.91  \\   \\

26.& IC 4329A  & 1.2 & 14.0   & 0.016 & 70.6    & 8.2  & 4.61   \\   \\

\hline \hline
\end{tabular} \\ 

\footnotetext{1}
{ The blackhole masses are obtained from \cite{2010A&A...524A..50D} and \cite{2012ApJ...745..107W}. The Redshift, V band magnitude, Luminosity distance, Seyfert classification are obtained from NED. The $\rm N_H^{Gal}$ are obtained from HEASARC.}\\

{The source identification numbers listed in column one are uniformly used in all places throughout this paper.}\\
}
\end{table*}


\begin{table*}
{\footnotesize
\centering
\caption[abc]{XMM-Newton observations of WAX sources.}
\label{obs-info}
  \begin{tabular}{llllllllllll}  \hline\hline

 No. &  Object&Obs & Date  & Total  & RA & Dec    \\ 
     &  name   &id  & of obs& exposure& (Degrees)& (Degrees)   \\ 
     &       &       & (DD-MM-YY) &  (Ks)   &     \\ 
     &       &       &         &              \\ \hline \\
 
 1.& NGC4593        & 0109970101 & 02-07-2000   & 28   & 189.914&$-5.344$   \\ \\

 2.& MRK 704        & 0502091601 & 03-05-2008 &  98    & 139.608 &16.305  \\ \\

 3.& ESO511-G030    & 0502090201 & 05-08-2007  &  112   & 214.843 & --26.644\\ \\

 4.& NGC 7213      & 0605800301 & 11-11-2009  &  132   & 332.317  & --47.166  \\ \\

 5.& AKN 564   & 0206400101 & 05-01-2005  & 101    & 340.663  & 29.725  \\   \\
   
6.& MRK 110        &0502090201 & 15-11-2004  & 47    &  141.303& 52.286   \\ \\

 7.& ESO198-G024   & 0305370101 & 04-02-2006  & 122    & 39.581& --52.192  \\  \\
 8.& Fairall-9    &  0605800401 & 09-12-2009  & 130 & 20.940 & --58.805 \\ \\
9.& UGC 3973          & 0502091001 & 26-04-2008  & 89     & 115.636 & 49.809  \\ \\
10.& NGC 4051       &0109141401 & 16-05-2001  & 122 & 180.790  & 44.531   \\ \\
  
11.& MCG-2-58-22    & 0109130701 & 01-12-2002 & 20 &346.181 & --8.685   \\ \\

12.& NGC 7469     & 0112170101  & 30-11-2000   &   18  & 345.815 & 8.874  \\  \\

13.& MRK 766 & 0109141301  & 20-05-2001   & 130   & 184.610 & 29.812    \\ \\
  
14.& MRK 590   & 0201020201  & 04-07-2004   & 112  &  33.639 & --0.766    \\ \\

15.& IRAS05078+1626      & 0502090501  & 21-08-2007   & 61   & 77.689 & 16.498  \\    \\
 
16.& NGC3227     & 0400270101  & 03-12-2006   & 107  & 155.877 & 19.865      \\ \\
   
17.& MR2251-178 & 0012940101  & 18-05-2002   & 65   & 343.524 & --17.581     \\    \\
18.& MRK 279    & 0302480401  & 15-11-2005   & 60   &   208.264 & 69.308   \\     \\
 
19.& ARK 120    & 0147190101  & 24-08-2003   & 112   & 79.047 & --0.149  \\  \\
20.& MCG+8-11-1 & 0201930201  & 09-04-2004   &  38 & 88.723 &46.439   \\ \\

21.& MCG-6-30-15& 0029740701    & 08-02-2001    & 130 &  203.973 & --34.295  \\ \\
  
22.& MRK 509    &  0306090201  &  19-10-2005    & 86 & 311.040  &--10.723    \\ \\

23.& NGC 3516   & 0107460701  & 09-11-2001    & 130  &166.697 & 72.568   \\ \\
24.& NGC 5548   & 0089960301  & 09-07-2001   & 96 & 214.498 & 25.136   \\ \\
   
25.& NGC 3783   & 0112210501  & 19-12-2001    & 138  & 174.757 & --37.738  \\ \\
26.& IC 4329A   & 0147440101  & 06-08-2003    & 136  & 207.330 & --30.309     \\ \\

\hline \hline
\end{tabular} \\ 

\footnotetext{1}{Obtained from Heasarc and XSA}.

\footnotetext{2}{}.

}
\end{table*}


\begin{table*}

{\footnotesize
\centering
  \caption{The Optical Monitor data and relevant parameters. The fourth column is the observed flux in the optical-UV bands, simultaneously observed with the X-ray data, using the OM camera. The sixth column lists the extinction corrected flux.\label{OM}}
  \begin{tabular}{llllllllllll} \hline\hline 

	  No. & Object      &  Filter$^1$     & Flux observed            &  $\rm ^{2}A_{\lambda}$      & Corrected flux           & $\rm ^{3}kT_{diskbb}$ & \\ 
     &   name      & used        & { $10^{-15}$ $\funit\AA^{-1}$} & Galactic          & { $10^{-15}$ $\funit\AA^{-1}$} &  $(\ev)$\\ 
     &             &             &   & extinction          &      &       \\ \hline

1.   & NGC4593     & UVW2        & 9.99                     &  0.191              &   11.91                  &   26.42     \\ \\

2.   & MRK 704     & UVW1        & 7.3                      &   0.150             &   8.3                    &  16.20 \\ \\   
3.   & ESO511-G030 & UVW2        & 14.58                    &  $0.533$            &   23.8                   & 10.34 \\  \\  
4.   & NGC 7213    &  UVW1       & 6.5                      &  0.079              &  6.99                    &   10.95 \\ \\   

5.   & AKN 564     &  UVW1       & 7.06                     &   0.312             &  9.4                     &  27.51\\  \\  
6.   & MRK 110     & UVW2        & 25                       &   0.100             & 27.4                     &  18.38 \\ \\  
 
7.   & ESO198-G024 &  UVM2       & 5.06                     &  0.270              & 6.49                     &  12.3 \\  \\
  
8.   & Fairall-9   & UVW1        & 17.74                    &  0.141              & 20.20                    &  9.21 \\ \\

9.   & UGC 3973    & UVW2        & 9.9                      &  0.549              & 16.4                     &  12.29\\   \\

10.  & NGC 4051    & UVW2        &  15.2                    & 0.101               & 16.6                     &  34.63 \\   \\ 

11.  & MCG-2-58-22 & UVW1        & 6.3                      &  0.218              & 7.7                      & 18.81   \\   \\

12.  & NGC 7469    & UVW2        & 33.5                     &  0.535              & 54.8                     & 21.85\\   \\

13.  & MRK 766     & UVW2        & 0.132                    & 0.152               & { 0.151}                 & 17.36 \\   \\

14.  & MRK 590     & UVW2        & 1.56                     &  0.290              & { 2.03 }                 & 9.21 \\   \\

15.  & IRAS05078+1626 & UVW2     & 0.32                      & 2.36               &{  2.7}                   & 14.11 \\   \\

16.  & NGC3227       & UVW1      & 9.0                      &  0.118              & 10.0                     & 18.38    \\   \\

17. & MR2251-178   & UVW2        & 8.05                     &  0.303              & 10.6                     & 9.76\\   \\

18.& MRK 279       & UVW2        & 24.32                    & 0.123               & 26                        &  17.36 \\   \\

19.& ARK 120       &  UVW2       & 35.8                     & 0.996               & 89.6                      &  10.34 \\   \\

20.& MCG+8-11-11   & UVW1        & 8.02                     & 1.136               & 22.8                       &  9.87 \\   \\

21.& MCG-6-30-15   & U Band      & 0.38                     & 0.267               & { 0.478 }                 & 23.15 \\   \\

22.& MRK 509       & UVW2        & 38                       & 0.446               & 57.3                       & 10.95   \\   \\

23.& NGC 3516      & UVW2        & 19.1                      & 0.329              & 25.8                        & 15.47 \\   \\

24.& NGC 5548      & U BAND      & 13.1                     & 0.088               &14.2                         & 12.29  \\   \\

25.& NGC 3783      & UVW2        & 26.15                    & 0.925               & 61.3                        &  14.60 \\   \\

26.& IC 4329A      & UVW1         & 1.38                    & 0.459                & { 2.1}                     & 11.60  \\   \\ 

 \hline \hline
\end{tabular} \\ 
{$^1$ U Band- $\rm 3440 \AA$, UVW1- $\rm 2910 \AA$, UVM2- $\rm 2310\AA$, UVW2- $\rm 2120 \AA$.}\\
{$^2$ $\rm A_{\lambda}$ is the Galactic extinction correction magnitude.}\\
{$^3$ $\rm kT_e$ is the blackbody temperature of the accretion disk emission as calculated in Section \ref{real-cont}}\\

}
\end{table*}


\begin{table*}
{\footnotesize
\centering
  \caption{The $2-10 \kev$ EPIC-pn continuum parameters.  \label{2-10}}
   \begin{tabular}{l l l l llllllllll} \hline\hline 

	   No. &Source  & $\Gamma$ &Fe K & Fe K                          & Fe K                     &    $\frac{\chi^2}{\rm dof}$\\ 
     &        &         &norm & Line centroid energy ($\kev$) & Line width $\sigma$($\ev$)&        \\ 

   &         &           & ($10^{-5}$)&           \\ \hline\\ \\

1. &NGC4593 & $1.76_{-0.02}^{+0.03}$ &  $4.16_{-0.90}^{+1.10}$ & $6.38_{-0.01}^{+0.03}$&  $\le 48$ & $\frac{178}{188}$\\ \\

2.& MRK 704   &  $1.76_{-0.02}^{+0.03}$  & $1.91_{-0.20}^{+0.31}$ & $6.39_{-0.05}^{+0.10}$  & $137_{-47}^{+50}$ & $\frac{173}{197}$\\ \\

3.&ESO511~G030 &$1.78_{-0.01}^{+0.02}$ &  $1.9_{-0.3}^{+0.2}$ & $6.33_{-0.05}^{+0.06}$&  $360_{-90}^{+110} $ &$\frac{293}{200}$\\ \\   

4.& NGC 7213   & $1.69_{-0.01}^{+0.02}$ &  $1.59_{-0.41}^{+0.41}$ & $6.39_{-0.02}^{+0.01}$& $ \le 60$ & $\frac{243}{193}$\\ 

   &           &                        & $1.62_{-0.5}^{+0.6}$ & $6.80_{-0.07}^{+0.08}$ &  $ 210_{-80}^{+90}$ \\ \\
5. &AKN564    & $2.47_{-0.02}^{+0.02}$ &  $8.0_{-2.0}^{+2.0}$ & $6.93_{-0.05}^{+0.03}$&  $\le 20$ & $\frac{214}{199}$\\ \\

6.&MRK110  & $1.75_{-0.01}^{+0.02}$ &  $2.13_{-0.6}^{+0.5}$ & $6.23_{-0.04}^{+0.04}$&  $100_{-30}^{+80} $ &$\frac{238}{197}$\\ \\   

7.& ESO198-G24 & $1.67_{-0.02}^{+0.02}$ &  $1.5_{-0.3}^{+0.4}$ & $6.42_{-0.03}^{+0.04}$& $130_{-50}^{+50}$ & $\frac{229}{199}$\\ 

   &       &                                  &  $4.07_{-1.61}^{+1.71}$ & $7.08_{-0.03}^{+0.05}$ &  $ \le 10$ \\ \\
8.& Fairall 9&  $1.68_{-0.01}^{+0.01}$ &  $14.0\pm 1.0$ & $6.42_{-0.01}^{+0.02}$& $120_{-20}^{+30}$ & $\frac{473}{314}$\\ 

   &       &                                  &  $3.15_{-0.8}^{+0.6}$ & $7.05_{-0.06}^{+0.02}$ &  $ 10$ \\ \\

9.&UGC3973    & $1.04_{-0.02}^{+0.03}$ &  $3.39_{-0.4}^{+0.6}$ & $6.37_{-0.02}^{+0.01}$&  $110_{-30}^{+40} $ &$\frac{277}{187}$\\ \\   

10. &NGC4051 & $1.88_{-0.02}^{+0.04}$ & $1.53_{-0.15}^{+0.15}$ &$6.37_{-0.01}^{+0.03}$ & $\le 20$ &$\frac{219}{194}$ \\

   &       &                                 & $4.5_{-0.51}^{+0.81}$ &$6.62_{-0.60}^{+0.21}$& $85$ &       \\ \\

11.& MCG-2-58-22 &  $1.70_{-0.04}^{+0.03}$ &  $5.5_{-3.0}^{+3.0}$ & $6.41_{-0.21}^{+0.17}$&  $300_{-17}^{+21} $ &$\frac{177}{165}$\\ \\

12.& NGC7469 & $1.87_{-0.02}^{+0.02}$ &  $2.9_{-0.3}^{+0.5}$ & $6.41_{-0.01}^{+0.01}$& $ 60 \pm 20$ & $\frac{260}{200}$\\ 

   &       &                                  &  $1.12_{-0.41}^{+0.31}$ & $7.00_{-0.01}^{+0.02}$ &  $\le 20$ \\ \\

13.&MRK766 & $2.066_{-0.02}^{+0.03}$ &  $4.53_{-1.21}^{+1.31}$ & $6.53_{-0.11}^{+0.07}$&  $380_{-100}^{+150}$ & $\frac{255}{199}$\\ \\

14. &MRK590 &  $1.59_{-0.05}^{+0.04}$ &  $1.14_{-0.41}^{+0.41}$ & $6.41_{-0.04}^{+0.05}$&  $270_{-40}^{+50} $ &$\frac{151}{172}$\\ \\

15.& IRAS050278 & $1.53_{-0.01}^{+0.01}$ &  $3.45_{-0.65}^{+0.55}$ & $6.39_{-0.02}^{+0.03}$&  $ 100_{-40}^{+50}$ &$\frac{201}{199}$\\ \\

16. &NGC3227 & $1.52_{-0.01}^{+0.01}$ &  $3.9_{-0.4}^{+0.6}$ & $6.41_{-0.01}^{+0.02}$&  $ 60\pm 20$ &$\frac{254}{202}$\\ \\

17.& MR2251-178& $1.41_{-0.01}^{+0.03}$ &  $1.25_{-0.50}^{+0.50}$ & $6.38_{-0.02}^{+0.04}$&  $\le 128 $ &$\frac{186}{197}$\\ \\

18.& MRK279  &  $1.80_{-0.01}^{+0.02}$& $3.0_{-0.41}^{+0.31}$ &  $6.43_{-0.02}^{+0.01}$ & $70 \pm 20$ & $\frac{250}{200}$\\ 

   &       &                                  &  $0.73_{-0.21}^{+0.21}$ & $6.99_{-0.03}^{+0.06}$ &  $\le 20$ \\ \\

19.& ARK 120 &  $1.95_{-0.02}^{+0.03}$ &  $4.7_{-0.9}^{+1.2}$ & $6.39_{-0.03}^{+0.03}$& $150_{-50}^{+50}$ & $\frac{339}{199}$\\ 

   &       &                           &  $90_{-10}^{+20}$ & $7_{-0.5}^{+0.01}$ &  $ 1000$ && . \\ \\

\hline \hline
\end{tabular} \\ 
}

\end{table*}

 
\begin{table*}
\centering
{\bf Table \ref{2-10} continued}
\begin{center}
\begin{tabular}{llllllllllllll}
\hline

No. &Source  & $\Gamma$ &Fe K & Fe K                          & Fe K                     &    $\cd$\\ 
     &        &         &norm & Line centroid energy ($\kev$) & Line width $\sigma$($\ev$)&        \\ 

   &         &           & ($10^{-5}$)&           \\ \hline\\ \\

20.& MCG+8-11-11& $1.64_{-0.01}^{+0.01}$ &  $5.5_{-0.7}^{+0.6}$ & $6.40_{-0.01}^{+0.02}$&  $70_{-30}^{+20} $ &$\frac{199}{196}$\\ \\  
 
 21.&MCG-6-30-15    & $1.85_{-0.01}^{+0.01}$ &  $7.6_{-0.8}^{+1.2}$ & $6.02_{-0.05}^{+0.08}$&  $470$ & $\frac{538}{200}$\\ 

   &       &                                &  $1.32_{-0.51}^{+0.31}$ & $6.39_{-0.05}^{+0.04}$ &  $\le 20$ \\ \\
                 
22. &MRK509 & $1.83_{-0.01}^{+0.01}$ &  $3.3_{-0.5}^{+0.45}$ & $6.39_{-0.01}^{+0.04}$&  $\le 20$ & $\frac{378}{199}$\\ 

   &       &                                &  $0.89_{-0.10}^{+0.15}$ & $7.10_{-0.01}^{+0.03}$ &  $\le 20$ \\ \\
23.& NGC 3516   & $0.94_{-0.02}^{+0.02}$ &$0.89_{-0.1}^{+0.2}$ & $6.84_{-0.05}^{+0.03}$ &  $ <100$ & $\frac{930}{197}$ \\ \\

24.& NGC5548 & $1.63_{-0.01}^{+0.03}$ &  $3.2_{-0.6}^{+0.6}$ & $6.39_{-0.03}^{+0.03}$&  $90_{-40}^{+50} $ &$\frac{187}{201}$\\ \\  
                                                   
25.&NGC3783  & $1.47_{-0.01}^{+0.01}$ &  $7.3_{-0.41}^{+0.31}$ & $6.38_{-0.01}^{+0.01}$&  $53 \pm 10$ & $\frac{1071}{314}$\\ 

   &       &                                &  $1.56_{-0.22}^{+0.22}$ & $7.02_{-0.01}^{+0.03}$ &  $\le 20$ \\ \\
                 
26.& IC4329A  & $1.63_{-0.01}^{+0.02}$ &  $5.3_{-0.4}^{+0.8}$ & $6.40_{-0.02}^{+0.01}$& $ \le 50$ & $\frac{276}{200}$\\ 

   &       &                                &  $9.2_{-1.6}^{+0.8}$ & $6.52_{-0.07}^{+0.07}$ &  $ 450 \pm 70$ \\ \\

\hline \hline
\end{tabular} \\ 
\end{center}

\end{table*}


\begin{table*}
{\footnotesize
\centering
\caption{The $0.3-10 \kev$ best fit parameters obtained using the EPIC-pn data.}
  \begin{tabular}{l l l l llllllllll} \hline\hline \label{EPIC-pn}

No. &Source   &  EPIC-pn & $^1$zwabs         &$\Gamma$  &Fe K     & Fe K          & $^3\rm kT_{BB}$ & $^4\rm kT_{BB}$& $^5$Reflection  &  $\cd$\\ 
    &        & counts    & ($\nh$)           &            &centroid energy& $^2$Eqw       &     &        &    (R)         &        \\ 

    &       &($10^5$)    & ($10^{20}\cmsqi$) &          &  ($\kev$)  & ($\ev$)           &($\ev)$   & ($\ev$)   \\ \hline\\ \\

1. &NGC4593   & $2.74$& ---&$1.74_{-0.08}^{+0.07}$  &  ---     & ---  & $86_{-1}^{+4}$ &  $268_{-20}^{+30}$  &  $0.44^{+0.21}_{-0.31}$  & $\frac{231}{241}$ \\ \\   

2.& MRK 704   &  $5.54$ & ---  & $1.86_{-0.03}^{+0.03}$& $6.39\pm 0.07$ & $ 79\pm 12.4$ & $89_{-5}^{+5}$ & $251\pm 20$ & $0.39\pm 0.21 $& $\frac{226}{244}$     \\  \\

3.&ESO511G030 &  $10$ & --- & $2.10 \pm 0.02$ & $--$ & & $53_{-3}^{+2}$ & $ 111_{-15}^{+13}$&$2.25_{-0.22}^{+0.32}$ & $\frac{268}{254}$    \\ \\

4.& NGC 7213  &  $5.06$ & --- & $1.69\pm 0.03$   & $6.80_{-0.08}^{+0.07}$  & $  113\pm 42.3$ &  $62_{-8}^{+8}$ & $177_{-8}^{+9}$  & $0.34_{-0.22}^{+0.12}$ &$\frac{284}{242}$ \\ \\

5. &AKN564    & $27.1$  & $<0.4$  &$2.49 \pm 0.02$     & $6.93\pm 0.04$            & $64\pm16$     & $66\pm 2$ & $149\pm 5$     & $0.15_{-0.13}^{+0.08}$& $\frac{317}{244}$        \\ \\

6.&MRK110  & $7.0$ & ---  & $1.92\pm 0.03$              & $6.23\pm 0.04$           & $70\pm 16.4$  &$ 62\pm 5$ & $140\pm 8$ & $1.07_{-0.31}^{+0.32}$  &$\frac{317}{253}$ \\ \\

7.& ESO198-G24 & $4.05$  & ---  & $1.79\pm 0.02$  & ---& ---& $149_{-5}^{+7}$& ---  & ${1.13\pm 0.21}$ & $\frac{265}{252}$   \\ \\

8.& Fairall 9 & $10$ & --- & $1.92\pm 0.04$ & $6.42\pm 0.02$ & $138\pm10$ & $74_{-3}^{+3}$ &$164\pm 9$  & $4_{+1}^{-3}$& $\frac{311}{255}$ \\ 

   &          &           &               &                    & $6.97_{-0.06}^{+0.02}$  & $48\pm 9.1$ \\ \\

9.&UGC3973    &$1.11$  &---  &  $1.15\pm 0.04$   & ---& & $102\pm 5$ & $225\pm 10$ & $0.4_{-0.4}^{+0.3}$ &$\frac{290}{236}$  \\ \\

10. &NGC4051& 18 & ---& $2.35_{-0.01}^{+0.04}$   & $---$ & & $101_{-3}^{+3}$ &---& $\sim 5$ & $\frac{315}{250}$   \\ \\

11.& MCG-2-58-22& $1.0$& ---  &  $1.76_{-0.04}^{ +0.04}$& $6.41\pm0.22$ & $ 40\pm24$ & $111_{-6}^{+7}$ & --- & $<0.46$ & $\frac{254}{225}$\\ \\  

12.& NGC7469 & $12$  & ---             & $1.99_{-0.07}^{+0.07}$   &$6.97\pm 0.02$  &  $33\pm8.8$ & $94_{-4}^{+1}$ & $234_{-19}^{+10}$ &  $1.17_{-0.41}^{+0.42}$ & $\frac{324}{255}$ \\ \\

13.&MRK766  & $18.9$  & ---&$ 2.08_{-0.01}^{+0.02}$ &  $6.58\pm0.07$     & $166\pm 47.6$&  $71_{-4}^{+3}$ &  $257_{-6}^{+6}$ & $<0.2$  & $\frac{305}{244}$ \\ \\

14. &MRK590  & $0.9$  & ---&$1.76_{-0.03}^{+0.03}$  &  ---    & --- & $13_{-1}^{+1}$ & ---              &  $1.65_{-0.52}^{+0.52}$  & $\frac{231}{231}$ \\ \\   

15.& IRAS050278 & $2.5$ & $21.9_{-0.8}^{+0.7}$   &   $1.77\pm 0.02$ &  $6.39\pm0.02$  & $64\pm 10.2$   & $67 \pm 5$ &  ---&$4.5\pm1.2$  & $\frac{246}{254}$\\ 
   & +1626      &    \\ \\

16. & NGC3227   & $11.5$&  $10.0_{-0.9}^{+1}$  & $1.57_{-0.07}^{+0.03}$     & --- &      & $55_{-2}^{+2}$  & --- & $<0.08$ & $\frac{360}{248}$ \\ \\

17.& MR2251-178 & $2.5$  &  ---          &  $1.43_{-0.03}^{+0.02}$ & $6.37\pm0.03$ & $34\pm13.6$   & --- & --- & $<0.1$ & $\frac{295}{245}$  \\ \\

18.& MRK279  &  $26$  & --- & $1.93\pm 0.02$  & $6.5\pm0.02$  & $42\pm4.2$  &  $84 \pm 1$ & $222_{-7}^{+6}$& $3_{-0.9}^{+1.5}$ & $\frac{276}{256}$   \\ 
        &                            &   &     &           & $6.95\pm 0.02$   & $15 \pm5.1$  &     \\ \\

19.& ARK 120 &   $17$ & --- &  $ 1.97_{-0.03}^{+0.02}$  & ---& &  $102\pm 2$  & $240\pm 4$ &  $0.75_{-0.11}^{+0.22}$& $\frac{314}{253}$   \\ \\

\hline \hline
\end{tabular} \\ 
}

{\begin{flushleft}

$^{1}${The intrinsic neutral absorption by the host galaxy}\\
$^2$The equivalent width of the Fe K emission line.\\
$^{3\, \&\, 4}$ The best fit blackbody temperatures.\\
$^5$ The Neutral reflection coefficient as defined in section \ref{sect_epic}.\\

\end{flushleft}
}

\end{table*}


\begin{table*}
\centering
{\bf Table \ref{EPIC-pn} continued}
\begin{center}
\begin{tabular}{llllllllllllll}
\hline

No. &Source   &  EPIC-pn & $^1$zwabs         &$\Gamma$  &Fe K     & Fe K          & $^3\rm kT_{BB}$ & $^4\rm kT_{BB}$& $^5$Reflection  &  $\cd$\\ 
    &        & counts    & ($\nh$)           &            &centroid energy& $^2$Eqw       &     &        &    (R)         &        \\ 

    &       &($10^5$)    & ($10^{20}\cmsqi$) &          &  ($\kev$)  & ($\ev$)           &($\ev)$   & ($\ev$)   \\ \hline\\ \\

20.& MCG+8-11-11 & $2.91$ &  $18.0_{-1.0}^{+0.7}$ & $1.68\pm 0.03$ & $6.99\pm 0.009$ & $32\pm 3$&  $268_{-16}^{+19}$  &---&  $<0.64$ & $\frac{240}{249}$ \\ \\

21.&MCG-6-30-15& $ 17.9$& $2.2\pm 2.1$ &$2.0\pm 0.02$       & $7.05\pm 0.05$    & $30\pm6.8$    & $89\pm 2$&    $277_{-11}^{+9}$   & $<5$   & $\frac{365}{237}$               \\ 
   &           &        &             &            &       & \\ \\

22. &MRK509 &  $15.2$& ---&  $ 2.13_{-0.02}^{+0.03}$ & --- & & $79_{-2}^{+2}$&   $161_{-7}^{+9}$  & $2.3_{-0.3}^{+0.5}$ & $\frac{342}{253}$\\  \\

23.& NGC 3516         &   & --- & $1.00_{-0.06}^{+0.07}$& $6.4\pm 0.005$&  $185\pm 10$      & $46\pm 12$ & --- & $<0.1$&  $\frac{302}{247}$ \\
   &           &                      &  &                         &  \\ \\

24.& NGC5548   & $11.6$ & ---            & $1.65\pm 0.02$   & ---  & --- & $87_{-1}^{+2}$ & $313_{-7}^{+16}$     & $<0.03$& $\frac{261}{245}$  \\ \\                                                           

25.&NGC3783 & $20$& --& $1.80 \pm 0.04$ & --- & &  $102_{-1}^{+3}$ & --- & $1.0_{-0.2}^{+0.3}$ & $\frac{433}{242}$ \\      &        &     &   &                 &  $6.3\pm 0.02$   &  $27\pm3.6$            &   \\ 

     &        &     &   &                 &  $6.98\pm 0.03$  & $20\pm 4$ \\ \\

26.& IC4329A & $23.3$  &$38_{-1}^{+1}$  & $ 1.80\pm 0.04$  & $6.87\pm 0.02$   & $32\pm4.8$  & $46_{-1}^{+4}$  & $286\pm 10$  & $1.67_{-0.12}^{+0.05}$ & $\frac{298}{247}$   \\ 

    &&&&& &  \\ \\

\hline \hline
\end{tabular} \\ 
\end{center}

{\begin{flushleft}

$^{1}${The intrinsic neutral absorption by the host galaxy}\\
$^2$The equivalent width of the Fe K emission line.\\
$^{3\, \&\, 4}$ The best fit blackbody temperatures.\\
$^5$ The Neutral reflection coefficient as defined in section \ref{sect_epic}.\\

\end{flushleft}
}


\end{table*}


\begin{table*}
{\footnotesize
\centering
\caption{The WA and narrow warm emission (WE) line parameters from combined fit of the EPIC-pn and the RGS data.} \label{RGS-table}

  \begin{tabular}{l l l l llllll} \hline\hline 

No.& Source  &   RGS              &  WE line         &  WE line         & WA                     & WA-$\rm log(\nh)$&WA-velocity &  $\Delta C$\\  
   &         &  Counts   &  centroid Energy & norm             & $\rm log(\xi/\xiunit)$ &                  & &            \\
   &         & ($10^3$)   &   ($\kev$)       &  ($10^{-4})$     &                        &                   & $\kms$  \\\hline \\

1.& NGC4593 &    $8.714$ & $0.652\pm 0.01$ &  $8.64_{-1.74}^{+1.22}$  &   $2.37_{0.10}^{+0.82}$ &  $20.93_{-0.11}^{+0.21}$  & $-510_{-30}^{+180}$ & -23  \\
  &         &               &               &                        & $3.24_{-0.10}^{+0.11}$  & $21.55^{+0.03}_{-0.09}$  &  $300_{-300}^{+300}$ & -42  \\ \\

2. & MRK 704  & $ 28.39 $  &$0.569\pm 0.003$  & $1.6_{-0.11}^{+0.05}$ & $ 0.25_{-0.51}^{+0.41}$ &  $20.3_{-0.12}^{+0.81}$  &  $-1500_{-300}^{+240}$ & -55\\  
    &         &            &         &                      & $ 2.31_{-0.15}^{+0.15}$  & $21.34_{-0.11}^{+0.11}$  & $-660_{-30}^{+30}$ & -45\\ 
   &          &            &         &                      &$3.13_{-0.51}^{+0.41}$ &  $21.33_{-0.11}^{+0.12}$ & $ -3540_{-90}^{+150}$  &  -53 \\ \\ 

	  3.&ESO511~G030 & $ 62.45$   &$0.547\pm 0.002$  & $0.39_{-0.13}^{+0.17}$             & {$ \ge 3.9$} &  { $<20$} &--- & ---    \\ 
  &           &            &$0.560\pm 0.002$  & $0.63_{-2.0}^{+2.4}$ &                        \\

  &           &            &$0.572\pm 0.003$ & $0.87_{-0.27}^{+0.13}$ & \\ \\

4.& NGC 7213 &   $30.91$     &  --- &   ---   &  { $<-0.75$} & { $ <20$} & --- & ---   \\ \\

5.& AKN 564  &  $161$      &$0.468\pm 0.001$        &$<0.41$        &    $-0.20_{-0.12}^{+0.21}$  & $20.31_{-0.21}^{+0.21}$  &$ -690_{-90}^{+150}$ & -163\\ 
  &          &             &$0.556\pm 0.001$        &  $2.05_{-0.40}^{+0.40}$           &     \\ 

  &          &             &$0.672\pm 0.003$        & $1.4_{-0.1}^{+0.3}$ & \\ \\

	  6.& MRK 110 &  $40.6$      & $0.558\pm 0.004$ & $3.73_{-0.61}^{+0.12}$ &{$>3.5$}  & {$<20.2$}  &  --- & ---    \\ \\

	  7.& ESO198-G24 & $27$      & $0.523\pm 0.003$ & 0 & { $1.2< \log\xi <1.93 $}  & { $<20$}  &  --- & ---   \\  \\

	  8.& Fairall 9& $41$        & $0.945_{-0.03}^{+0.028}$ & $3.07_{-1.31}^{+1.22}$ &{ $-1< \log\xi <4 $}  &  { $18.8< \log\nh <22.7 $}  &  --- & ---   \\ \\

9.&UGC3973   &  $7.44$& $0.579\pm 0.002$  & $1.7_{-0.21}^{+0.21}$ & $2.04_{-0.23}^{+0.57}$ & $21.29_{-0.06}^{+0.14}$ & $<-2400$ & -93\\ 
   &          &     &       \\ \\

10.& NGC 4051 &  $111.3$& $0.561_{-0.002}^{+0.001}$ & $1.49\pm0.33$ &    $ 0.28_{-0.12}^{+0.12}$ &$20.43_{-0.21}^{+0.04}$    &  $-600_{-30}^{+30}$ & -161 \\

   &         & &   $0.597_{-0.001}^{+0.001}$  &  $1.04_{-0.04}^{+0.5}$ &  $2.87_{-0.21}^{+0.12} $   &   $22.39_{-0.22}^{+0.22} $ &  $-688_{-30}^{+30}$ & -458\\ \\

	  11.& MCG-2-58-22 & $ 7.40$& ---  &  --- &{ $-1< \log\xi <4 $}  &  { $18.8< \log\nh <22.7 $}  &  ---&---  \\ \\

12.& NGC 7469 &    $ 78.69$  & $0.538\pm 0.2$   & $ 0.45_{-0.36}^{+0.21}$ &  $2.8_{-0.1}^{+0.2}$    &  $20.96_{-0.15}^{+0.21}$& $-1590_{-90}^{+90}$ & -117\\ 
   &          &              &$0.554 \pm 0.0001$ & $0.33_{-0.11}^{+0.32}$  \\
   &          &              &$0.672 \pm 0.003$  &  $ 0.3_{-0.11}^{+0.12}$ \\ \\

13.& MRK 766 &   $120$     & $0.538_{-0.002}^{+0.006}$    &  $ 10\pm 0.2$     &   $1.35_{-0.21}^{+0.22}$ & $20.53_{-0.05}^{+0.05}$ & $ -810_{-60}^{+60}$ & -256\\
  &          &              &$0.655\pm 0.001$             & $13.8_{-0.5}^{+0.6}$ &   $-0.94_{-0.12}^{+0.15}$ & $20.46_{-0.08}^{+0.07} $ &$-1020_{-30}^{+90}$  & -22  \\ 

&            &              & $0.847\pm 0.003$            &  $0.44_{-0.07}^{+0.10}$ & $-0.70_{-0.22}^{+0.15}$ & $20.6_{-0.2}^{+0.2} $ & $0_{-120}^{+150} $  & -120  \\ 

&            &              &$0.965\pm 0.001$             & $1.5_{-0.2}^{+0.1}$                            \\ \\

	  14.&  MRK 590   & $ 18.8$ & $0.556\pm 0.0001$ & $<0.12$ &  { $-1< \log\xi <4 $}  &  { $18.8< \log\nh <22.7 $}  &     --- & --- \\ \\

15.& IRAS050278&  $10.8$      &  $0.534\pm 0.0007$  & $2.8_{-1.8}^{+1.7}$ &  $-0.47_{-0.21}^{+0.21}$  & $21.16_{-0.15}^{+0.12}$  & $-900_{-30}^{+600}$ &-17  \\
     &         &               &  $0.561\pm 0.004$ & $4.2_{-1.0}^{+1.3}$ &  $2.18_{-1.01}^{+0.52} $  & $20.66_{-0.31}^{+0.22}$ & $300_{-300}^{+30}$ & -18\\ \\

16.& NGC 3227  &  $41.8$ &$0.564\pm 0.001$      & $3.1_{-0.5}^{+0.6}$ &   $ 0.04_{-0.08}^{+0.08}$ &  $20.45_{-0.07}^{+0.09}$  &  $ <-1290$ & -240\\ 

  &           &     & $0.584\pm 0.001$      & $ 0.51\pm 0.05$     &   $ 2.97_{-0.08}^{+0.12}$ &  $21.50_{-0.08}^{+0.11}$ &  $ <-2460$ & -32\\

  &           &     &$0.670\pm 0.003$    & $1.8_{-0.2}^{+0.1}$& $2.24_{-0.07}^{+0.07} $  &  $21.08_{-0.22}^{+0.22}$   &   $<-1377$  & -161\\ \\

17.& MR2251-178 &  $17.29$ &$0.564\pm 0.001$ & $3.02_{-0.5}^{+1}$  &  $1.60_{-0.02}^{+0.03}$ & $20.96_{-0.03}^{+0.41}$  &  $ -3150_{-30}^{+60}$ & -12\\ 

   &            &  &  &      &  $2.92_{-0.12}^{+0.51} $ & $21.70_{-0.05}^{+0.21}$  & $-3090_{-1200}^{+210} $ & -185
 \\ \\

18.& MRK 279  &   $ 55.3$  &---         & ---        &{ $ \ge 3.9$} & { $ <20.0 $} &  ---& ---    \\ \\

 19.& ARK 120 &    $123$   & $0.554\pm 0.001$ & $2.33_{-0.41}^{+0.42}$ & { $ \ge 3.5$} &  { $<20.0 $} & --- & ---  \\

\hline \hline
\end{tabular} \\ 
}

\end{table*}


\begin{table*}
\centering
{\bf Table 6 continued}
\begin{center}
\begin{tabular}{llllllllllllll}
\hline
No.& Source  &   RGS              &  WE line         &  WE line         & WA                     & WA-$\rm log(\nh)$&WA-velocity &  $\Delta C$\\  
   &         &  Counts   &  centroid Energy & norm             & $\rm log(\xi/\xiunit)$ &                  & &            \\
   &         & ($10^3$)   &   ($\kev$)       &  ($10^{-4})$     &                        &                   & $\kms$  \\\hline \\

20.&  MCG-8+11+11&  $18.7$ &$0.547\pm 0.002$  & $7.9_{-0.6}^{+0.6}$ &   $ 3.18_{-0.12}^{+0.12}$  &  $21.17_{-0.21}^{+0.22}$  & $-2340_{-30000}^{+900}$& -23 \\ 
   &    &            &$0.729\pm 0.001$ & $0.19\pm 0.03$ &    \\ \\

21.& MCG-6-30-15&  $ 132.1$  &$0.593_{-0.002}^{+0.001}$       & $13.0\pm 2.0$         &    $-0.36_{0.05}^{+0.05}$ & $20.55_{-0.15}^{+0.15}$  & $ -870_{-30}^{+150}$  & -116  \\
  &            &    &$0.654\pm 0.001$       & $2.31\pm 0.2$            &    $1.25_{-0.15}^{+0.15}$ & $ 20.82\pm 0.03$ &  $-450_{-30}^{+90}$ & -187\\
 &            &     &$0.677\pm 0.002$ &  $2.4_{-0.25}^{+0.4}$      &    $2.53_{-0.05}^{+0.03}$ & $ 21.17\pm 0.05$ &  $-2370_{-60}^{+60}$ & -162\\ \\

22.& MRK 509 &   $98$ &$0.562\pm 0.001$         & $2.6\pm 0.2$     &    $3.24_{-0.21}^{+0.05}$ & $20.8_{-0.14}^{+0.15}$  & $-6500_{-120}^{+150}$ & -181    \\ 
   &         &        & $0.598\pm 0.001$        & $0.23\pm 0.08$\\ \\

23.& NGC 3516  & $16.69$&  $0.556\pm 0.001$  & $0.41\pm 0.01$  & $2.69_{-0.14}^{+0.12}$ & $21.37_{-0.05}^{+0.12}$ &$-2280_{-90}^{+252}$ & -230\\ 
   &           &   &$0.572\pm 0.002$  &  $0.39\pm 0.01$  \\
   &           &   &$0.884$  &  $<0.04$    \\ \\

24.& NGC 5548 &   $ 89.2$&$0.562\pm 0.0001$  & $1.1\pm 0.2$  &  $2.92_{-0.07}^{+0.03}$ & $21.53_{-0.05}^{+0.11}$  & $-3609_{-270}^{+180} $ &-347      \\ 
   &          &    &$0.664\pm 0.002$&   $1.6_{-0.3}^{+0.2}$  &  $1.86_{-0.05}^{+0.08} $ & $ 21.13_{-0.07}^{+0.07}$ & $-1236_{-30}^{+42}$  &-836       \\ 
   &          &     &$0.769\pm 0.001$  & $<0.05$\\

\\

25.& NGC 3783  &  $65.2$ &$0.566\pm 0.001$        & $6.2\pm 1.2$           &  $ 1.55_{-0.15}^{+0.15}$ & $21.69_{-0.02}^{+0.03}$ & $-1650_{-60}^{+90}$ & $>1000$\\ 
  &     &        & $0.824\pm 0.001$      &  $0.6\pm 0.02$     &  $ 2.91_{-0.02}^{+0.02}$ & $22.18_{-0.08}^{+0.10}$ &  $-1635_{-60}^{+120}$ & $>1000$\\   \\

26.& IC4329A  &  $ 92.5$&$0.528\pm 0.003$  & $2.5\pm 0.8$ & $-0.58_{-0.11}^{+0.22}$  & $20.96_{-0.31}^{+0.21}$   & $ -1020_{-120}^{+150}$ & -51\\ 
   &          &   &$0.649\pm 0.005$  &  $0.17\pm 0.05$ & $1.87_{-0.21}^{+0.22}$  & $20.54_{-0.08}^{+0.05}$  &$-660_{-120}^{+120}$ & -46\\
   &          &   &        &          & $ 3.33_{-0.03}^{+0.07}$  & $21.27_{-0.32}^{+0.22}$   & $-990_{-30}^{+120}$   & -235\\ \\

\hline \hline
\end{tabular} \\ 
\end{center}


\end{table*}


\begin{table*}
{\footnotesize
\centering
  \caption{The diskline parameters for the detected broad Fe K$\alpha$ emission lines. \label{diskline}}
  \begin{tabular}{l l l l llllll} \hline\hline 

 Source      & Profile   & Line centroid       & Eqw       &$^1$Betor10            &$^1$ Index     & Inclination \\  
             & (model)&   ($\kev$)             & $\ev$     & (diskline)            & (Laor)    & $\theta^{\circ}$           \\\hline \\

ESO~511-G030 & Laor      &$6.86_{-0.32}^{+0.11}$    &$294\pm31$    &---                   & $6.86_{-1.21}^{+1.22}$  & $59_{-10}^{+10}$\\ \\

UGC~3973     & Diskline  & $6.5_{-0.2}^{+0.2}$      &$434\pm43$  & $-3.97_{-0.51}^{+0.71}$   &---                     & $19_{-6}^{+6}$ \\ \\
NGC~4051     & Laor      &$7_{-0.20}^{+0.01}$       & $461\pm52$ & ---                  & $7.6\pm 0.3$           &  $<41$ \\ \\
NGC~3227     & Diskline  &$6.60_{-0.11}^{+0.05}$    & $80\pm 11 $    & $-1.36_{-0.22}^{+0.32}$ & ---                    &  $<5$ \\ \\
ARK~120      & Diskline  &$6.9_{-0.10}^{+0.06}$     &$92\pm 10$   & $-3.8^{+0.4}_{-0.4}$    &---                     &  $<50$\\ \\
MCG-6-30-15  & Diskline  & $6.43_{-0.1}^{+0.1}$     & $123 \pm 20$   & $-2.44^{+0.23}_{-0.3}$  &---                     & $<10$\\ \\
MRK~509      & Laor      &$6.63_{-0.05}^{+0.05}$    &$310 \pm 25$  & ---                     &$4.53\pm0.32$           & $48_{-20}^{+1}$ \\ \\

NGC~3516     &Laor       & $6.4_{-0.1}^{+0.1}$      &$1560\pm 86$   &---                      &$4.2_{-0.2}^{+0.2}$     & $50_{-5}^{+2}$\\ \\

NGC~3783     & Laor      &$6.4_{-0.1}^{+0.1}$       &$110 \pm 12$    & ---                  & $3.07_{-0.22}^{+0.21}$& $<11$  \\ \\
IC~4329A     & Diskline  &$6.30_{-0.05}^{+0.15}$    &$125 \pm 24$  & $-2.19_{-0.21}^{+0.22}$                & ---                   & $30_{-5}^{+10}$ \\ \\

\hline \hline
\end{tabular} \\ 
}

$^1$ The emissivity profiles for the broad emission lines.

\end{table*}


\begin{table*}
{\footnotesize
\centering
  \caption{The fluxes and luminosities of the sources obtained from the broad band spectral analysis.}. \label{Luminosity}

  \begin{tabular}{l l l l llllll} \hline\hline 

No.& Source &  $\rm F_{2\kev} $ & $\alpha_{OX}$ & $ \rm L_{0.3-10\kev}$&$^1\rm L_{ion}$ & $\rm L_{ion}/L_{Edd}$ &    \\ \\
   &        &  $(10^{-12})\funit\, \kev^{-1}$ &              & $ \lunit$            & $\lunit$      &   \\  \hline\\ \\

1&NGC4593    &   $11.5\pm0.01$  &  $0.989\pm 0.01$ & $4.34_{-0.06}^{+0.06}\times 10^{42}$ & $1.37\times 10^{44}$ & $0.180$ \\ \\

2&MRK704     &   $2.94\pm 0.02$   & $1.64 \pm 0.05$ &$4.66_{-0.04}^{+0.04}\times 10^{43}$  & $2.18\times 10^{44}$ &  $0.040$\\    \\

3&ESO511-G030 &  $5.55\pm 0.04$   & $1.22\pm 0.01$ &$5.08_{-0.02}^{+0.03}\times 10^{43}$ & $1.23\times 10^{44}$ & $0.003$ \\ \\

4&NGC7213     & $61.5 \pm0.11$   & $0.726\pm 0.007 $&$1.14_{-0.005}^{+0.005}\times 10^{42}$  & $5.74\times 10^{42}$ &  $ 0.0002$\\ \\

5&AKN564     &  $7.6\pm 0.03$    & $1.126\pm 0.011$ &$1.12_{-0.03}^{+0.03}\times 10^{44}$ & $7.92\times 10^{44}$ & $1.21^{\rm 2}$\\ \\

6&MRK110     &  $7.65\pm 0.08$   & $1.196\pm 0.011$ &$1.66_{-0.03}^{+0.03}\times 10^{44}$ & $2.37\times 10^{45}$ & $ 0.72$\\ \\

7&ESO198-G24 &  $2.61\pm 0.03$   & $1.166\pm 0.013$ &$8.21_{-0.04}^{+0.04}\times 10^{43}$ & $4.21\times 10^{44}$ & $0.025$\\ \\

8&Fairall 9  &  $69.5\pm 0.1$   & $0.883\pm 0.01$ &$2.53_{-0.009}^{+0.009}\times 10^{44}$ & $5.68\times 10^{45}$ & $0.109$ \\ \\

9&UGC 3973   &  $1.18\pm 0.08$   & $1.424\pm 0.012$ &$1.43_{-0.02}^{+0.02}\times 10^{43}$& $2.368\times 10^{44}$ &  $0.0144$\\ \\

10& NGC 4051  & $7.55\pm 0.02$   & $1.115\pm 0.011$ &$1.31_{-0.005}^{+0.005}\times 10^{42}$ & $2.20\times 10^{43}$ & $0.085$\\ \\

11&MCG-2-58-22 & $7.6\pm 0.02$    &$1.09 \pm 0.009$ &$2.43_{-0.008}^{+0.008}\times 10^{44}$ & $1.13\times 10^{45}$ & $0.381$\\ \\

12 &NGC 7469  & $8.75\pm 0.33$   & $0.91\pm  0.009$  &$3.499_{-0.01}^{+0.02}\times 10^{43}$& $1.06\times 10^{45}$ & $0.649$ \\ \\

13&MRK766     &  $7.9\pm 0.04$    &$0.32\pm 0.002$  &$2.80_{-0.007}^{+0.007}\times 10^{43}$ & $4.96\times 10^{43}$ & $0.012$\\ \\

14&MRK 590    &  $1.52\pm0.1$   &$1.03\pm 0.009$ &$1.54_{-0.04}^{+0.04}\times 10^{43}$ & $4.15\times 10^{43}$ & $0.0008$\\ \\

15&IRAS05078  &  $5.925\pm 0.03$  & $0.853\pm 0.008$ &$3.03_{-0.03}^{+0.03}\times 10^{43}$ & $7.46\times 10^{43}$ & $0.008$\\ \\

16&NGC 3227   &  $7.64\pm 0.03$   & $1.13\pm 0.011 $ &$3.23_{-0.01}^{+0.01}\times 10^{42}$ & $1.915\times 10^{43}$ & $0.005$\\ \\

17&MR2251-178 &  $4.15\pm 0.04$   & $1.14\pm 0.011$ &$2.65_{-0.01}^{+0.01}\times 10^{44}$ & $1.40\times 10^{45}$ &  $0.034$\\ \\

18&MRK 279   &   $7.5\pm 0.03$    & $1.192\pm 0.012$ &$1.15_{-0.006}^{+0.006}\times 10^{44}$ & $1.54\times 10^{45}$ & $0.37$ \\ \\

19&ARK 120    &  $11.85\pm 0.13$  & $1.322\pm 0.011$ &$2.14_{-0.002}^{+0.002}\times 10^{44}$ & $3.489\times 10^{45}$ & $0.107$ \\ \\

20&MCG+8-11-11 & $10.9\pm 0.04$   & $1.213\pm 0.010$ &$6.63_{-0.008}^{+0.008}\times 10^{43}$ & $3.23\times 10^{44}$ & $0.008$\\ \\

21&MCG-6-30-15 &  $14.05\pm 0.04$  & $0.44\pm 0.005$& $1.739_{-0.03}^{+0.03}\times 10^{43}$ & $1.411\times 10^{43}$ & $0.011$\\ \\

22&MRK 509    &  $10.42\pm 0.03$  & $1.27 \pm 0.011$&$2.06_{-0.007}^{+0.007}\times 10^{44}$ & $2.42\times 10^{45}$ & $0.093$ \\ \\

23&NGC 3516   &  $1.75\pm 0.02$   & $1.54 \pm 0.013$ &$3.32_{-0.06}^{+0.06}\times 10^{42}$ & $2.97\times 10^{43}$ & $0.004$\\ \\

24&NGC 5548   &  $9.74\pm 0.02$   & $1.30\pm 0.02$ &$4.87_{-0.002}^{+0.002}\times 10^{43}$ & $2.10\times 10^{44}$ & $0.013$\\ \\

25&NGC 3783   &  $14.685\pm 0.12$ & $1.22\pm 0.012$ &$2.74_{-0.05}^{+0.05}\times 10^{43}$ & $3.45\times 10^{44}$ & $0.042$ \\ \\

26&IC 4329A   &  $25.5\pm 0.03$   & $0.673\pm 0.007$ &$1.837_{-0.005}^{+0.005}\times 10^{44}$ & $1.88\times 10^{45}$ & $0.091$  \\ \\

\hline \hline
\end{tabular} \\ 
}
\begin{flushleft}
$^1$ The ionising luminosity is calculated over the energy range of $13.6 \ev-30 \kev$.\\

$^{ 2}${This super Eddington rate for AKN~564 has also been seen by other authors like \cite{2012ApJ...753...75C, 2012ApJ...745..107W}}

\end{flushleft}

\end{table*}


\begin{table*}
{\footnotesize
\centering
  \caption{The Spearman rank correlations for WA parameters. The first quantity in the bracket is the Spearman correlation coefficient while the second term is the correlation probability.\label{corr}}
  \begin{tabular}{llllllllllll} \hline\hline 

 Quantity & WA-$\log\xi$& WA-$\log\nh$ & WA-velocity    \\ \hline \\ 

WA-$\log\xi$ & 1 &  (0.64,{\bf $>99.99\%$})  & (0.33,$94\%$)   \\ \\

WA-$\log\nh$ & (0.64,{\bf $>99.99\%$})  & 1 & (0.36,{\bf $96\%$})   \\ \\ 

WA-velocity & (0.33,{\bf $94\%$})         &  (0.36,{\bf $96\%$}) & 1    \\ \\

 \hline \hline
\end{tabular} \\

}

\end{table*}


\begin{table*}

{\footnotesize
\centering
  \caption{The linear regression analysis for warm absorber parameters ($y=a\,x+b$). \label{Table:lin-reg}}
  \begin{tabular}{llllllllllll} \hline\hline 

 $x$ & $y$ & $a$ &  \hspace{1cm}Dev$(a)$ & \hspace{1cm} $b$ & \hspace{1cm} Dev$(b)$ & \hspace{1cm}$R_S$ &\hspace{1cm} $P_{null}$\\ \hline \\ 

$\log\xi$ &$\log\nh$        &  $0.31$ &\hspace{1cm} $0.06$ & \hspace{1cm}$20.46$  & \hspace{1cm}$0.11$ & \hspace{1cm}$0.64$ & \hspace{1cm} $>99\%$\\ \\
$\log\xi$ & $\log v_{out}$  & $0.12$ &\hspace{1cm} $0.03$ & \hspace{1cm}$2.97$  & \hspace{1cm}$0.05$ & \hspace{1cm}$0.33$ & \hspace{1cm} $>93\%$ \\ \\
$\log\nh$ & $\log v_{out}$  & $0.8$ &\hspace{1cm} $0.7$ & \hspace{1cm}$-13$  & \hspace{1cm}$18$ & \hspace{1cm}$0.36$ & \hspace{1cm} $>96\%$ \\ \\

 \hline \hline
\end{tabular} \\ 

\footnotetext{1}{ $R_S$ stands for the Spearman rank correlation coefficient.}

}
\end{table*}


\begin{table*}

{\footnotesize
\centering
\caption{The Spearman rank correlations for chosen subsets of WA parameters and the continuum parameters. The first quantity in the bracket is the Spearman correlation coefficient while the second term is the correlation probability. See section \ref{subsec:corr-analysis} for details. \label{Tab:corr-high-and-low}}
  \begin{tabular}{llllllllllll} \hline\hline 

 Quantity           & MBH            & $\rm L_{Xray}$  &$\rm L_{ion}$  & $\Gamma$    & $\alpha_{OX}$   \\ \hline \\ 

WA-highest$\log\xi$ & (0.05,0.84)    &  (0.40,0.11)    & (0.23,0.37)   & (0.04,0.87) & (-0.01,0.97)  \\ \\
WA-lowest$\log\xi$  & (-0.34,0.174)  &  (-0.35,0.15)   & (-0.48,0.05)  &(0.10,0.68)  & (0.11,0.66)  \\ \\

WA-highest$\log\nh$ & (0.10,0.68)    & (0.40,0.10)     & (0.43,0.08)   &(0.16,0.53)  & (0.018,0.94)    &    \\ \\ 
WA-lowest$\log\nh$  & (-0.03,0.89)   & (-0.19,0.45)    & (-0.32,0.19)  &(0.09,0.71)  & (-0.11,0.65)      &    \\ \\ 

WA-highest-velocity & (0.25,0.33)    & (0.23,0.36)     & (0.16,0.53)   &(-0.14,0.56) & (0.14,0.58)   \\ \\
WA-lowest-velocity  & (0.29,0.25)    & (0.036,0.88)    & (-0.08,0.75)  &(-0.29,0.24) & (-0.04,0.87)   \\ \\

 \hline \hline
\end{tabular} \\

}
\end{table*}





\begin{figure*}
  \centering
  \hbox{
\includegraphics[width=7cm,angle=0]{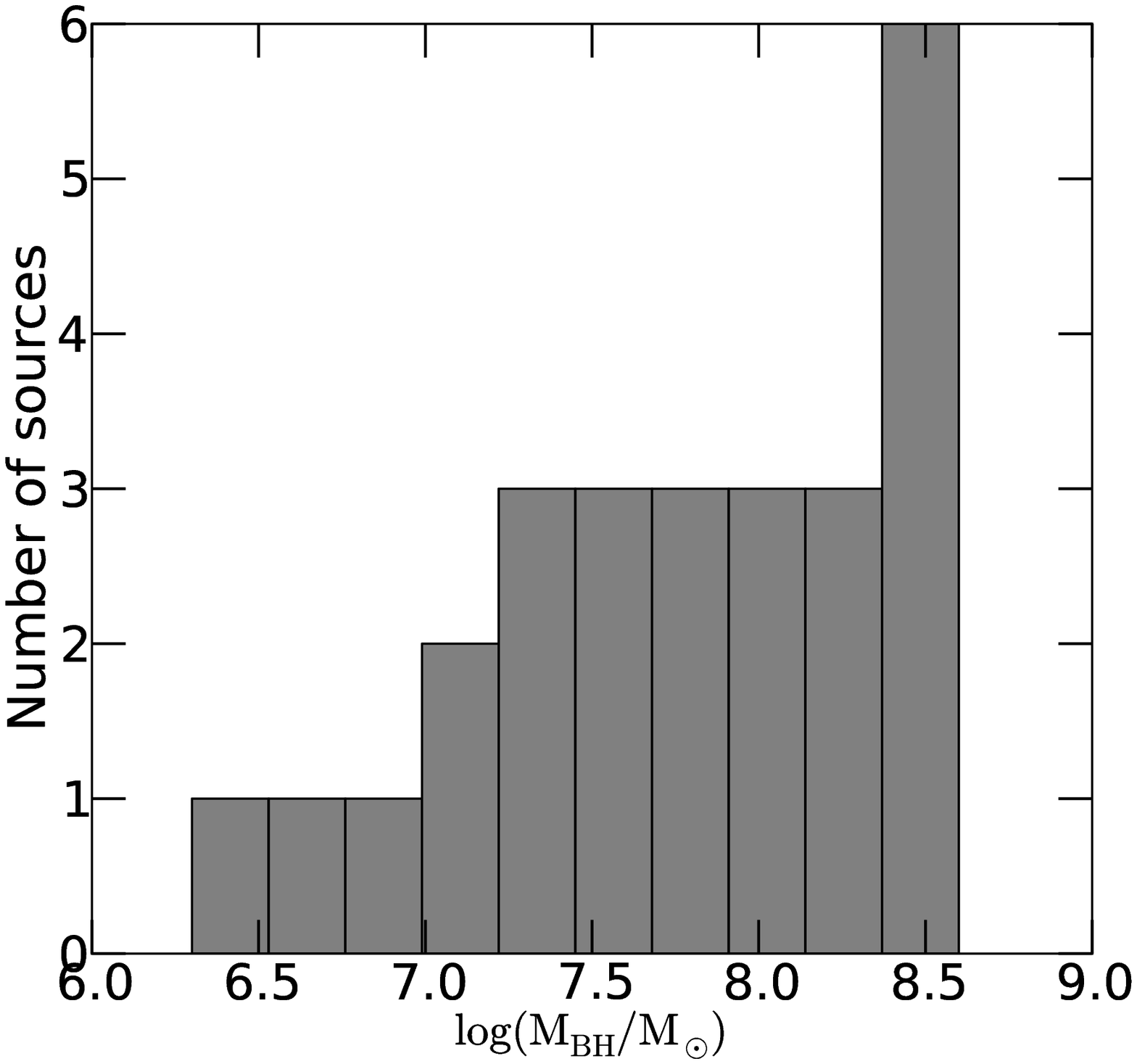} 
\includegraphics[width=7cm,angle=0]{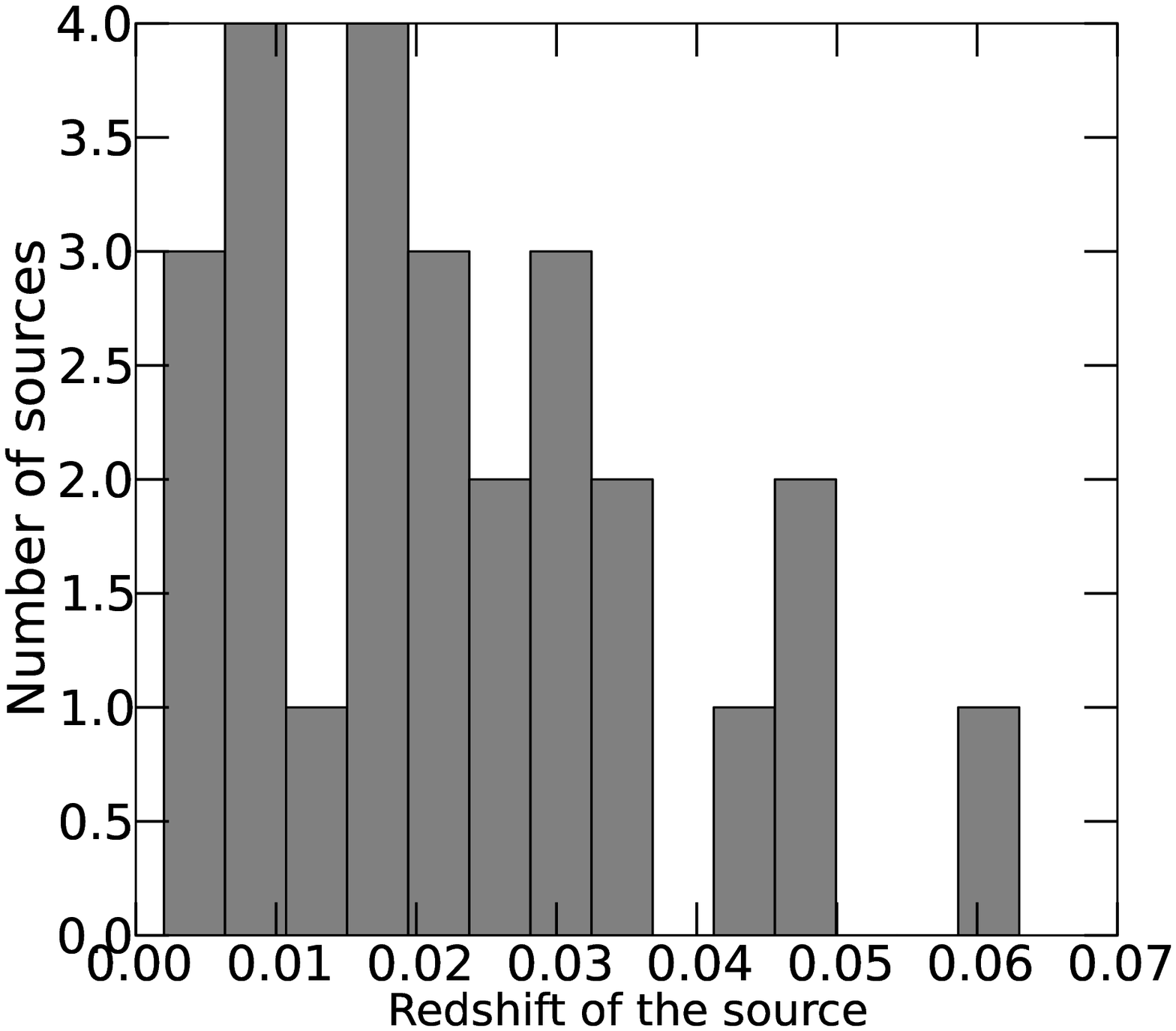} 
}
\caption{Left: The distribution of black hole mass (left panel) and redshift (right panel) for the WAX sample.}
\label{sample_2}
\end{figure*}

\begin{figure*}
  \centering
\hbox{
\includegraphics[width=5.5cm,angle=0]{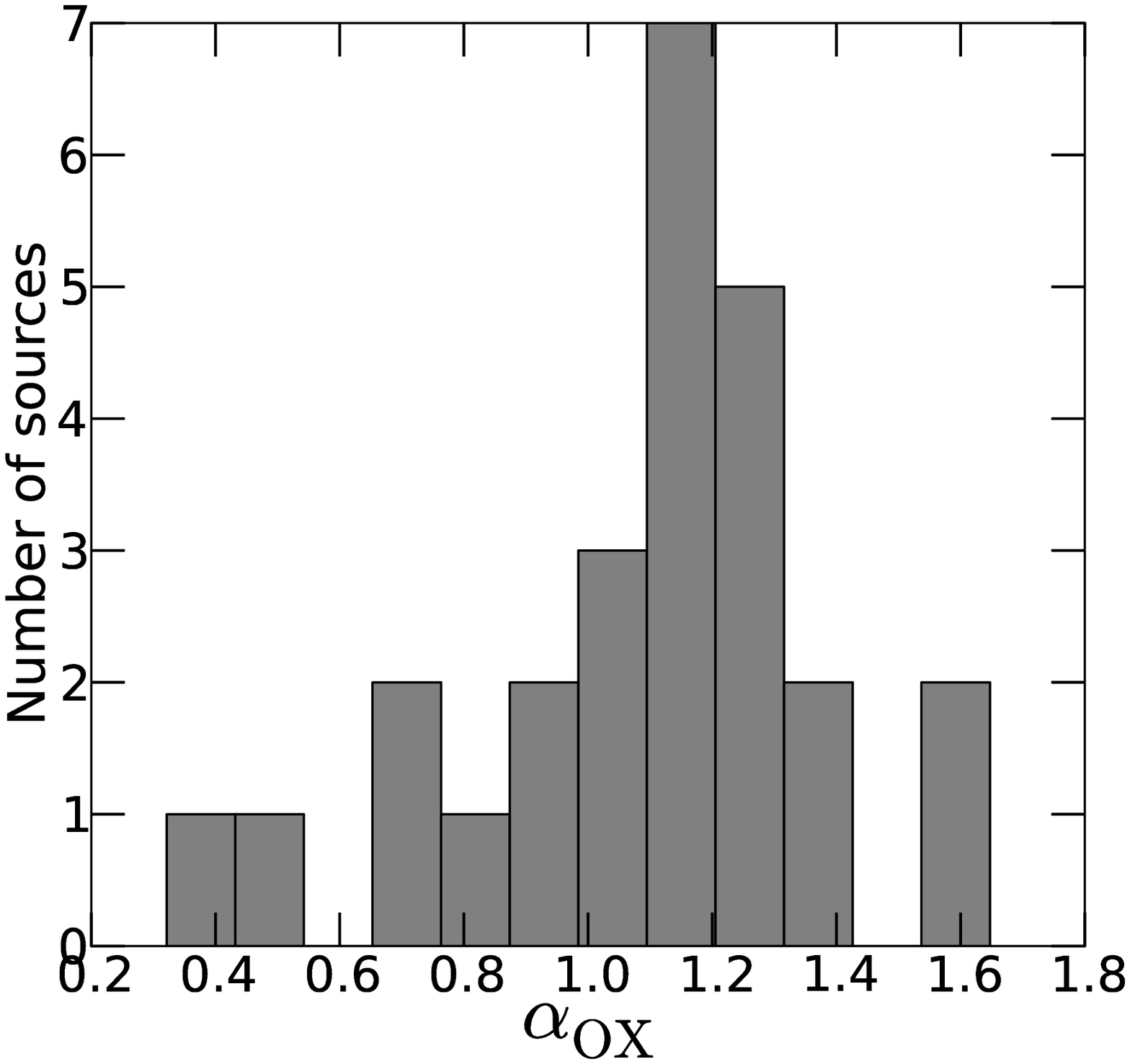} 
\includegraphics[width=5.5cm,angle=0]{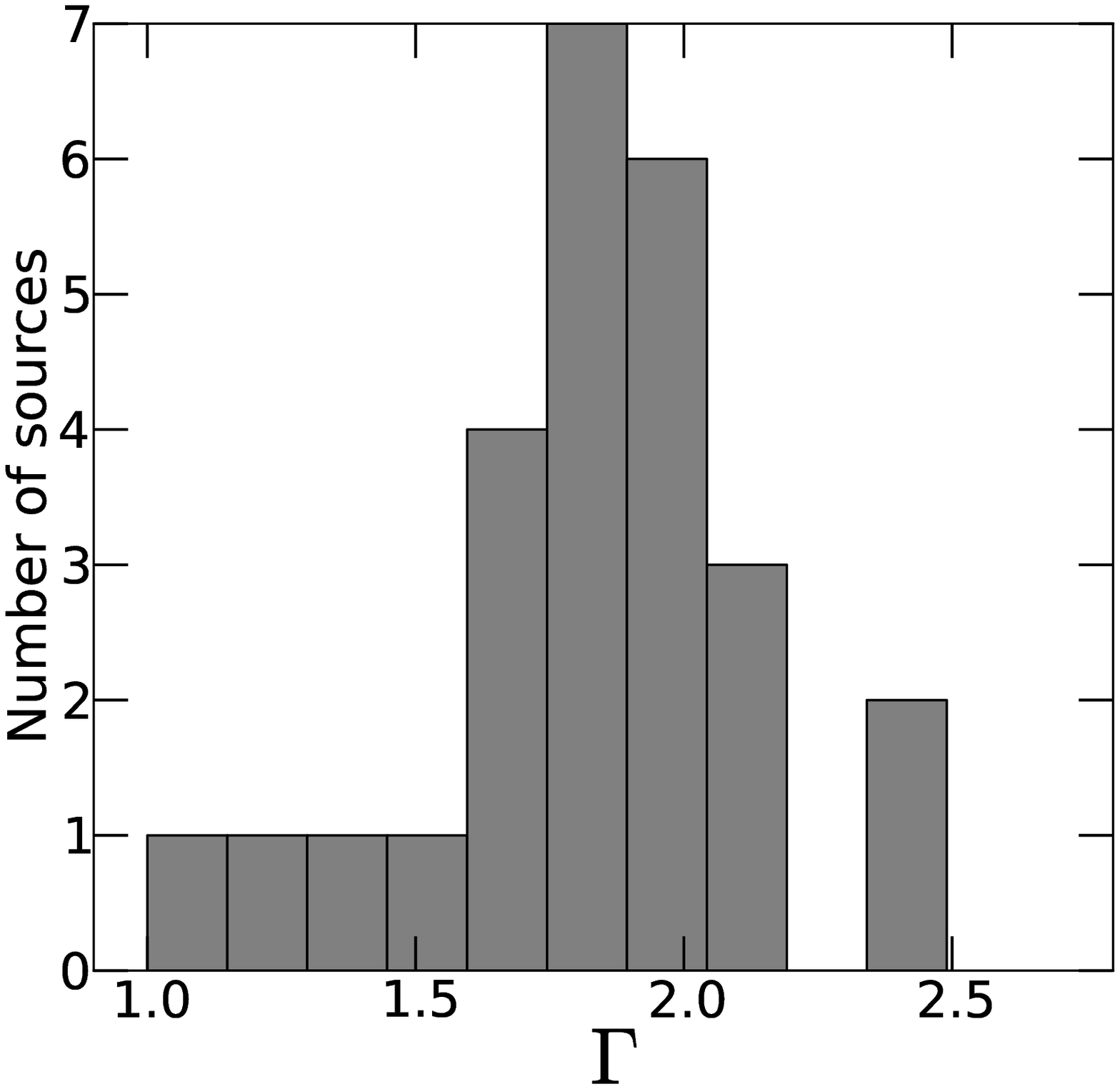} 
\includegraphics[width=5.5cm,angle=0]{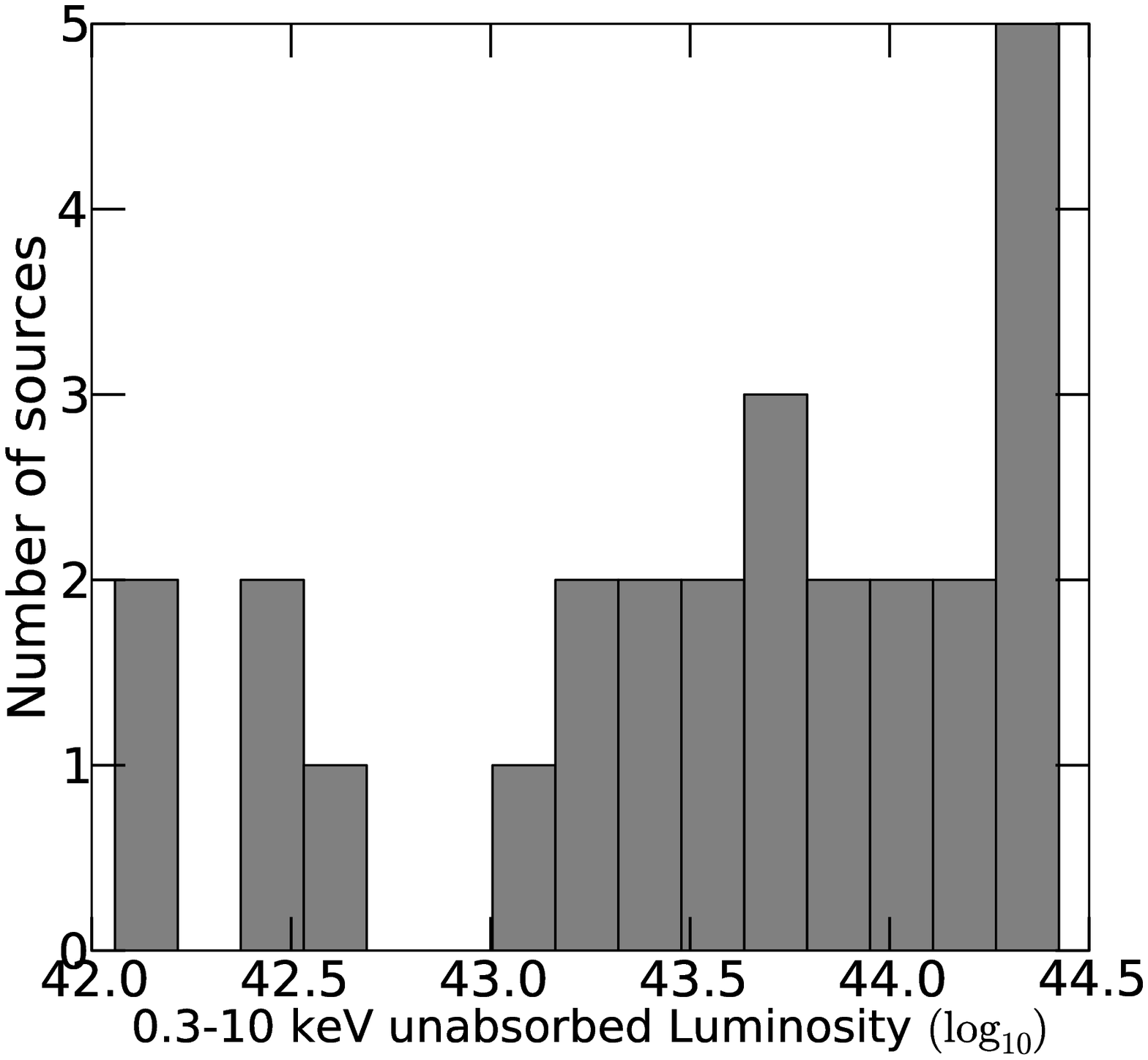} 
} 
\caption{From left to right: The distributions of $\alpha_{OX}$, power-law photon index $\Gamma$, and the $0.3-10\kev$ X-ray unabsorbed luminosity for the WAX sample. }
\label{sample_1}
\end{figure*}

\clearpage

\begin{figure}
  \centering
  
\hbox{
\includegraphics[width=8cm,angle=0]{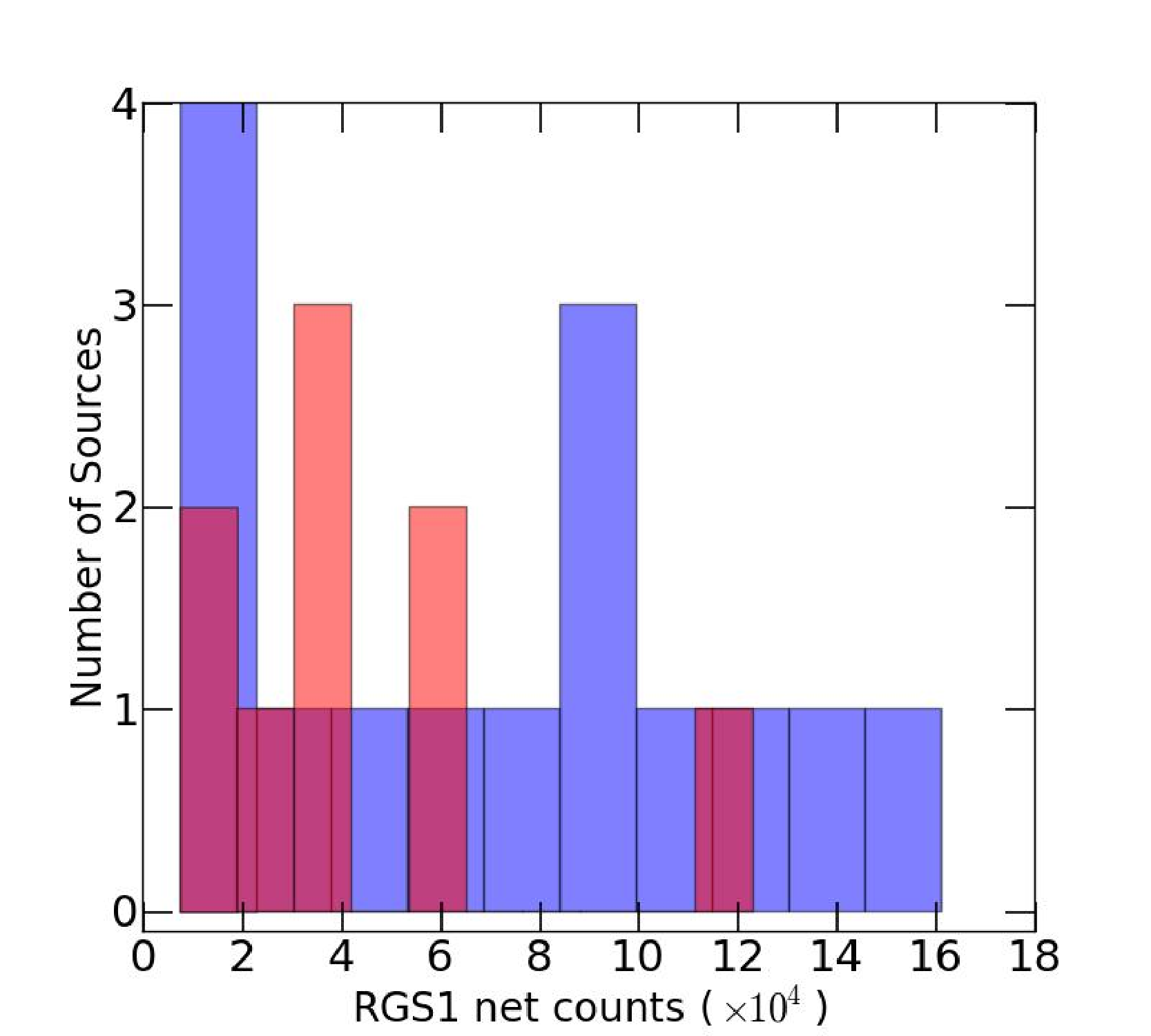} 
\includegraphics[width=8cm,angle=0]{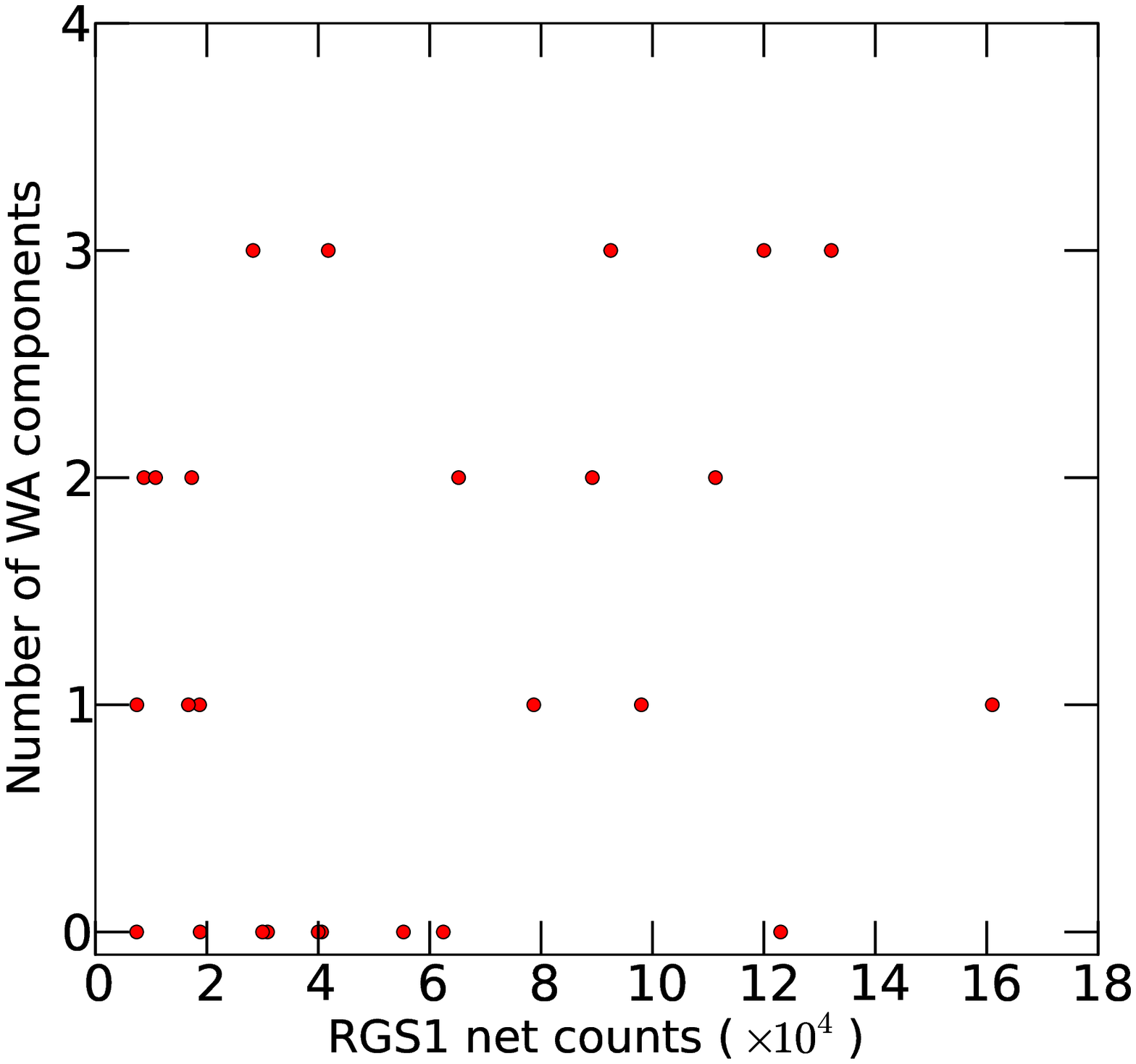} 
} 

\caption{Left: Histogram of the number of sources as a function of net 0.4--2~keV RGS counts. The blue histogram is for sources with confirmed WA detections while the red histogram is for sources with no WA detections; Right: Number of detected WA components as a function of RGS1 net counts}\label{fig:SNR-RGS}

\end{figure}

\begin{figure*}
  \centering 
  
\hbox{
\includegraphics[width=8cm,angle=0]{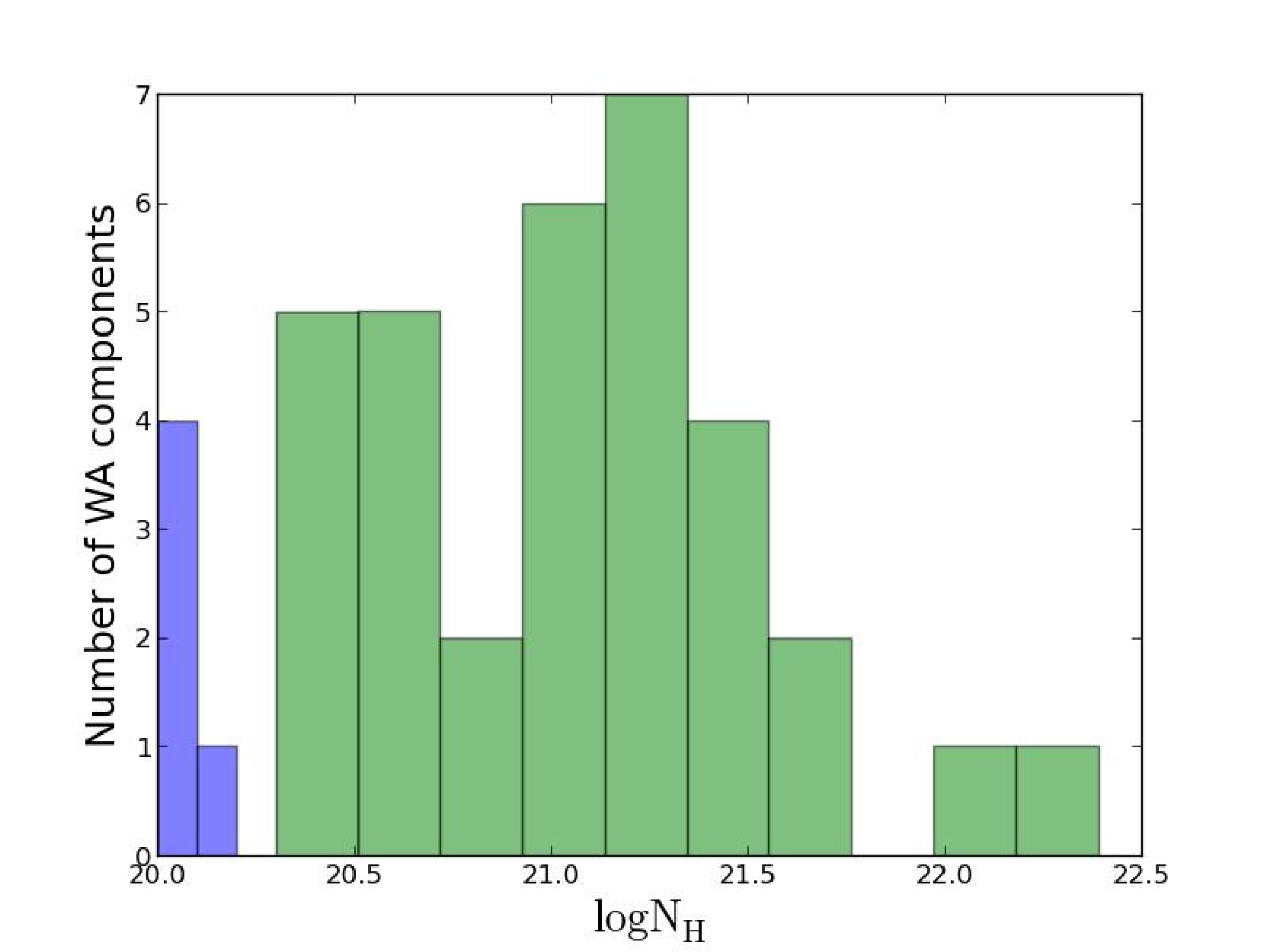} 
\includegraphics[width=8cm,angle=0]{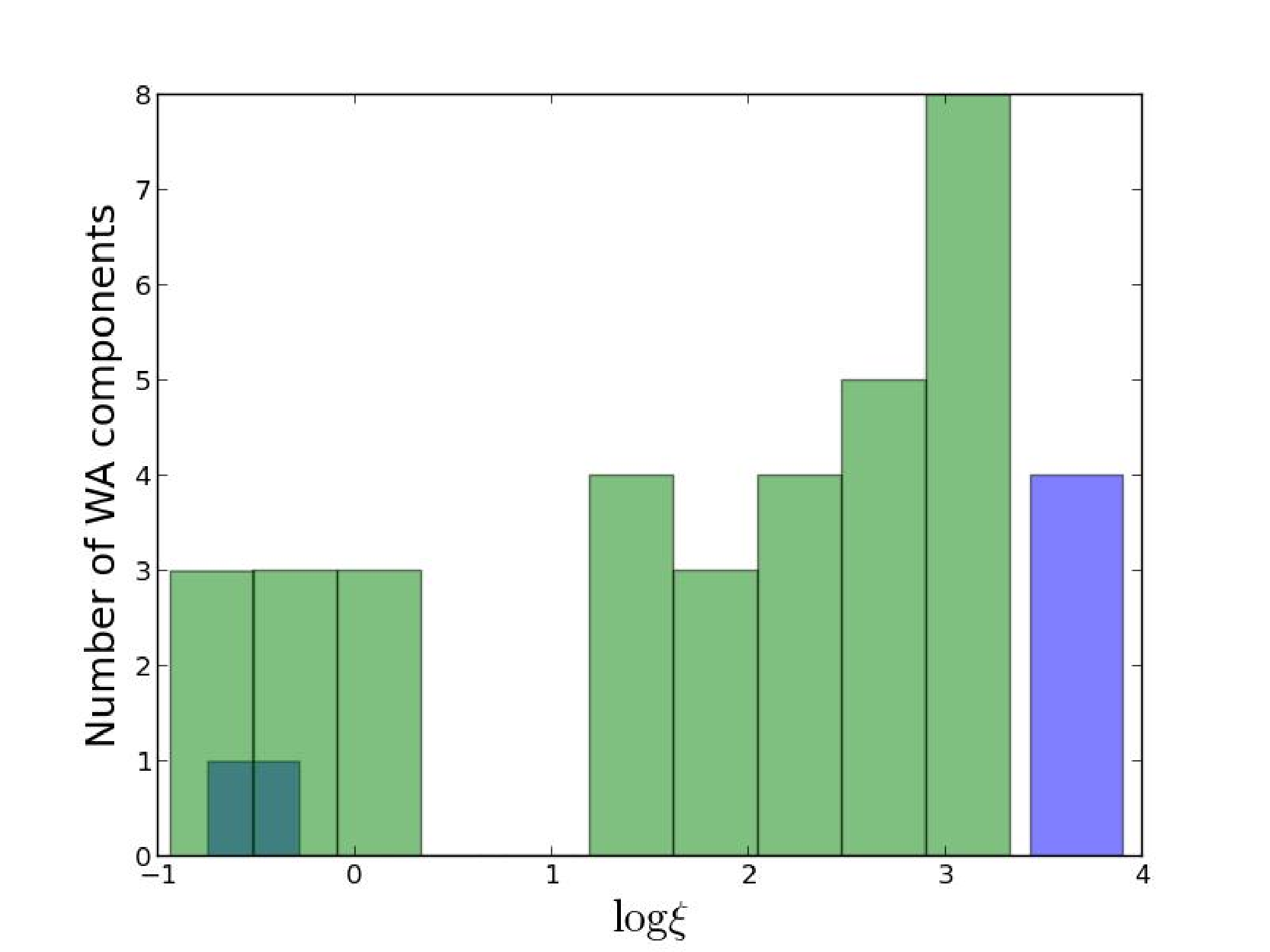} 
} 

\caption{Left: The distribution of column density for detected WA components shown in green. The upper limit on the column densities for sources without detectable WA are shown in blue. Right:The distribution of the ionisation parameter for warm absorbers with the same color coding as on the {\it left}.}
\label{xi-NH-distribution}
\end{figure*}


\clearpage
\begin{figure*}
  \centering
  \hbox{
 \includegraphics[width=8cm,angle=0]{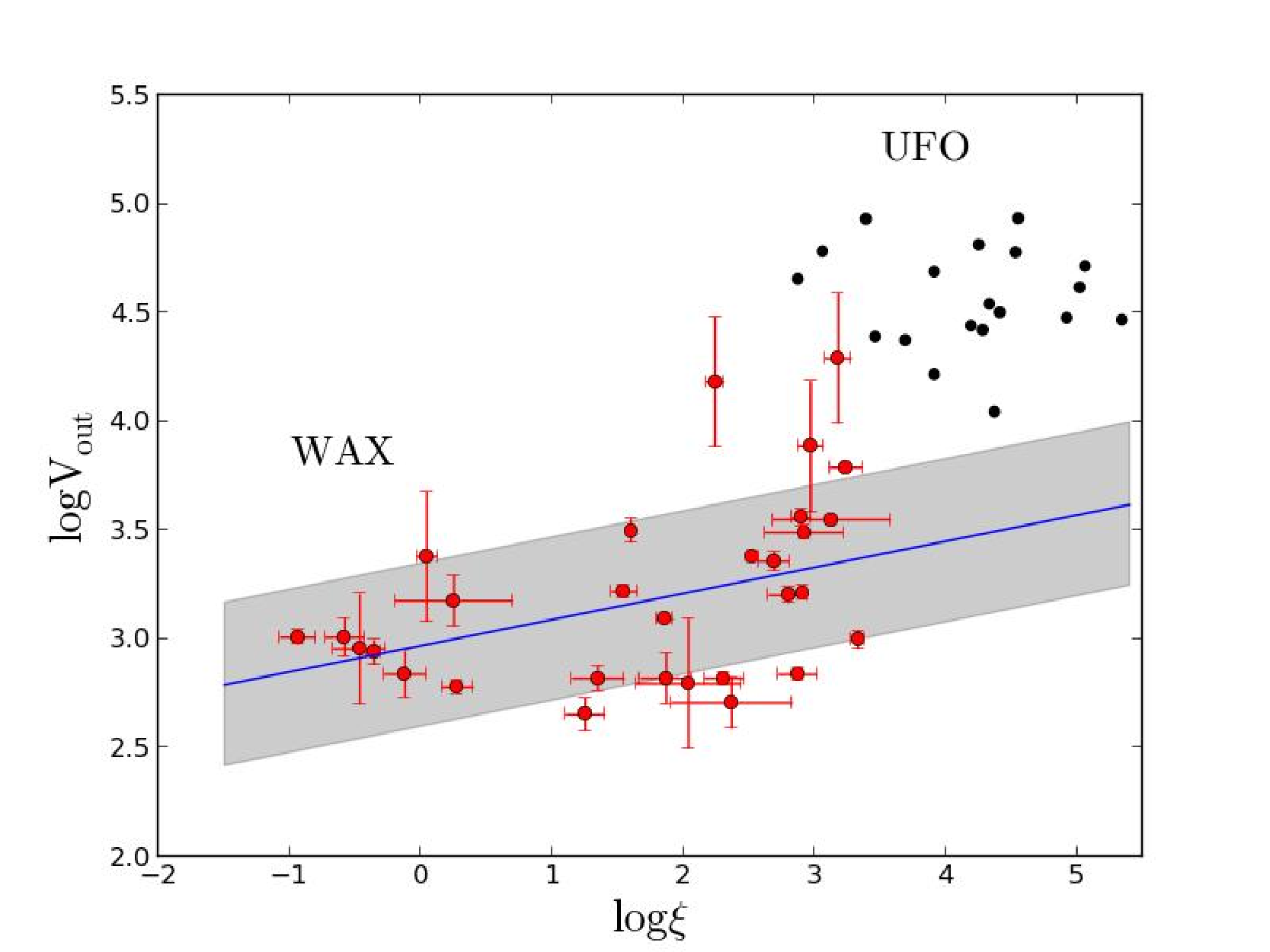}
 \includegraphics[width=8cm,angle=0]{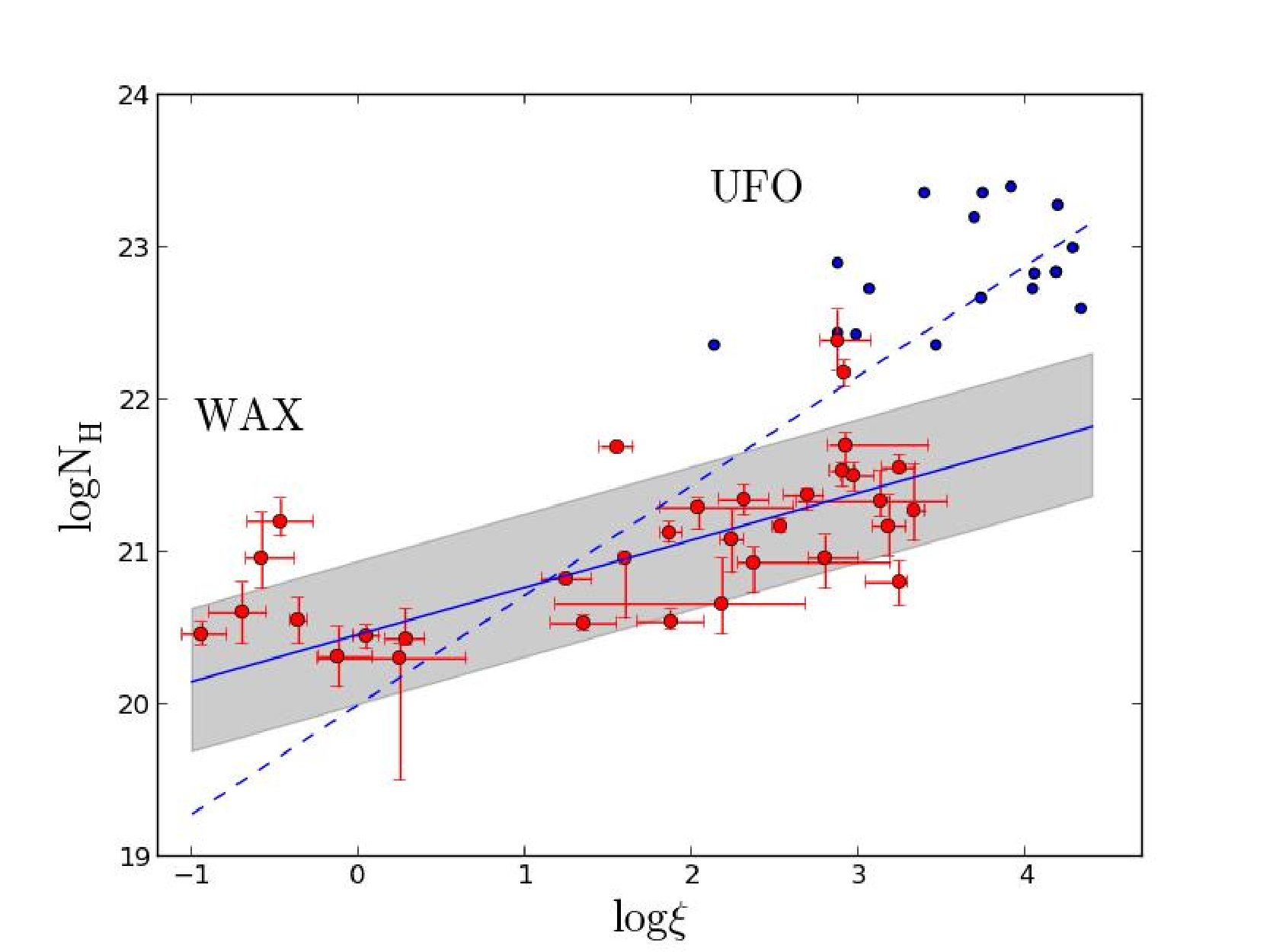}
}
\caption{{\tt Left:} Correlation between ionisation parameter and outflow velocity of the WAs in the WAX sample (large dots) and the UFO (small dots). The shaded regions correspond to the root mean square (RMS) deviation of the data points from the best fit linear regression line (solid line). The UFOs do not lie in the defined band. {\tt Right:} the same for the ionisation parameter versus the column density. The dotted line is the best fit linear regression line for the UFOs obtained by T13.}

\label{fig:linear-reg-xi-V}
\end{figure*}


\begin{figure*}
  \centering
 {
 \includegraphics[width=10cm,angle=90]{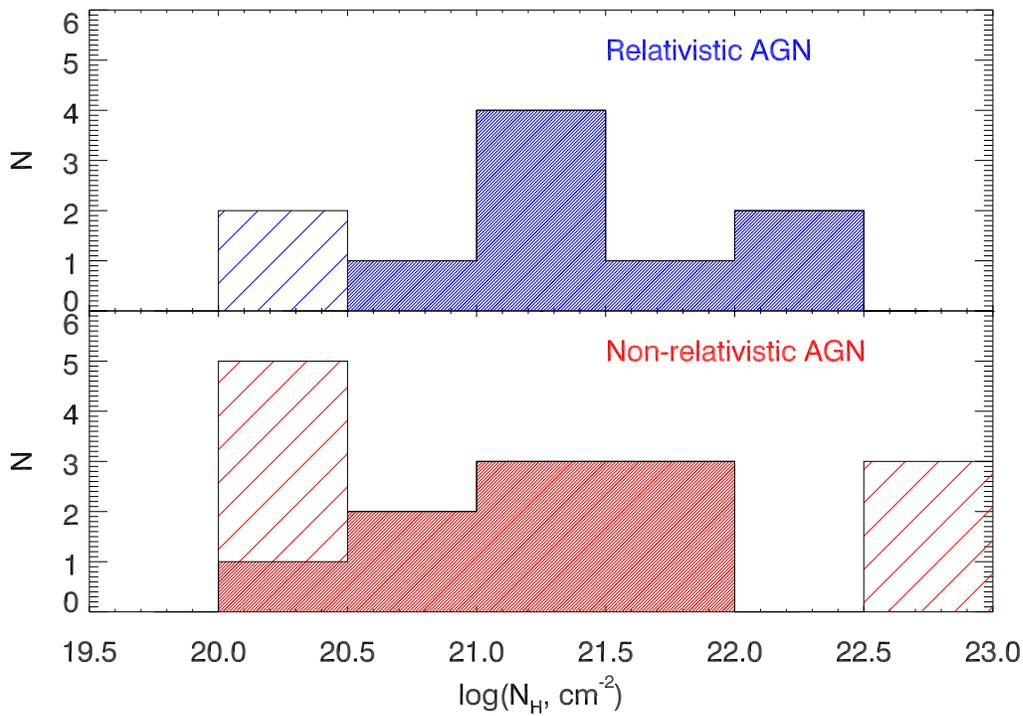}
 }
\caption{ Distribution of the column density of the detected WA components. The top panel is for the sources which have relativistic Fe K$\alpha$ emission lines while the bottom panel is for sources without such a detection. The shaded areas correspond to upper limits.}

\label{Fe-line-NH-distrib}
\end{figure*}


\begin{figure*}
  \centering
 {
 \includegraphics[width=10cm,angle=0]{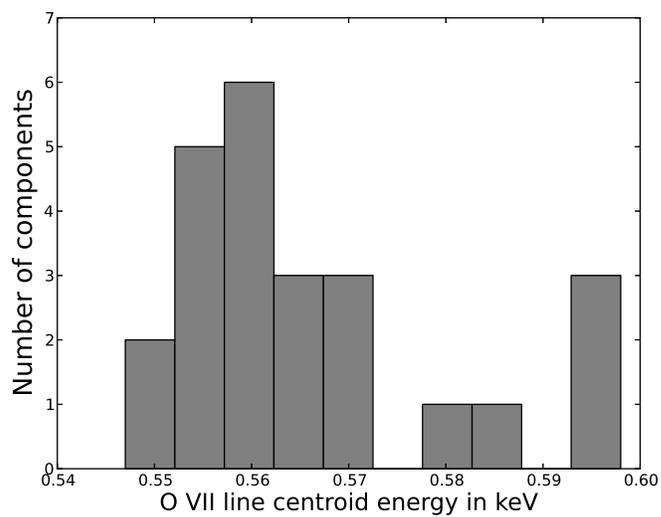}
 }
\caption{Distribution of the O VII emission line centroid energies. The resonant line of the O VII He-like triplet emission has a rest frame energy of $0.574\kev$ while the forbidden line has a rest frame energy of $0.561\kev$. We detect more forbidden line emission in the WAX sample compared to resonant lines. This implies a photo-ionisation origin of the emission lines from distant clouds.}

\label{OVII}
\end{figure*}





\clearpage
\appendix 
\section{Figures.}
We show in this Section: a) the SED of the WAX sample (cf. \S~\ref{real-cont}) in Figs.~\ref{sed} to ~\ref{sed_3}; b) the EPIC-pn spectra ({\it upper panels}) and the
residuals in units of data/model ratio ({\it lower panels}) in Figs.~\ref{epicpn} to ~\ref{epicpn_2}; c) the RGS spectra ({\it upper panels}) and the
residuals in units of contribution to the Cash statistics  ({\it lower panels}) in Figs.~\ref{rgs} to ~\ref{rgs_2}; d) the best-fit WA components superposed to
the RGS spectra in Figs.~\ref{wa} to ~\ref{wa_2}

\clearpage

\begin{figure}
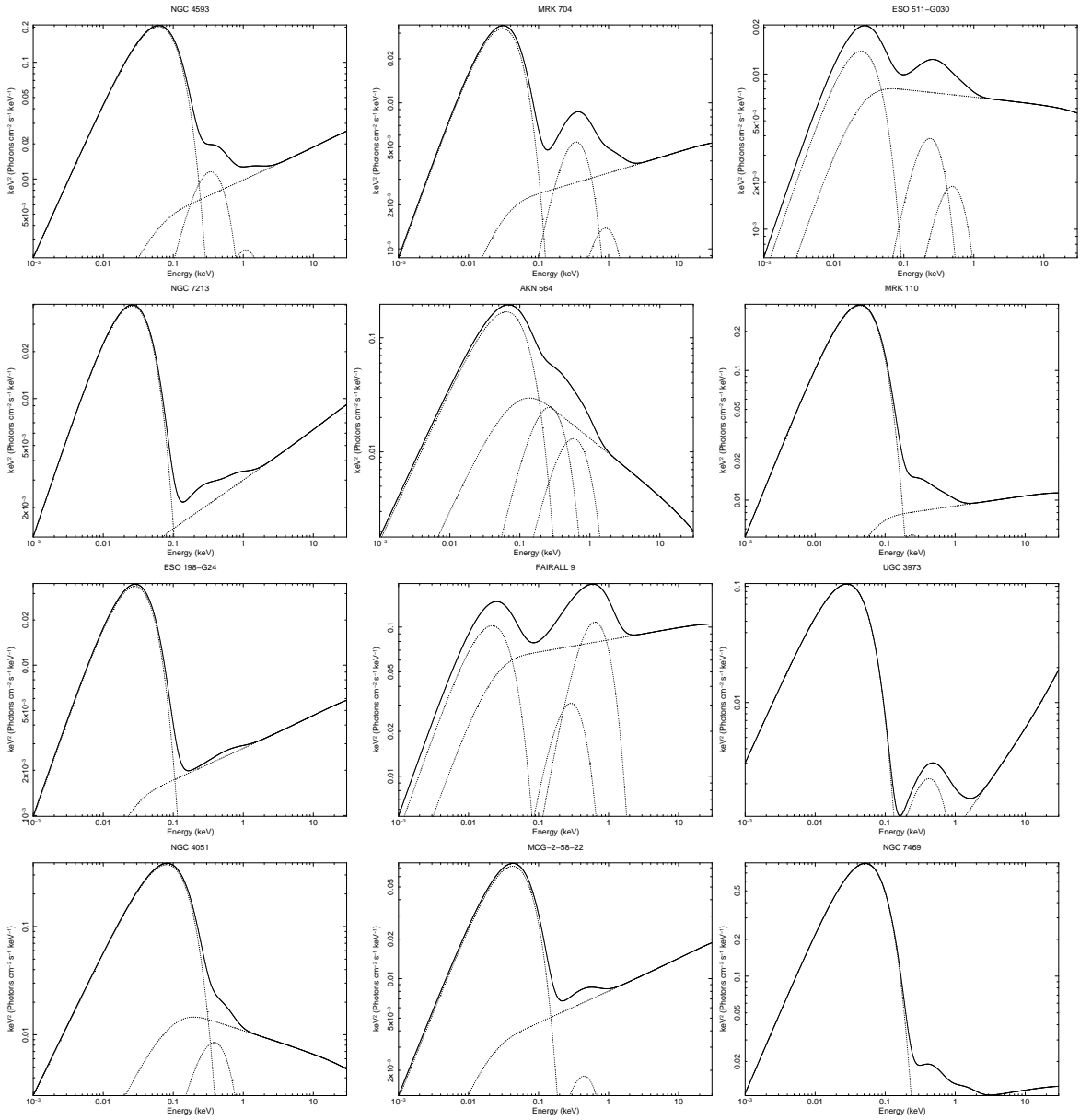

  \centering
  \vbox{
\hbox{
\includegraphics[width=4cm,angle=-90]{SEDs_July13/NGC4593.ps} 
\includegraphics[width=4cm,angle=-90]{SEDs_July13/MRK704.ps} 
\includegraphics[width=4cm,angle=-90]{SEDs_July13/ESO511-G030.ps} 
} 

\hbox{
\includegraphics[width=4cm,angle=-90]{SEDs_July13/NGC7213.ps} 
\includegraphics[width=4cm,angle=-90]{SEDs_July13/AKN564.ps} 
\includegraphics[width=4cm,angle=-90]{SEDs_July13/MRK110.ps} 
} 

\hbox{
\includegraphics[width=4cm,angle=-90]{SEDs_July13/ESO198-G24.ps} 
\includegraphics[width=4cm,angle=-90]{SEDs_July13/Fairall9.ps} 
\includegraphics[width=4cm,angle=-90]{SEDs_July13/UGC3973.ps} 
} 

\hbox{
\includegraphics[width=4cm,angle=-90]{SEDs_July13/NGC4051.ps} 
\includegraphics[width=4cm,angle=-90]{SEDs_July13/MCG-2-58-22.ps} 
\includegraphics[width=4cm,angle=-90]{SEDs_July13/NGC7469.ps} 
} 

}
\caption{The $1\ev-30 \kev$ SEDs for the sources in the sample. The SED consists of three components: The UV bump, the soft X-ray excess and the powerlaw. }
 \label{sed}
\end{figure}


\clearpage

\begin{figure}
  \centering
  \vbox{
\hbox{
\includegraphics[width=4cm,angle=-90]{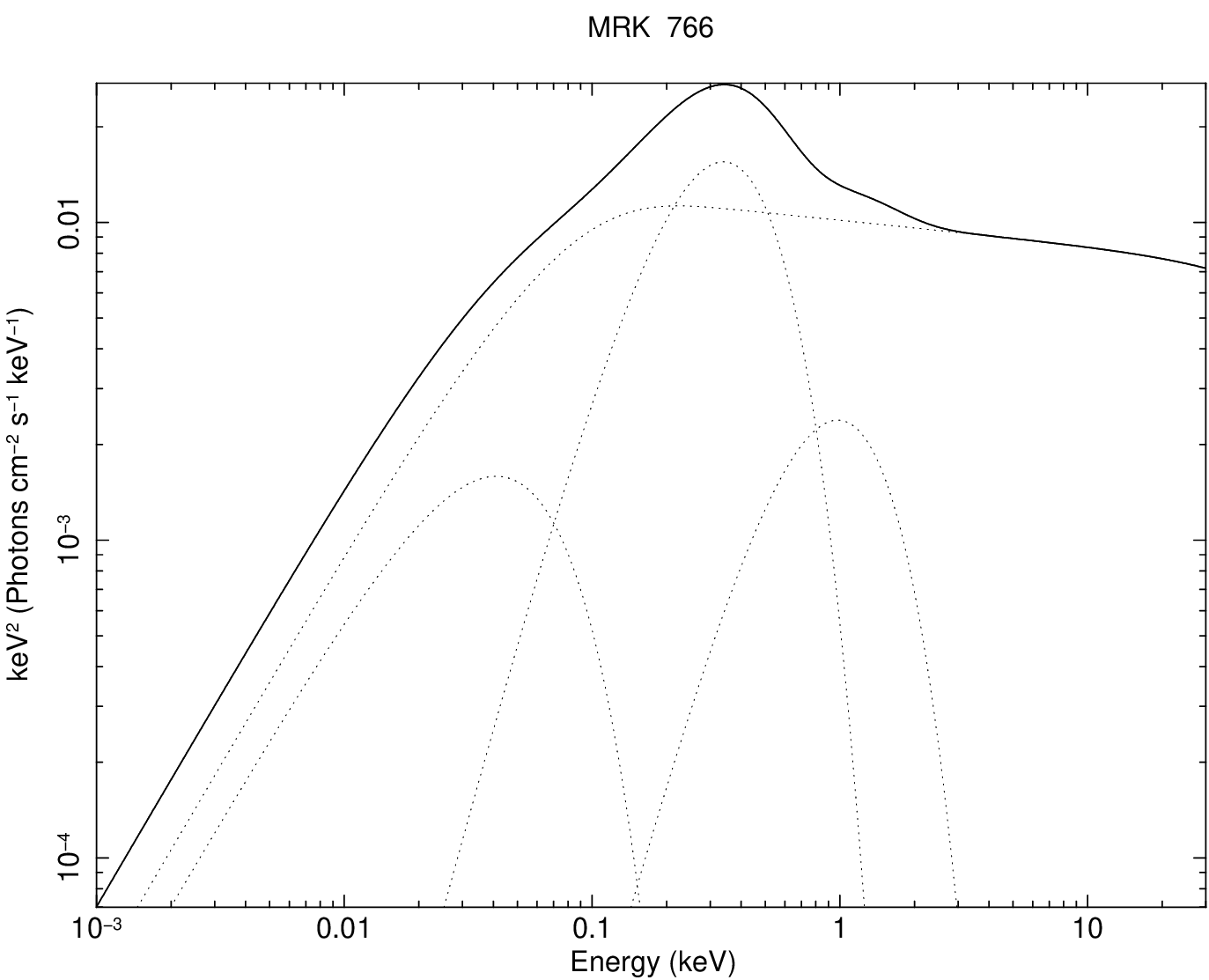} 
\includegraphics[width=4cm,angle=-90]{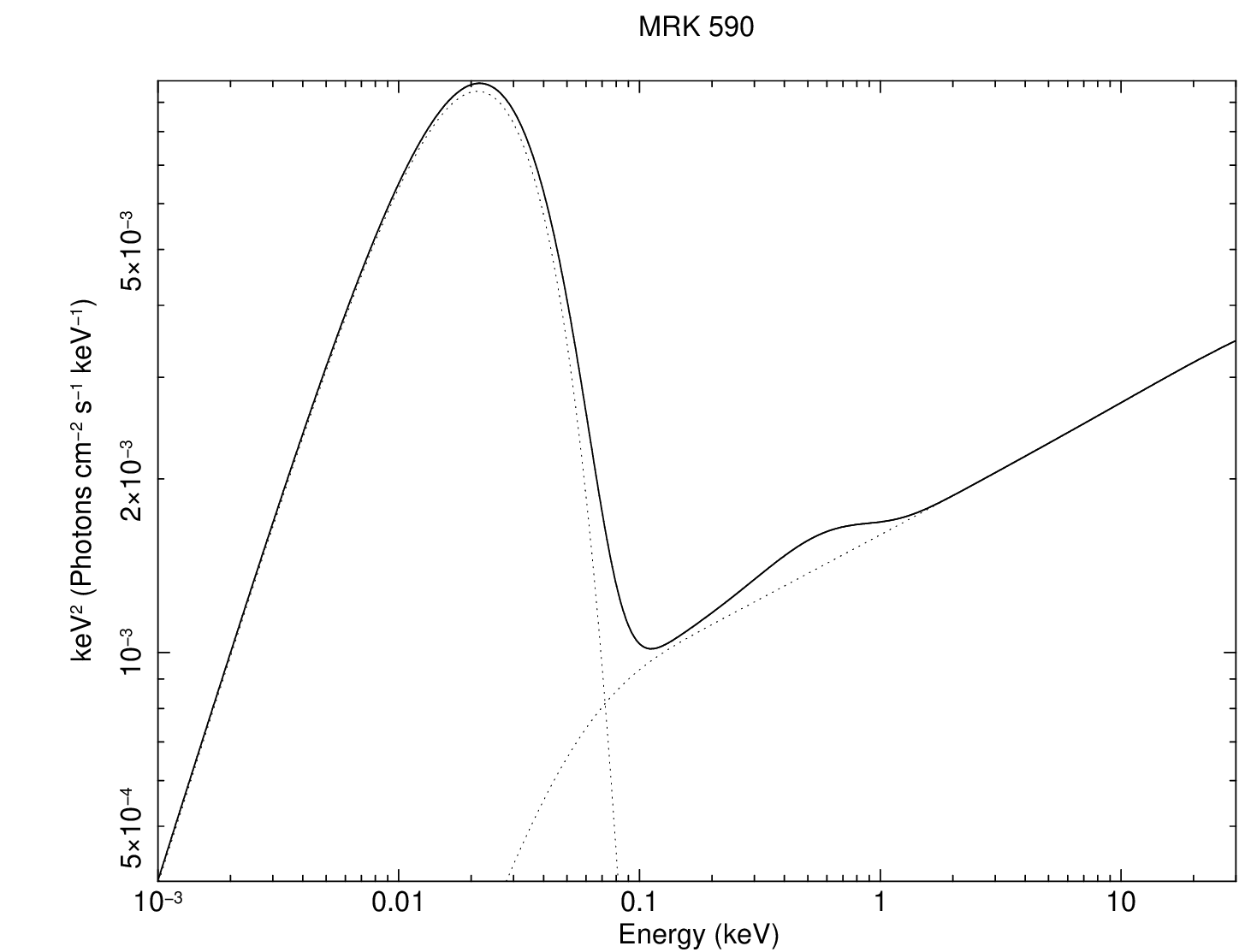} 
\includegraphics[width=4cm,angle=-90]{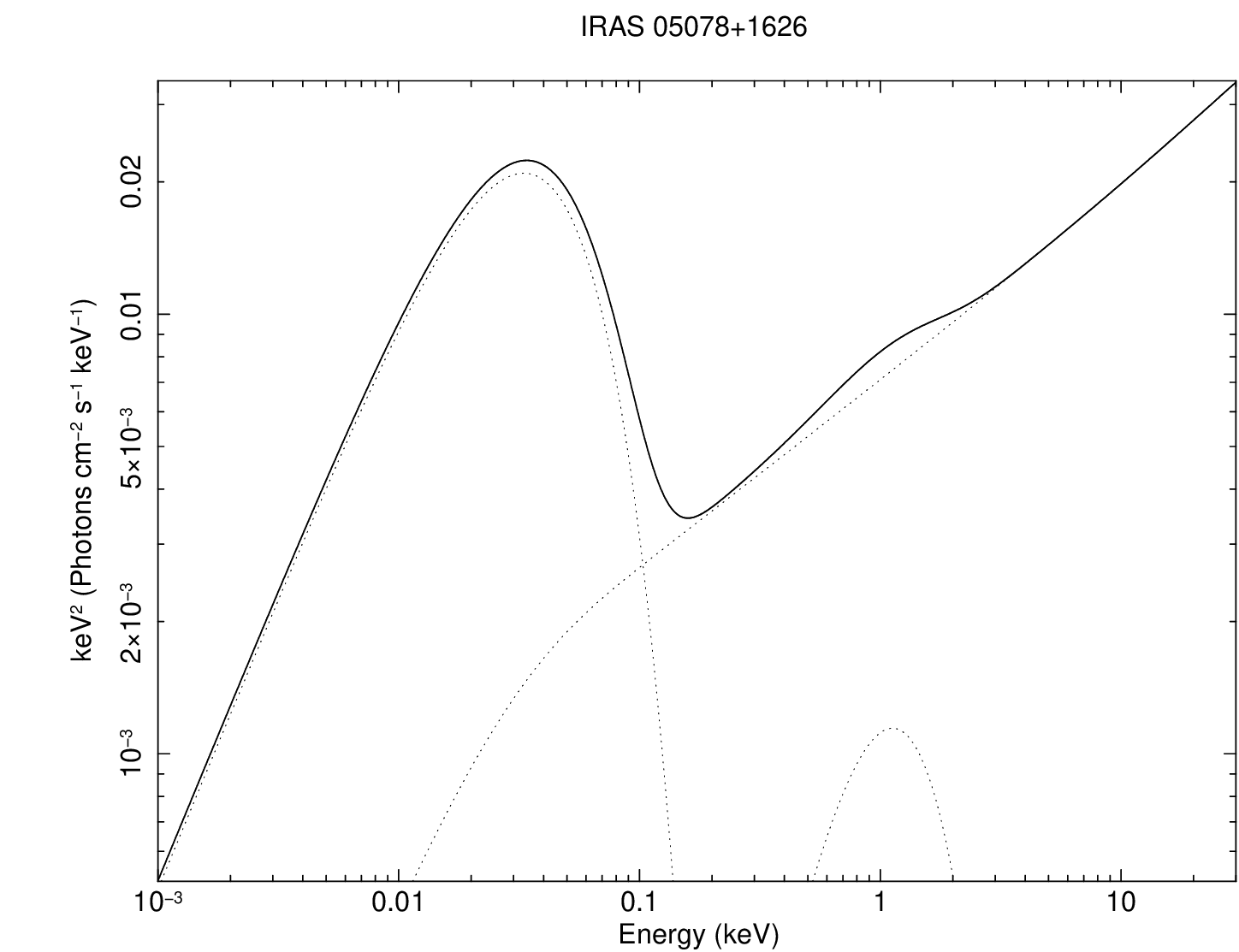} 
} 

\hbox{
\includegraphics[width=4cm,angle=-90]{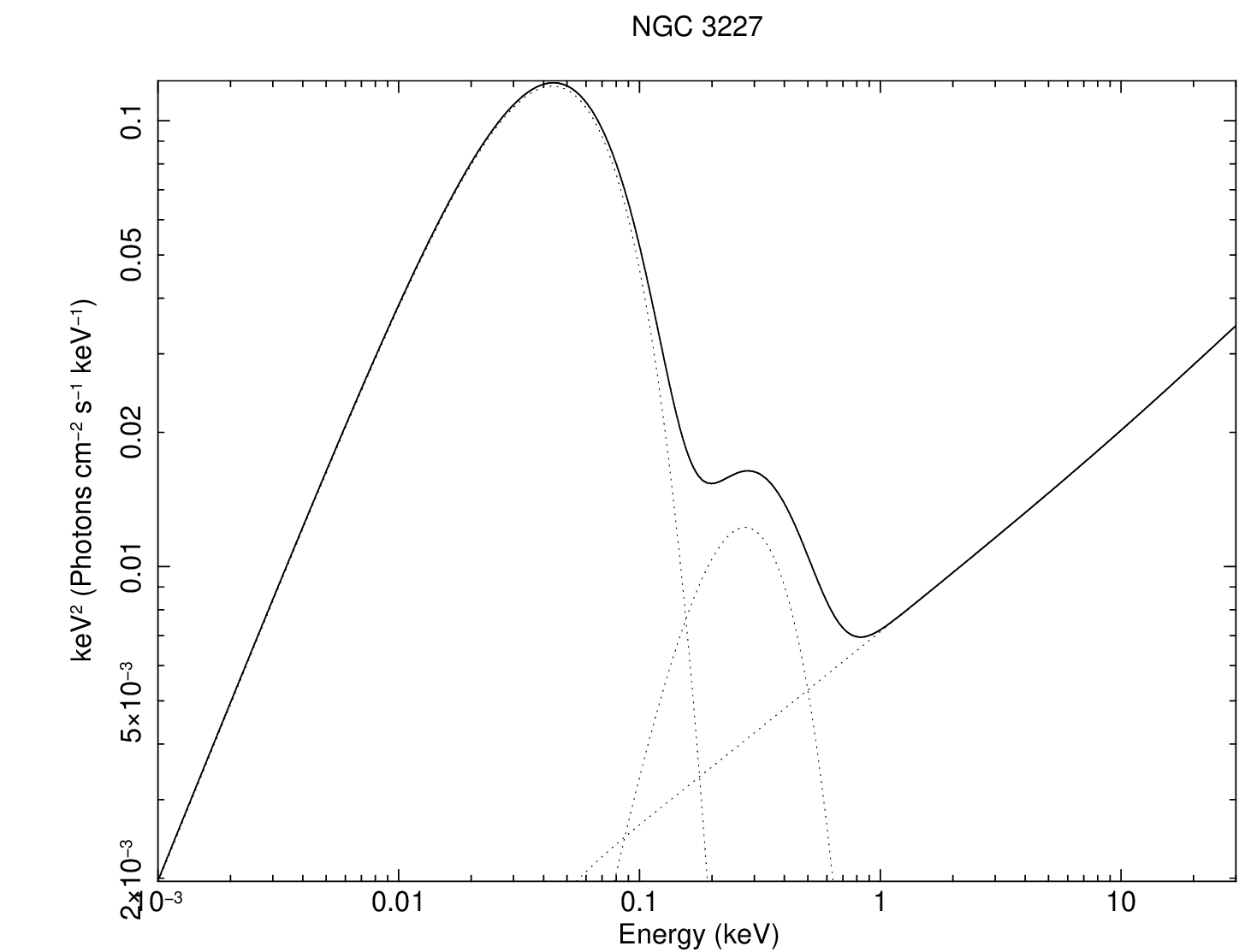} 
\includegraphics[width=4cm,angle=-90]{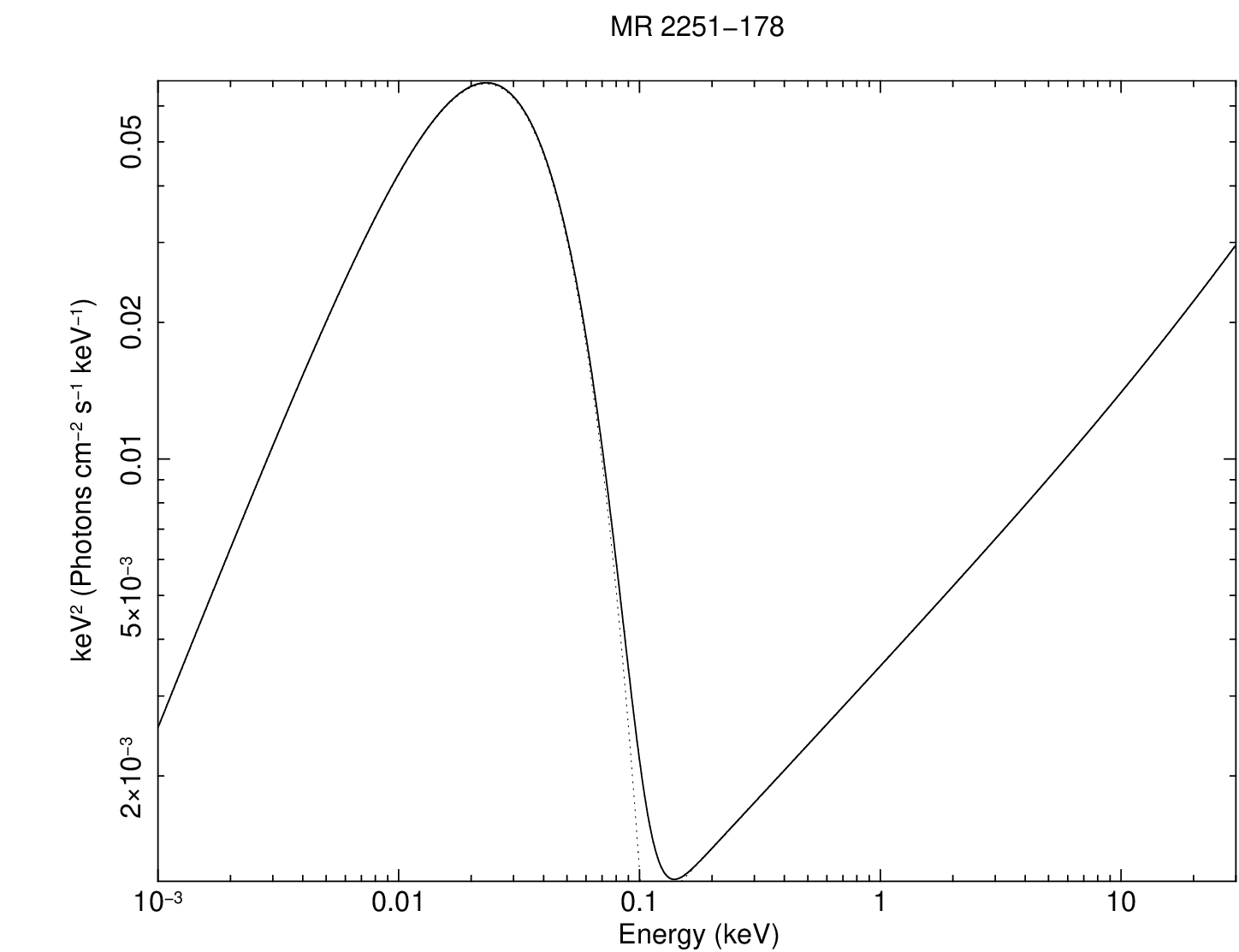} 
\includegraphics[width=4cm,angle=-90]{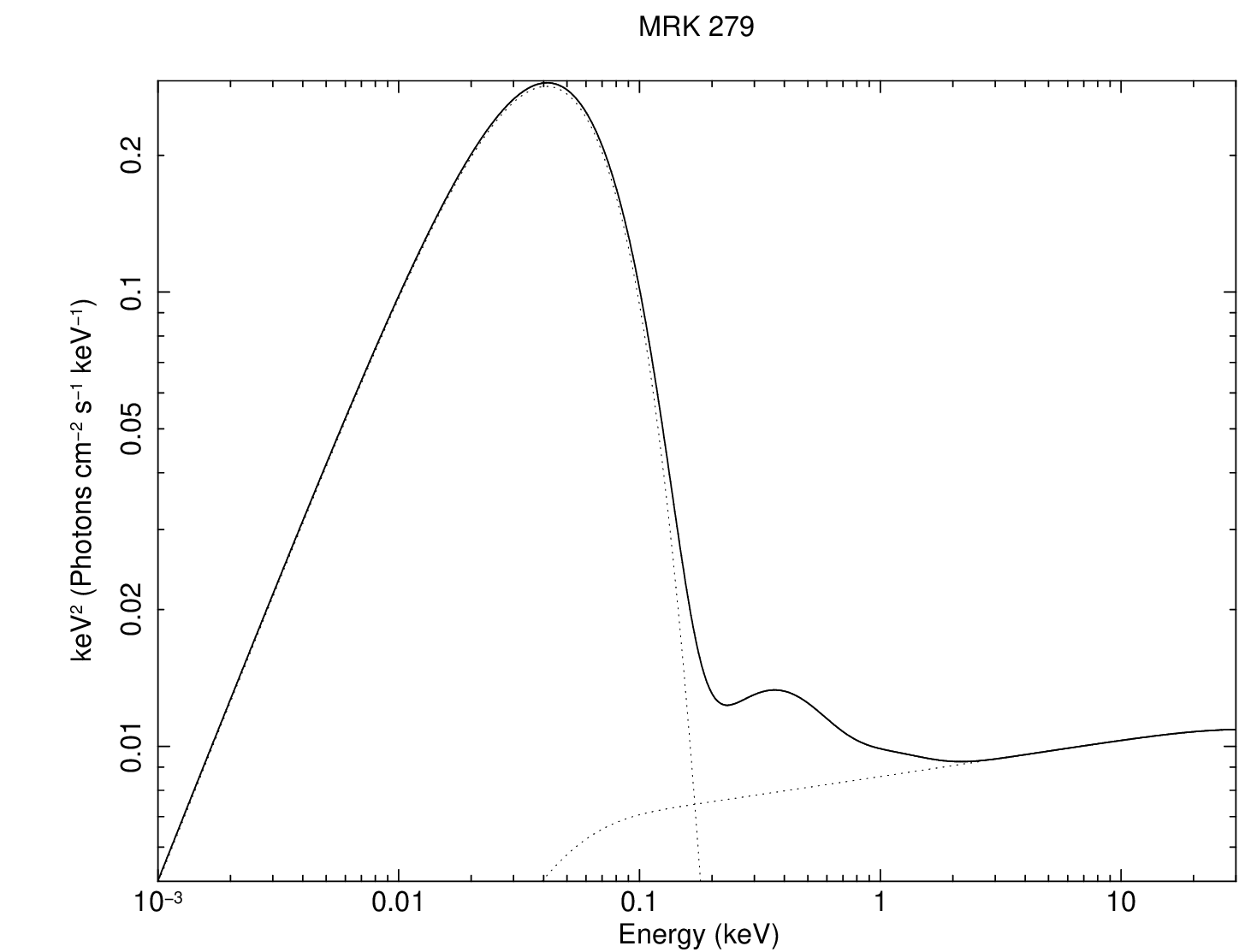} 
} 

\hbox{
\includegraphics[width=4cm,angle=-90]{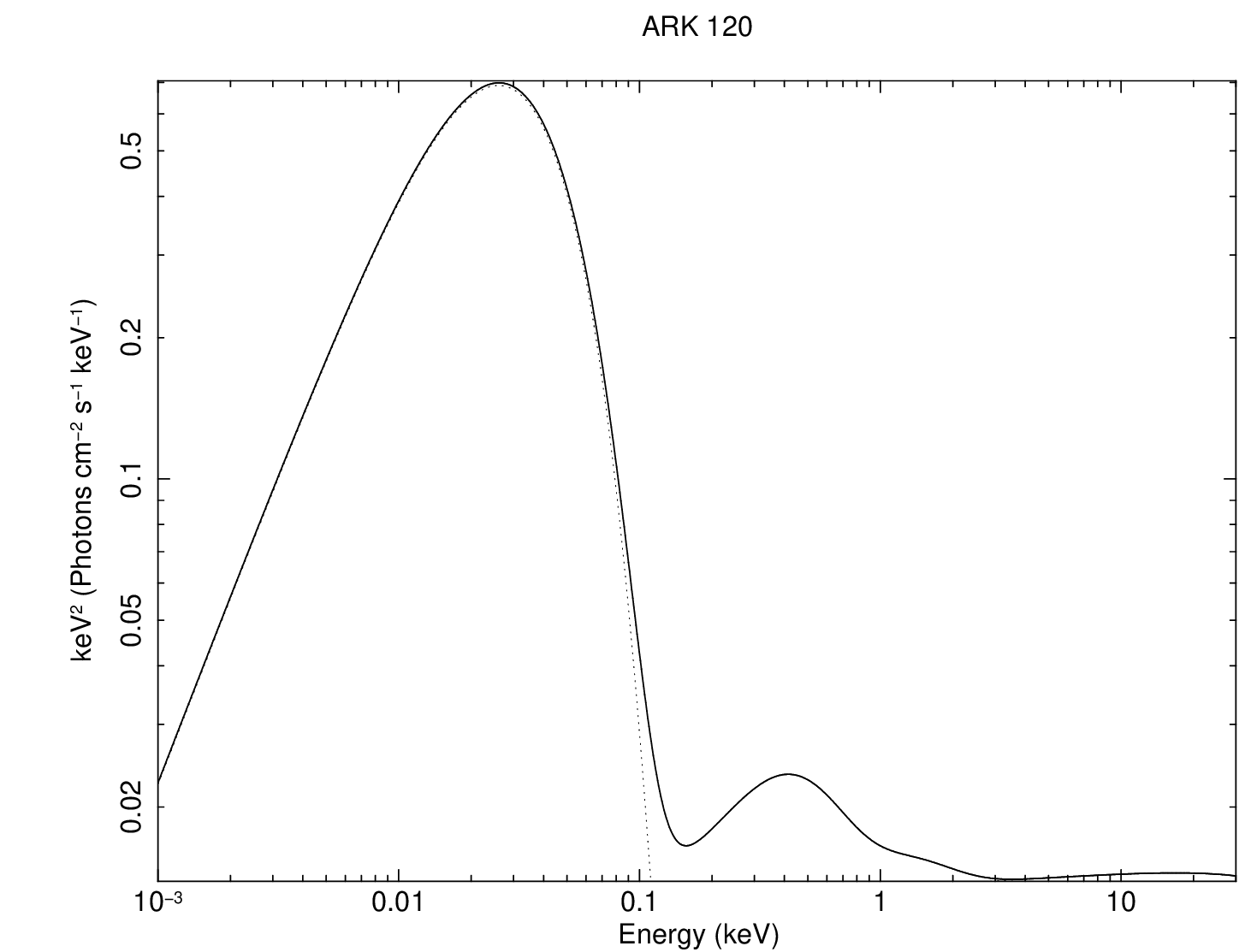} 
\includegraphics[width=4cm,angle=-90]{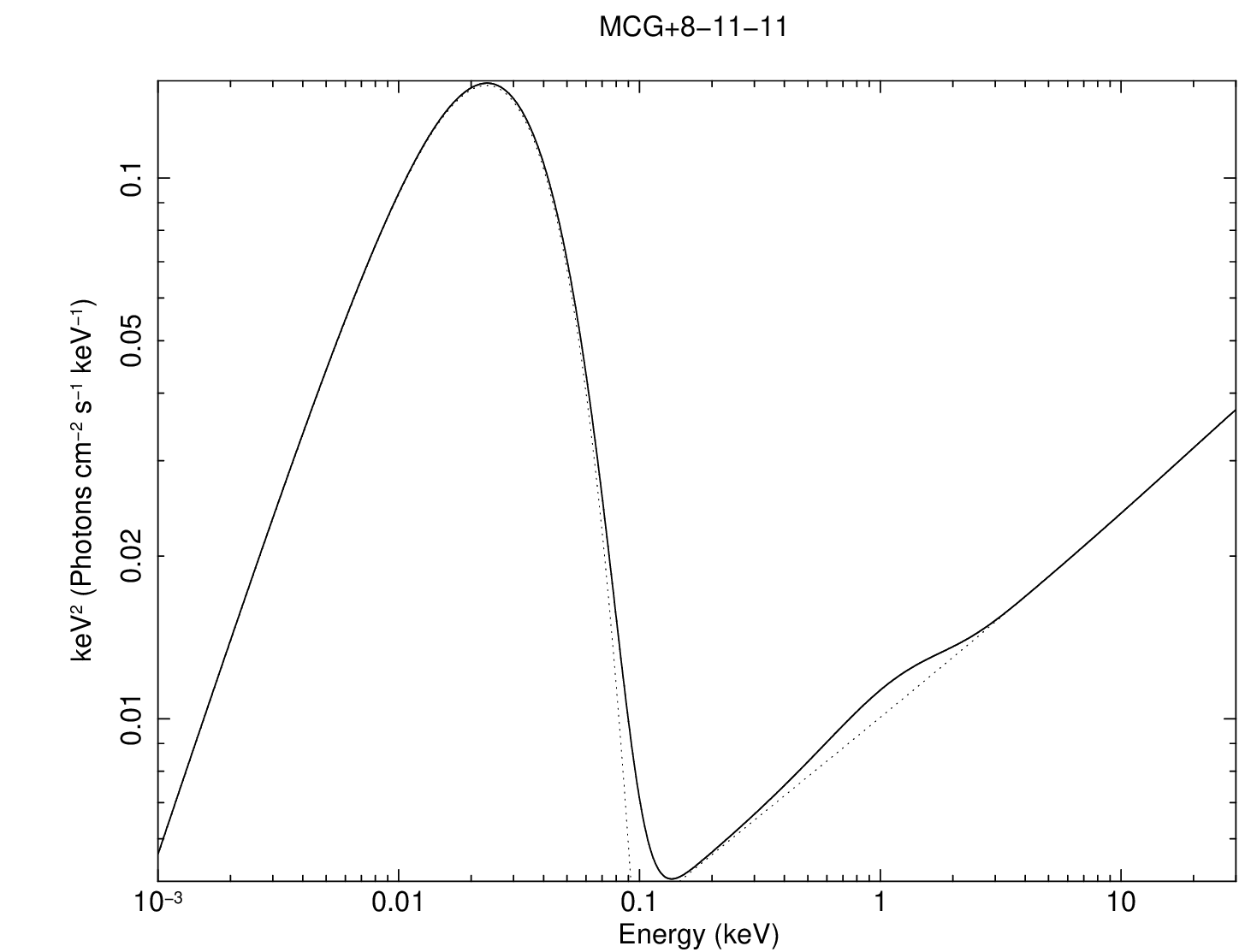} 
\includegraphics[width=5cm,angle=-90]{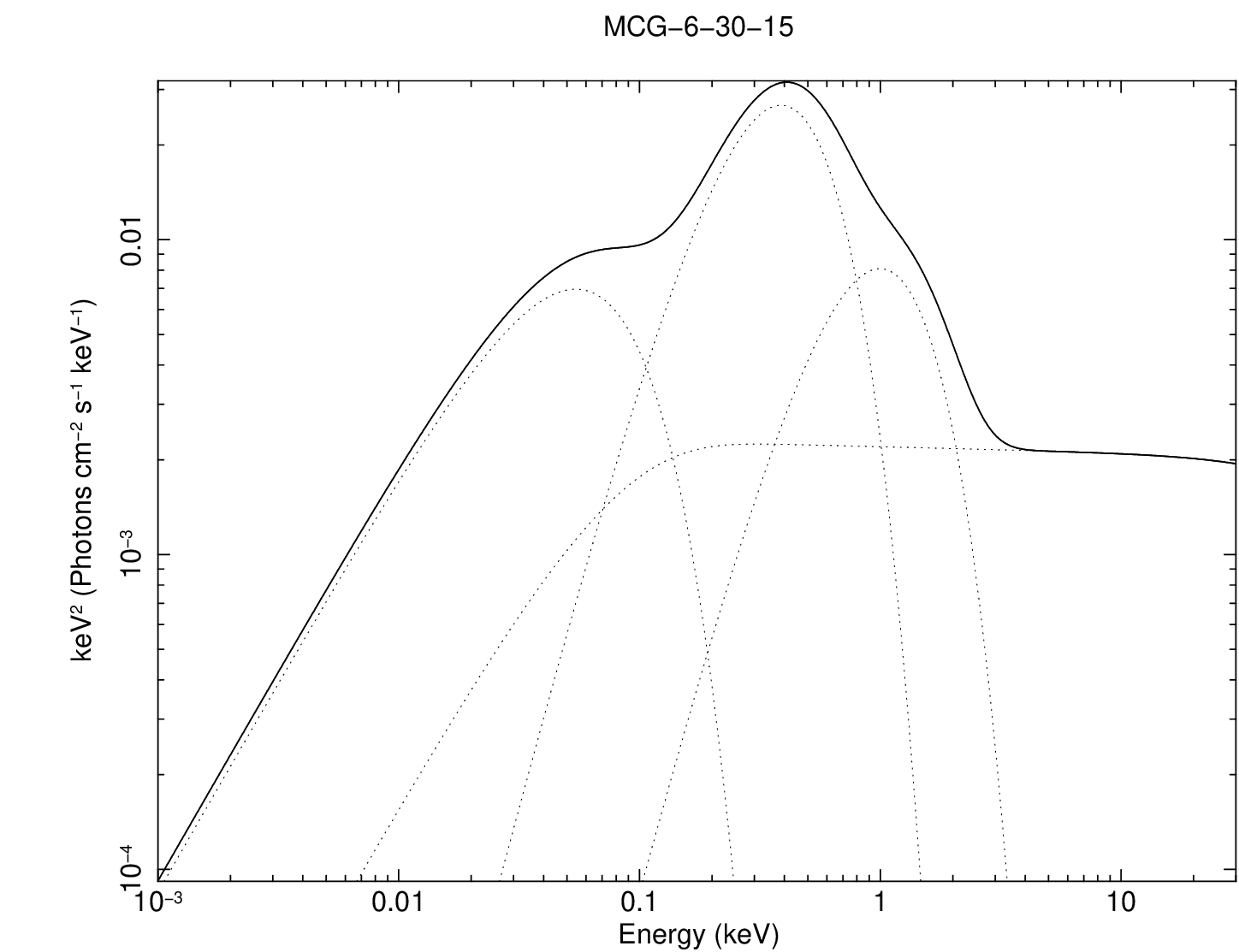} 
} 

\hbox{
\includegraphics[width=4cm,angle=-90]{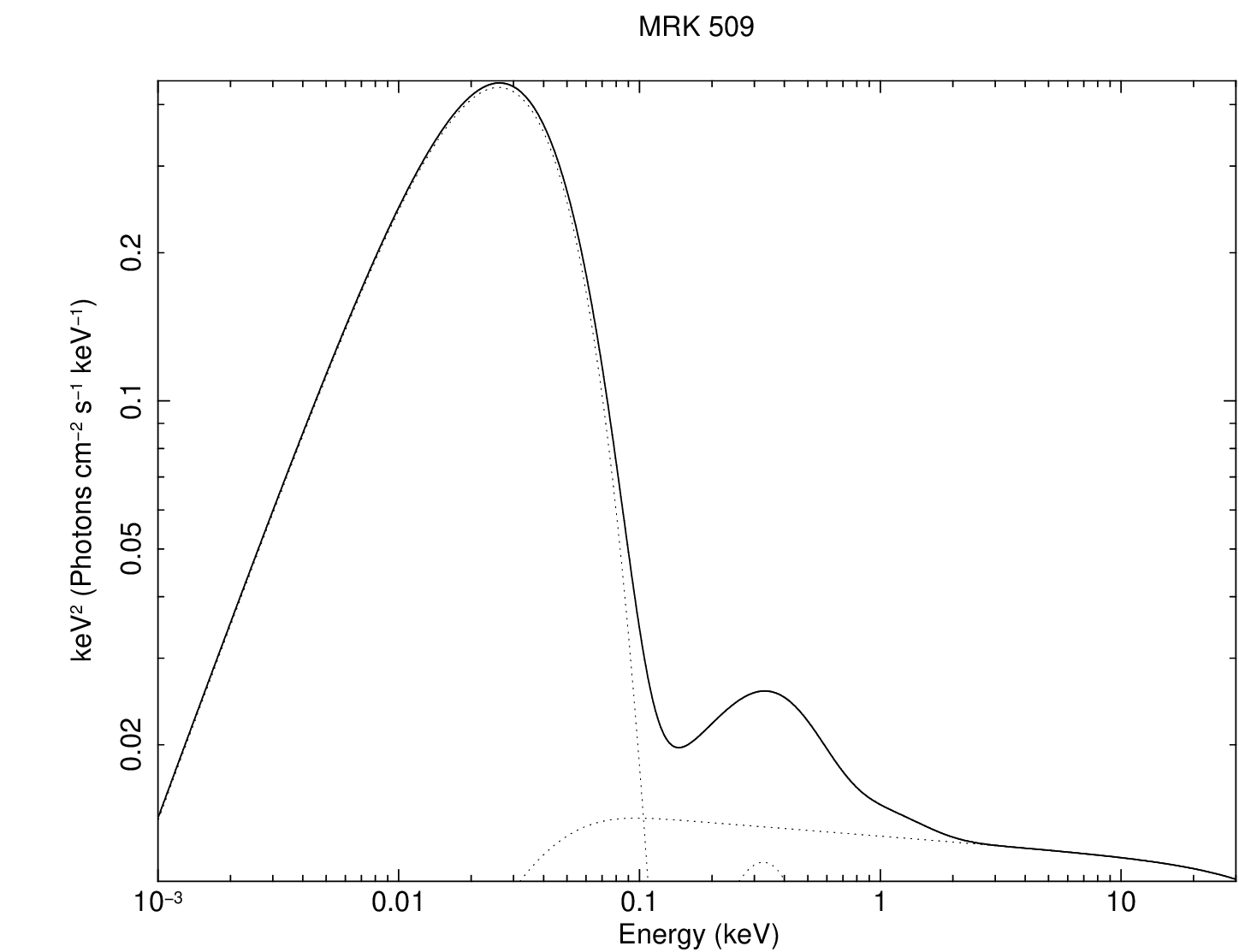} 
\includegraphics[width=4cm,angle=-90]{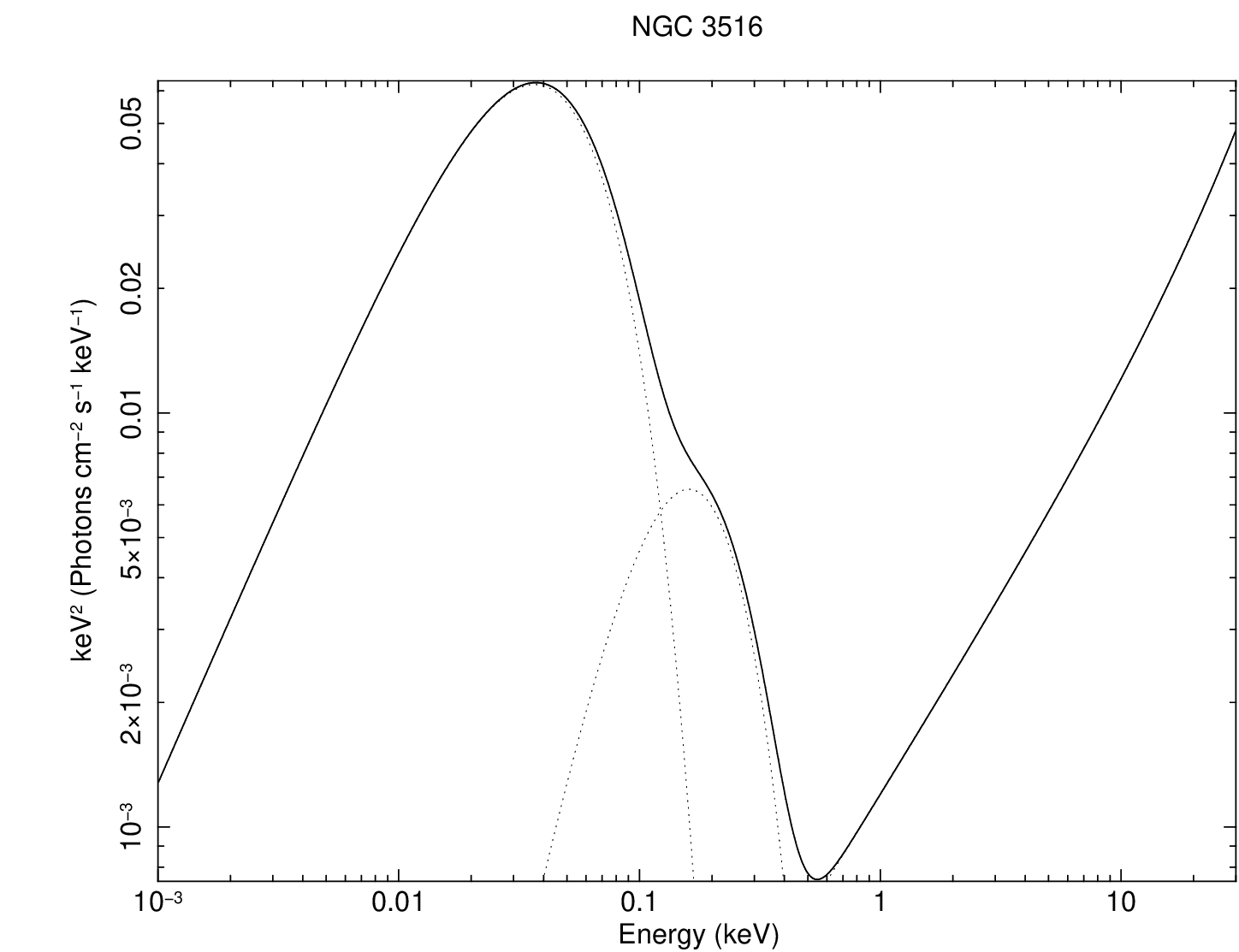} 
\includegraphics[width=4cm,angle=-90]{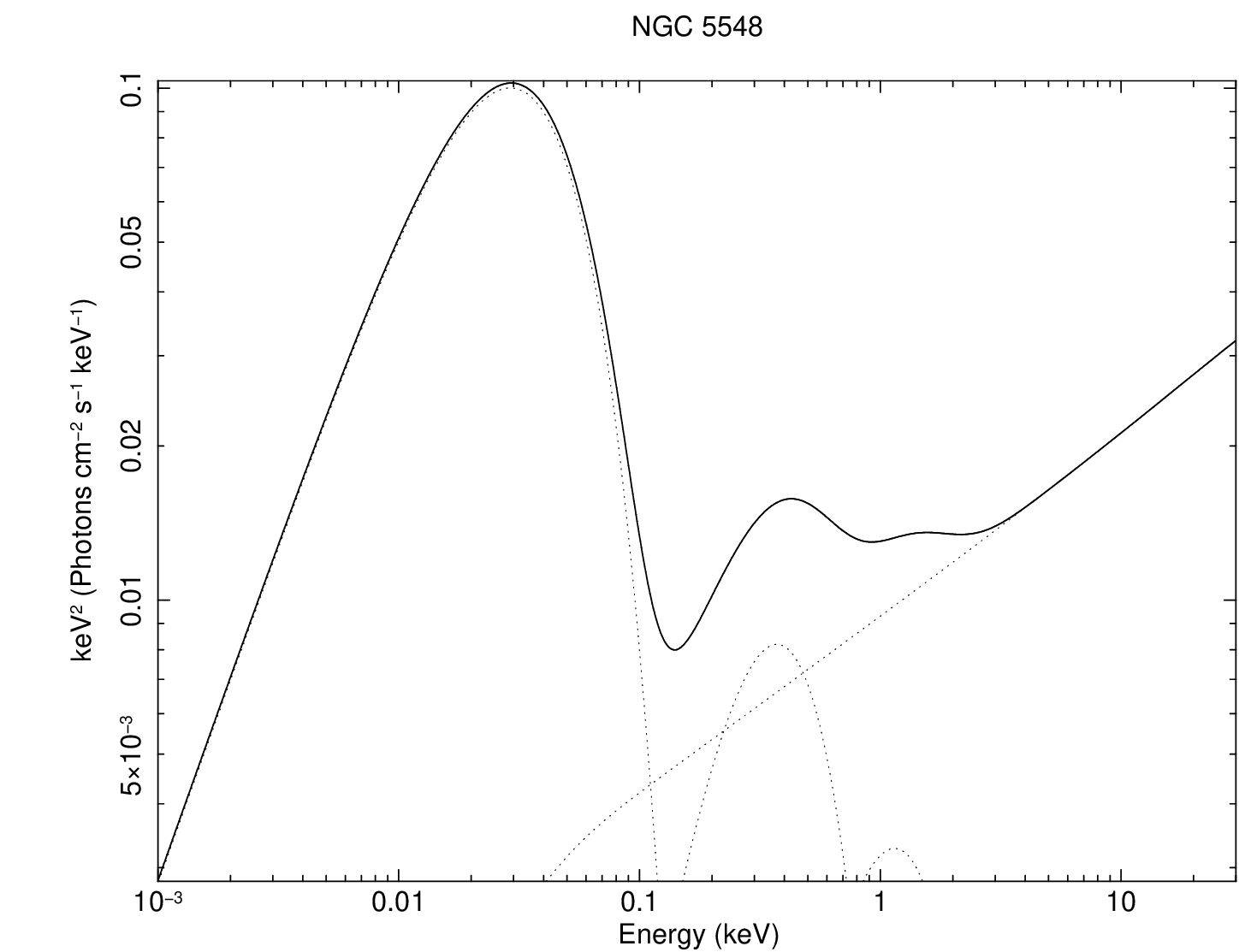} 
} 

}
\caption{Continued from Fig.~\ref{sed}}
 \label{sed_2}
\end{figure}


\clearpage

\begin{figure}
  \centering
  \vbox{
\hbox{
\includegraphics[width=4cm,angle=-90]{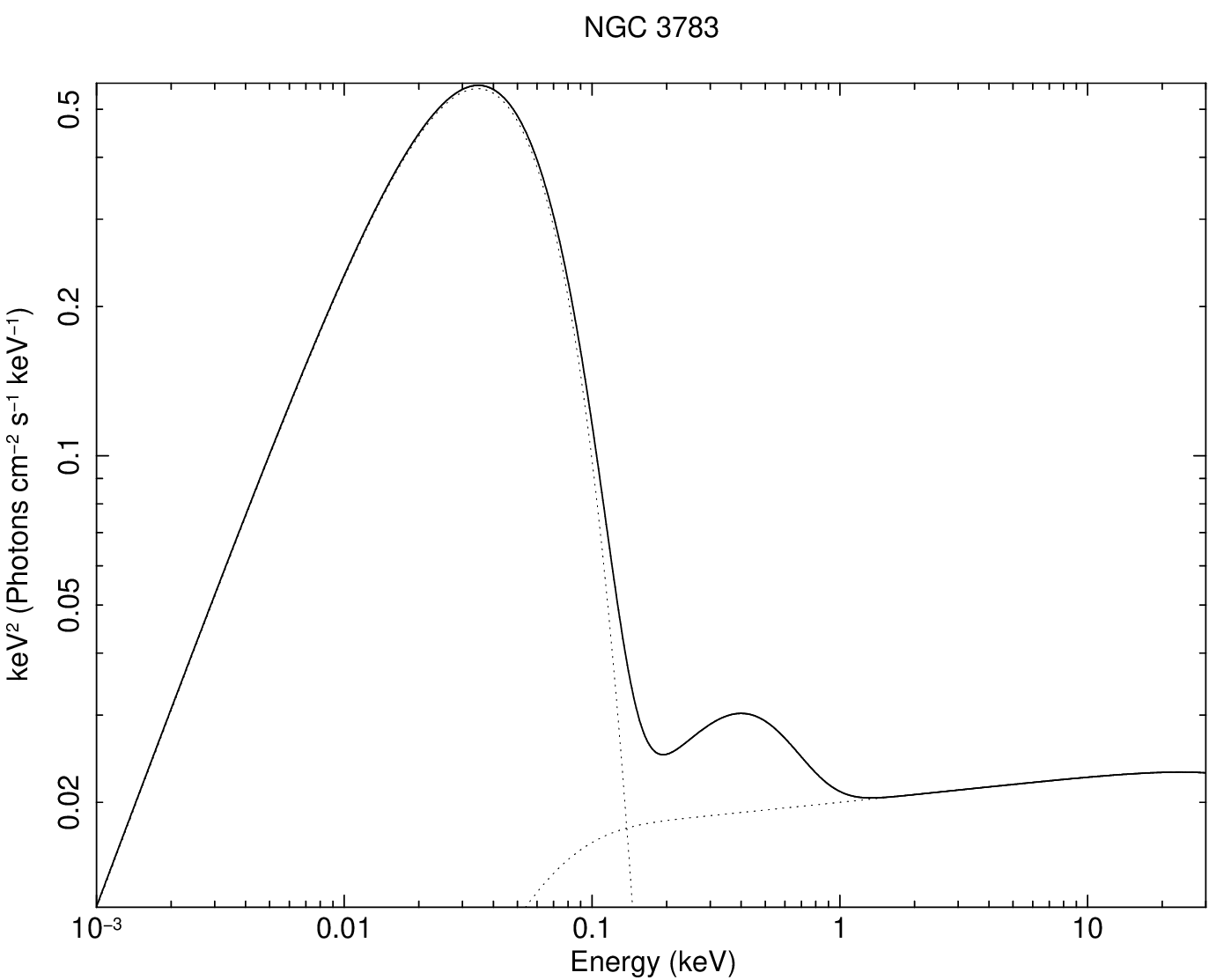} 
\includegraphics[width=4cm,angle=-90]{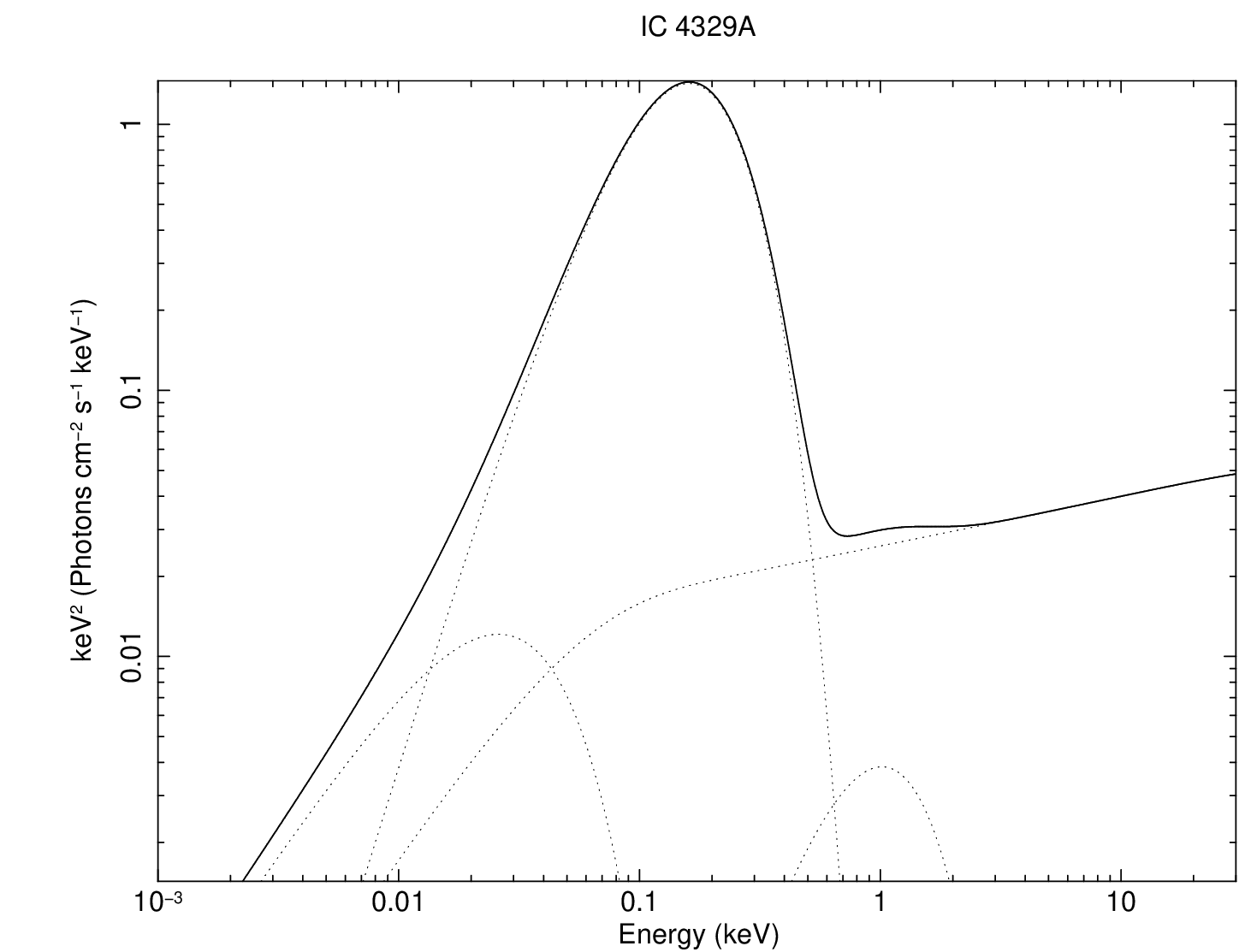} 
} 
}
\caption{Continued from Fig.~\ref{sed}}
 \label{sed_3}
\end{figure}



\clearpage

\begin{figure}
  \centering

  \vbox{
\hbox{
\includegraphics[width=4cm,angle=-90]{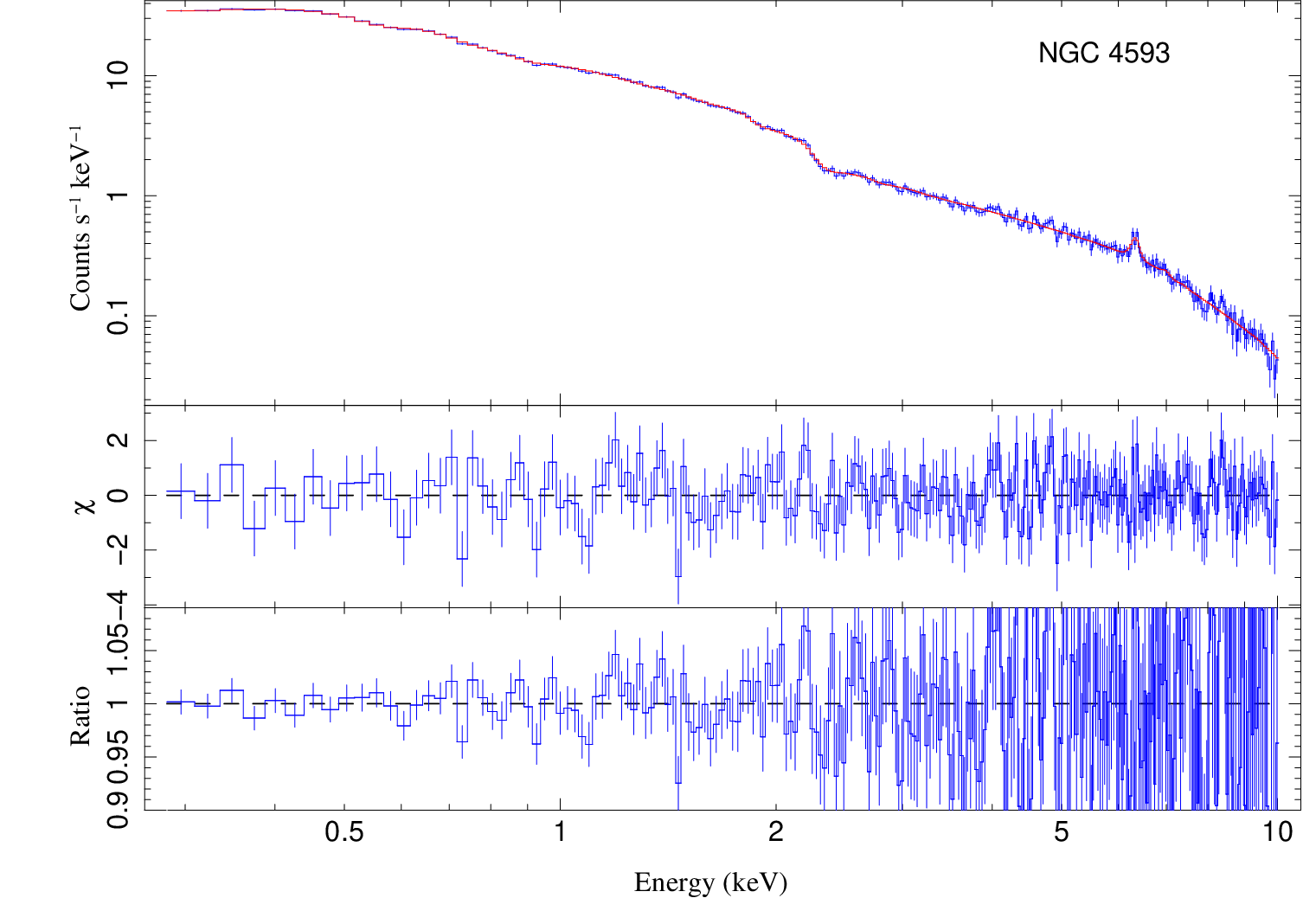}
\includegraphics[width=4cm,angle=-90]{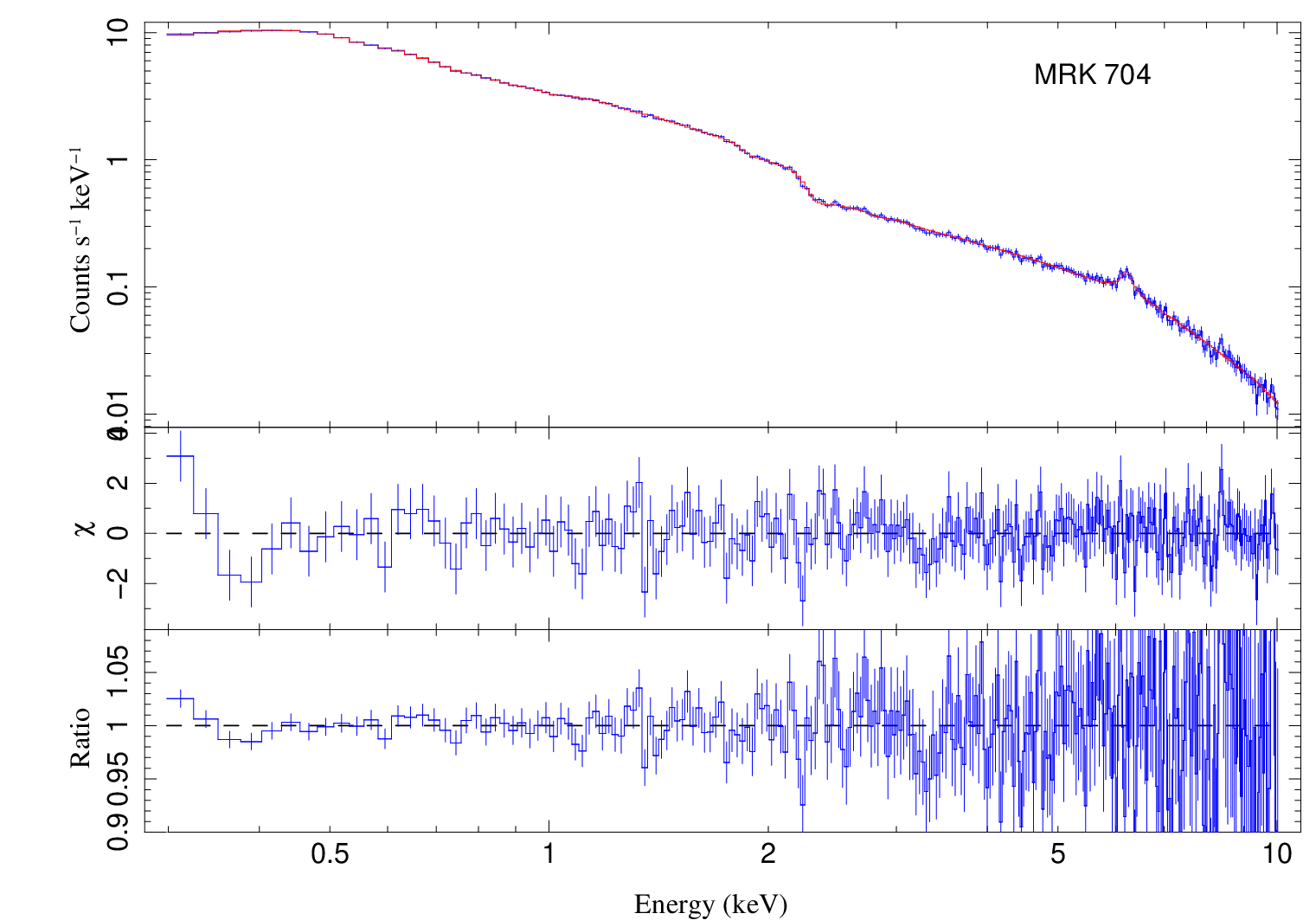}
\includegraphics[width=4cm,angle=-90]{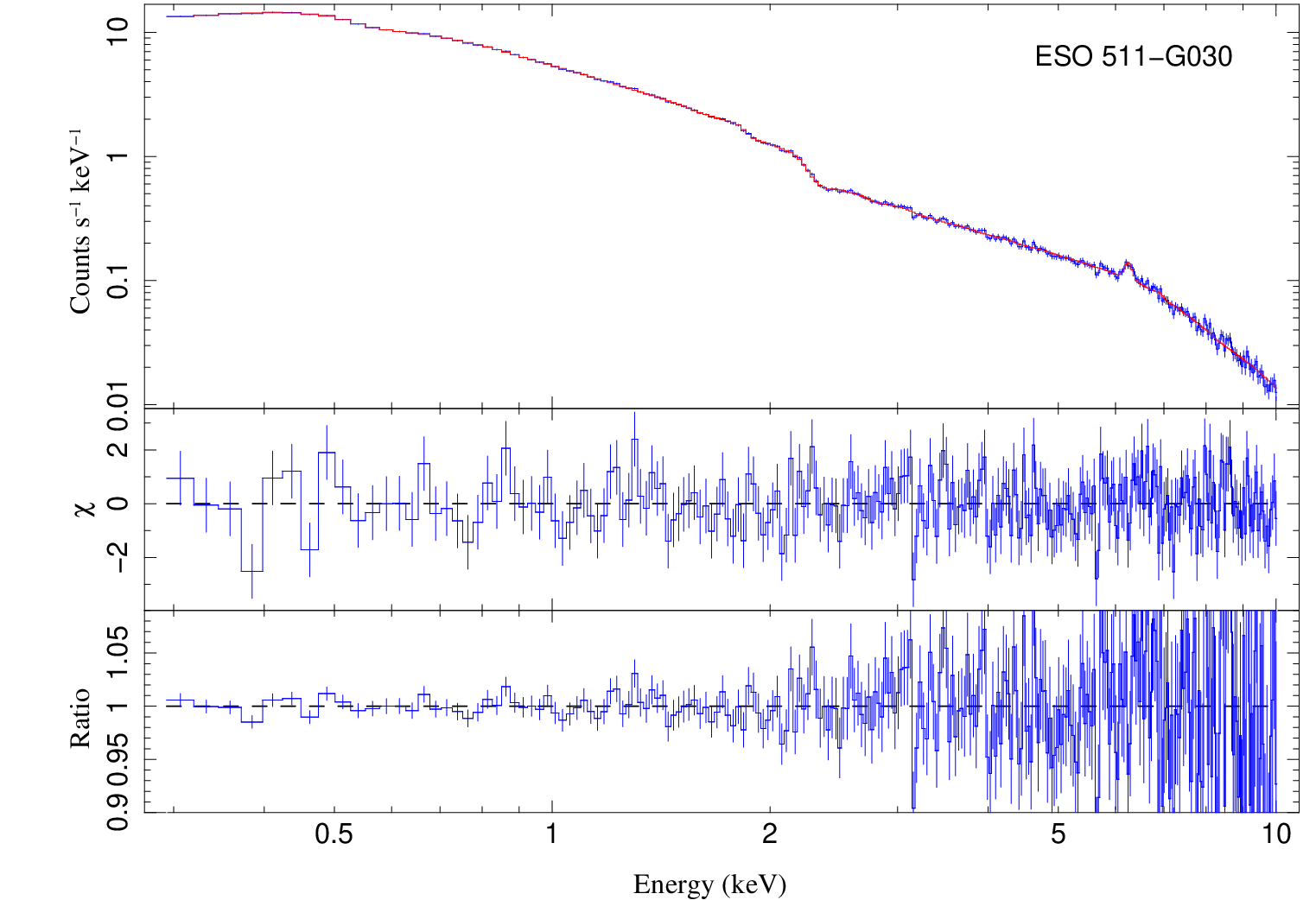}

} 

\hbox{
\includegraphics[width=4cm,angle=-90]{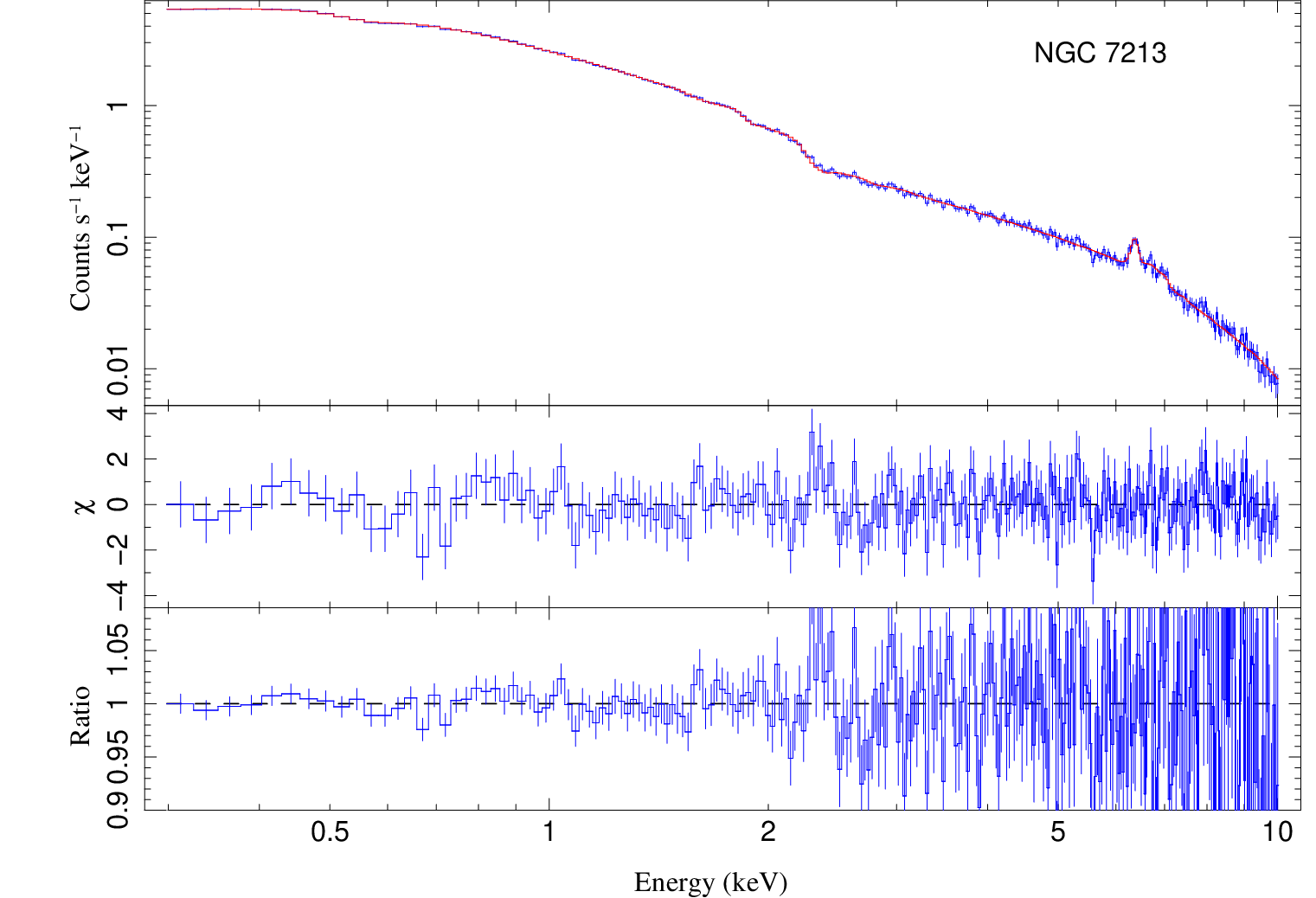}
\includegraphics[width=4cm,angle=-90]{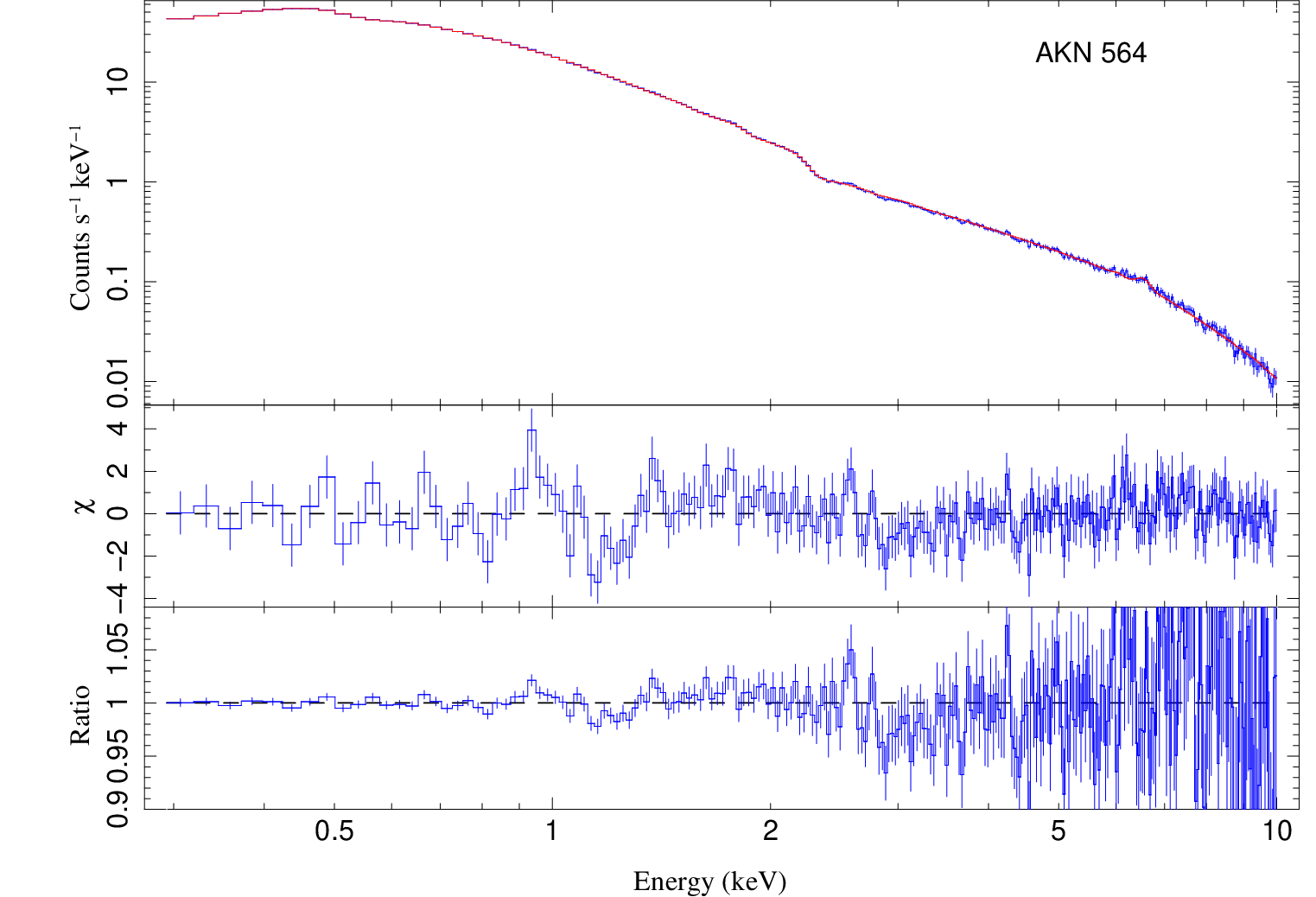}
\includegraphics[width=4cm,angle=-90]{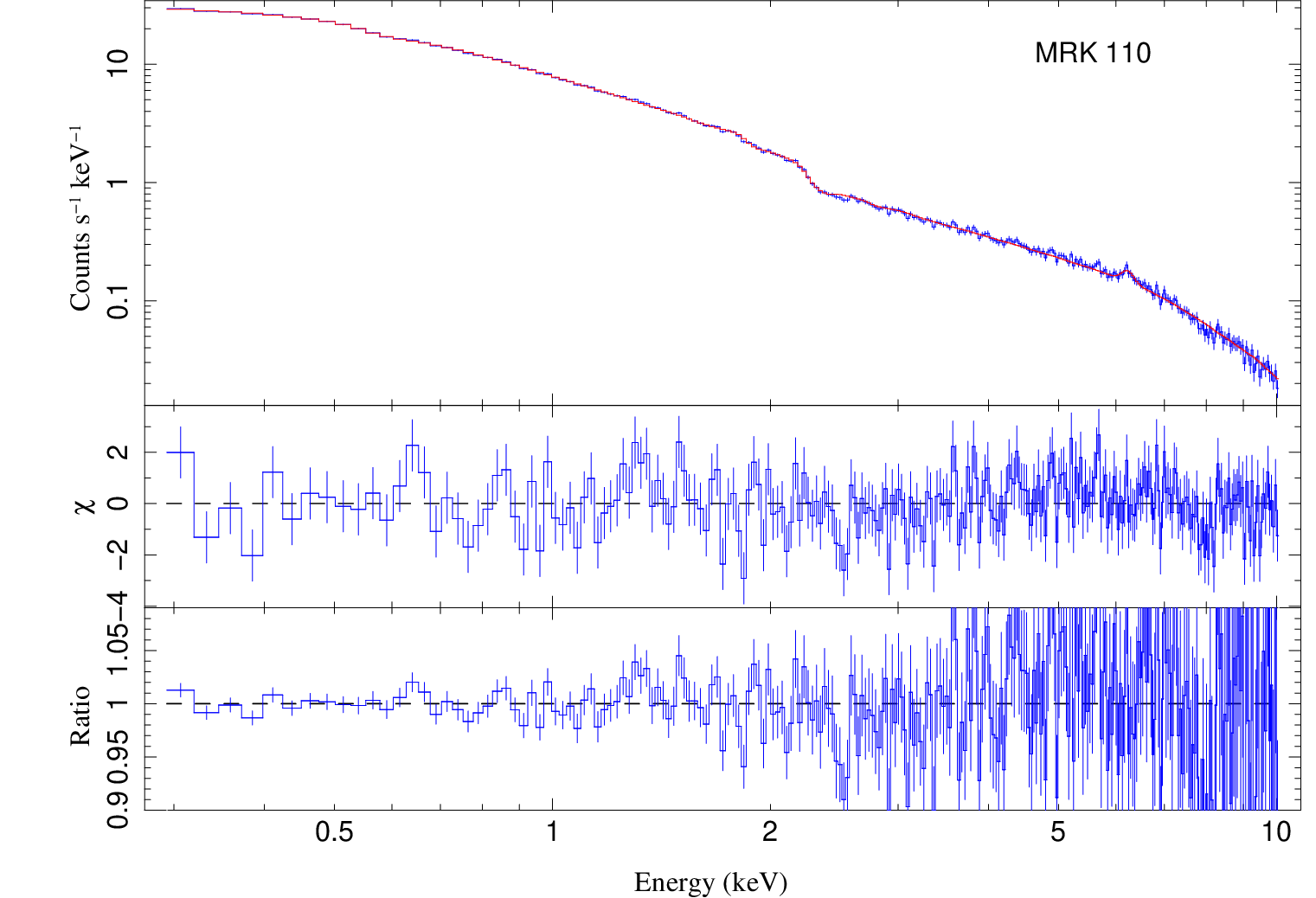}

} 

\hbox{
\includegraphics[width=4cm,angle=-90]{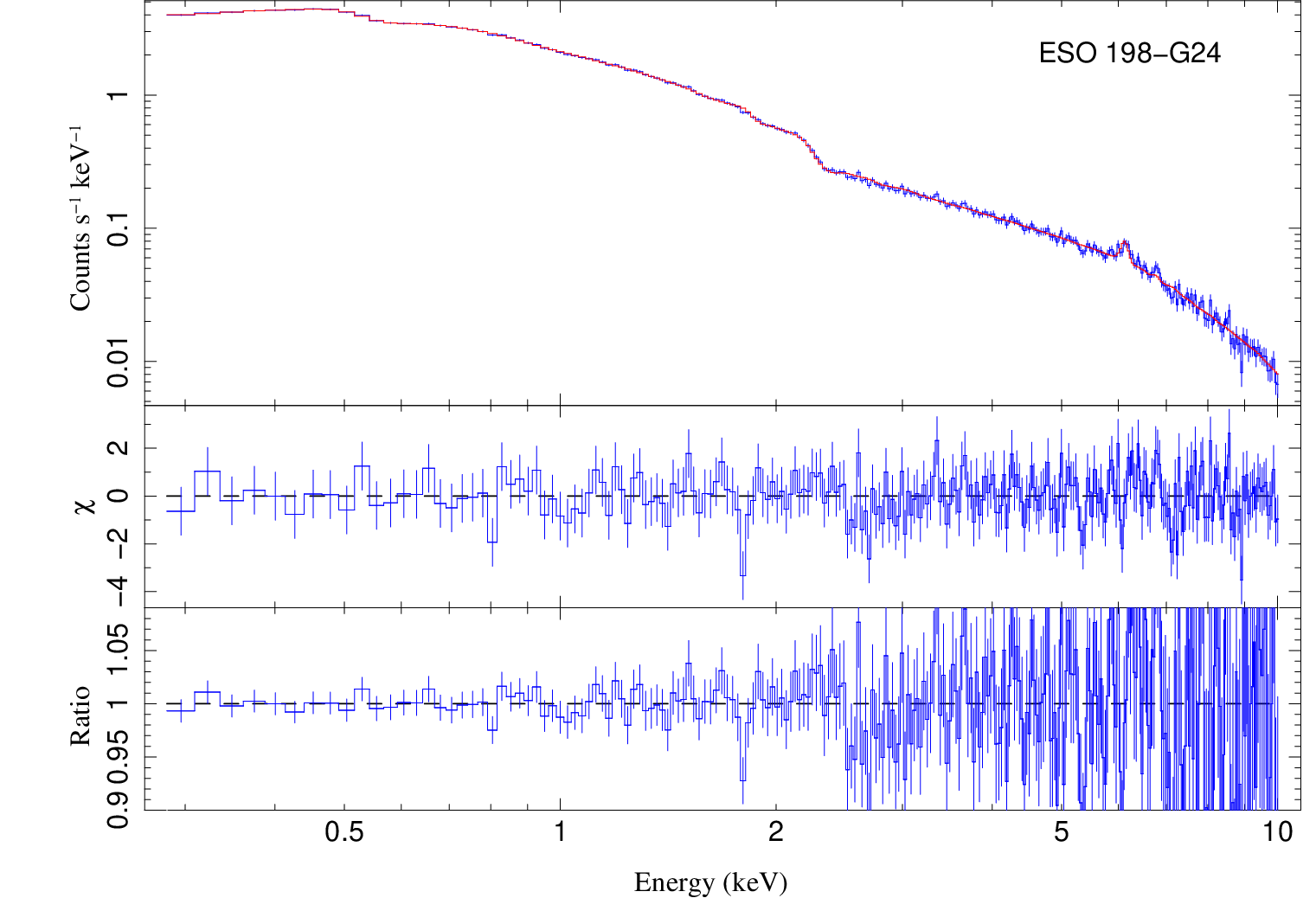}
\includegraphics[width=4cm,angle=-90]{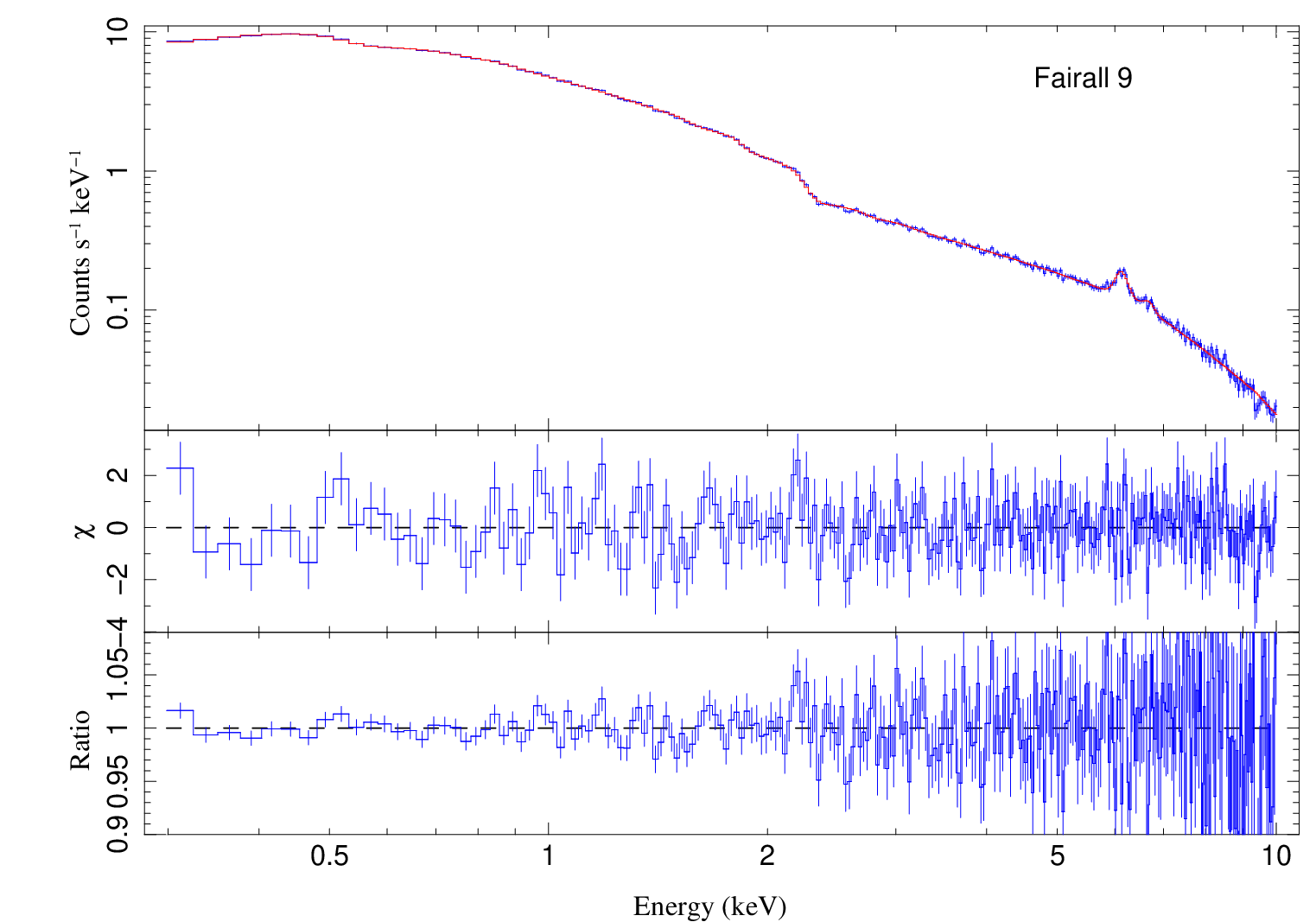}
\includegraphics[width=4cm,angle=-90]{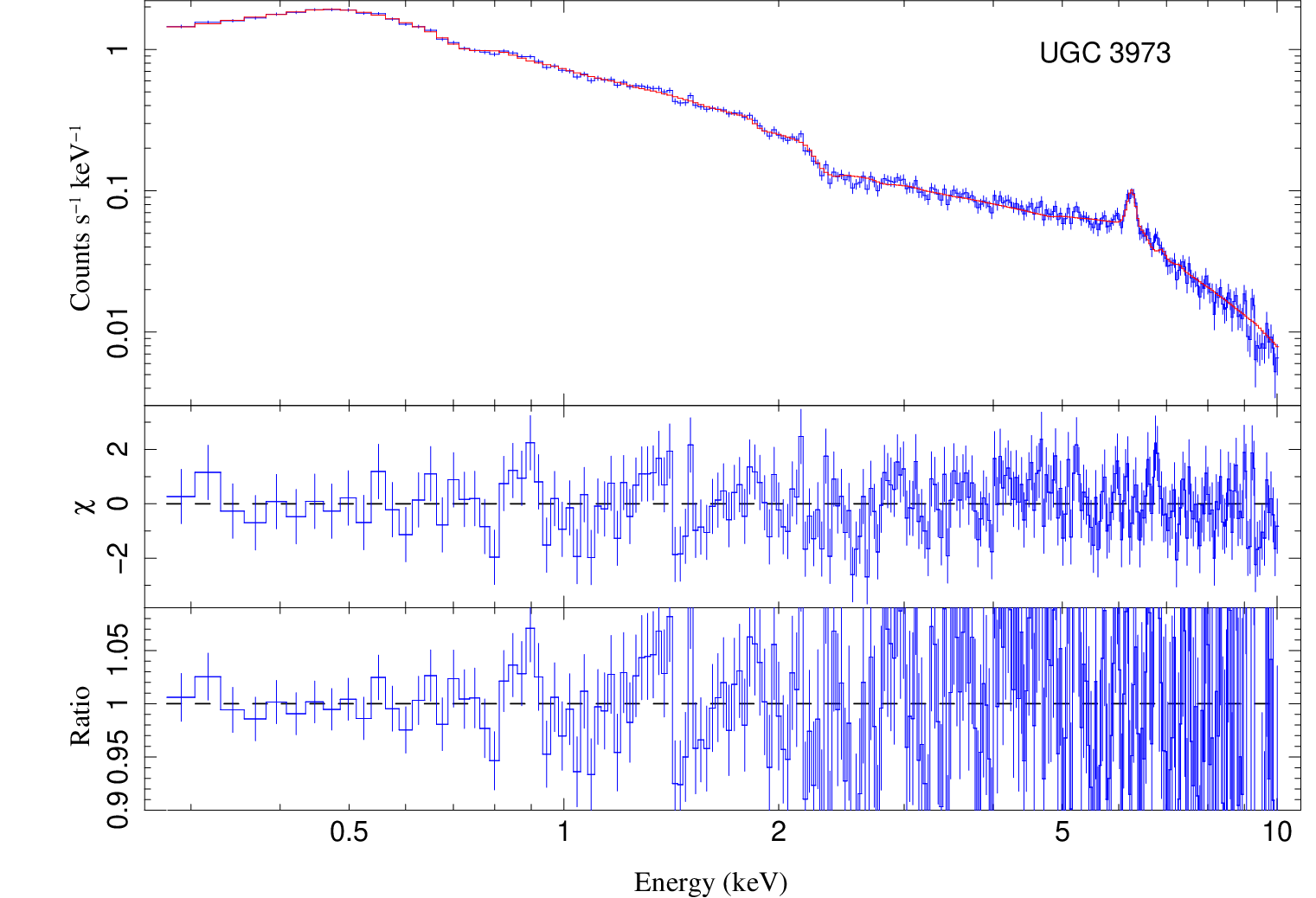}
} 
\hbox{
\includegraphics[width=4cm,angle=-90]{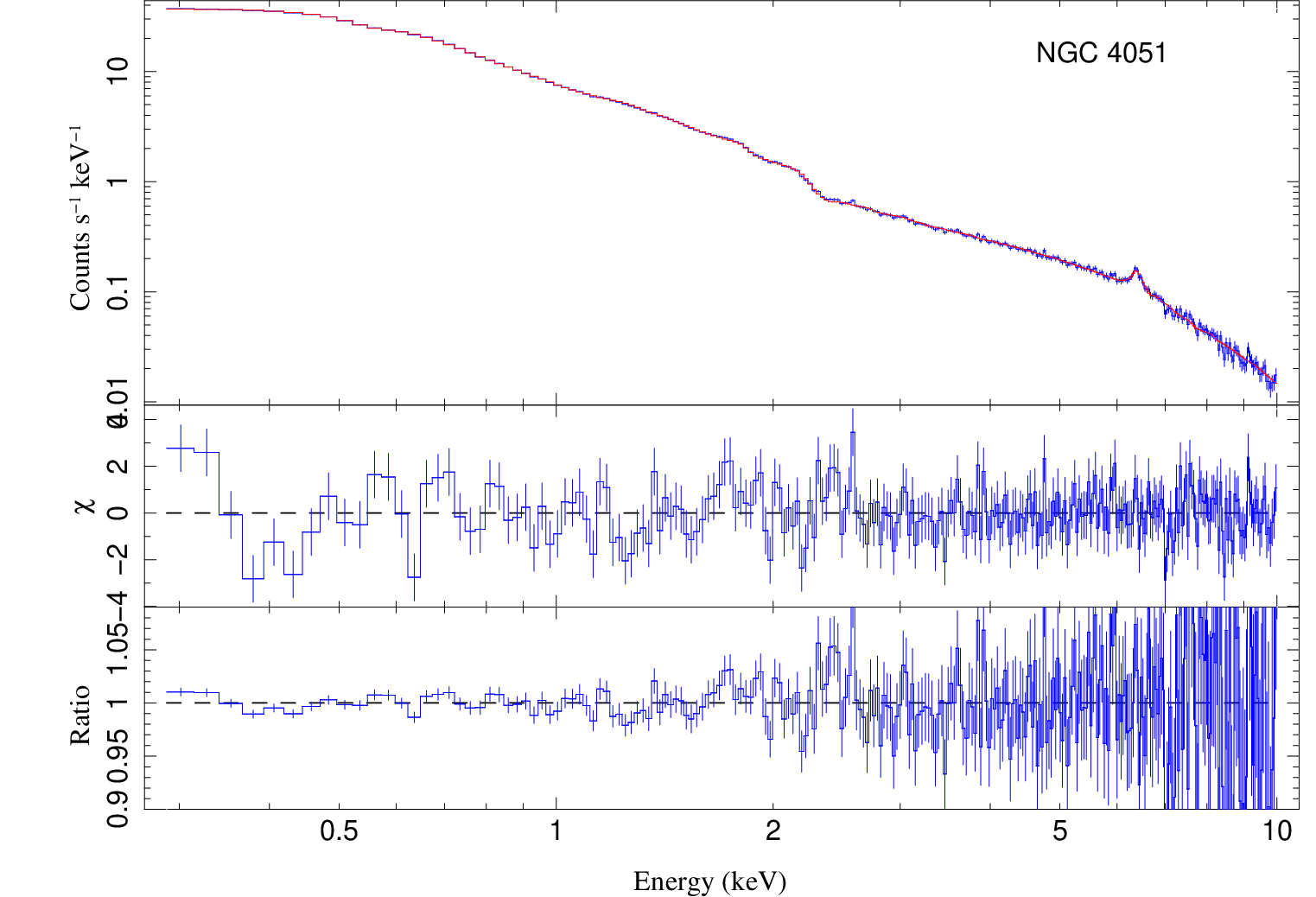}
\includegraphics[width=4cm,angle=-90]{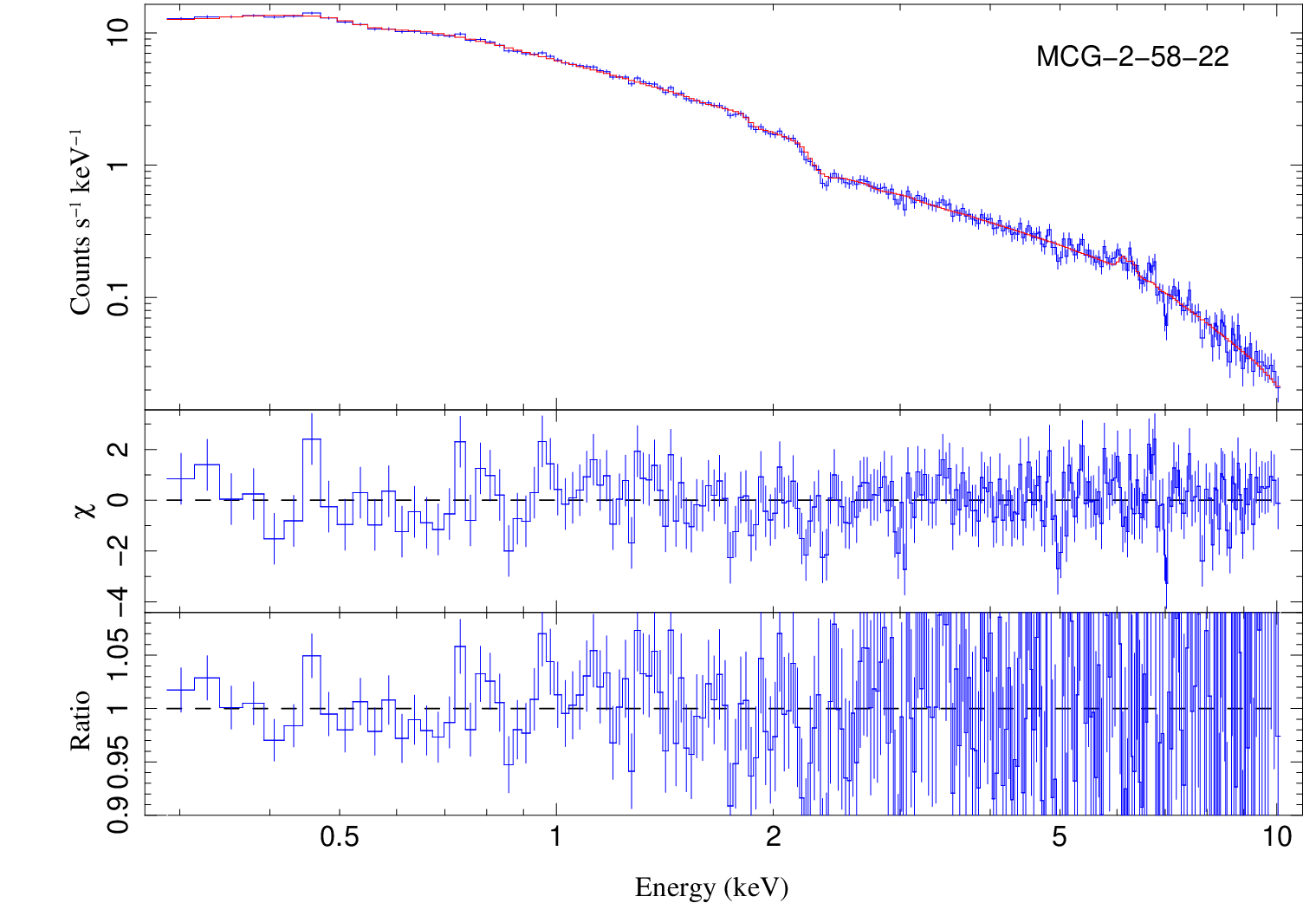}
\includegraphics[width=4cm,angle=-90]{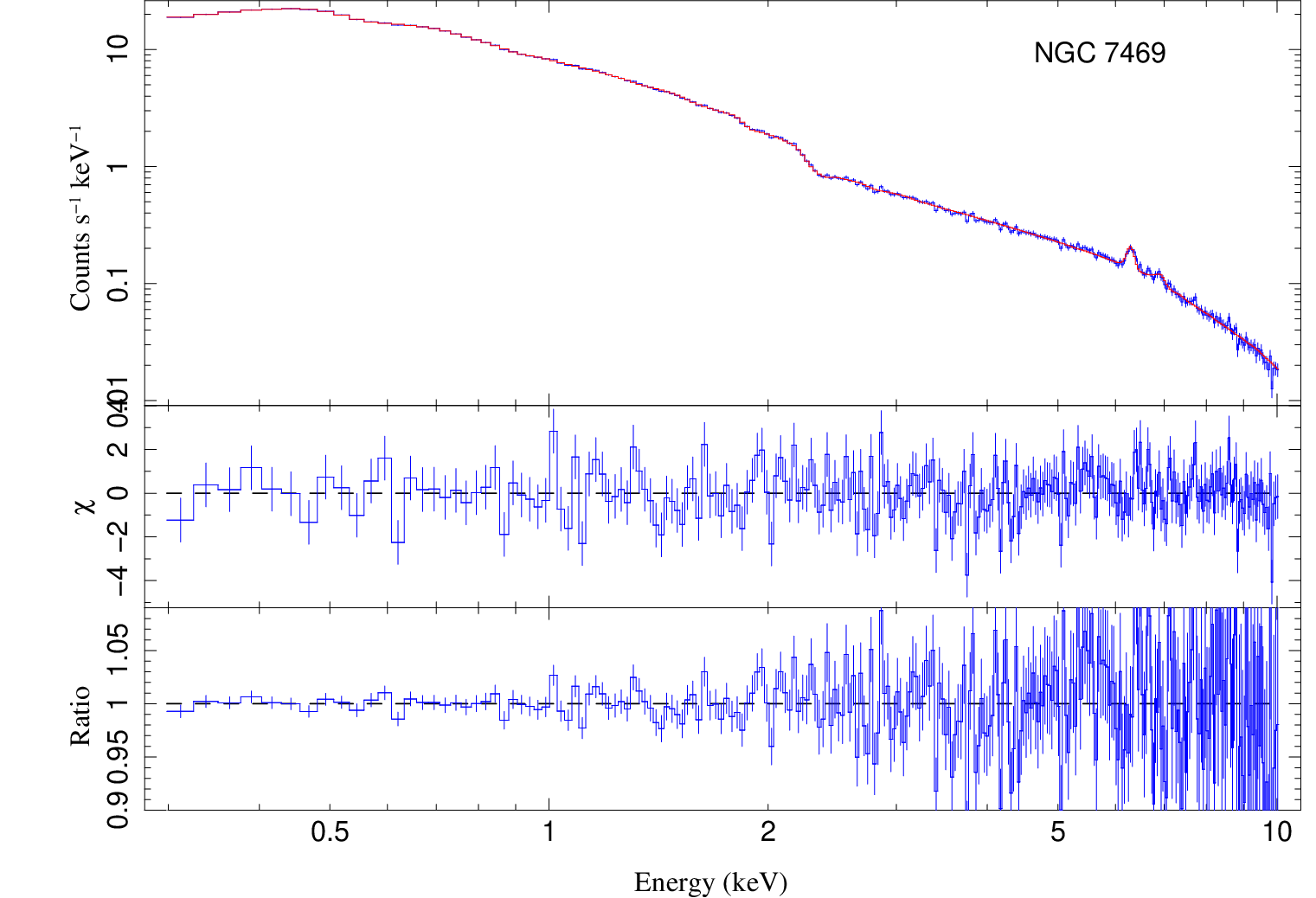}
} 

\hbox{
\includegraphics[width=4cm,angle=-90]{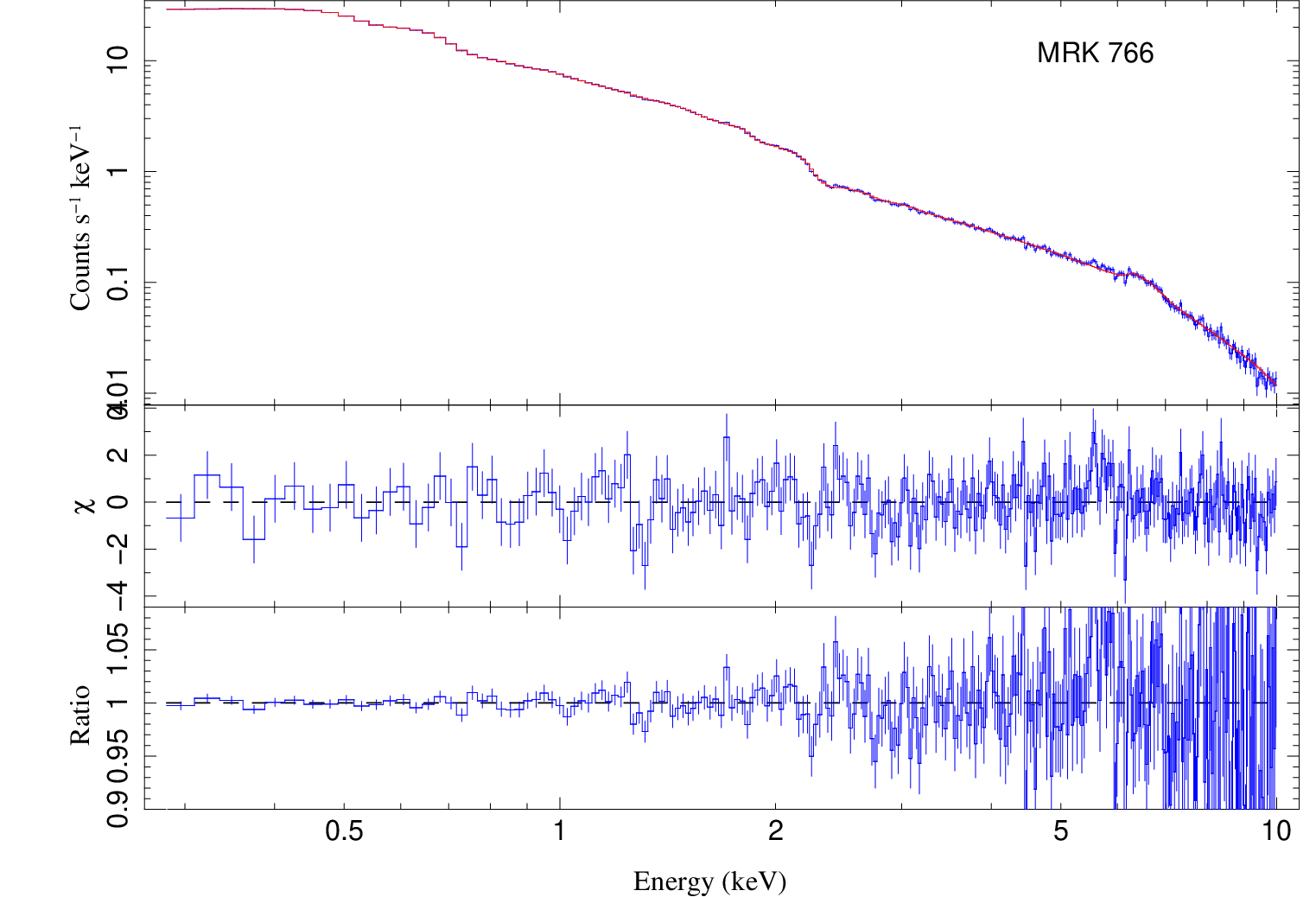}
\includegraphics[width=4cm,angle=-90]{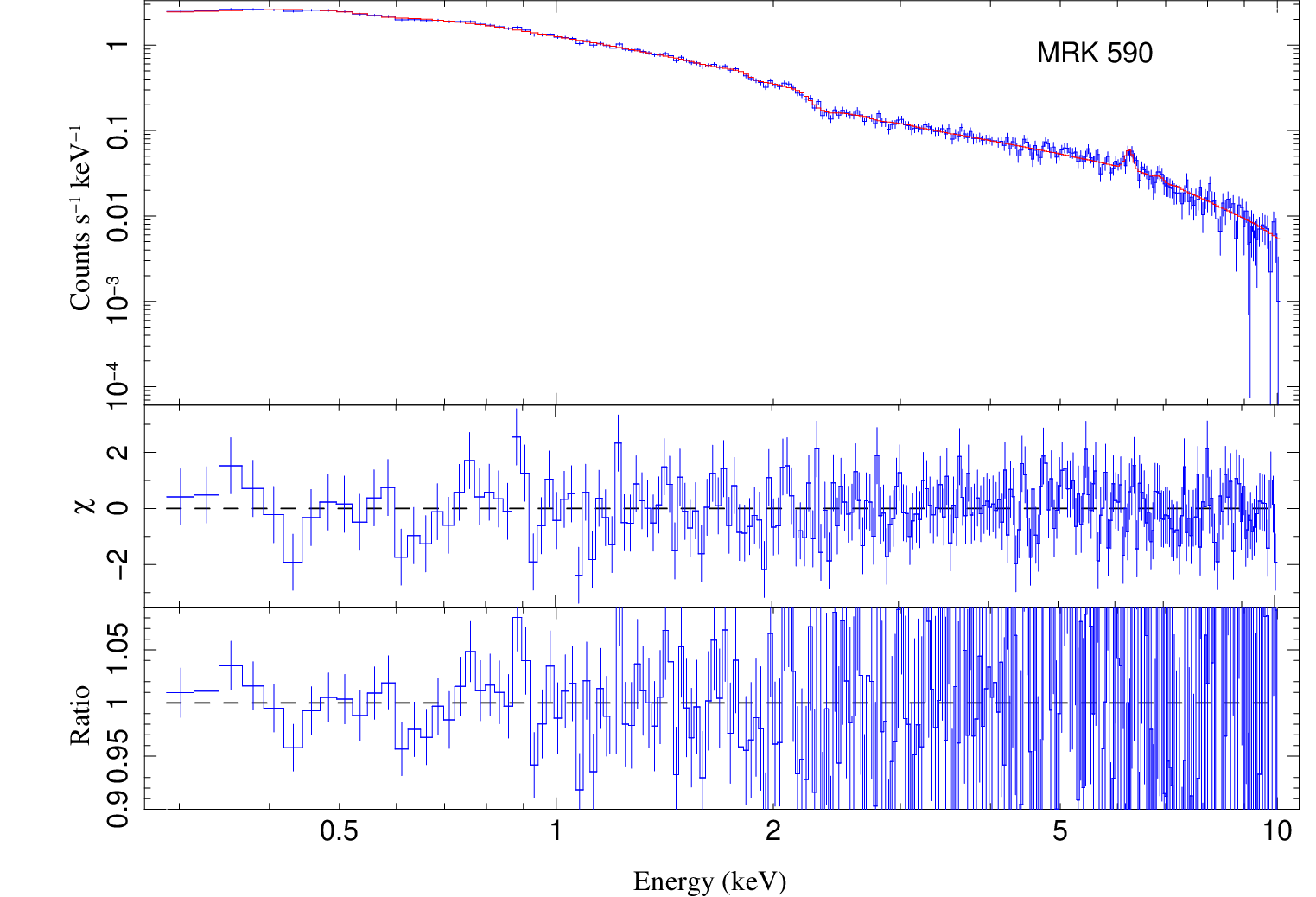}
\includegraphics[width=4cm,angle=-90]{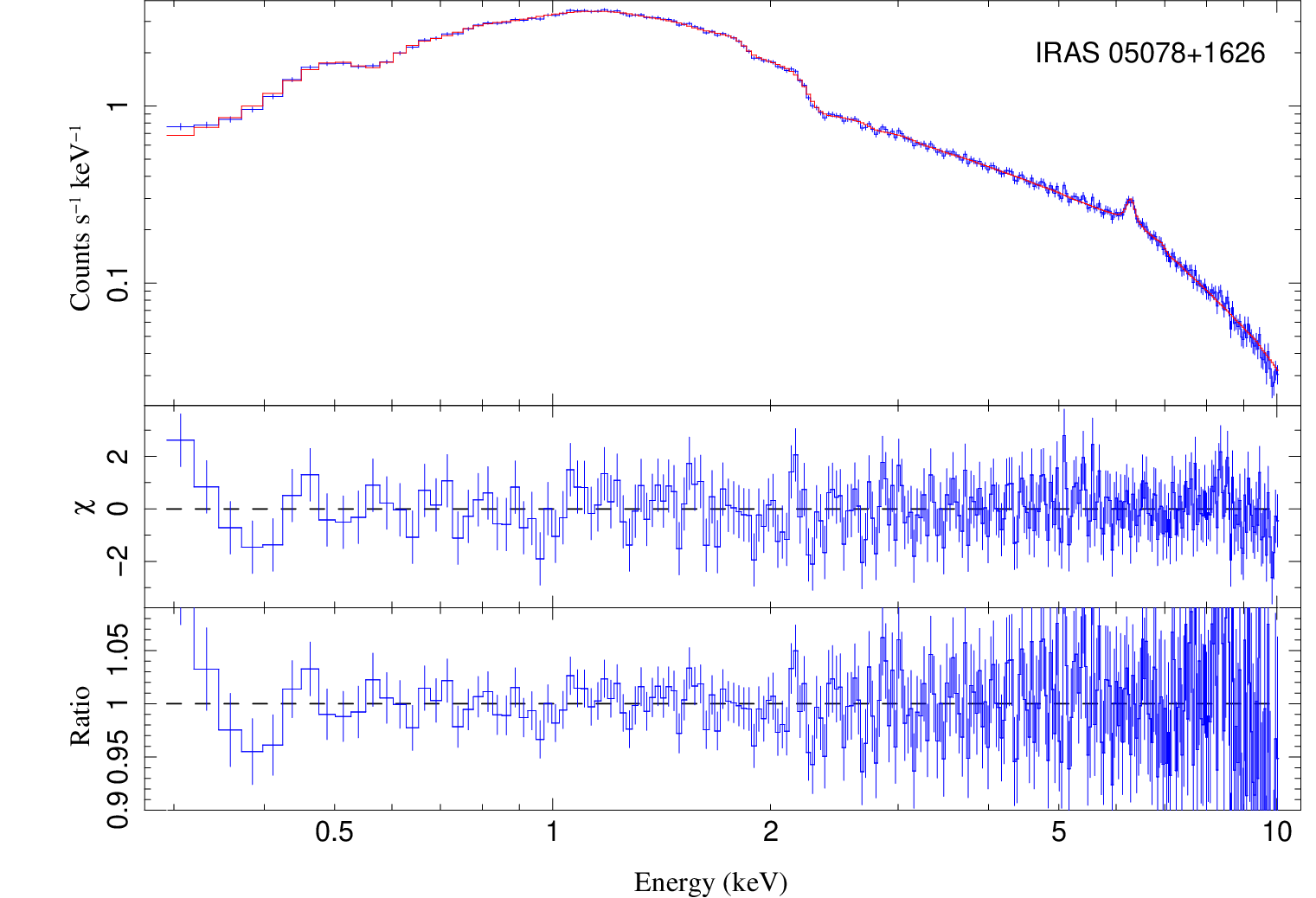}
}

}

\caption{The best fit EPIC-pn data and the model plotted in red. The lower two panels are the residuals between the data and the best-fit model expressed as $\Delta\chi=(data-model)/\sigma$ and the data to model ratio respectively.}
 \label{epicpn}
\end{figure}


\clearpage

\begin{figure}
  \centering

  \vbox{
\hbox{
\includegraphics[width=4cm,angle=-90]{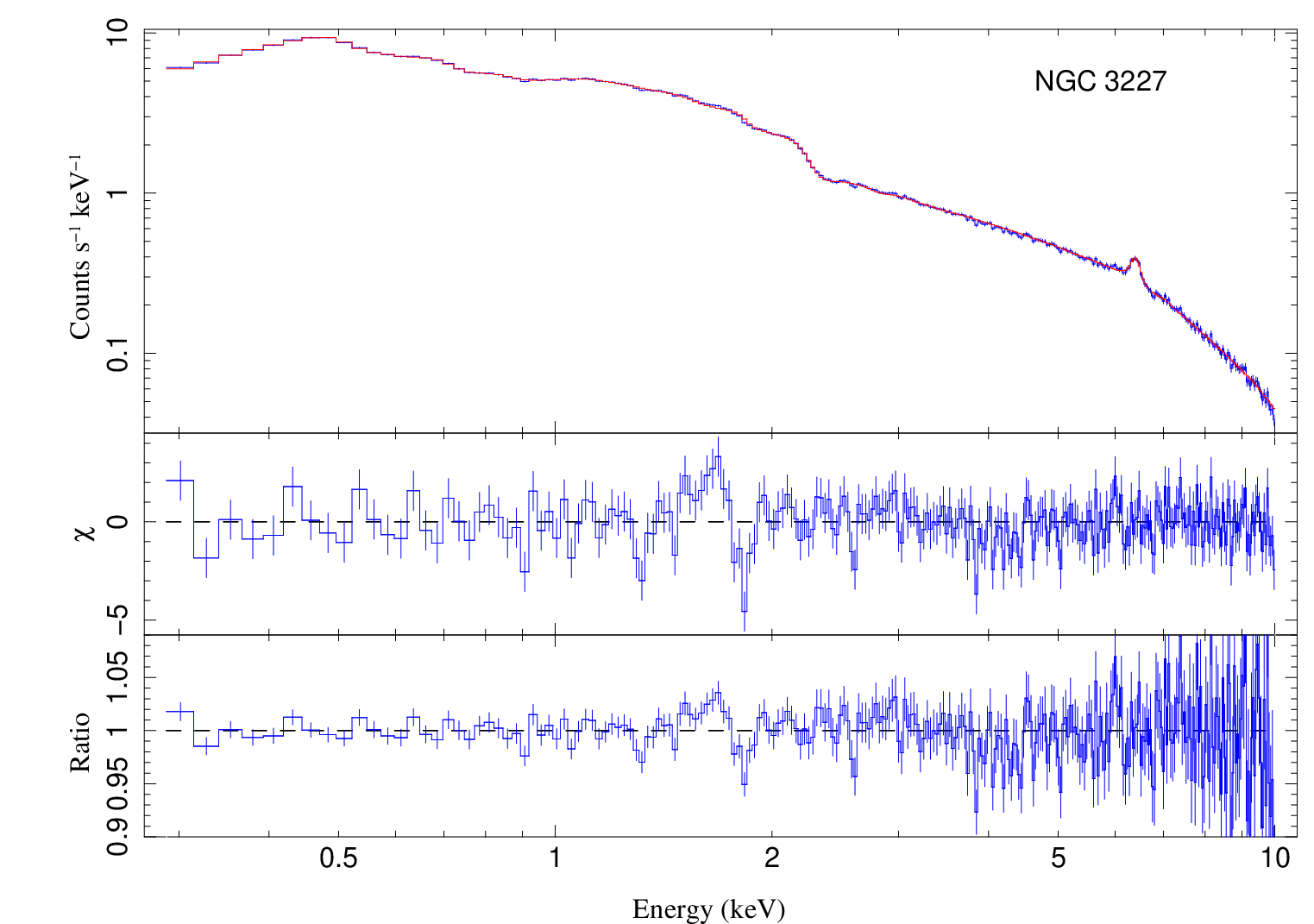}
\includegraphics[width=4cm,angle=-90]{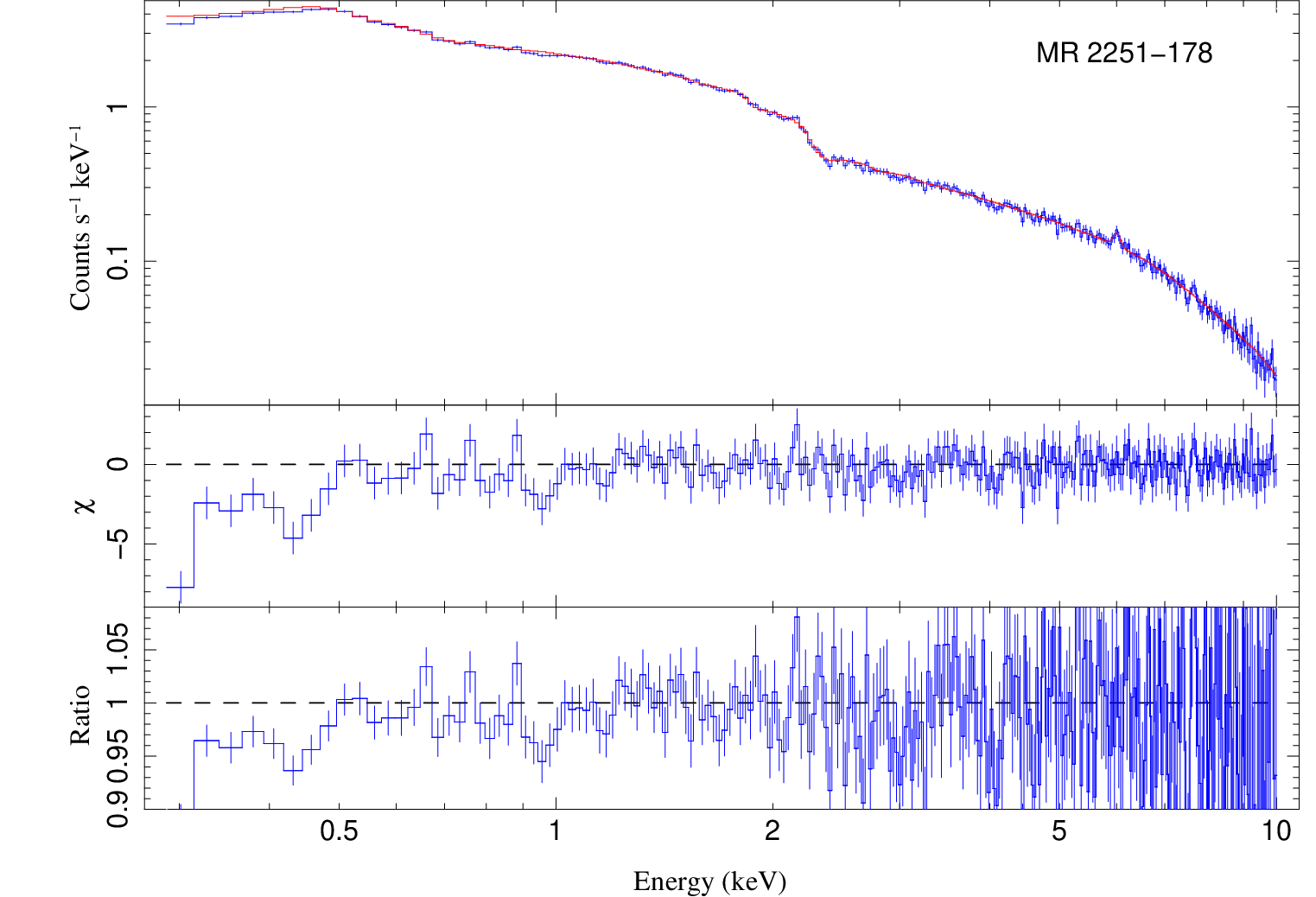}
\includegraphics[width=4cm,angle=-90]{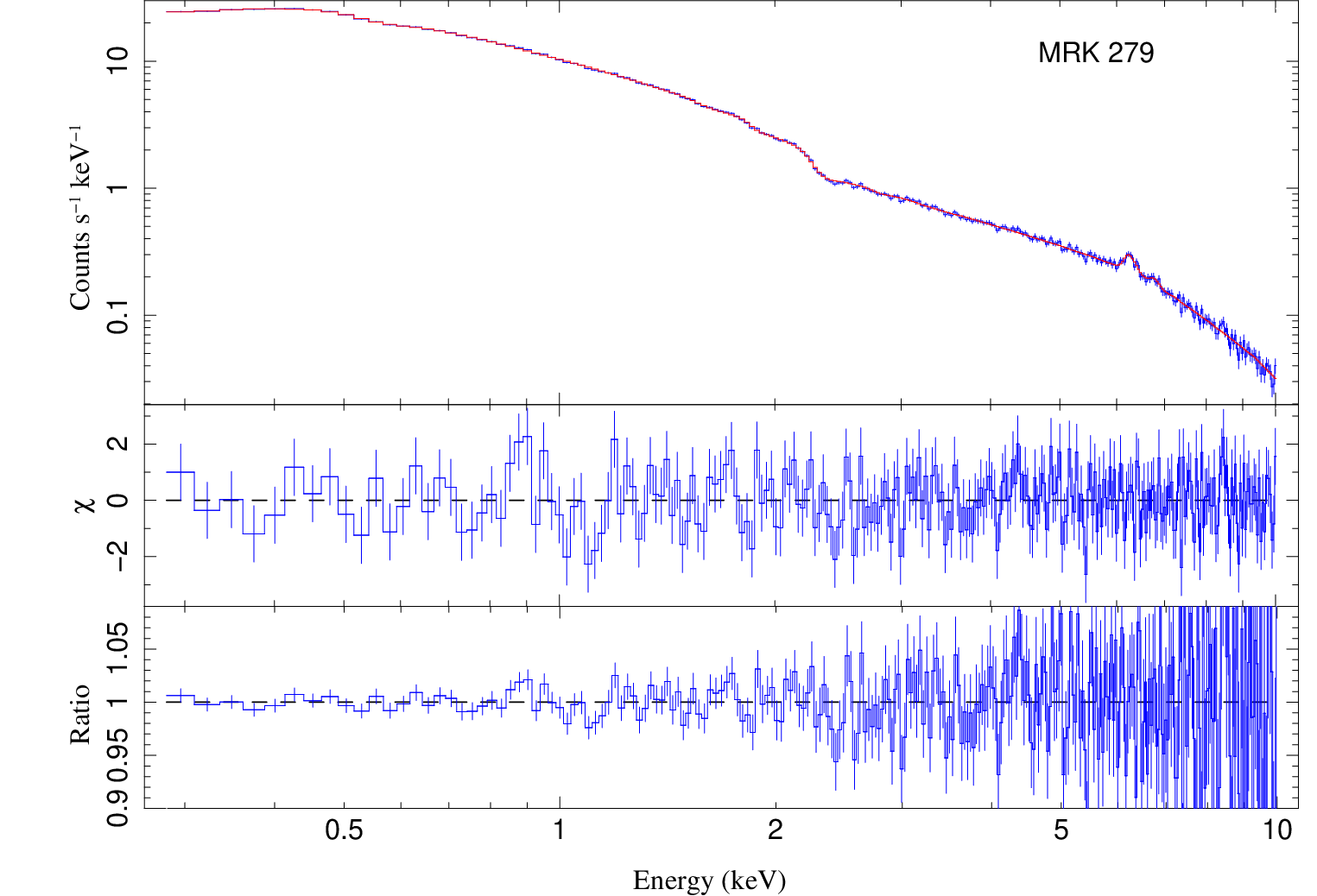}

} 

\hbox{
\includegraphics[width=4cm,angle=-90]{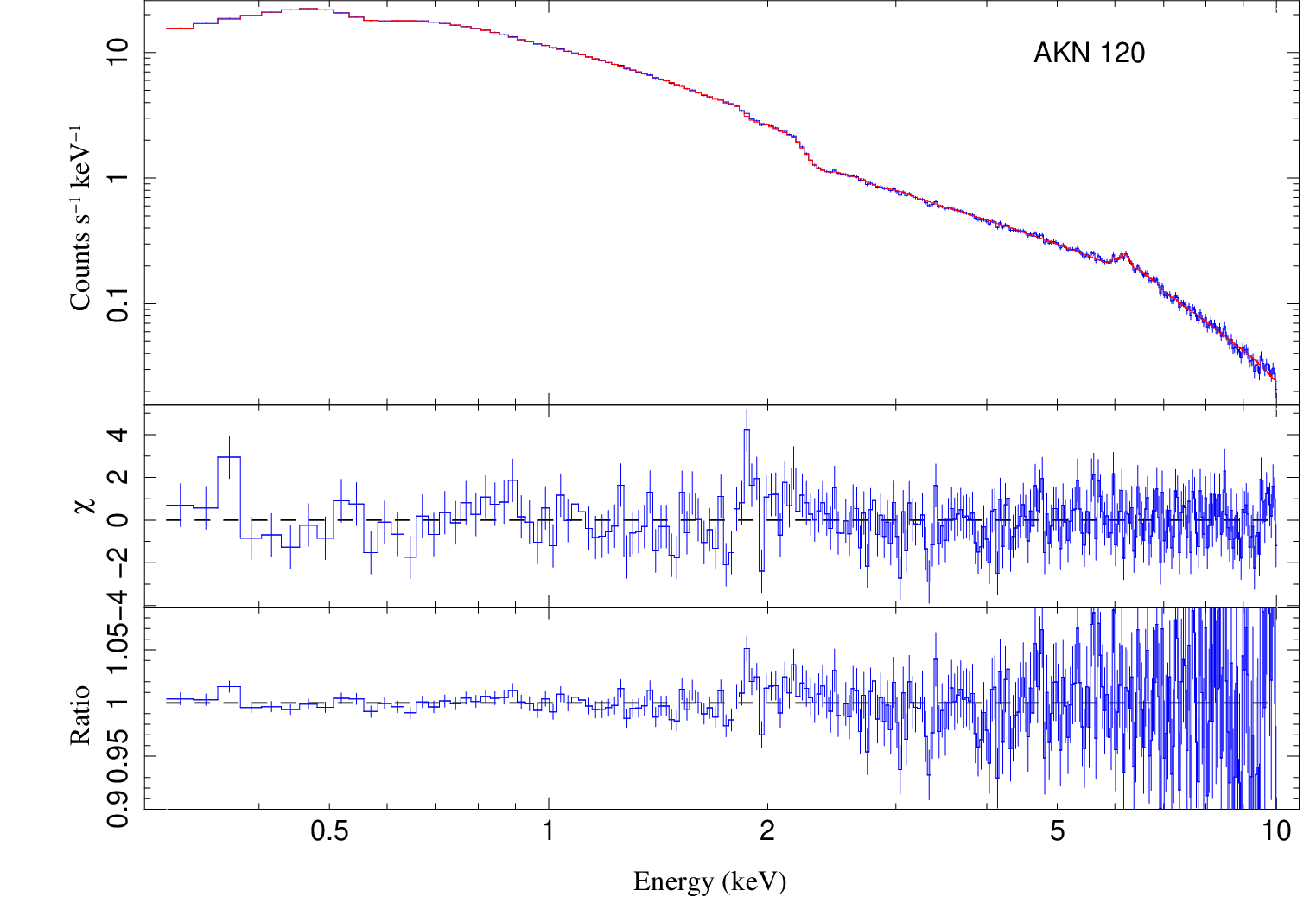}
\includegraphics[width=4cm,angle=-90]{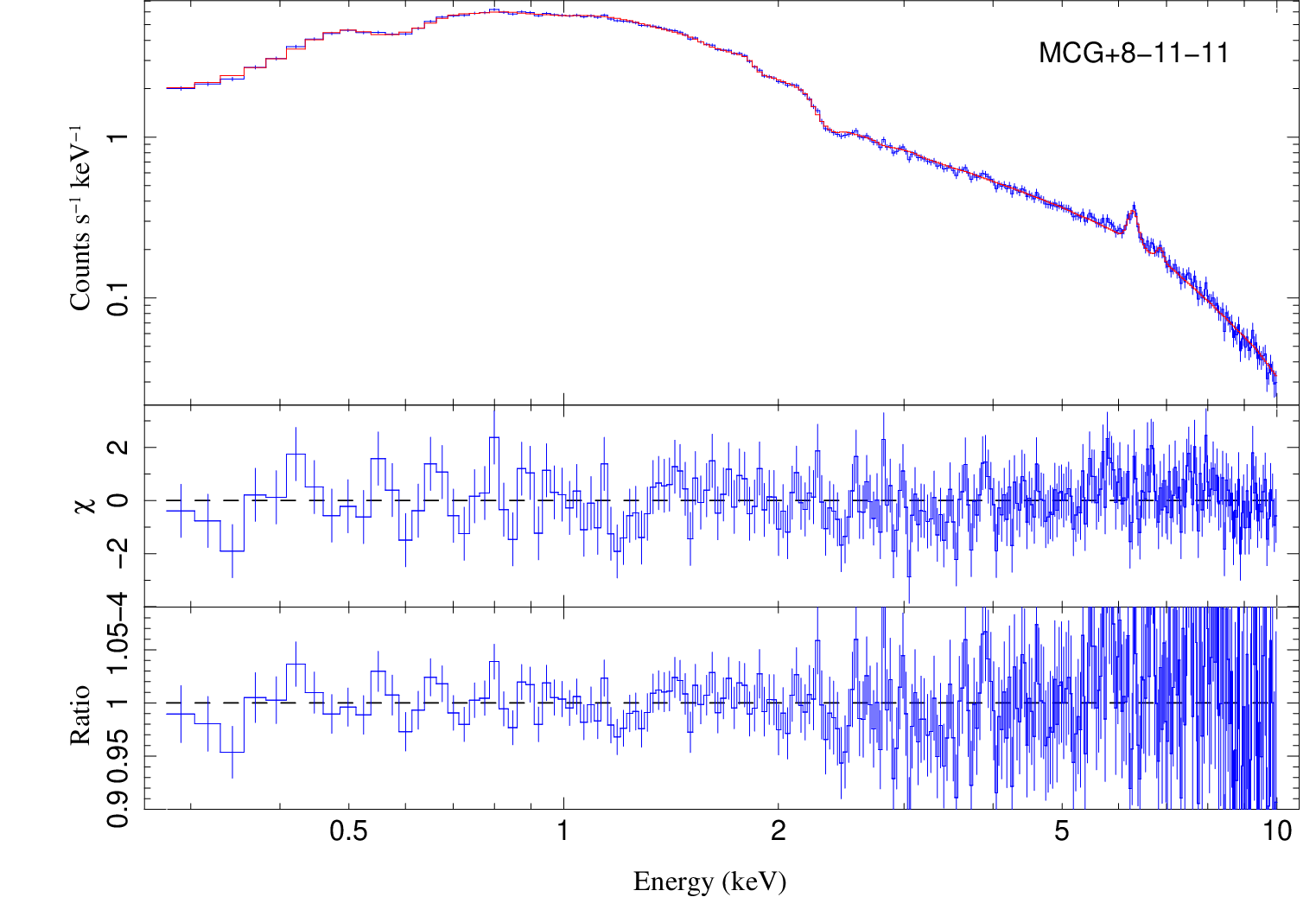}
\includegraphics[width=4cm,angle=-90]{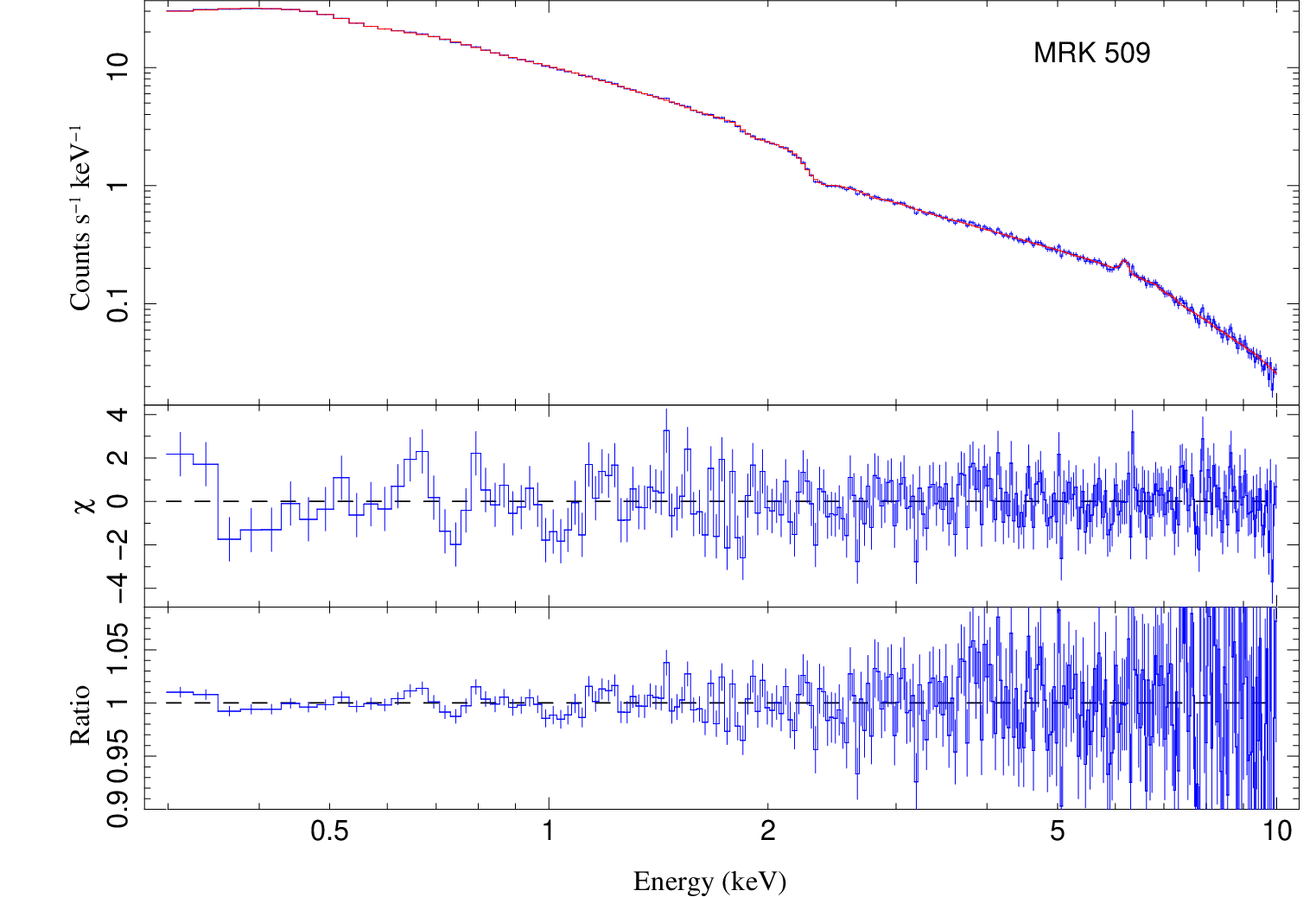}

} 

\hbox{
\includegraphics[width=4cm,angle=-90]{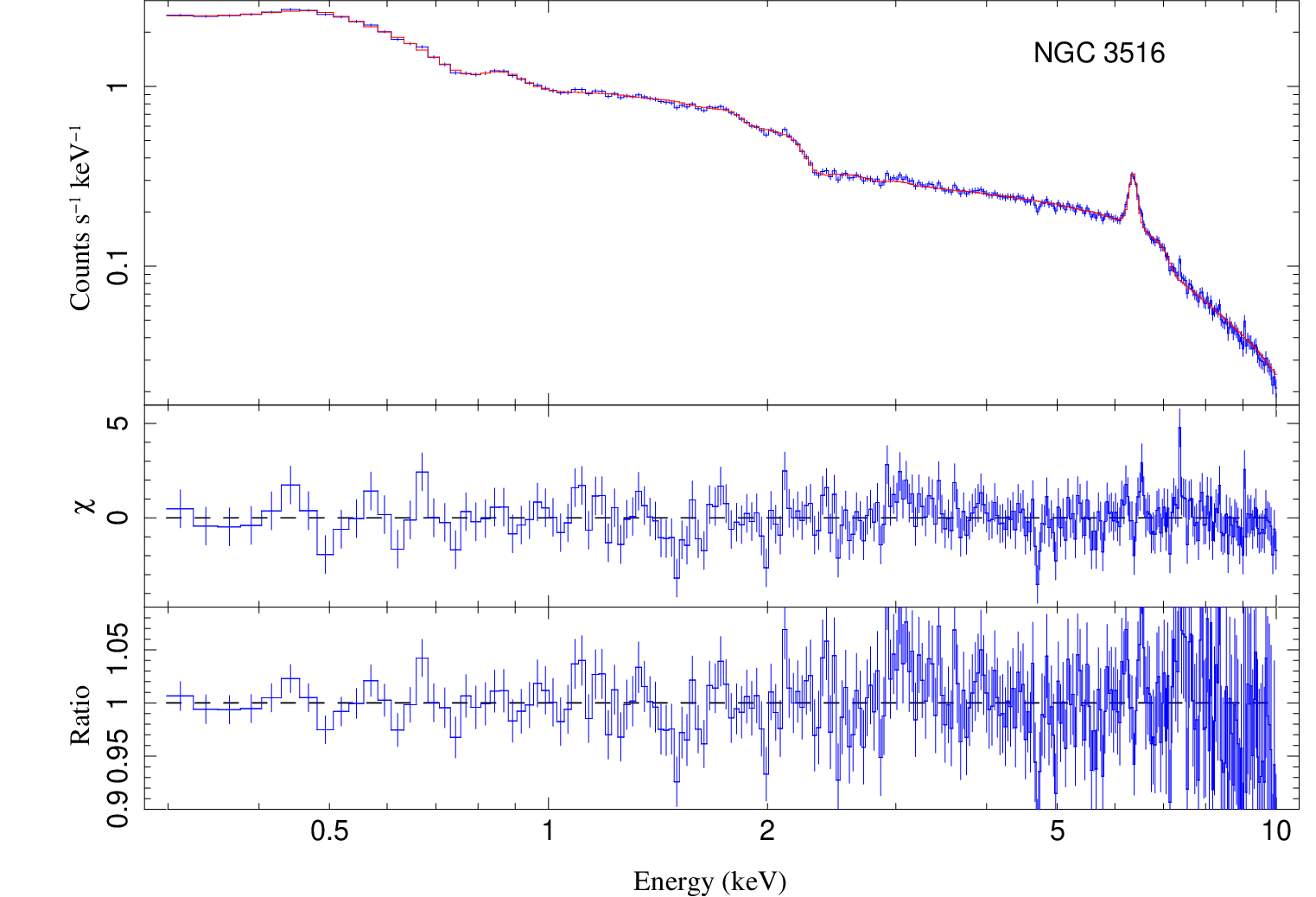}
\includegraphics[width=4cm,angle=-90]{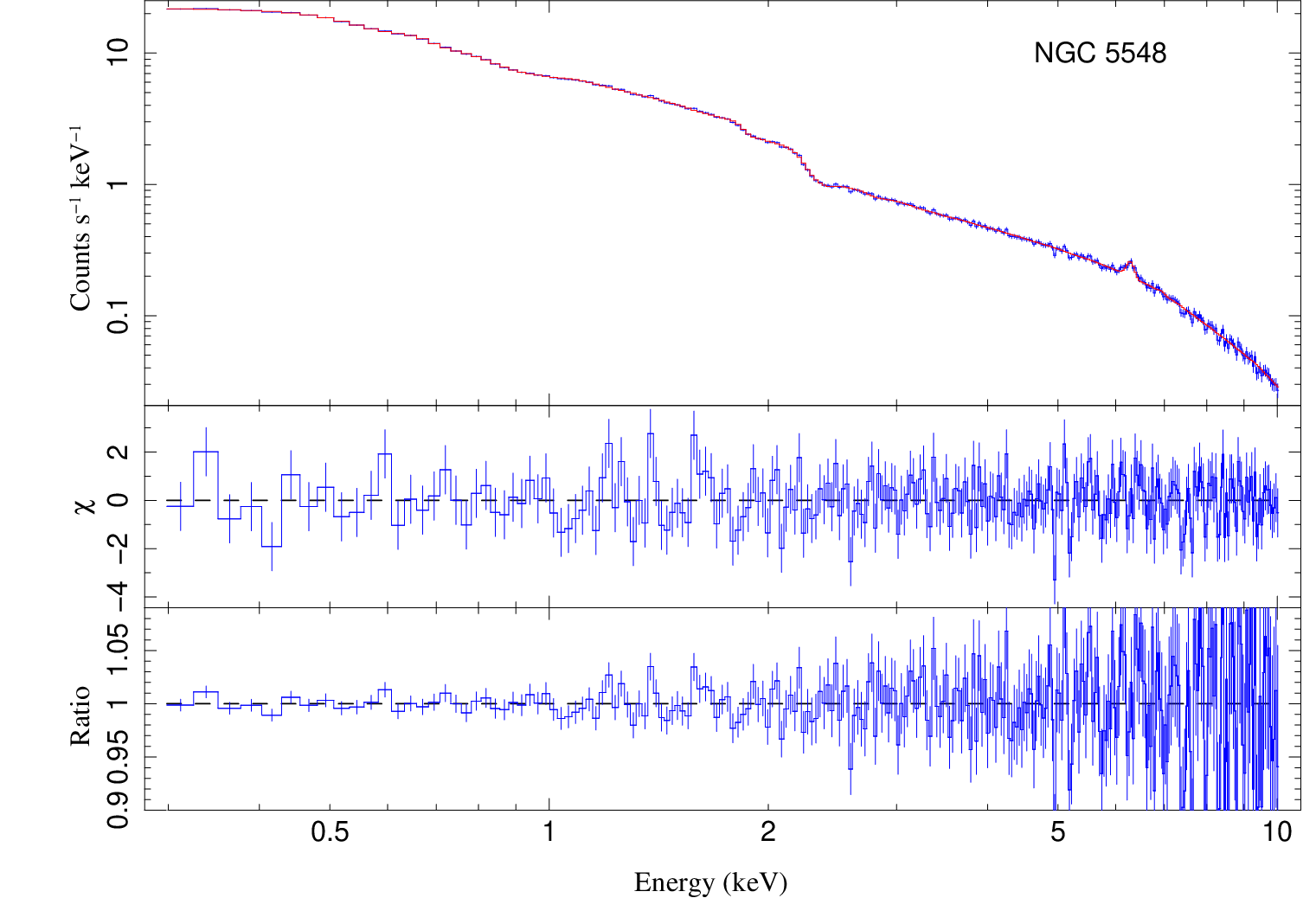}
\includegraphics[width=4cm,angle=-90]{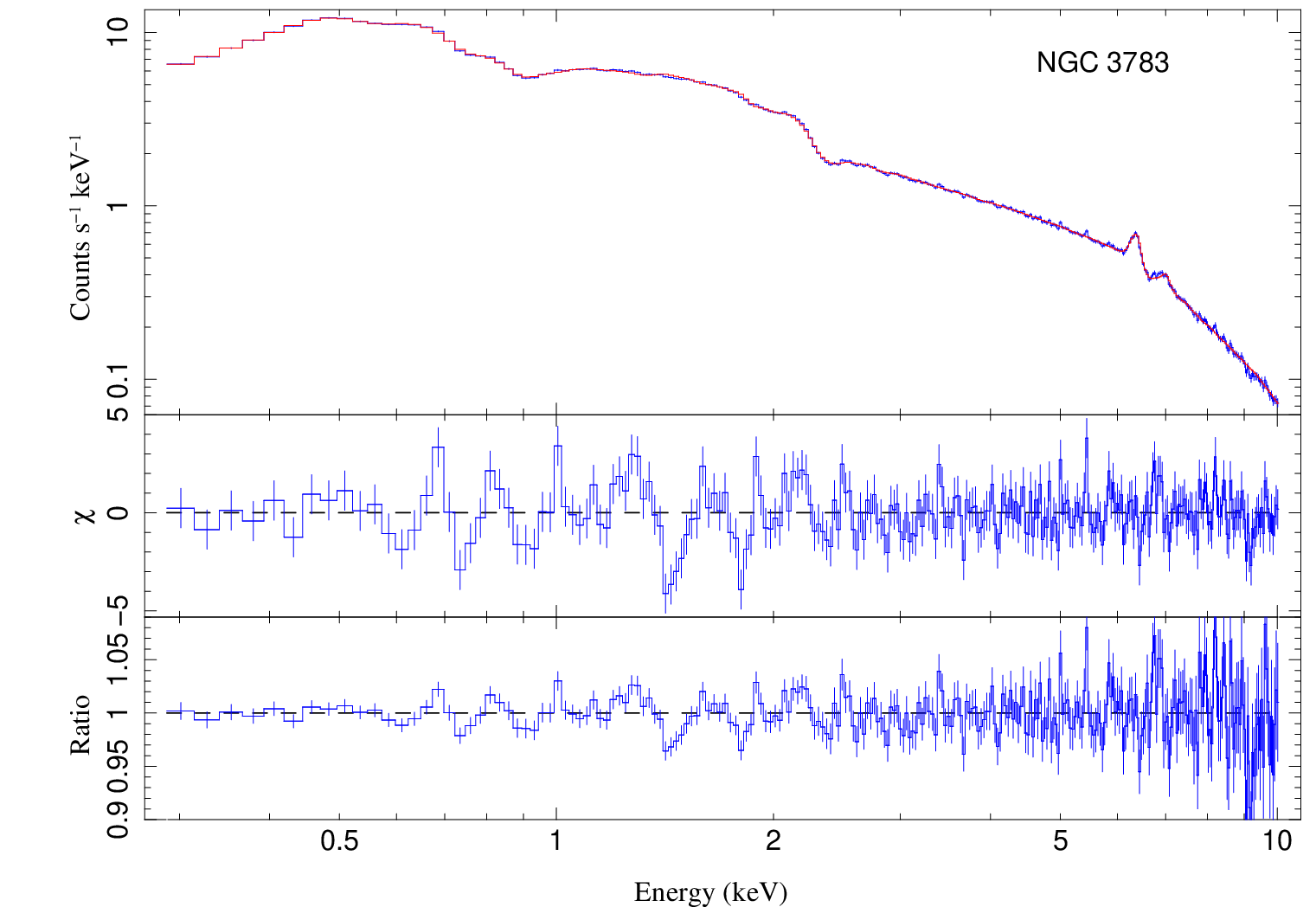}
} 
\hbox{
\includegraphics[width=4cm,angle=-90]{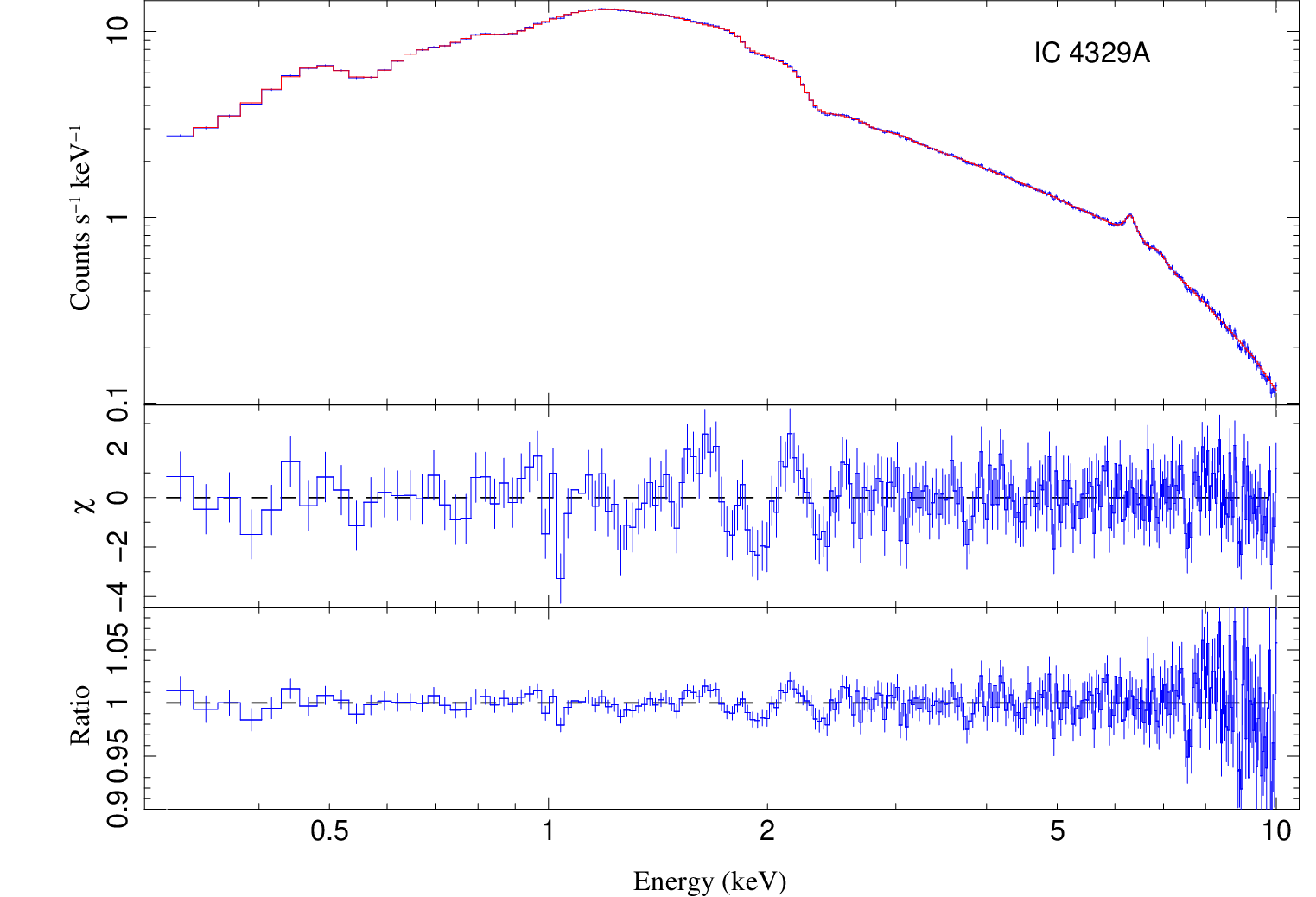}
}

}

\caption{Continued from Fig.~\ref{epicpn}}
 \label{epicpn_2}
\end{figure}


\clearpage

\begin{figure}
  \centering

  \vbox{
\hbox{
\includegraphics[width=6.5cm,angle=-90]{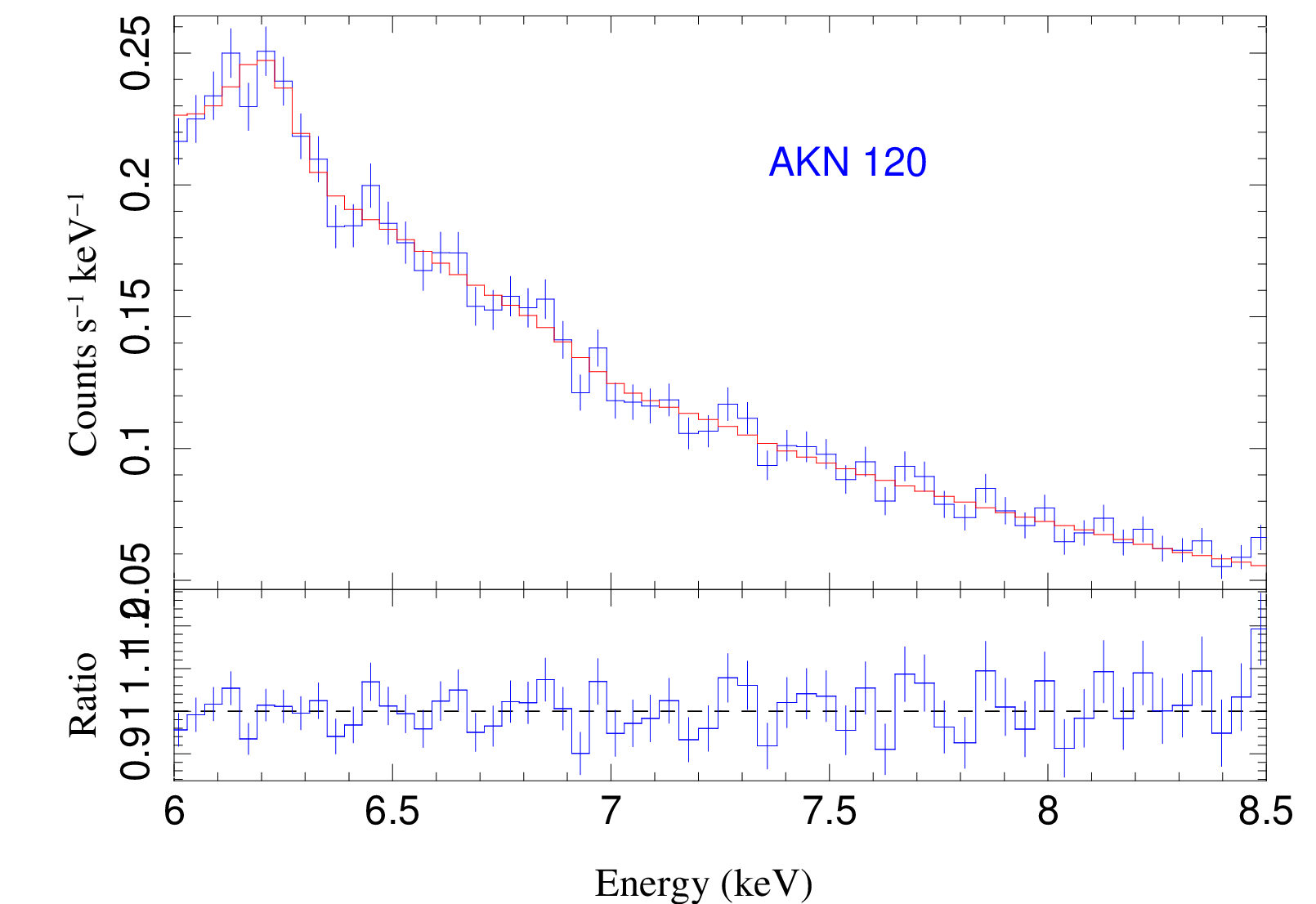}
\includegraphics[width=6.5cm,angle=-90]{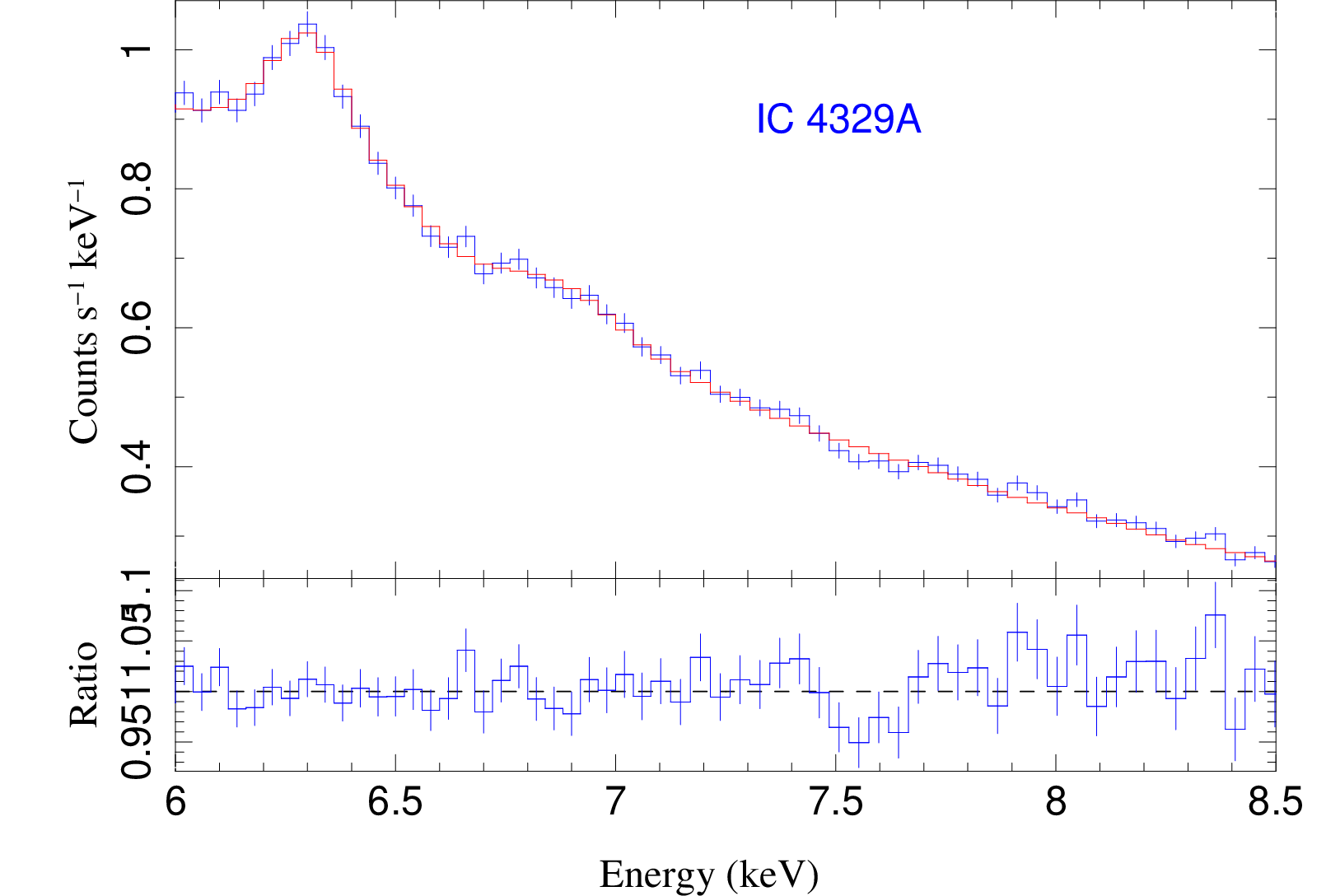}

} 

\hbox{
\includegraphics[width=6.5cm,angle=-90]{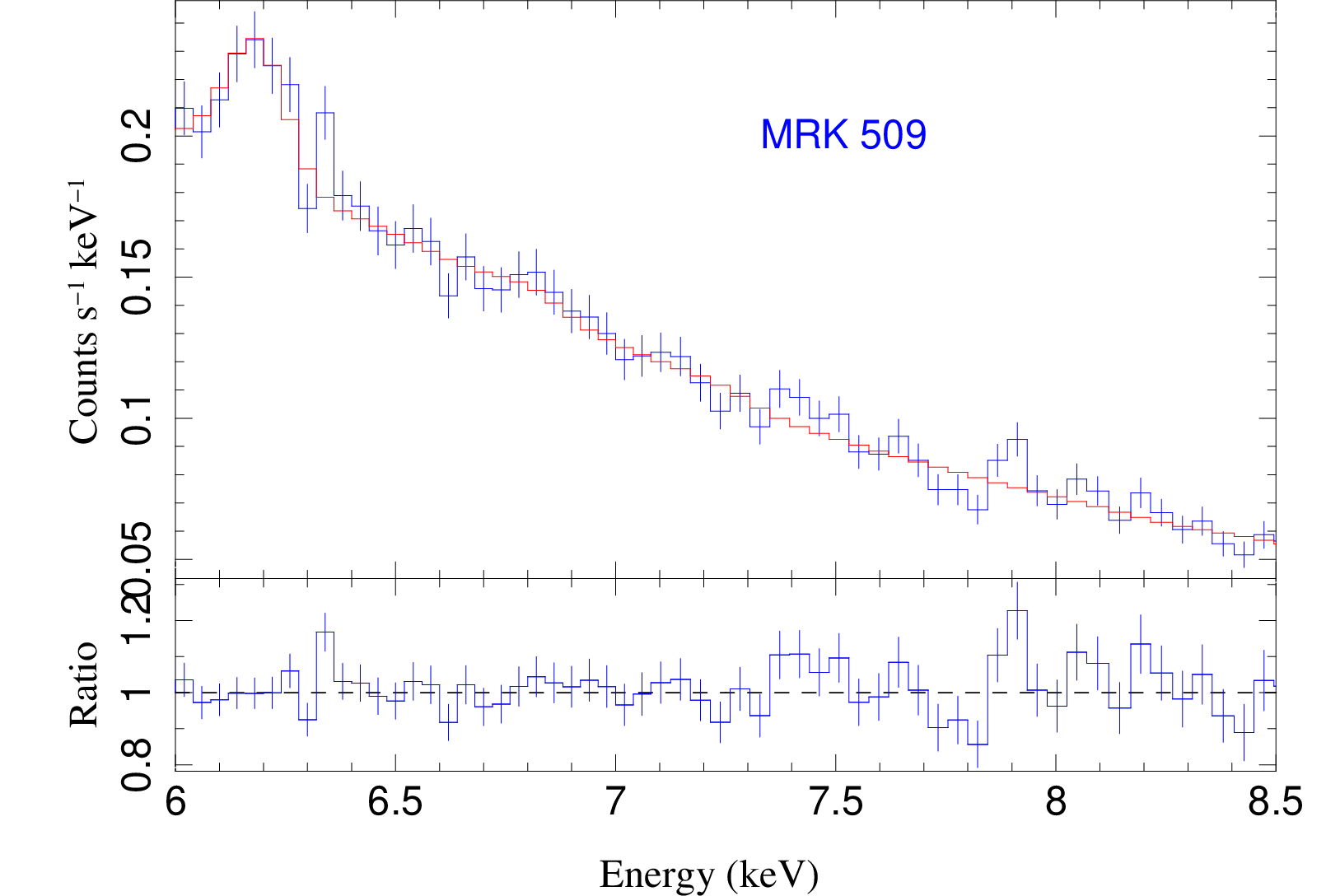}
\includegraphics[width=6.5cm,angle=-90]{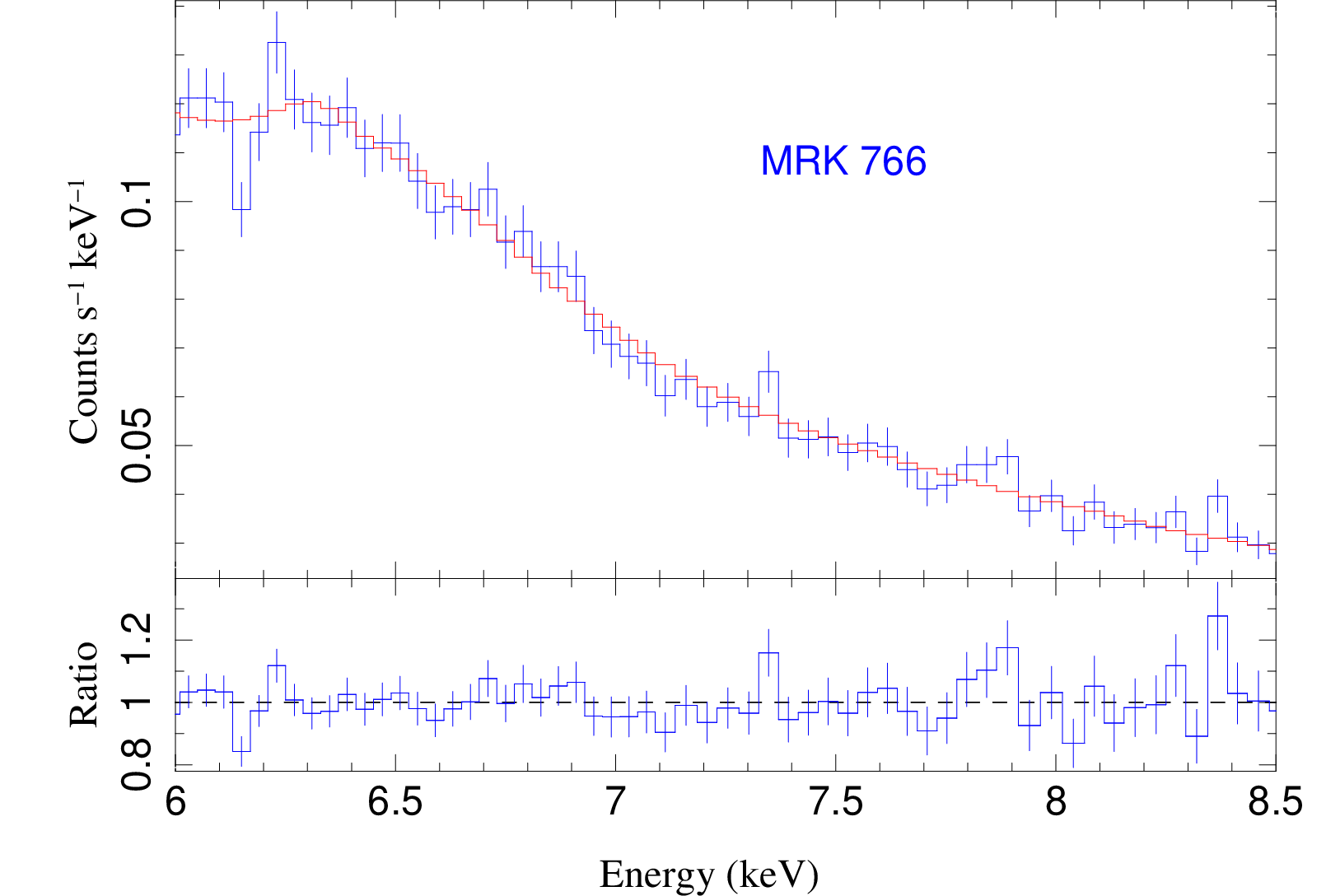}
} 

\hbox{
\includegraphics[width=6.5cm,angle=-90]{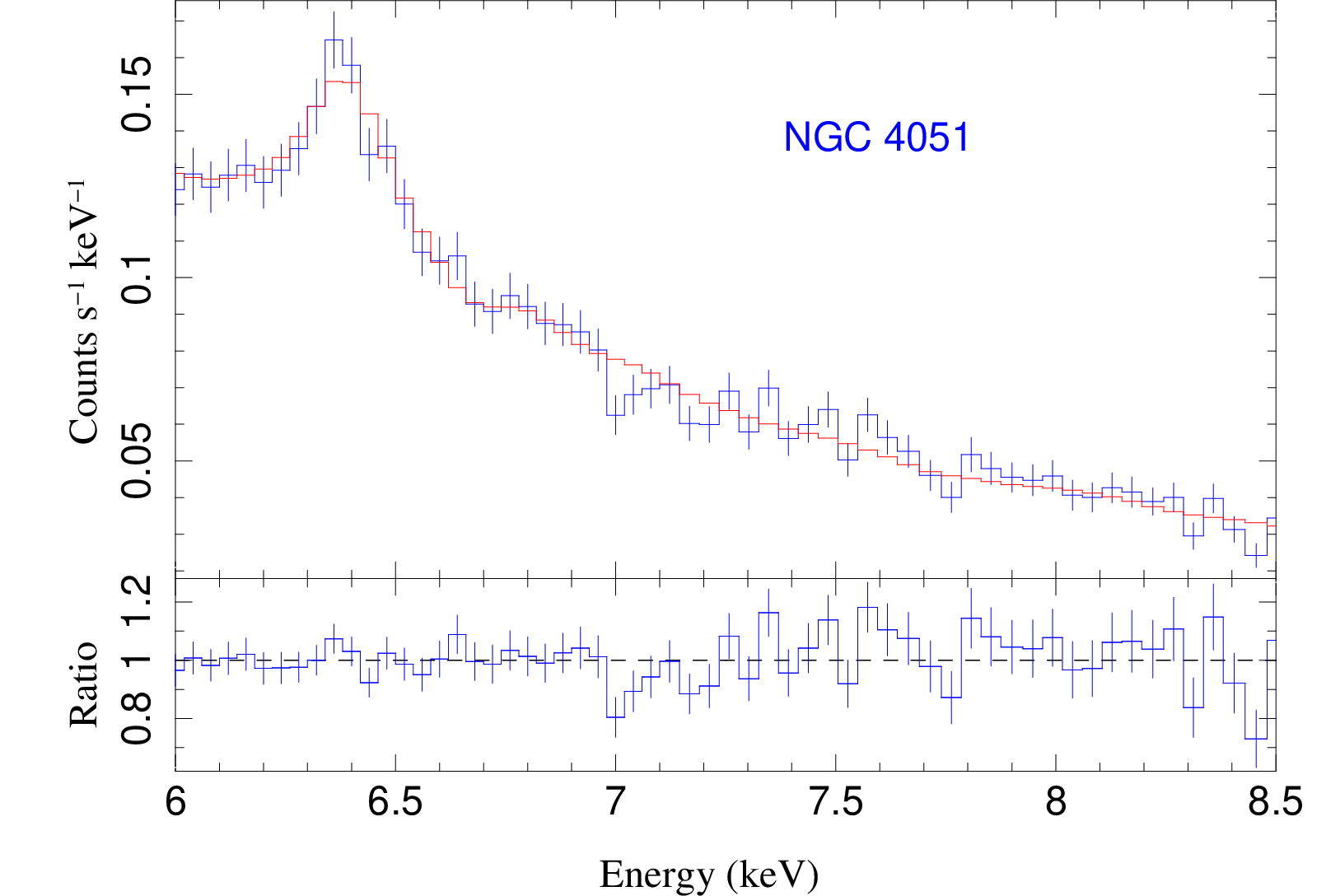}
\includegraphics[width=6.5cm,angle=-90]{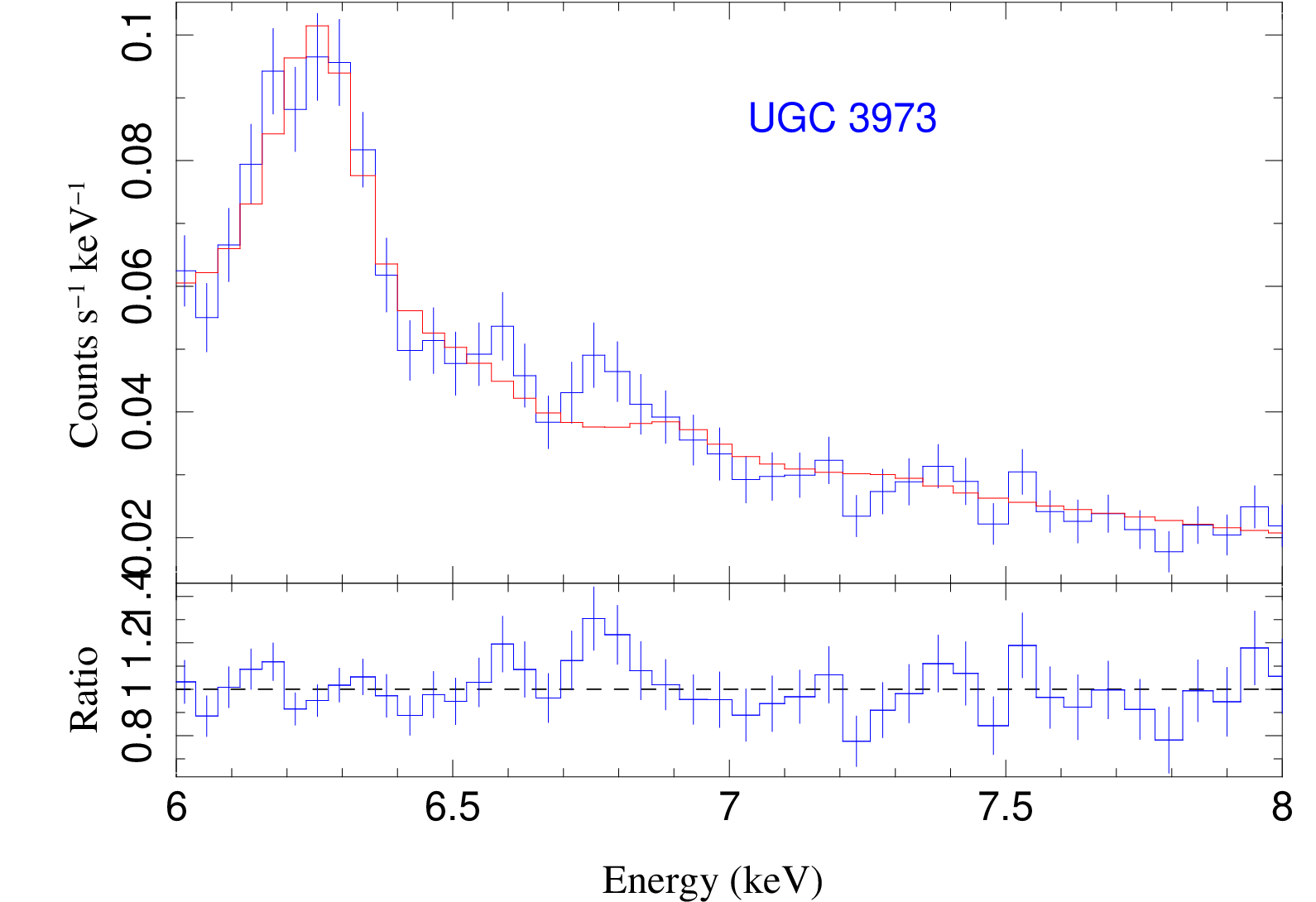}
}

}

\caption{The zoomed fits of the EPIC-pn data in the energy range of $6-8.5\kev$ for the sources in the WAX sample where T10 have found UFO. We do not statistically detect any negative residuals after we model the data as enumerated in Section \ref{section:RGS-analysis}.}
 \label{epicpn_zoomed}
\end{figure}



\clearpage

\begin{figure}
  \centering

  \vbox{
\hbox{
\includegraphics[width=4cm,angle=-90]{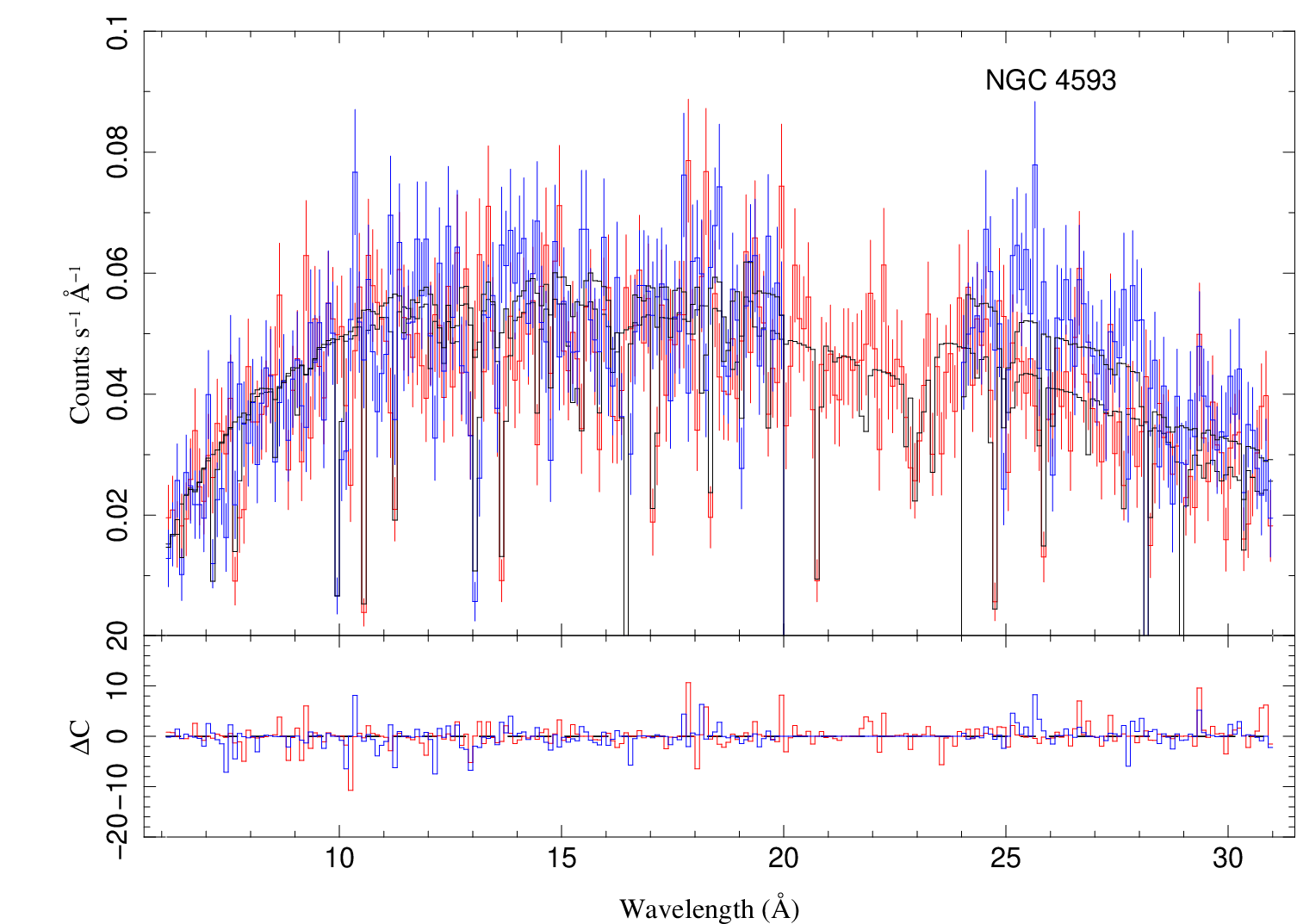}
\includegraphics[width=4cm,angle=-90]{WAX_RGS_plot_30June13/MRK704-RGScompare.ps}
\includegraphics[width=4cm,angle=-90]{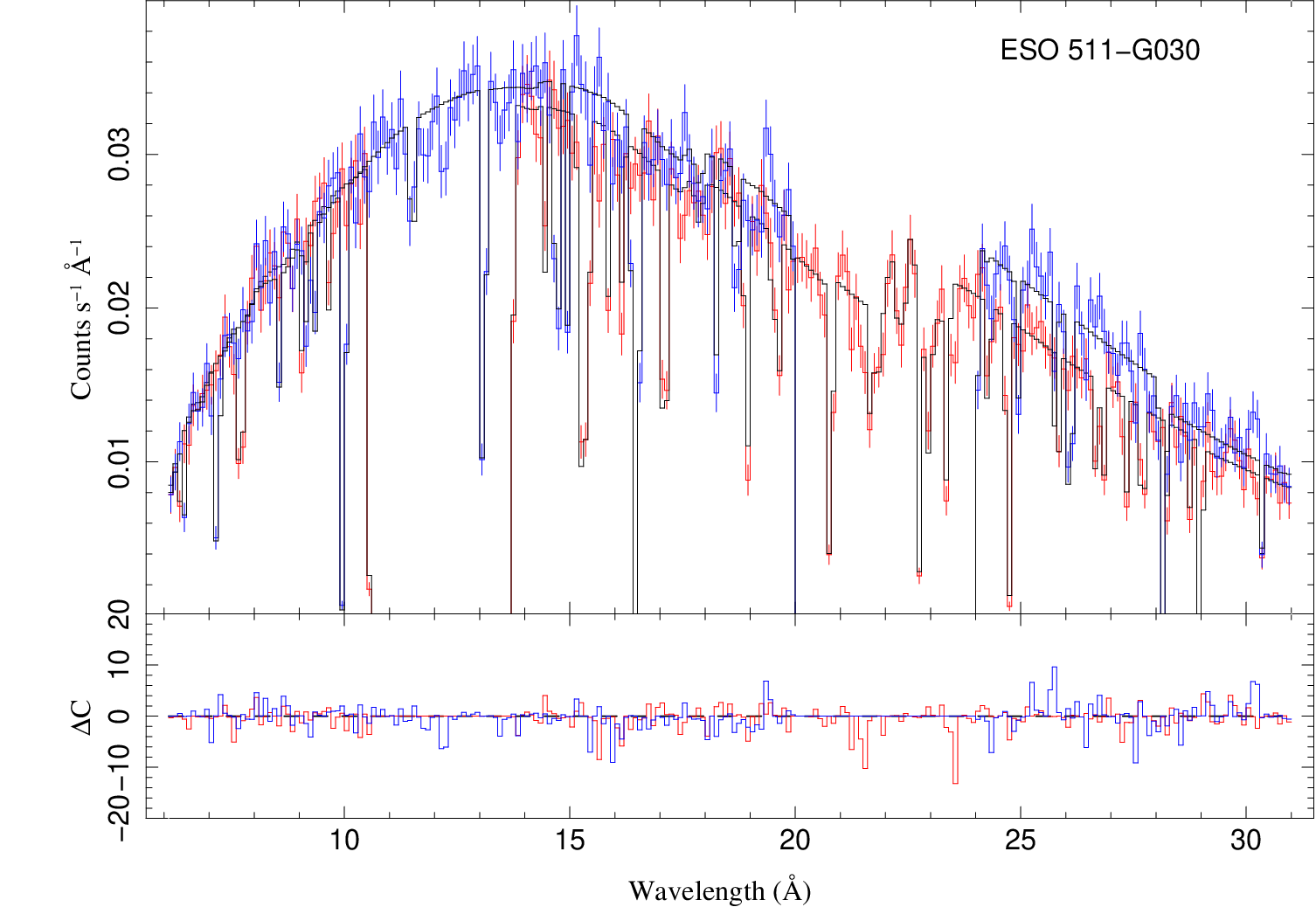}

} 

\hbox{
\includegraphics[width=4cm,angle=-90]{WAX_RGS_plot_30June13/NGC7213-RGScompare.ps}
\includegraphics[width=4cm,angle=-90]{WAX_RGS_plot_30June13/AKN564-RGScompare.ps}
\includegraphics[width=4cm,angle=-90]{WAX_RGS_plot_30June13/MRK110-RGScompare.ps}

} 

\hbox{
\includegraphics[width=4cm,angle=-90]{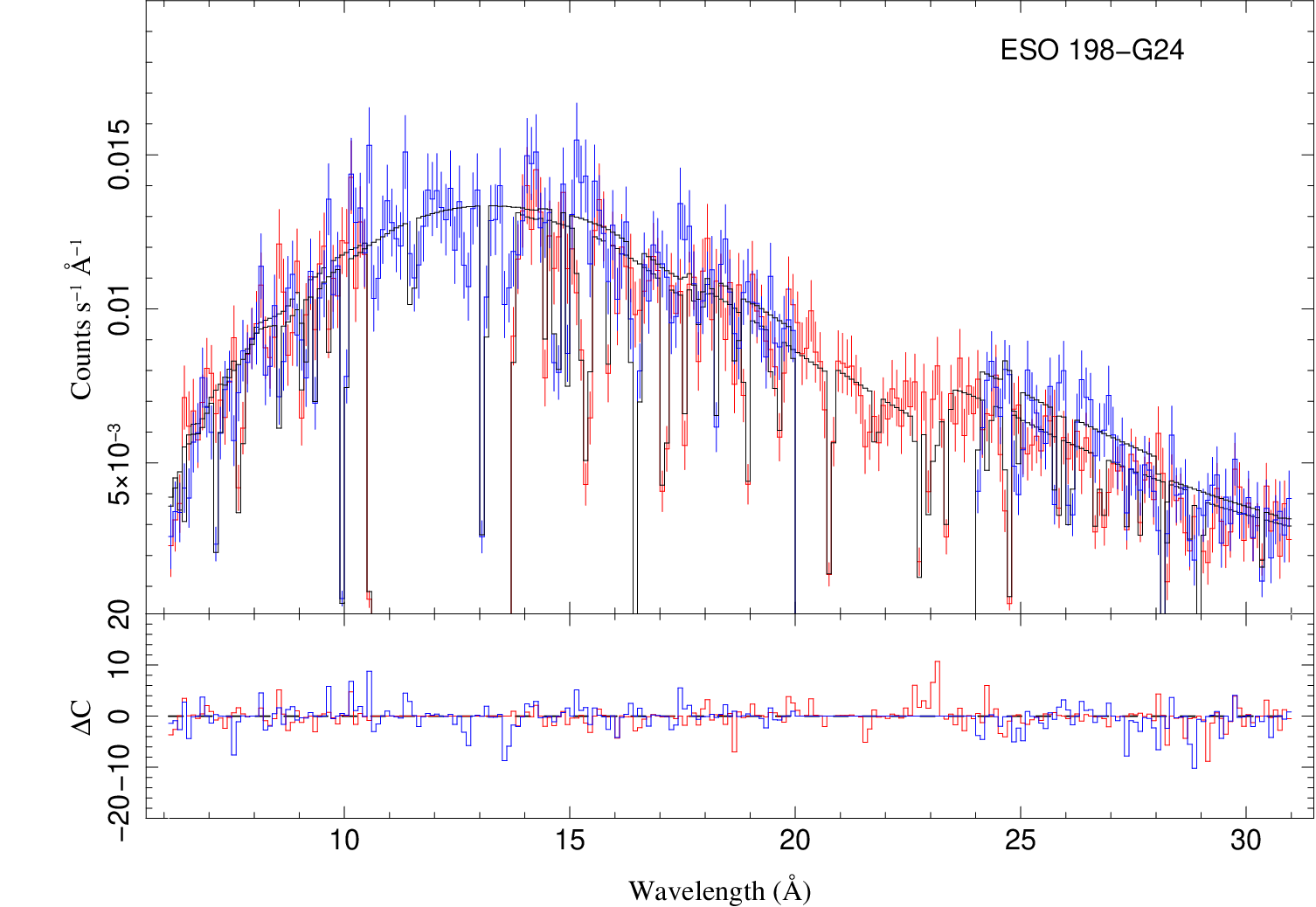}
\includegraphics[width=4cm,angle=-90]{WAX_RGS_plot_30June13/UGC3973-RGScompare.ps}
\includegraphics[width=4cm,angle=-90]{WAX_RGS_plot_30June13/NGC4051-RGScompare.ps}
}

\hbox{
\includegraphics[width=4cm,angle=-90]{WAX_RGS_plot_30June13/MCG-2-58-22-RGScompare.ps}
\includegraphics[width=4cm,angle=-90]{WAX_RGS_plot_30June13/NGC7469-RGScompare.ps}
\includegraphics[width=4cm,angle=-90]{WAX_RGS_plot_30June13/MRK766-RGScompare.ps}
}

\hbox{
\includegraphics[width=4cm,angle=-90]{WAX_RGS_plot_30June13/MRK590-RGScompare.ps}
\includegraphics[width=4cm,angle=-90]{WAX_RGS_plot_30June13/IRAS05078-RGScompare.ps}
\includegraphics[width=4cm,angle=-90]{WAX_RGS_plot_30June13/NGC3227-RGScompare.ps}
}

}

\caption{The best fit RGS data with the best fit model and the residuals .}
 \label{rgs}
\end{figure}


\clearpage

\begin{figure}
  \centering

  \vbox{
\hbox{
\includegraphics[width=4cm,angle=-90]{WAX_RGS_plot_30June13/MR2251-RGScompare.ps}
\includegraphics[width=4cm,angle=-90]{WAX_RGS_plot_30June13/MRK279-RGScompare.ps}
\includegraphics[width=4cm,angle=-90]{WAX_RGS_plot_30June13/AKN120-RGScompare.ps}

} 

\hbox{
\includegraphics[width=4cm,angle=-90]{WAX_RGS_plot_30June13/MCG8-11-11-RGScompare.ps}
\includegraphics[width=4cm,angle=-90]{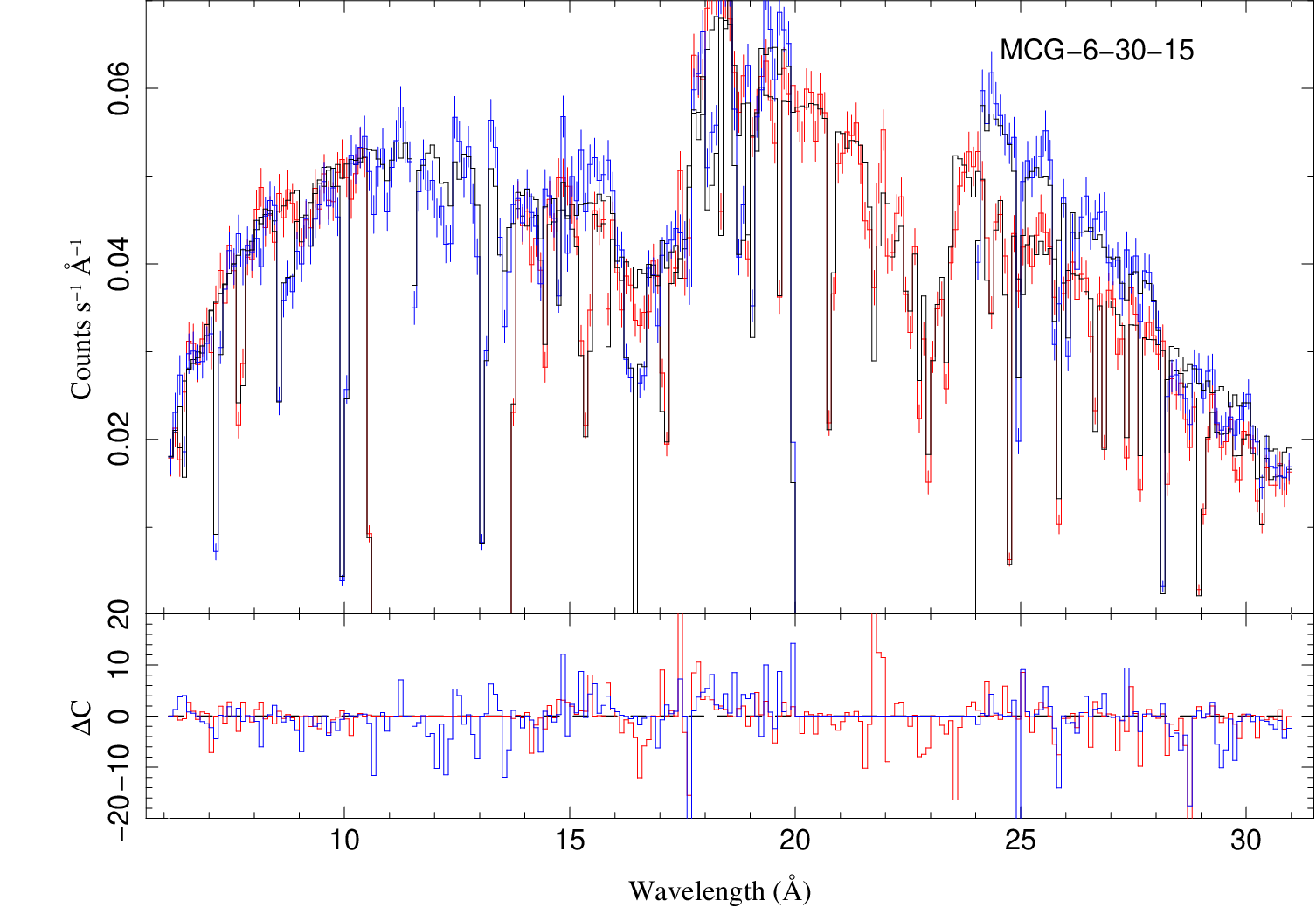}
\includegraphics[width=4cm,angle=-90]{WAX_RGS_plot_30June13/MRK509-RGScompare.ps}

} 

\hbox{
\includegraphics[width=4cm,angle=-90]{WAX_RGS_plot_30June13/NGC3516-RGScompare.ps}
\includegraphics[width=4cm,angle=-90]{WAX_RGS_plot_30June13/NGC5548-RGScompare.ps}
\includegraphics[width=4cm,angle=-90]{WAX_RGS_plot_30June13/NGC3783-RGScompare.ps}
}

\hbox{
\includegraphics[width=4cm,angle=-90]{WAX_RGS_plot_30June13/IC4329A-RGScompare.ps}
}

}

\caption{Continued from Fig.~\ref{rgs}}
 \label{rgs_2}
\end{figure}


\clearpage

\begin{figure}
  \centering

  \vbox{
\hbox{
\includegraphics[width=4cm,angle=-90]{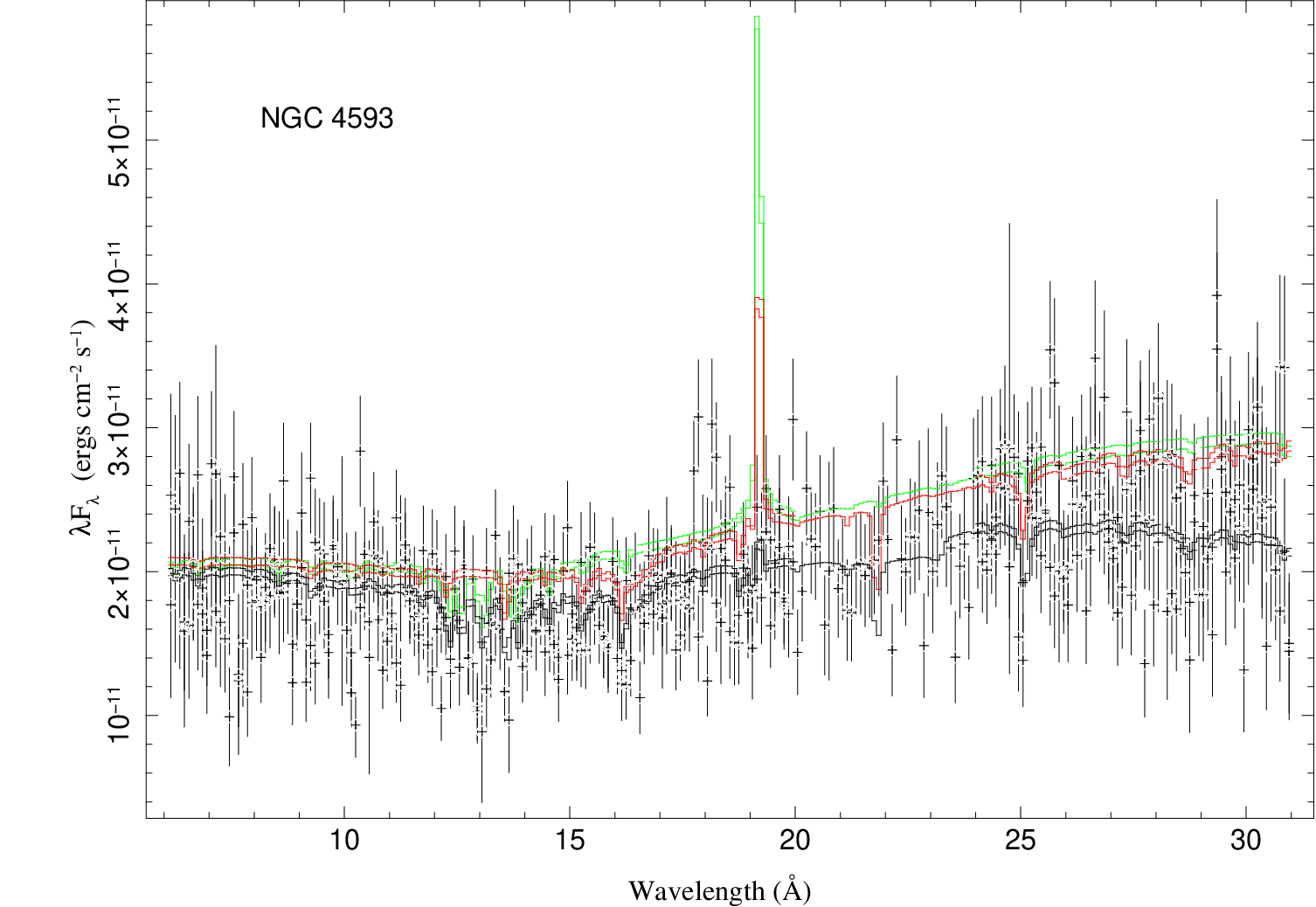} 
\includegraphics[width=4cm,angle=-90]{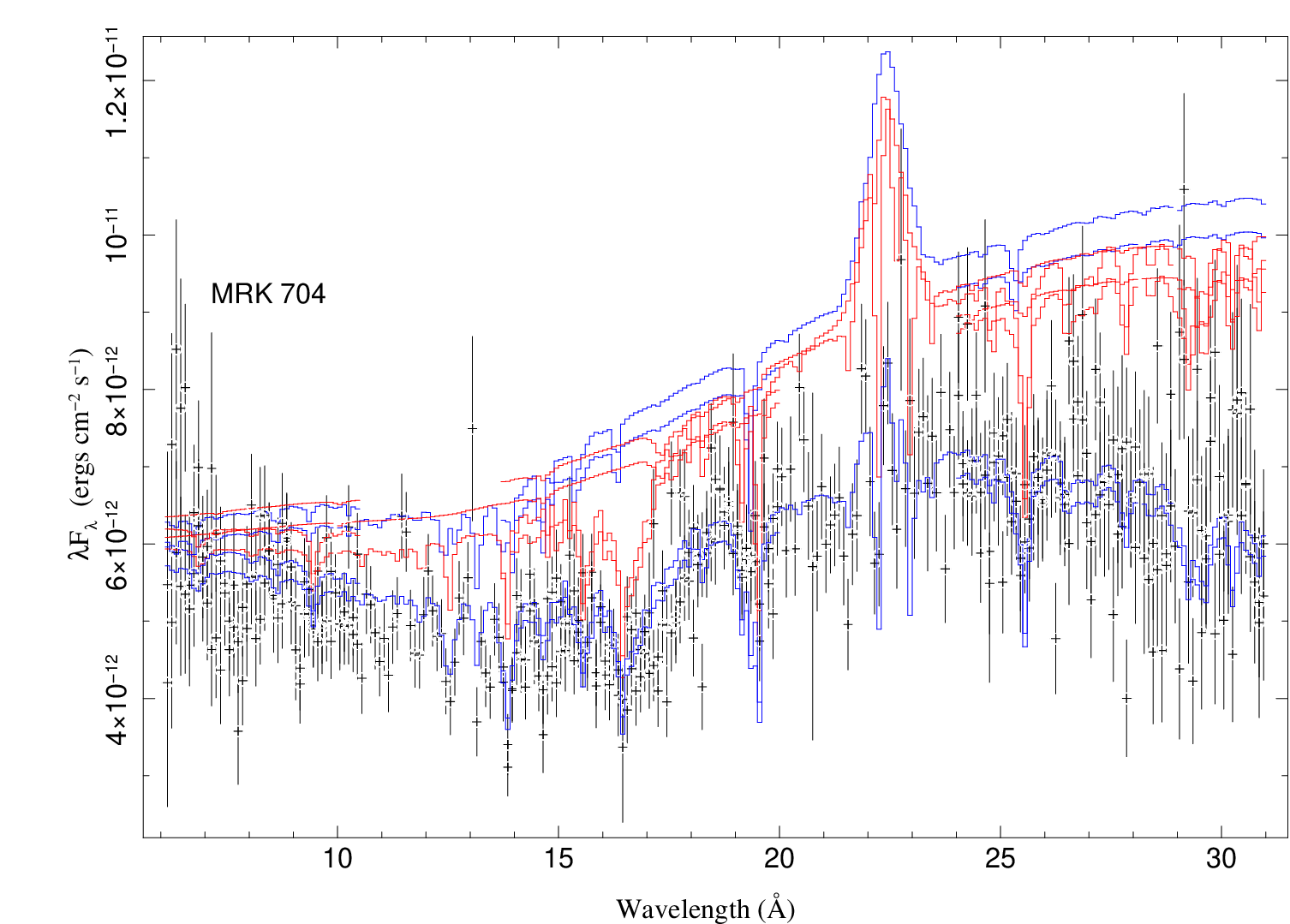} 
\includegraphics[width=4cm,angle=-90]{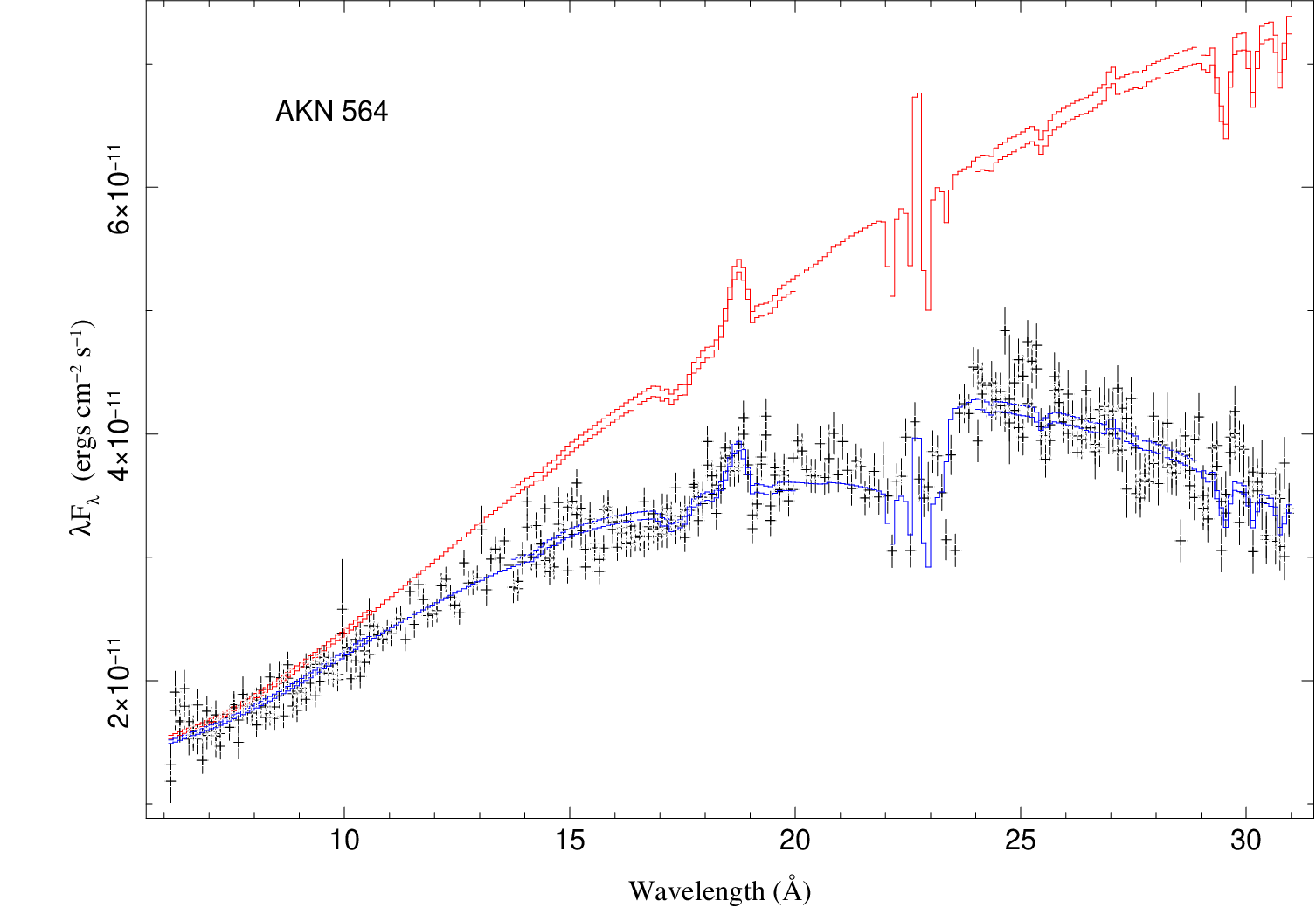} 
}

\hbox{
\includegraphics[width=4cm,angle=-90]{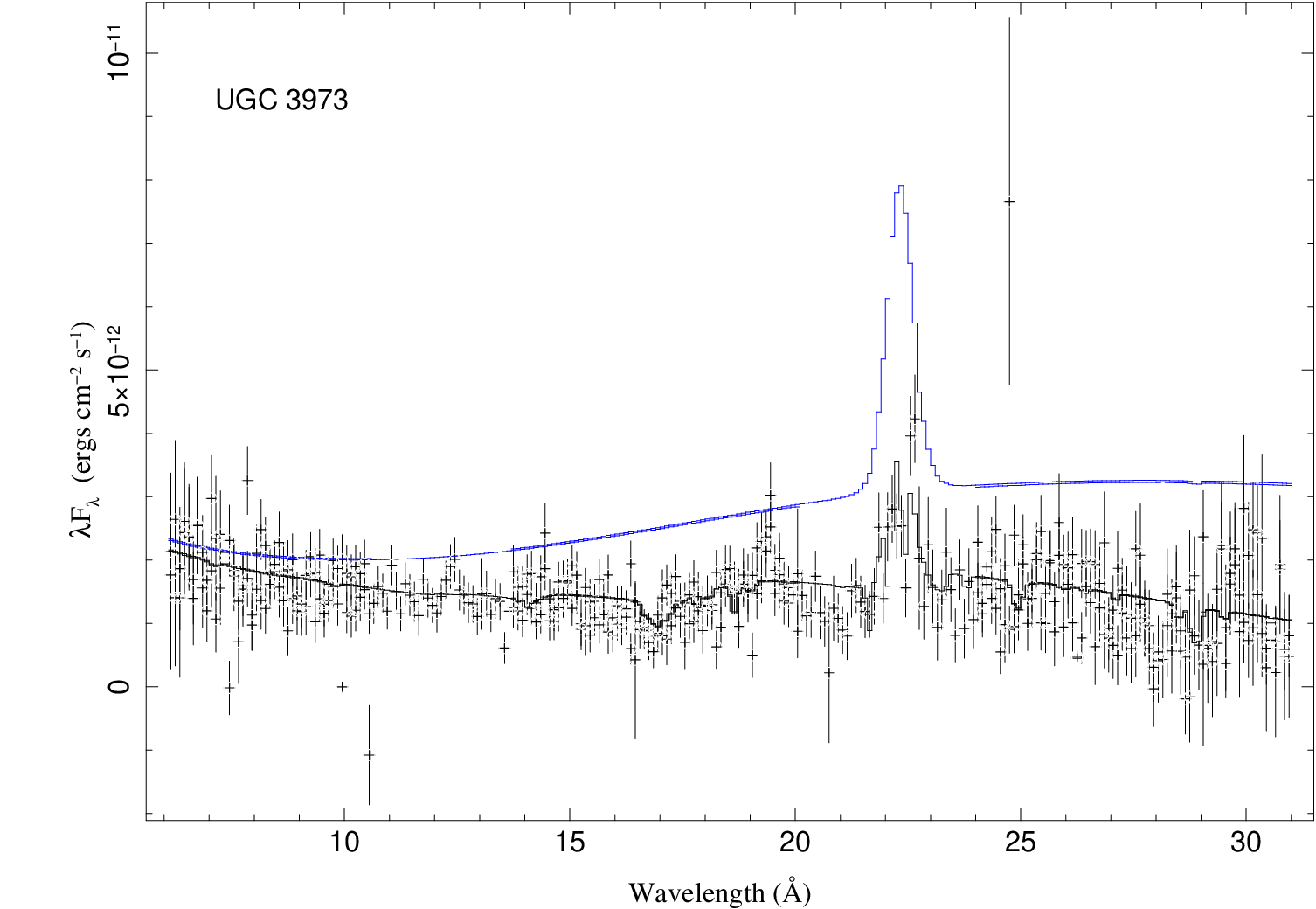} 
\includegraphics[width=4cm,angle=-90]{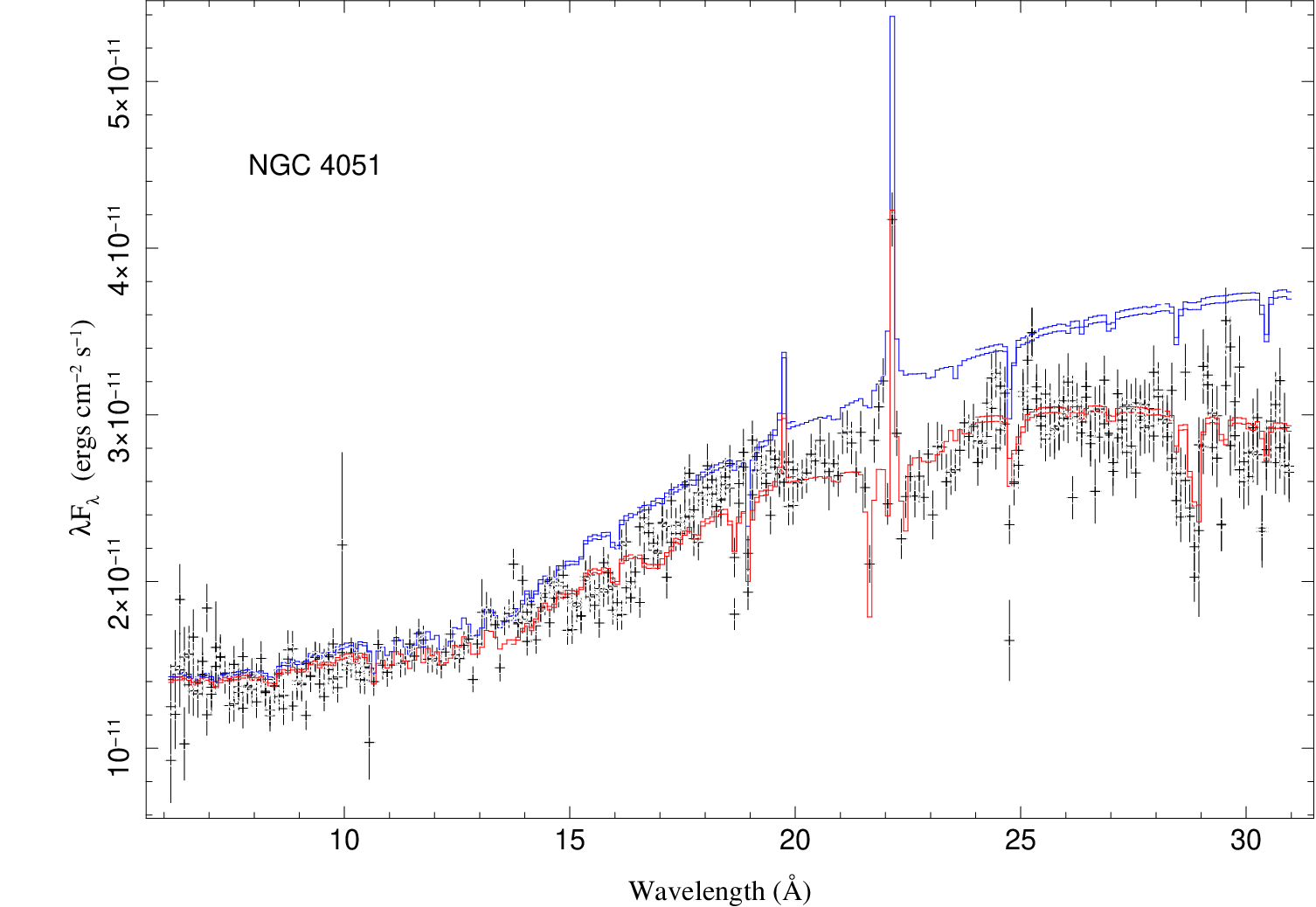} 
\includegraphics[width=4cm,angle=-90]{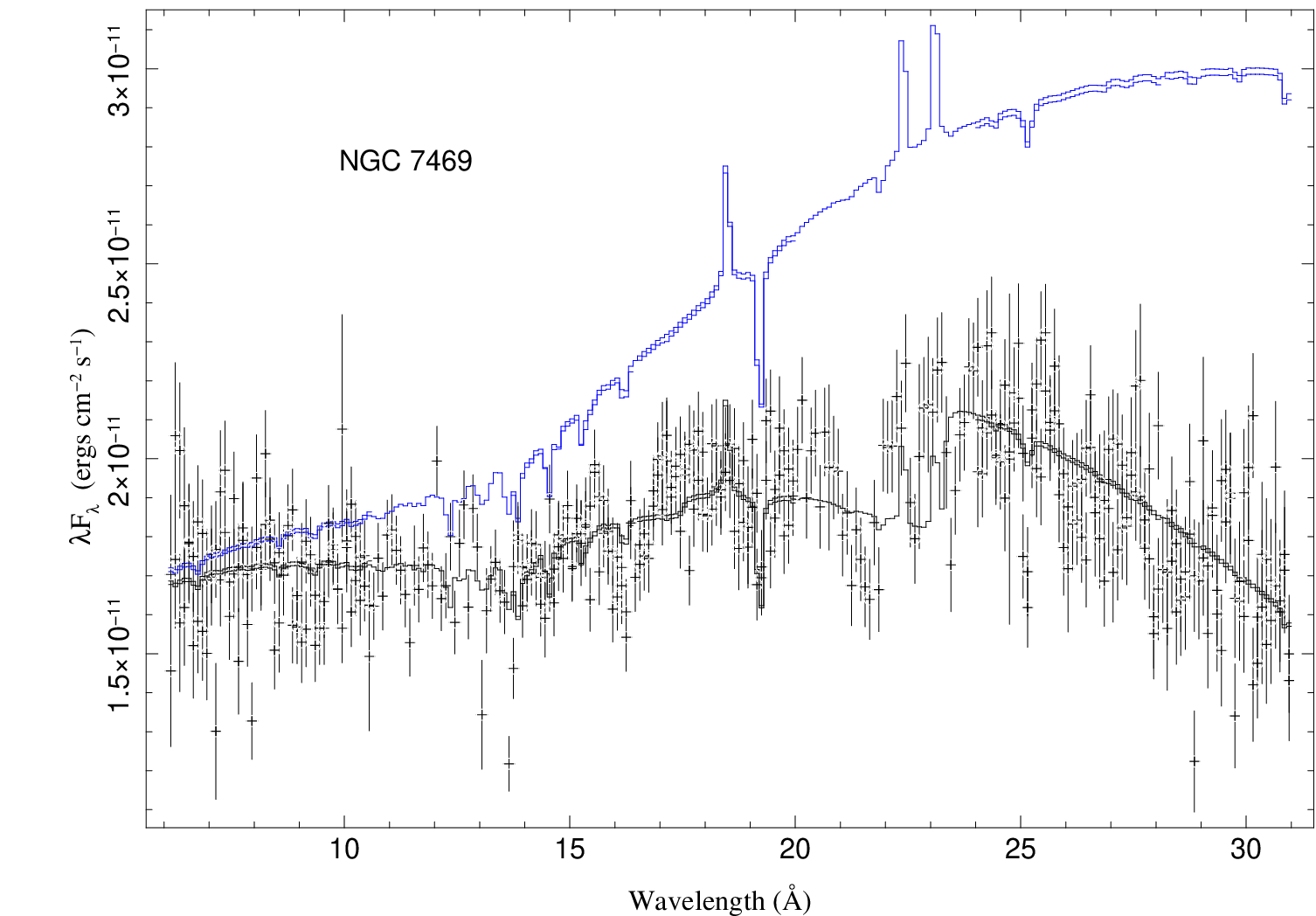} 
} 

\hbox{
\includegraphics[width=4cm,angle=-90]{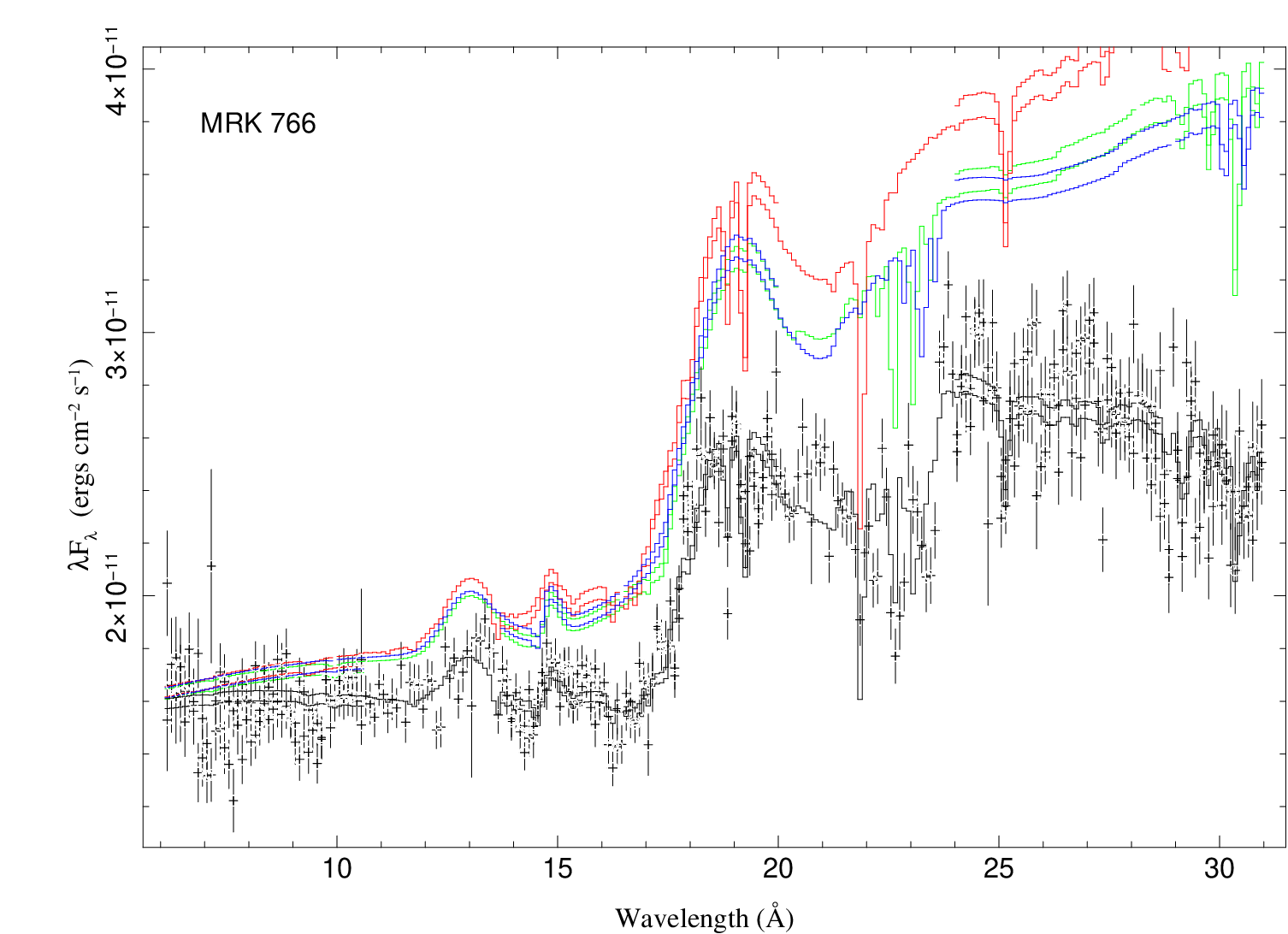} 
\includegraphics[width=4cm,angle=-90]{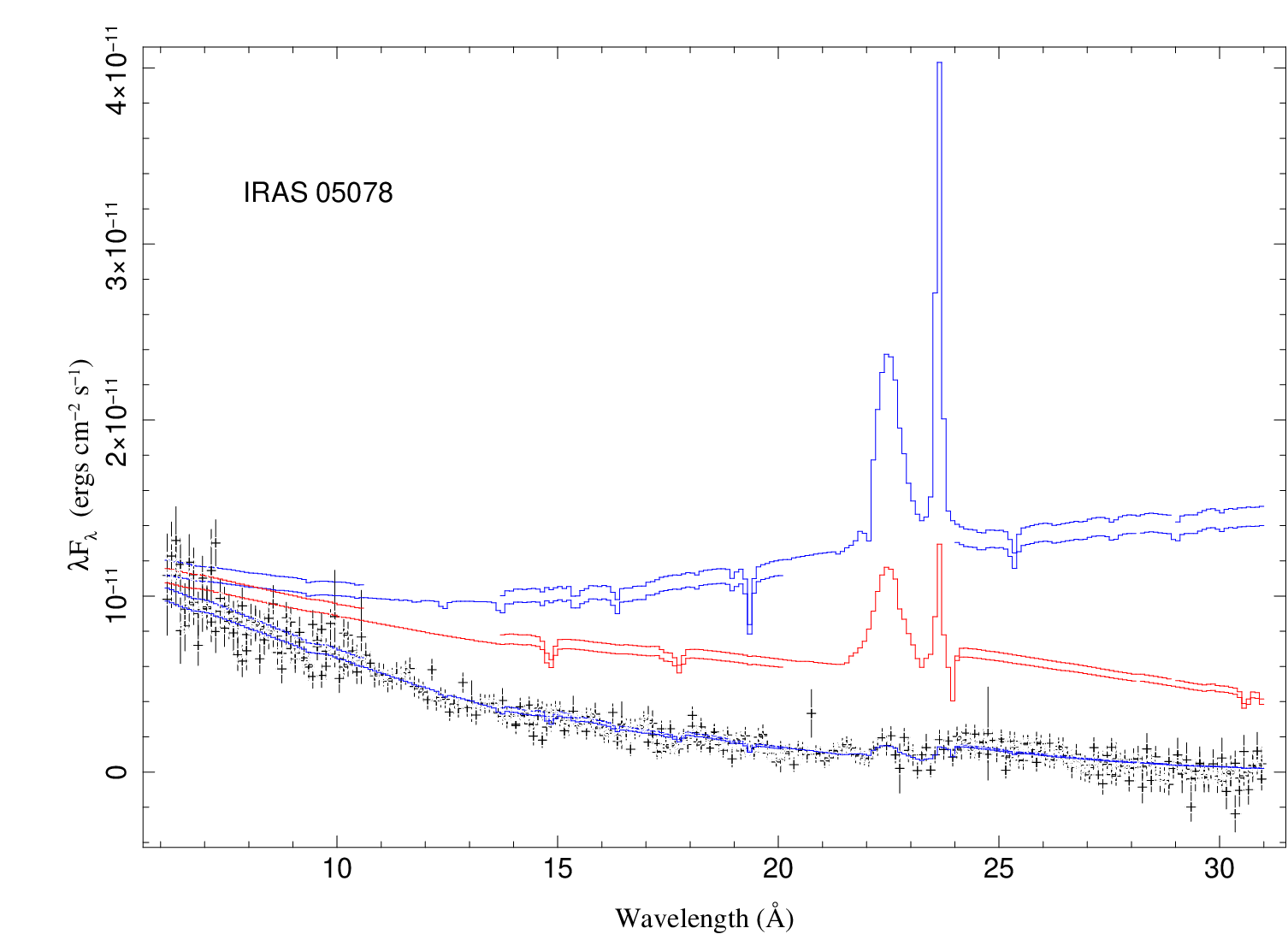} 
\includegraphics[width=4cm,angle=-90]{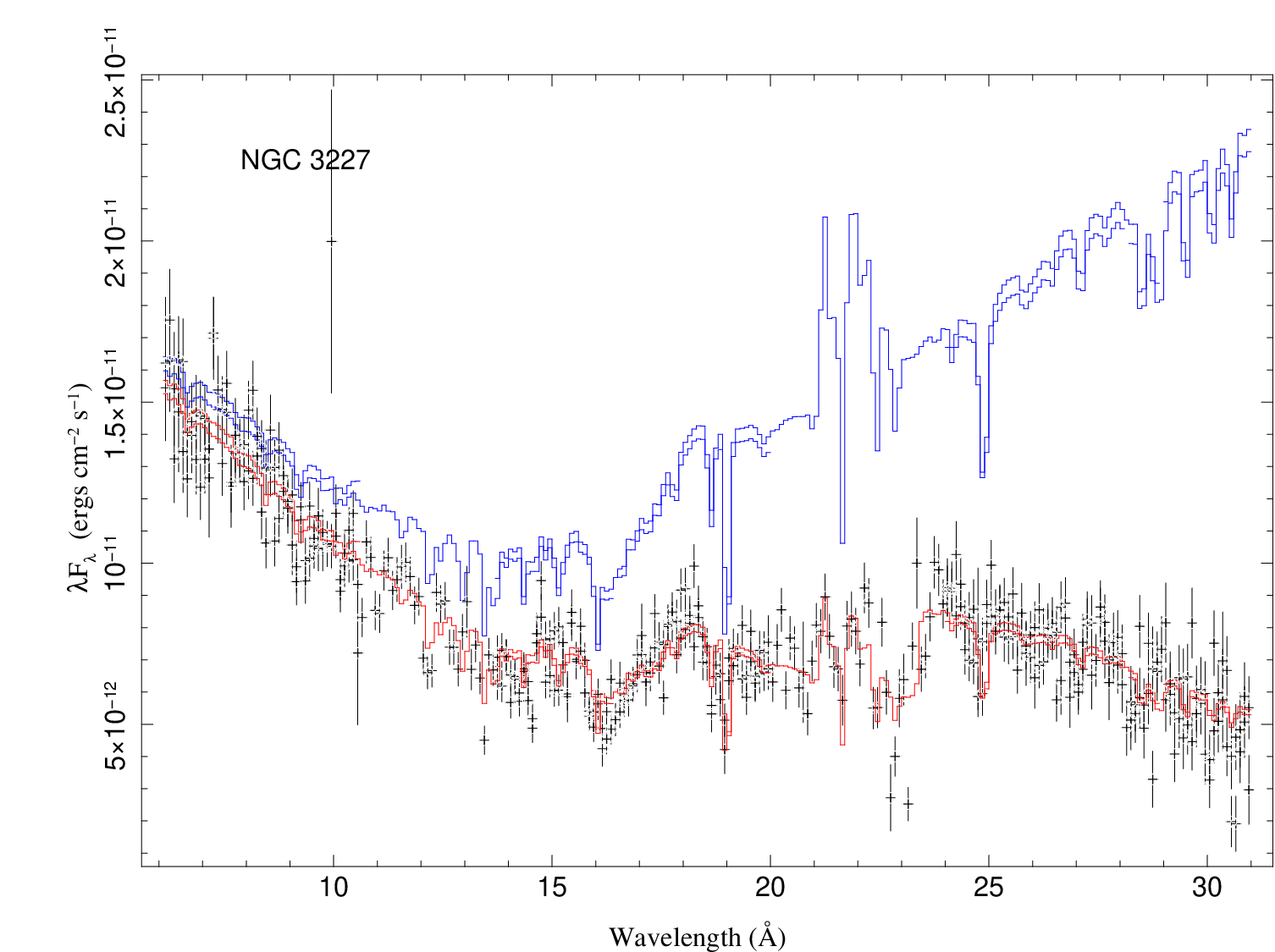} 
} 

\hbox{
\includegraphics[width=4cm,angle=-90]{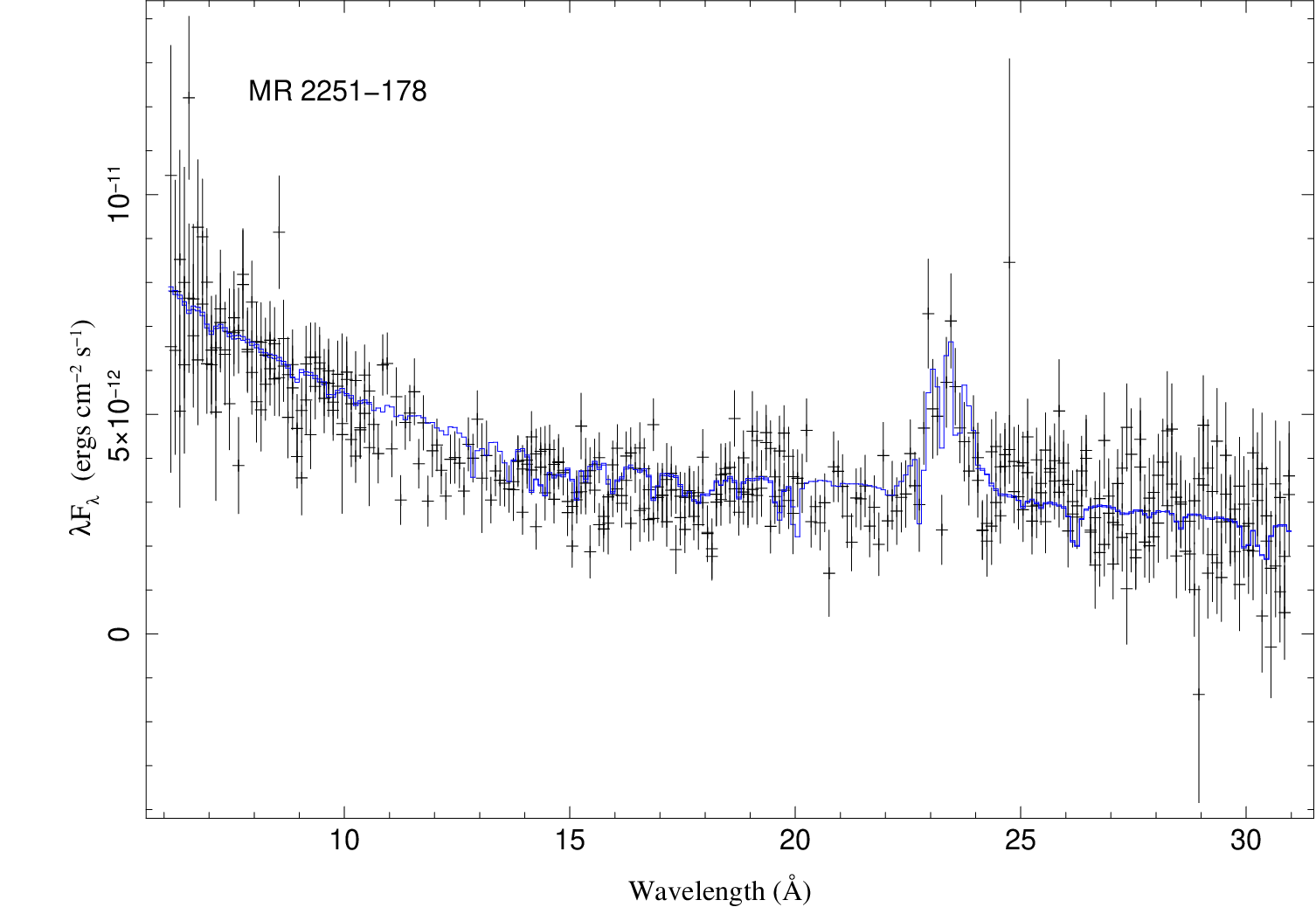} 
\includegraphics[width=4cm,angle=-90]{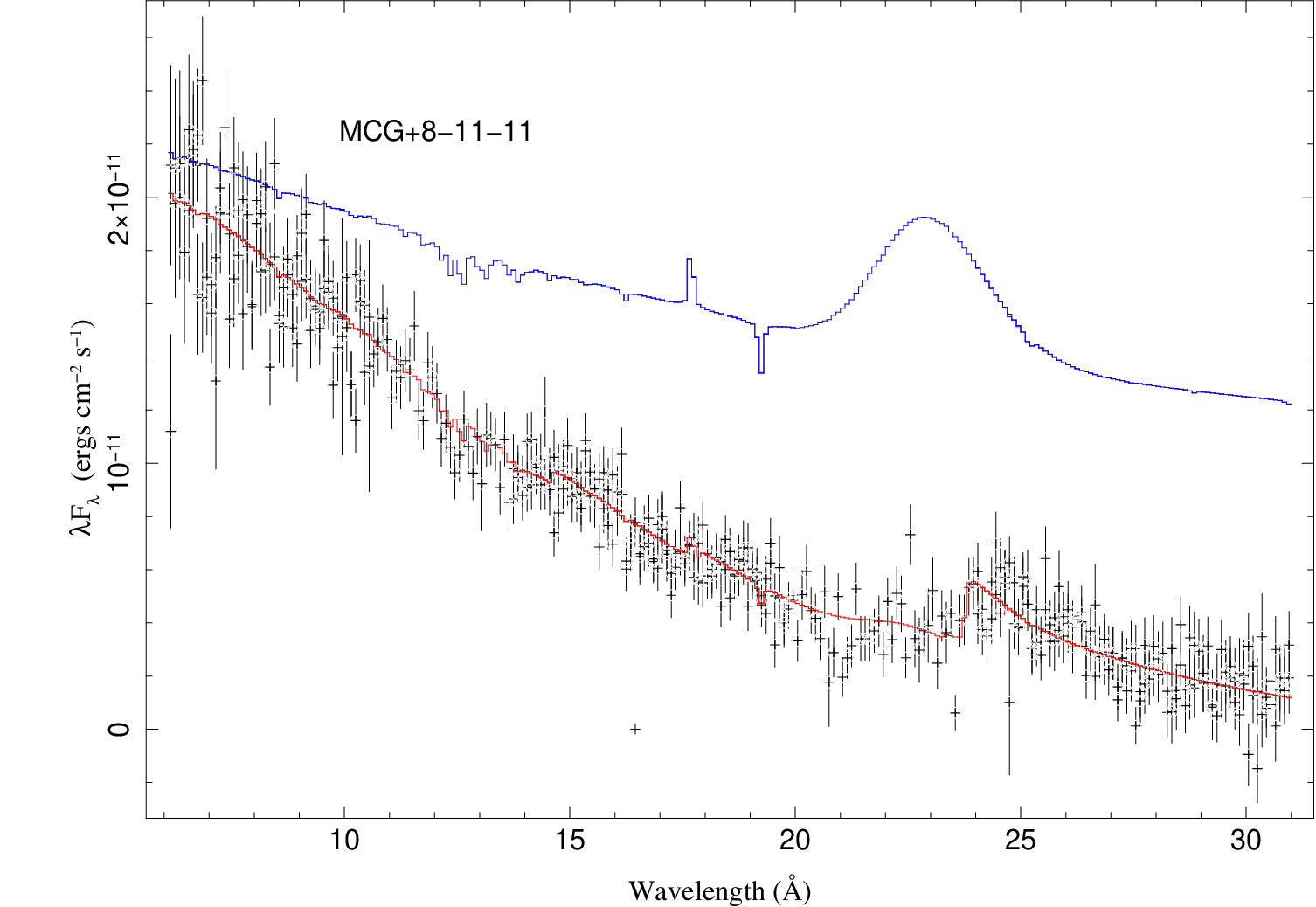} 
\includegraphics[width=4cm,angle=-90]{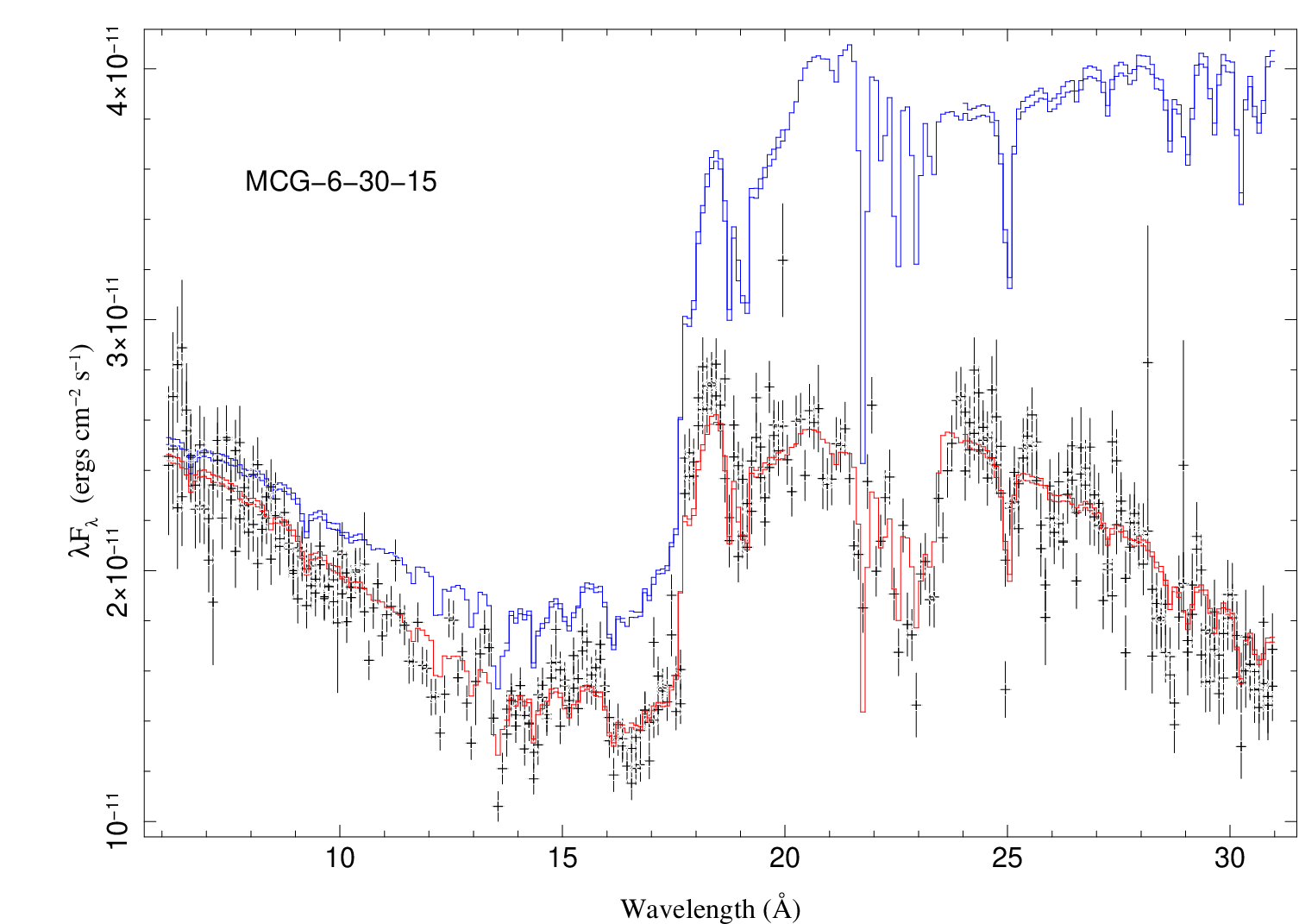} 
}

}
\caption{Figures showing the different WA components fitted to the data. The topmost curve in all the cases is the best fit model unabsorbed by any WA. The bottom-most curve along with the data is the best fit model. The multiple curves in certain cases are due to absorption due to multiple WA components.}
 \label{wa}
\end{figure}

\clearpage

\begin{figure}
  \centering

  \vbox{
\hbox{
\includegraphics[width=4cm,angle=-90]{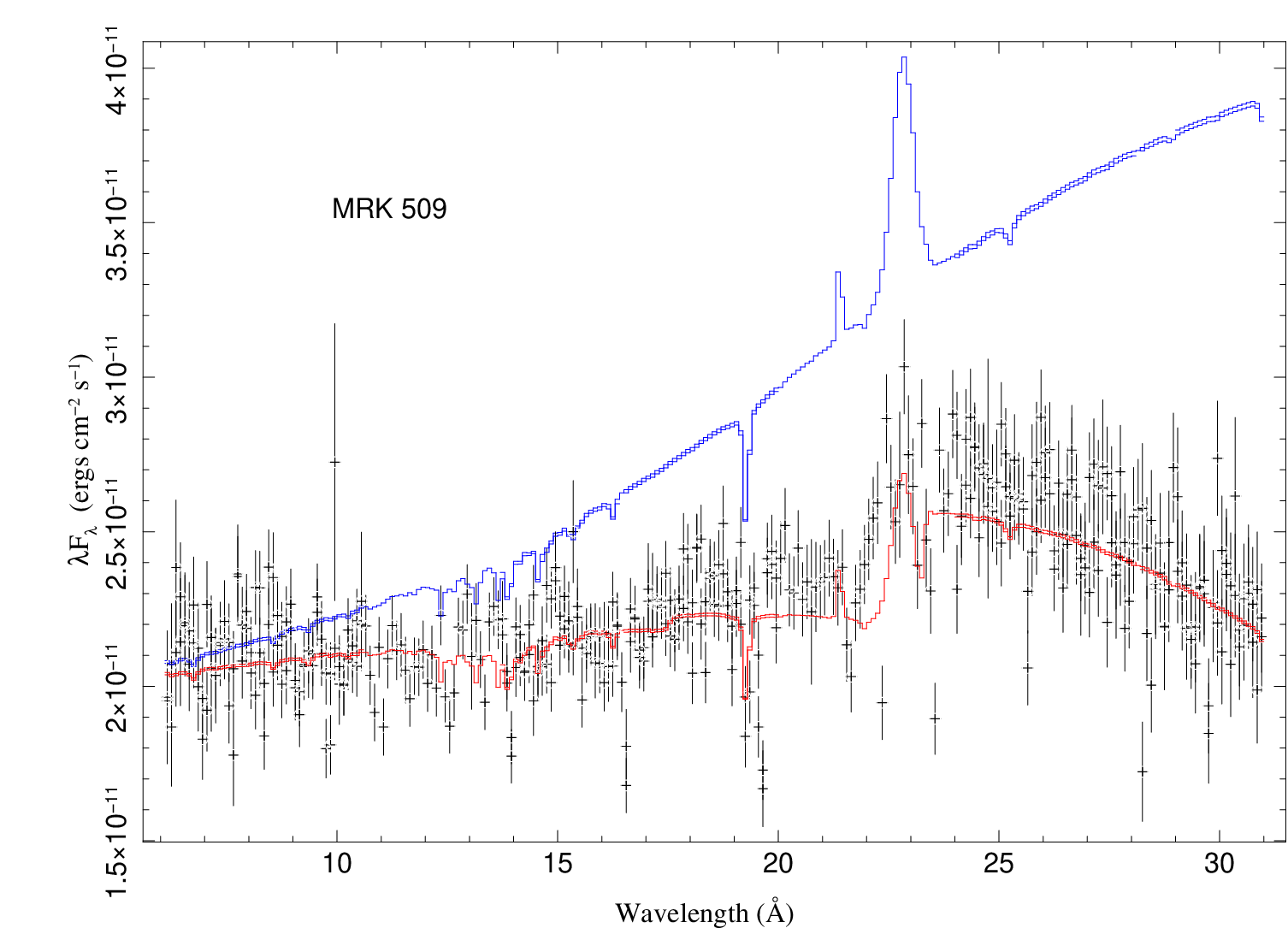} 
\includegraphics[width=4cm,angle=-90]{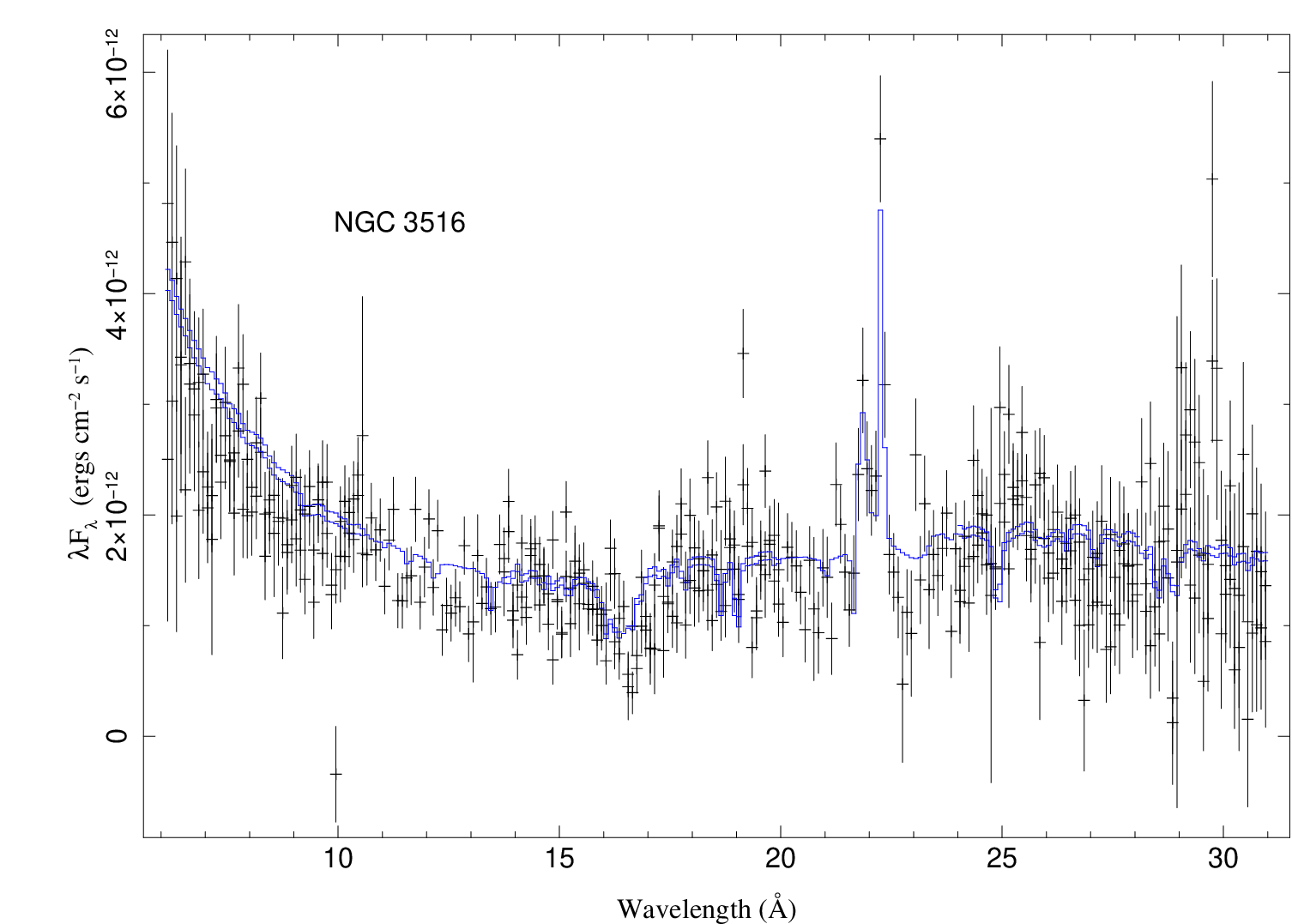} 
\includegraphics[width=4cm,angle=-90]{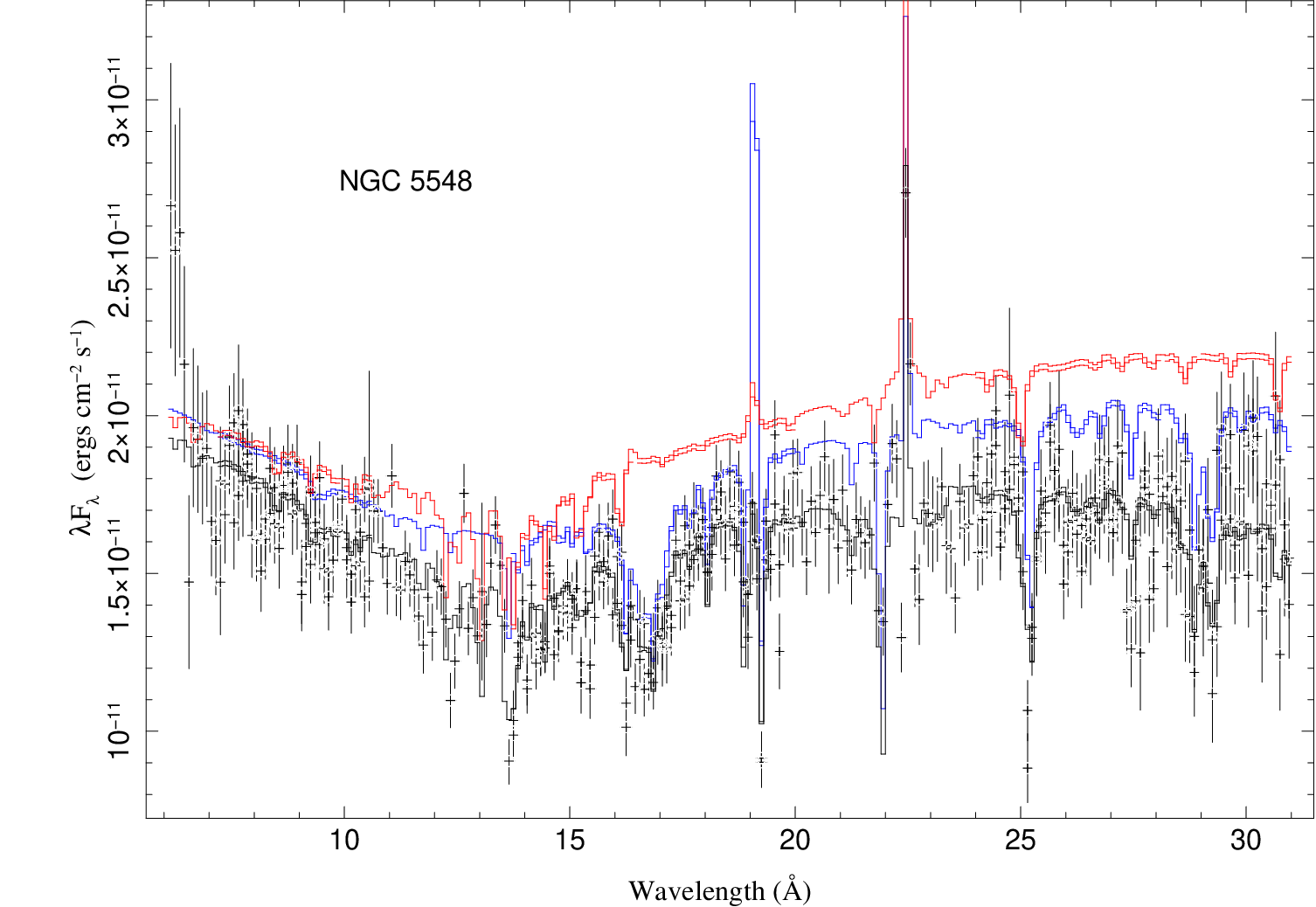} 
}

\hbox{
\includegraphics[width=4cm,angle=-90]{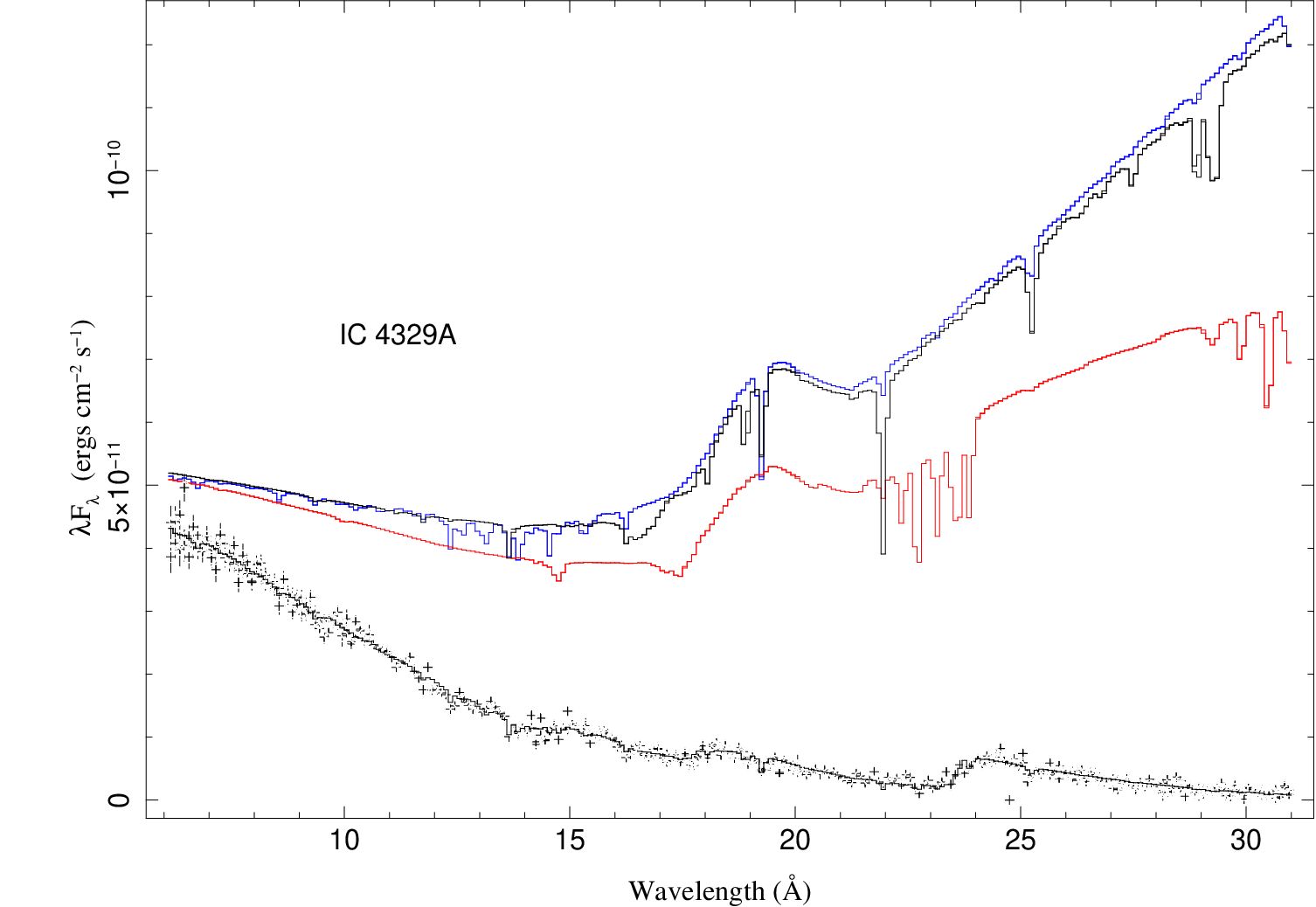} 
}

}
\caption{Continued from Fig.~\ref{wa}}
 \label{wa_2}
\end{figure}


\clearpage

\section{Effects of turbulent velocity on WA parameters.}
\label{subsec:turbulent-velocity}

{\color{black}The micro-turbulent velocity $v_{turb}$ is an important parameter of the WA clouds. The thermal motions of the electrons as well as contributions from other broadening effects yield a turbulent velocity of the cloud. Typically, in most studies the turbulent velocity is fixed to $v_{\rm turb}=100\kms$, as the limited data statistics do not allow us to constrain the intrinsic width of the absorption lines. However, in some cases with enough SNR, individual line widths have yielded constraints on the turbulent velocity typically to the range $100-1000\kms$ \citep{2001A&A...365L.168S,2013MNRAS.430.2650L}. The turbulent velocity sometimes affects the column density measurements of the WA.

\citet{2013MNRAS.430.2650L} in their study of a bright Seyfert galaxy IRAS~13349+2438 using high resolution {\it Chandra} data found that the turbulent velocities measured for different ionic species were different. However, they did not find any statistical difference in their measured values of ionic column densities when the turbulent velocity was frozen to $100\kms$ compared to that when it was left free to vary during the fit. We have carried out an exercise to verify how changes in turbulent velocity affect the measured ionisation parameter and the column density. We have considered three sources (AKN~564, NGC~5548, MCG-6-30-15) from the WAX sample for the purpose, which have one, two and three WA components respectively, and also sufficient SNR. Our intention was also to check whether the turbulent velocities affect multi-component WAs similarly as they affect a single WA component. We created WA table models with the respective SEDs as enumerated in Section \ref{subsec-CLOUDY}. While creating each of the models we froze the turbulent velocities to $50\kms$, $100\kms$, $200\kms$ and $600\kms$. With these WA models we fitted the EPIC-pn and RGS datasets simultaneously as enumerated in Section \ref{section:RGS-analysis}. Table \ref{Table:turbulent-vel} shows that for all cases the column densities are consistent within errors. However, the best fit statistic is different mostly for the case of MCG-6-30-15. For MCG-6-30-15 the statistic becomes worse gradually as the turbulent velocity is increased which means that the data prefer a lower turbulent velocity. As the column density and the ionisation parameter remain unchanged with velocity, a turbulent velocity of $v_{\rm turb}=100\kms$ is a reasonable assumption for the sample study.}

\begin{table*}
{\footnotesize
\centering
\caption{The WA parameters for the different CLOUDY models developed using the three turbulent velocities: $50\kms$, $100\kms$, $200\kms$, and $600\kms$.}\label{Table:turbulent-vel}
  \begin{tabular}{llllllllllll} \hline\hline 
	  
	  Source 	& WA component		& 	WA parameters 	&  $v_{turb}=50\kms$  & $v_{turb}=100\kms$  & $v_{turb}=200\kms$ & $v_{turb}=600\kms$  \\ \hline \\ 

	  AKN~564	&  WA-1	  		&    $\log\xi$    	 &$-0.20^{+0.11}_{-0.10}$&	$-0.20^{+0.12}_{-0.09}$& $-0.20^{+0.11}_{-0.11}$&$-0.1^{+0.11}_{-0.11}$  \\ 

			&		  	&    $\log\nh$      	& $20.48^{+0.05}_{-0.13}$&$20.31_{-0.2}^{+0.2}$	&$20.36_{-0.1}^{+0.12}$&$20.29_{-0.1}^{+0.12}$	\\ \\
	  		&   C/dof		&    			&$8097/6885$		 & $8105/6885$		& $8110/6885$		& $8128/6885$                \\ \hline \\


	  NGC~5548	&WA-1			&$\log\xi$  		&$1.81_{-0.05}^{+0.06}$	&$1.86_{-0.05}^{+0.08}$	&$1.85_{-0.05}^{+0.05}$&$1.90_{-0.05}^{+0.05}$		\\
				 &			&$\log\nh$ 		&$21.16_{-0.03}^{+0.03}$&$21.13_{-0.07}^{+0.07}$&$21.13_{-0.04}^{+0.04}$&$21.05_{-0.06}^{+0.05}$	\\ \\	
				  		
			 &WA-2			&$\log\xi$  		&$2.85_{-0.03}^{+0.05}$&$2.92_{-0.07}^{+0.03}$&$2.86_{-0.05}^{+0.01}$&$2.83_{-0.05}^{+0.01}$		\\
			 &			&$\log\nh$ 		&$21.70_{-0.11}^{+0.02}$	&$21.53_{-0.05}^{+0.11}$&$21.55_{-0.07}^{+0.03}$&$21.43_{-0.07}^{+0.03}$ \\ \\			   
			 & C/dof		   &   			& $7468/6893$		&$7448/6893$		& $7463/6893$		&$7672/6893$                 	 	\\	\hline \\

	  MGC-6-30-15	&WA-1			&$\log\xi$  		&$-0.31_{-0.02}^{+0.02}$&$-0.36_{-0.03}^{+0.02}$&$-0.33_{-0.02}^{+0.02}$ & $-0.35_{-0.02}^{+0.02}$   \\
				    &			&$\log\nh$ 		&$20.68_{-0.05}^{+0.12}$&$20.55_{-0.15}^{+0.15}$&$20.65_{-0.03}^{+0.06}$ &$20.82_{-0.03}^{+0.06}$	\\ \\	
			
			     &WA-2			&$\log\xi$  		&$1.24_{-0.02}^{+0.02}$	&$1.25_{-0.02}^{+0.02}$	&$1.25_{-0.02}^{+0.01}$	&$1.27_{-0.02}^{+0.01}$		\\
			&			&$\log\nh$ 		&$20.90_{-0.03}^{+0.05}$&$20.82_{-0.03}^{+0.03}$&$20.65_{-0.06}^{+0.06}$ &$20.55_{-0.06}^{+0.06}$	\\ \\	
						
		      &WA-3			&$\log\xi$  		&$2.54_{-0.03}^{+0.02}$&$2.52_{-0.02}^{+0.02}$	&$2.52_{-0.02}^{+0.02}$	&$2.50_{-0.02}^{+0.02}$	\\
			  &			&$\log\nh$ 		&$21.24_{-0.05}^{+0.03}$&$21.17_{-0.05}^{+0.05}$&$21.08_{-0.05}^{+0.05}$ & $21.07_{-0.05}^{+0.05}$\\	\\	
			
			   & C/dof		   &    			&  $8195/6875$		& $8257/6885$		&$8273/6885$	&$8329/6885$             	 	\\	


	  \hline \hline
\end{tabular} \\ 

}

The three sources were chosen so that they have one, two and three components of WA respectively.
\end{table*}


\section{Multiple observations of the WAX sources}
\label{subsec:multiple-obs}

{\color{black} Most of the sources in the WAX sample have multiple observations at various epochs with different exposure times. To characterise the spectral properties of the sources better, one should use as many datasets as possible in the analysis. However, combining the datasets corresponding to different X-ray spectra may hamper the detection of weak warm absorber features, because in principle a) the underlying continuum cannot any longer be described by the phenomenological model adopted in this paper, b) different WA states may mix, confusing the interpretation of the results\footnotemark.

	Most of the AGN in the WAX sample have exhibited spectral variability as shown in Figures \ref{Fig:varied} and hence cannot be added. The criterion used to determine whether a source has varied between the observations is as follows: we have simultaneously fitted the EPIC-pn datasets of all the observations of a given source, in the energy band $3-5\kev$ with a powerlaw and relative-normalisations between the datasets which are free to vary. The relative-normalisation takes into account the flux variability, if any, between the observations. If the datasets have different spectral shapes (powerlaw slope or soft-excess shape) a common $\Gamma$ will not be able to describe them which will lead to deviations between the datasets in the residual plots (see Fig. \ref{Fig:varied}). We consider a source to exhibit spectral variability whenever we find the deviations in the data to model ratios between the datasets is $\ge 10\%$. Figure \ref{Fig:varied2} show that the sources AKN~564, MRK~279, MRK~509, NGC~7469, and MR~2251-178 have at least two observations where the data have not shown spectral variability. 
	
	For each source exhibiting spectral variability, we have decided to use the EPIC and RGS spectra extracted from the longest observation, so as not to introduce any bias associated with the source flux or spectral state. Each source is therefore associated with a unique set of warm absorber components in the discussion of the global properties of the warm absorber present in the paper. 

	For the five sources mentioned above, which do not show spectral variability, we have combined the non-varied datasets. Table \ref{Table:comb-data} shows the datasets which have been combined for the five sources. It should be noted that AKN~564 and MR~2251-178 have also been observed in 2011, i.e after the WAX sample was finalised. The EPIC-pn datasets of these sources are combined using the commands {\tt addascaspec} and the RGS datasets are combined using {\tt rgscombine} and are simultaneously fitted as enumerated in Section \ref{section:RGS-analysis}. Table \ref{Table:WA-par-comb-data} shows the best fit WA parameters of these five sources for the two cases: First, the longest observation case, and second, the new combined datasets. We see that the best fit WA parameters are significantly different in the two cases. However, our aim is to find out whether the newly derived parameters alter our global conclusions inferred from the correlation results derived in Section \ref{subsec:corr-analysis} and Fig. \ref{fig:linear-reg-xi-V}. Table \ref{appendix-C-corr} and \ref{Table:appendix-C-lin-reg} show that that the correlation and linear regression results are similar to those obtained earlier in Tables \ref{corr} and \ref{Table:lin-reg}. Our global inferences are therefore not affected by our choice of the datasets of these five non-varied sources. So we have considered the present (longest) datasets of the sources as a representative sample (statistical realisation) of the parent population of the datasets of the sources in our sample.}

	\footnotetext{ This applies also for the variability {\it during} an observation. We believe that this effect does not have a major impact on the global WA properties mentioned in the paper. A discussion of the inter-observation WA variability is deferred to a future paper.}

\begin{table*}
{\footnotesize
\centering
\caption{The datasets of the sources which have been combined for our analysis.}\label{Table:comb-data}
  \begin{tabular}{llllllllllll} \hline\hline 
	  
	  Number		&Source 	& observation id	& observation date 	&  Total exposure  & Combined exposure \\
			 &              &         &   (dd/mm/yyyy)        &   in ks           & in ks \\ \hline \\ 

	  1			&AKN~564	&   0670130201		&24-05-2011		&60		&	366  \\
				   &		&0670130501		&11-06-2011		&67		&		\\
				   &		&0670130601		&17-06-2011		&61		&		\\
				   &		&0670130701		&25-06-2011		&64		&		\\
				   &		&0670130801		&29-06-2011		&58		&		\\
				   &		&0670130901		&01-07-2011		&56		&		\\ \\
	  
	  2			&MR~2251-178	&0670120201		&11-11-2011		&133		&	394	 \\
				   &		&0670120301		&13-11-2011		&128		&		\\
				   &		&0670120401		&15-11-2011		&133		&		\\ \\

	  3			&NGC~7469	&0112170301		&26-12-2000		&25		&	110	\\
	      			&		&0207090101		&30-11-2004		&85			&		\\ \\

	  4			& MRK~279	&0302480401		&15-11-2005		&60		& 98		\\
				   &		&0302480601		&19-11-2005		&38		&		\\ \\

	  5			&MRK~509	&0601390701		&07-06-2009		&63		&240		\\
				   &		&0601390801		&10-11-2009		&61		&		\\
				   &		&0601390401		&23-10-2009		&61		&		\\
				   &		&0601391001		&18-11-2009		&66		&		\\

	  \hline \hline
\end{tabular} \\ 

}

For the three sources AKN~564, MR~2251-178 and MRK~509 the observations quoted above are near simultaneous and hence added together. For the first two cases, the datasets were not available publicly when the WAX project had started.
\end{table*}

\begin{table*}
{\footnotesize
\centering
\caption{The WA parameters for the sources which have non-varied datasets. The combined datasets refer to those enumerated in Table \ref{Table:comb-data}. }\label{Table:WA-par-comb-data}
  \begin{tabular}{llllllllllll} \hline\hline 
	  
	  Number 	&		 Source 	& WA component		& 	WA parameters 	&  Longest observation  & Combined datasets  \\ \hline \\ 

	  1		&	AKN~564	&  WA-1	  	&    $\log\xi$    	 &$-0.20^{+0.11}_{-0.10}$&	$1.87^{+0.01}_{-0.08}$  \\ 

			   &		 & 		&    $\log\nh$      	& $20.48^{+0.05}_{-0.13}$&     $20.40^{+0.01}_{-0.05}$ 	\\ 
			   &		&		&$v_{\rm out}$		& $-690_{-90}^{+150}$	& $-850\pm 30$	\\	\\


	  2		&	NGC~7469		&  WA-1	  	&    $\log\xi$    	 &$2.80^{+0.20}_{-0.10}$	&$2.78^{+0.20}_{-0.10}$ \\ 

			 &		  		&		&    $\log\nh$      	& $20.96^{+0.20}_{-0.15}$	&$20.85^{+0.20}_{-0.15}$	\\ 
			 &				&		&$v_{\rm out}$		&$-1590_{-90}^{+90} $		&$-1650_{-90}^{+90} $	\\	\\


	3		&	MR~2251-178	&  WA-1	  	&    $\log\xi$    	 &$1.60^{+0.03}_{-0.02}$&	$-0.75^{+0.01}_{-0.05}$  \\ 

			 &		  	&		&    $\log\nh$      	& $20.96^{+0.40}_{-0.03}$&      $20.68_{-0.01}^{+0.01} $	\\ 
			 &			&		&$v_{\rm out}$		&$-3150_{-30}^{+60}$	&	$-800_{-30}^{+40}$\\	\\

			&			&  WA-2	  	&    $\log\xi$    	 &$2.92^{+0.50}_{-0.10}$&	$2.61^{+0.01}_{-0.02}$  \\ 

			&		  	&		&    $\log\nh$      	& $21.698^{+0.2}_{-0.05}$&      $21.76_{+0.01}^{-0.01}$	\\ 
			&			&		&$v_{\rm out}$		&$-3150_{-30}^{+60}$	&	$-840_{-30}^{+40}$\\	\\


	4		&	MRK~279		&  WA-1	  	&    $\log\xi$    	 &---& --- \\ 

			 &		  	&		&    $\log\nh$      	& ---& ---		\\ 
			 &			&		&$v_{\rm out}$		&  ---& ---	\\	\\


	5		&	MRK~509	&  WA-1	  	&    $\log\xi$    	 &$3.24^{+0.05}_{-0.2}$		&$2.84^{+0.01}_{-0.02}$	  \\ 

			 &		&  			&    $\log\nh$  & $20.8^{+0.15}_{-0.14}$	&$20.84^{+0.03}_{-0.01}$	\\ 
			 &		&			&$v_{\rm out}$	&$-6500_{-120}^{+150}$		&$-900_{-50}^{+50}$	\\	\\

	  \hline \hline
\end{tabular} \\ 

}

\end{table*}


\begin{table*}
{\footnotesize
\centering
  \caption{The Spearman rank correlations for WA parameters. The first quantity in the bracket is the Spearman correlation coefficient while the second term is the correlation probability.\label{appendix-C-corr}}
  \begin{tabular}{llllllllllll} \hline\hline 

 Quantity & WA-$\log\xi$& WA-$\log\nh$ & WA-velocity    \\ \hline \\ 

WA-$\log\xi$ & 1 &  (0.62,{\bf $>99.99\%$})  & (0.33,$94\%$)   \\ \\

WA-$\log\nh$ & (0.62,{\bf $>99.99\%$})  & 1 & (0.36,{\bf $96\%$})   \\ \\ 

WA-velocity & (0.33,{\bf $94\%$})         &  (0.36,{\bf $96\%$}) & 1    \\ \\

 \hline \hline
\end{tabular} \\

}

\end{table*}


\begin{table*}

{\footnotesize
\centering
  \caption{The linear regression analysis for warm absorber parameters ($y=a\,x+b$). \label{Table:appendix-C-lin-reg}}
  \begin{tabular}{llllllllllll} \hline\hline 

 $x$ & $y$ & $a$ &  \hspace{1cm}Dev$(a)$ & \hspace{1cm} $b$ & \hspace{1cm} Dev$(b)$ & \hspace{1cm}$R_S$ &\hspace{1cm} $P_{null}$\\ \hline \\ 

$\log\xi$ &$\log\nh$        &  $0.31$ &\hspace{1cm} $0.06$ & \hspace{1cm}$20.46$  & \hspace{1cm}$0.11$ & \hspace{1cm}$0.64$ & \hspace{1cm} $>99\%$\\ \\
$\log\xi$ & $\log v_{out}$  & $0.15$ &\hspace{1cm} $0.03$ & \hspace{1cm}$2.97$  & \hspace{1cm}$0.05$ & \hspace{1cm}$0.33$ & \hspace{1cm} $>93\%$ \\ \\
$\log\nh$ & $\log v_{out}$  & $0.8$ &\hspace{1cm} $0.7$ & \hspace{1cm}$-13$  & \hspace{1cm}$18$ & \hspace{1cm}$0.36$ & \hspace{1cm} $>96\%$ \\ \\

 \hline \hline
\end{tabular} \\ 

\footnotetext{1}{ $R_S$ stands for the Spearman rank correlation coefficient.}

}
\end{table*}


\clearpage

\begin{figure}
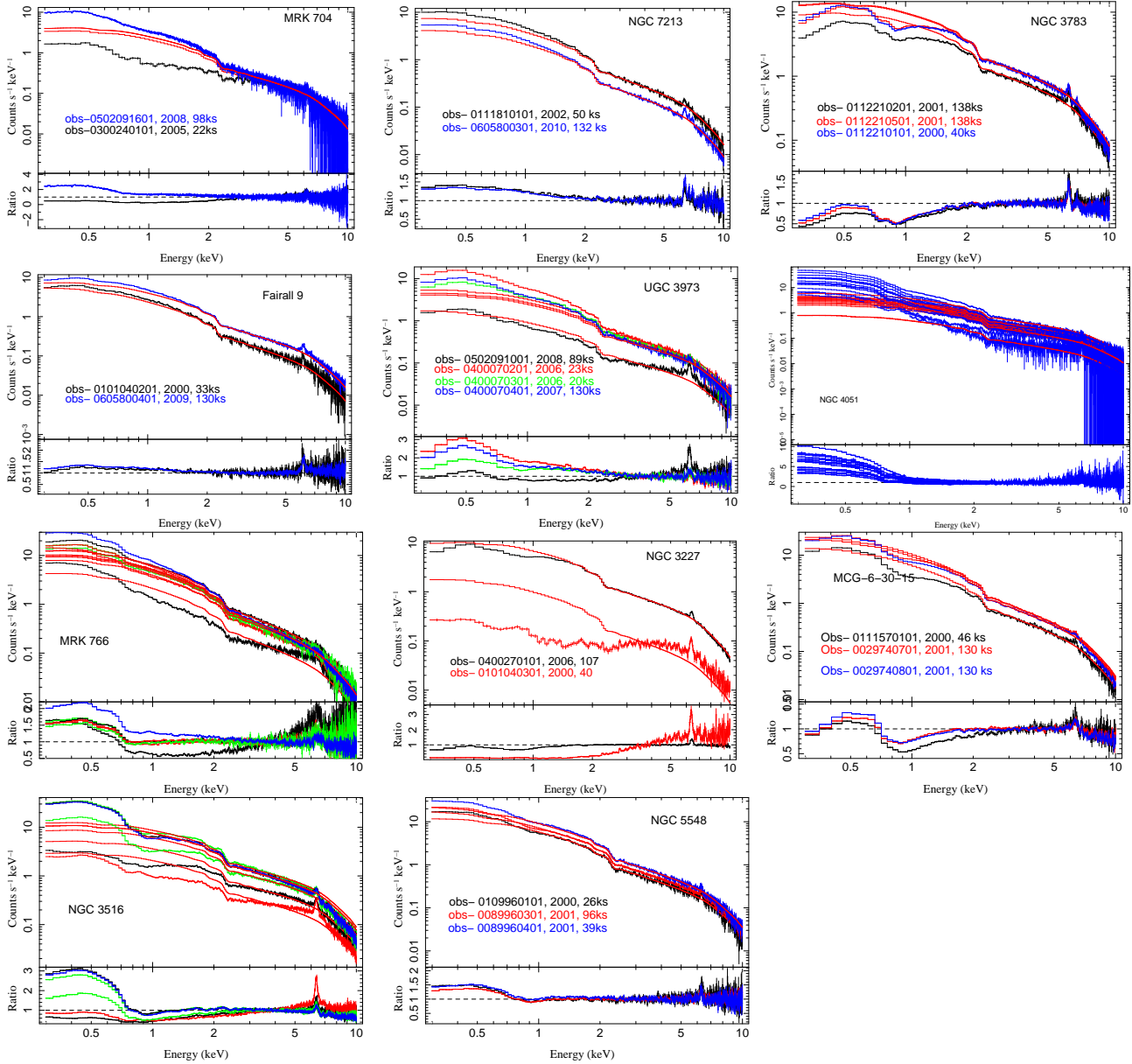

  \centering

  \vbox{
\hbox{
\includegraphics[width=4cm,angle=-90]{Final_compare_plots_12Jan14/MRK704-compare.ps}
\includegraphics[width=4cm,angle=-90]{Final_compare_plots_12Jan14/NGC7213-compare.ps}
\includegraphics[width=4cm,angle=-90]{Final_compare_plots_12Jan14/NGC3783-compare.ps}

} 

\hbox{
\includegraphics[width=4cm,angle=-90]{Final_compare_plots_12Jan14/Fairall9-compare.ps}
\includegraphics[width=4cm,angle=-90]{Final_compare_plots_12Jan14/UGC3973-compare.ps}
\includegraphics[width=4cm,angle=-90]{Final_compare_plots_12Jan14/NGC4051-comparing.ps}

} 

\hbox{
\includegraphics[width=4cm,angle=-90]{Final_compare_plots_12Jan14/MRK766_comparing.ps}
\includegraphics[width=4cm,angle=-90]{Final_compare_plots_12Jan14/NGC3227-compare.ps}
\includegraphics[width=4cm,angle=-90]{Final_compare_plots_12Jan14/MCG-6-30-15-comparing.ps}
}

\hbox{

\includegraphics[width=4cm,angle=-90]{Final_compare_plots_12Jan14/NGC3516-compare.ps}
\includegraphics[width=4cm,angle=-90]{Final_compare_plots_12Jan14/NGC5548-compare.ps}

}

}

\caption{The variability between the different observations for the different sources in the WAX sample which have multiple datasets. These are the sources which have shown spectral variability from one observation to the other.}
 \label{Fig:varied}
\end{figure}

\clearpage

\begin{figure}
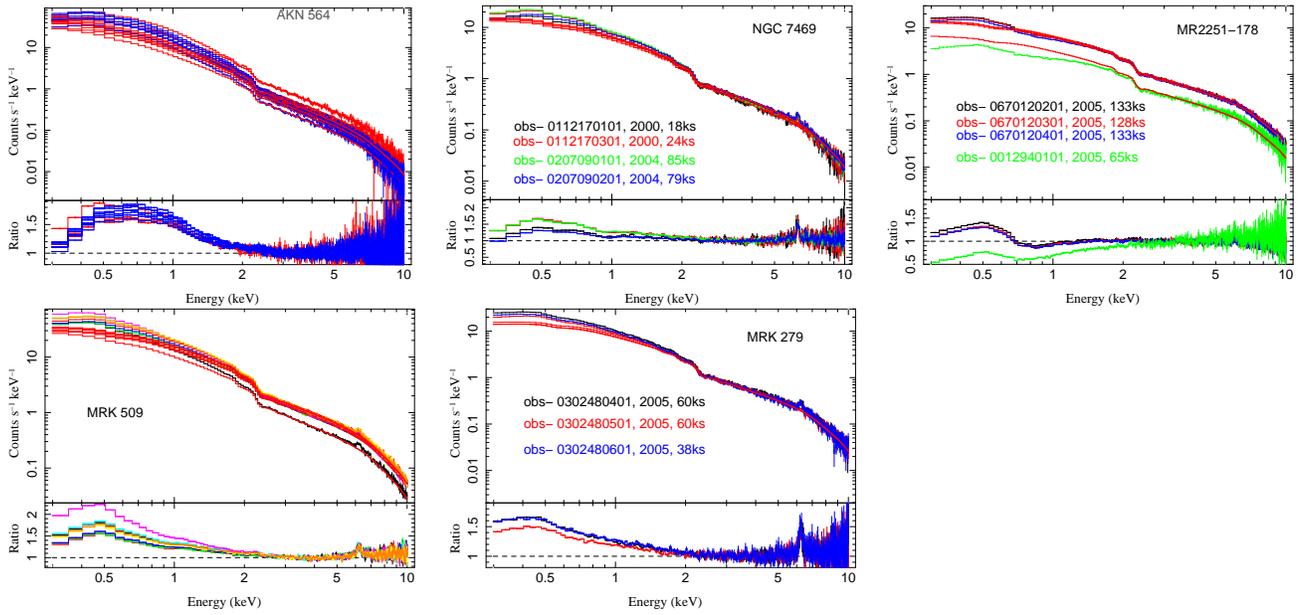

  \centering

  \vbox{

  \hbox {
\includegraphics[width=4cm,angle=-90]{Final_compare_plots_12Jan14/AKN564-compare.ps} 
\includegraphics[width=4cm,angle=-90]{Final_compare_plots_12Jan14/NGC7469-compare.ps}
\includegraphics[width=4cm,angle=-90]{Final_compare_plots_12Jan14/MR2251-178-compare.ps}

}

\hbox {
\includegraphics[width=4cm,angle=-90]{Final_compare_plots_12Jan14/MRK509-compare.ps}
\includegraphics[width=4cm,angle=-90]{Final_compare_plots_12Jan14/MRK279-compare.ps}

}

}
\caption{Sources which have not shown spectral variability in at least two of their observations. See Table \ref{Table:comb-data} for the observations whose data have been combined and analysed.}
 \label{Fig:varied2}
\end{figure}

\clearpage
\section{Comparison with previous studies}
\label{subsec:comparison}
 Our spectral results are in general agreement with the previous studies as far as the broadband continuum as well as the WA properties are concerned. Below we list individually each source wherever we find any differences in the measured properties of soft X-ray spectra and the WA compared to our studies.

\begin{itemize}

\item NGC~4593- \cite{2003A&A...408..921S} mentioned that the ionisation state is hard to constrain using the RGS dataset for this source. They have found a large spread in the ionisation parameter using the {\it slab} model in XSTAR. In our analysis we could detect two discrete components of WA.

\item Mrk~704- \cite{2011ApJ...734...75L} have found two discrete WA components with $\xi\sim 20$ and $\xi\sim 500$. Both these ionisation states and the ouflow velocities are comparable with our detections. In our analysis we detect an additional component of WA with $\log\xi\sim 3.13$. \cite{2011ApJ...734...75L} had detected a partially covering neutral absorber.

\item AKN~564- This source was stuided by \cite{2008A&A...490..103S}. All the four datasets from 2001 and one from 2005 (100 ks) were combined to obtain the final RGS spectrum. We have used only 2005 (100 ks) data set. They detected 5 emission lines and 3 WA components. $\lgxi=-0.3\pm0.5$, $\nhwa=0.9\pm0.7\times 10^{20} \cmsqi$, and $\lgxi=1.0\pm0.4$, $\nhwa=1.1\pm0.6\times 10^{20} \cmsqi$, and $\lgxi=2.2\pm0.4$, $\nhwa=2.2\pm2.1\times 10^{20} \cmsqi$. We however detect only one WA component.

\item NGC~4051- This same dataset has been studied multiple times \citep{2008RMxAC..32..123K,2006MNRAS.368..903P,2004ApJ...606..151O}. \cite{2006MNRAS.368..903P} find that the ions cover nearly 4 decades of ionisation parameter. They also find relativistic \ion{O}{8} emission lines. We detect two discrete WA components of ionisation parameters $\log\xi \sim 0.28$ and $\log\xi \sim 2.87$.

\item NGC~7469 - \cite{2003A&A...403..481B} have combined two datasets to obtain a total of $160 \ks$ exposure. They found a continuous distribution of WA ionisation parameter with $-2 \le\lgxi\le 2$. We detect a single WA component of ionisation parameter of $\log\xi \sim 2.8$.

\item MRK~766- Studied by \cite{2003ApJ...582...95M}. They found WA with a large range in $\xi$ as well as broad relativistic emission lines in the soft X-rays at $0.5006 \kev$, $0.65 \kev$, and $0.367 \kev$. We have detected similar emission lines in the soft X-rays. We detect WA primarily in low ionisation states from $\log\xi \sim -0.94- 1.35$.

\item IRAS05078+1626- \cite{2010A&A...512A..62S} found a single component of WA with an ionisation parameter of $\lgxi \sim 2.5\pm 1.0$, and $\nhwa \ge 10^{24} \cmsqi$. We found two components of WA and the highest column density is that of $\log\nh\sim 21.16$.

\item NGC~3227-  This is a Seyfert 1.5 galaxy and have been studied by \cite{2009ApJ...691..922M}. The $\Gamma= 1.57$ and the intrinsic neutral absorption column is $\nh= 3\times 10^{21} \cmsqi$. There are two WA components. $\lgxi=1.2$, $\nhwa=1\times 10^{21} \cmsqi$, and $\lgxi=2.9$, $\nhwa=2\times 10^{21} \cmsqi$. We also detect a similar neutral absorption column but we find three discrete WA components.

\item MR2251-178 -This same dataset had been studied by \cite{2004ApJ...611...68K}. However they have combined it with other datasets. The $\Gamma=1.54\pm 0.024$. They detected two warm absorber components $\log U=-1.78$, $\nhwa=1\times 10^{21.51} \cmsqi$, and $\log U=-4$, $\nhwa=2\times 10^{20.06} \cmsqi$. Note that they have used Tartar's definition of ionisation parameter (represented by U) and there is no direct conversion until we know the appropriate SED. We also detected two WA components.

\item MRK~279-\cite{2010A&A...512A..25C} studied the same dataset ($\sim 60 \ks$) but they have combined 3 other datasets to make $\sim 160 \ks$. Very poor quality of the RGS dataset. They found two WA. $\lgxi=0.8\pm0.3$, $\nhwa=0.7\times 10^{20} \cmsqi$, and $\lgxi=2.6$, $\nhwa=3\times 10^{20} \cmsqi$. We have not detected any WA in this source and have found strict upper and lower limits on column density and ionisation parameters.

\item MCG+8-11-11 - This source have been studied by \cite{2006A&A...445..451M}. The slope $\Gamma=1.8$. The single component warm absorber $\xi\sim 460$. There is an absorption edge in soft-Xrays with $\tau \sim 0.041$. We too detect one WA component but with a much higher ionisation parameter. 

\item MRK~509 - \cite{2007A&A...461..135S} have studied this dataset and found three WA components, with $\lgxi=0.89$, $\lgxi=2.14$, and $\lgxi=3.26$. The slope $\Gamma=2.14$. We detected two WA components.

\item NGC~3516 - \cite{2003ApJ...594..128T} detected large range of ionisation states for WA in this source. We detected just one WA component.

\item NGC~3783 -\cite{2003ApJ...598..232B} studied the same dataset as ours but also added another $138 \ks$ dataset taken almost simultaneously. The slope $\Gamma=1.7$ and the WA $\xi \sim 30$. We detected two WA components.

\item IC~4329A - \cite{2005A&A...432..453S} studied the same dataset. The source is heavily absorbed. The slope $\Gamma=1.71$. There are four components of WA, $\lgxi=-1.37$, $\lgxi=0.56$, $\lgxi=1.92$, and $\lgxi=2.70$. We detected three WA components, which are nearly similar to these.

\end{itemize}


$Acknowledgements:$  The authors are grateful to the anonymous referee for insightful comments which improved the quality of the paper. Author SL is grateful to Ehud Behar, Gerard Kriss, Smita Mathur, Gary Ferland and Ryan Porter for insightful discussion on various aspects of this paper. Author SL is grateful to the ESAC faculty of the European Space Agency for its grants to allow him to visit ESAC for collaboration work. MG acknowledges financial support from the IUCAA (Pune). Author SL is grateful to CSIR, Government of India for supporting this work. SC wants to acknowledge the use of facilities at Harvard University, Cambridge, MA, USA and Max Planck Institute for Radio Astronomy, Bonn, Germany. This research has made use of the NASA/IPAC Extragalactic Database (NED) which is operated by the Jet Propulsion Laboratory, California Institute of Technology,
under contract with the National Aeronautics and Space Administration.

\bibliographystyle{mn2e} 
\bibliography{mybib.bib}

\end{document}